\documentclass[11pt, a4paper, oneside]{Thesis} 

\graphicspath{{./Figures/}} 

\usepackage[T1]{fontenc}
\usepackage{color,graphicx,Arrayeq,amsmath,amssymb,mathrsfs,braket,cite}
\usepackage{cancel,tcolorbox,mdframed,framed}
\usepackage{bbold}
\usepackage{quotchap,slashed,wrapfig,wasysym}
\definecolor{mybrown}{RGB}{234,26,29} 
\definecolor{mycol}{RGB}{40,23,137}
\def\order(#1){{\cal O} \left(#1 \right)}  
\def\Eqn#1{Eq.\ (\ref{#1})}                
\def\Eqs#1#2{Eqs.\ (\ref{#1}) and (\ref{#2})}
\long\def\rpl#1!!#2!!{\textcolor{red}{#1} \textcolor{blue}{#2}} 
\def \Slash{\slash\!\!\!\!}
\def \cm{\cal{M}}

\def \ml{\mathscr{L}}
\def \ma{\mathcal{A}}
\def\m{\scriptstyle}
\def\l{\lambda}
\def\tb{\tan\beta}

\newcommand{\specialcell}[2][c]{%
  \begin{tabular}[#1]{@{}c@{}}#2\end{tabular}}

\hypersetup{urlcolor=mycol, colorlinks=true} 
\title{\ttitle} 

\allowdisplaybreaks  
\begin{document}

\frontmatter 

\setstretch{1.3} 

\fancyhead{} 
\rhead{\thepage} 
\lhead{} 

\pagestyle{fancy} 

\newcommand{\HRule}{\rule{\linewidth}{0.5mm}} 

\hypersetup{pdftitle={\ttitle}}
\hypersetup{pdfsubject=\subjectname}
\hypersetup{pdfauthor=\authornames}
\hypersetup{pdfkeywords=\keywordnames}


\begin{titlepage}
\begin{center}
\includegraphics[scale=0.35]{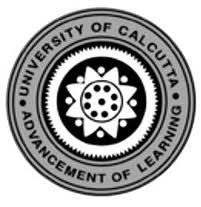} \\
\textsc{\LARGE \univname}\\[1.5cm] 
\textsc{\Large Doctoral Thesis}\\[0.5cm] 

\HRule \\[0.4cm] 
{\huge \bfseries \ttitle}\\[0.4cm] 
\HRule \\[1.5cm] 

\begin{minipage}{0.35\textwidth}
\begin{center} \large
\emph{Author:}\\
\href{http://www.saha.ac.in/web/?option=com_content&view=article&id=955&Itemid=1332&mid=609&tab=tab1}{\authornames}\\ 
\end{center}
\end{minipage}\hspace*{2.8cm}
\begin{minipage}{0.45\textwidth}
\begin{center} \large
\emph{Supervisor:} \\
\href{http://www.saha.ac.in/web/theory-personal-page?mid=70&tab=tab1}{\supname} \\
\end{center}
\end{minipage}\\[5mm]
\addressnames \\[3cm]
 
\large \textit{A thesis submitted in fulfilment of the requirements\\ for the degree of \degreename}\\[0.3cm] 
\textit{in the}\\[0.4cm]
\groupname\\\deptname\\[2cm] 
 
{\large \today}\\[2cm] 
 
\vfill
\end{center}

\end{titlepage}


\pagestyle{empty} 

\null\vfill 

{\fontfamily{pzc}\selectfont
`` What I have done is to show that it is possible for the way the universe began to be determined by the laws of science. In that case, it would not be necessary to appeal to God to decide how the universe began. This doesn't prove that there is no God, only that God is not necessary. "}

\begin{flushright}
{\fontfamily{pag}\selectfont Stephen Hawking}
\end{flushright}

\vfill\vfill\vfill\vfill\vfill\vfill\null 

\clearpage 


\addtotoc{Abstract} 

\abstract{\addtocontents{toc}{\vspace{1em}} 

In this thesis, we investigate the implications of the LHC Higgs data on different BSM scenarios. 
Since the data seem to agree with the SM expectations, any nonstandard couplings will be strongly constrained.
First we investigate, in a model independent way, the constraints on the nonstandard Higgs couplings with
the fermions and the vector bosons in view of high energy unitarity and the measured value of the Higgs to
diphoton signal strength. Then we concentrate on a particular BSM scenario, namely, the two Higgs-doublet 
models (2HDMs). Consistency of the Higgs data with the corresponding SM predictions strongly motivates
us to work in the {\em alignment limit}. In this limit, including the informations of Higgs mass and its SM-like
nature, we find many new constraints on the nonstandard masses and $\tan\beta$. We also study the constraints
on the charged scalar mass arising from the $h\to \gamma\gamma$ signal strength measurements and observe
that the charged scalar does not necessarily decouple from the diphoton decay width. We then move on to some
particular variants of 2HDMs, known as BGL models, and study the flavor constraints on these models. Here we
find that lighter than conventionally allowed nonstandard scalars can successfully negotiate the stringent bounds
coming from flavor physics data and can leave unconventional decay signatures that can be used as distinctive 
features of these models. We also analyze the stability and unitarity constraints in a three Higgs-doublet model
(3HDM) with $S_3$ symmetry and find that there must be many more nonstandard particles below 1~TeV. We 
also observe that the nondecoupling feature of the charged scalar in the context of $h\to \gamma\gamma$ is 
not unique to the 2HDMs only, instead it is a general property of the multi doublet extensions of the SM
with an exact discrete symmetry. 
}

\clearpage 


\setstretch{1.0} 

\acknowledgements{\addtocontents{toc}{\vspace{1em}} 

First I thank my supervisor, Prof. Gautam Bhattacharyya, who has been the {\em guide} during my PhD career in the truest
sense. It has been a pleasure to work with him and I learned a lot of things from him apart from physics.
I am grateful to Prof. Palash B. Pal for his helps with the basics of physics as well as \LaTeX. I am also thankful to my
university professors, specially Anirban Kundu and Anindya Datta, for exposing me to the interesting and
dynamic field of modern high energy physics.

But most of the time during the PhD, you don't talk or think about academics. That's when friends become important.
I thank all my batchmates of the P.MSc year 2010 for a wonderful start of my PhD life. Due to the lack of my social skills,
I did not have many acquaintances. Despite that I was very lucky to meet some great young people who could endure
me and with time became close friends. Among them, Aminul interacted the most primarily because we shared the
same office. In the first two years when my desk was beside him, he used to monitor my internet activities which,
I must admit, has influenced my research in a positive way. He also tried to teach me how to maintain a balance between
arrogance and modesty during social encounters. I don't know if I have learned anything or not but I thank him for
his efforts. Next I thank Rajani for allowing me into his hostel room whenever I felt like going to Ruby and disturb him!
Although he has a good strong heart, I apologize for any {\em joke} of mine that might have made him feel like
I was actually testing that goodness and strength of his heart. I thank Amit (Dutta Banik) for being such a nice and supportive
friend. I also thank Avik and Vijay for their friendships. I thoroughly enjoyed the companionship of Amit (Dey) and
Debashis (Saha) during our many short P.MSc adventures. I thank KD for sharing his vast knowledge on different topics.
I acknowledge the efforts of Arindam-da to educate me politically. I thank Mayukh and Kousik for the `adda' sessions in
their lab. I thank all members of G11, 363, MSA-I (Hostel) for many colorful memories.

My special thanks to Sadhan with whom I shared an apartment for four years. I fondly remember the weekends when
I woke up at 11.30 in the morning (?) to find that he had already done all the hard works -- bought the chicken, cleaned
it and started cooking. Week after week with a toothbrush in my mouth I watched him cooking and wondered how he
can live with the laziest possible flatmate and never complain! In fact, Sadhan is the reason why I had the good fortune
of tasting some delicious dishes prepared by Indrani-di (now Sadhan's wife) on some lucky weekends. I express my sincere
gratitude to Sadhan for keeping me alive during the weekends and holidays.

Next I thank Sir (Ujjal) for introducing me to many good movies. I also thank Ipsita for showing me how to `act smart' at
the airport and win a window seat. I am thankful to Amit-da (Chakraborty) for helps at different stages of my PhD.

Finally I thank my family for the constant support that has helped me to remain focused.

\vfill
\begin{flushright}
Dipankar Das \\ Kolkata, India
\end{flushright}
}
\clearpage 


\pagestyle{fancy} 

\lhead{\emph{Contents}} 
\tableofcontents 

\lhead{\emph{List of Figures}} 
\listoffigures 

\lhead{\emph{List of Tables}} 
\listoftables 



\setstretch{1.5} 

\lhead{\emph{Abbreviations}} 
\listofsymbols{ll} 
{
\textbf{SM} & \textbf{S}tandard \textbf{M}odel \\
\textbf{BSM} & \textbf{B}eyond \textbf{S}tandard \textbf{M}odel \\
\textbf{CM} & \textbf{C}enter of \textbf{M}omentum \\
\textbf{LHC} & \textbf{L}arge \textbf{H}adron \textbf{C}ollider \\
\textbf{IVB} & \textbf{I}ntermediate \textbf{V}ector \textbf{B}oson \\
\textbf{QED} & \textbf{Q}uantum \textbf{E}lectro-\textbf{D}ynamics \\
\textbf{QCD} & \textbf{Q}uantum \textbf{C}hromo-\textbf{D}ynamics \\
\textbf{LH} & \textbf{L}eft \textbf{H}anded \\
\textbf{RH} & \textbf{R}ight \textbf{H}anded \\
\textbf{2HDM} & Two \textbf{H}iggs-\textbf{D}oublet \textbf{M}odel \\
\textbf{3HDM} & Three \textbf{H}iggs-\textbf{D}oublet \textbf{M}odel \\
\textbf{VEV} &  \textbf{V}acuum \textbf{E}xpectation \textbf{V}alue \\
\textbf{BGL} &  \textbf{B}ranco \textbf{G}rimus \textbf{L}avoura \\
\textbf{NFC} &  \textbf{N}atural \textbf{F}lavor \textbf{C}onservation \\
}


\setstretch{1.3} 

\pagestyle{empty} 

\dedicatory{Dedicated to my Parents} 

\addtocontents{toc}{\vspace{2em}} 


\mainmatter 

\pagestyle{fancy} 


\begin{savequote}[60mm]  
 Two roads diverged in a wood, and I, \\
 I took the one less traveled by ...    
 \qauthor{Robert Frost in ``The Road Not Taken"}    
\end{savequote}
%

\lhead{Chapter 1. \emph{A bottom-up approach to the Standard Electroweak Theory}}

\chapter[A bottom-up approach to the Standard Electroweak Theory]{A bottom-up approach to the \\ Standard Electroweak Theory} 

\label{Chap1} 

The original theory was constructed fifty years ago from gauge theoretical point of view \cite{Weinberg:1967tq}. Here, however, we shall take a different route to reconstruct the essential virtues of the theory from the consideration of `tree-unitarity' \cite{Joglekar:1973hh,Horejsi:1993hz}. First of all, we need to note that every scattering amplitude can be expanded in terms of the partial waves \cite{Bhattacharyya:2009gw}:
\begin{eqnarray}
 {\cal M}(\theta) = 16 \pi \sum_{l=0}^{\infty}(2l+1) a_l  P_l(\cos\theta) \,,
\label{uni_amp}
\end{eqnarray}
where, $\theta$ is the angle of scattering. Every partial wave amplitude is bounded from the `unitarity' condition:
\begin{eqnarray}
|a_l| \le 1 \,.
\label{unitarity}
\end{eqnarray}
If we trust in perturbative calculations then the unitarity condition must be satisfied order by order \cite{Chanowitz:1978mv}, \emph{i.e.}, it must hold in tree level also. If the tree level amplitude grows with energy then unitarity is bound to be violated at large energies. In view of this, let us postulate the following:
\begin{quotation}
 Any remnant energy growth in a scattering amplitude must be canceled either by tuning the couplings suitably or, if the first option is unavailable, by introducing new particles.
\end{quotation}
In the following sections we shall see how this hypothesis along with the experimental facts and the requirement of `minimality' lead us to the correct description of nature.

\section{A quantum mechanical prelude}
\begin{figure}
\centering
\includegraphics[scale=0.6]{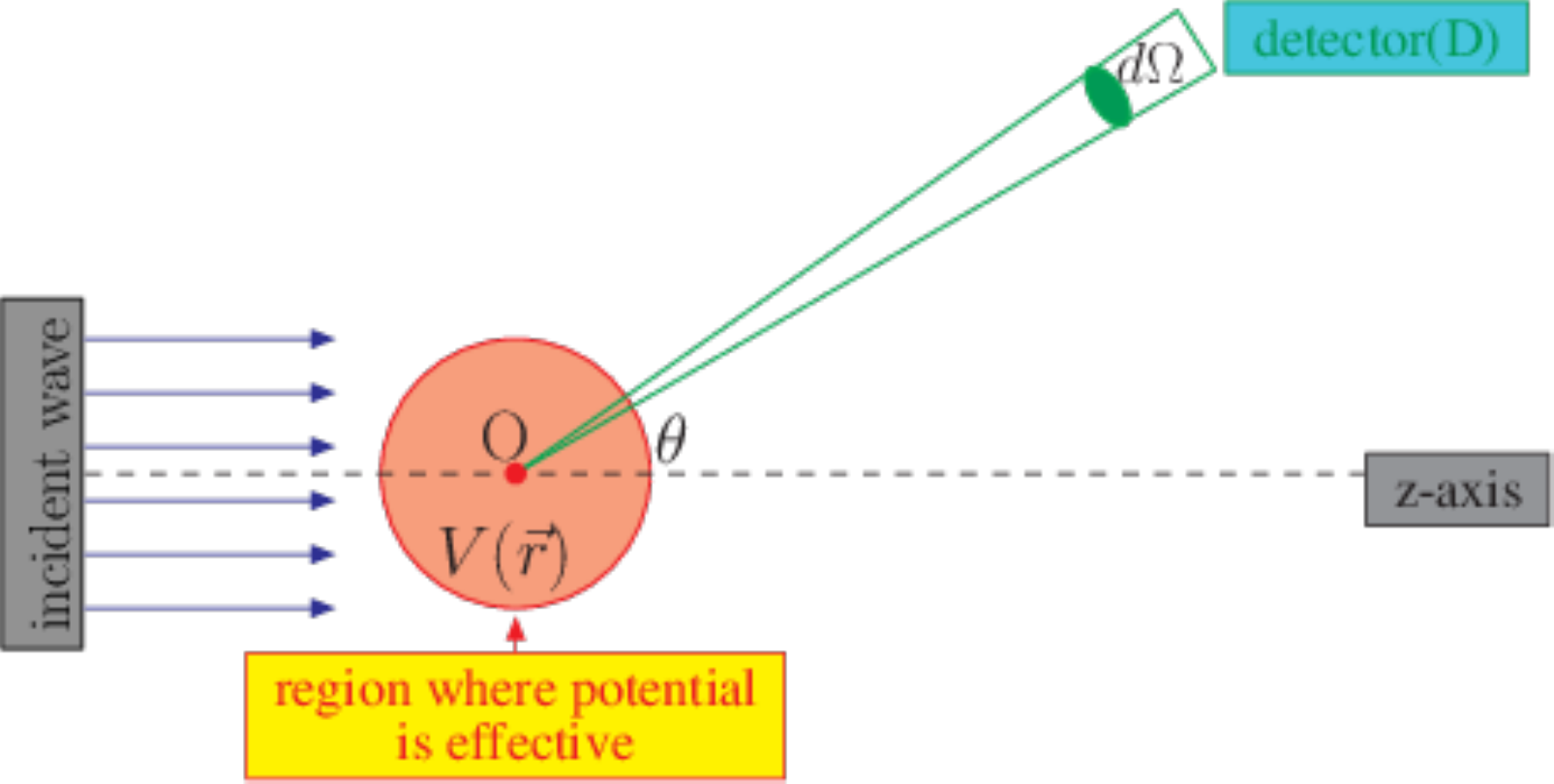}
\caption[Scattering in QM]{\em Schematic diagram for scattering by a potential $V(\vec{r})$.}
\label{qm-scat}
\end{figure}
A good way to start will be to make a sense of \Eqs{uni_amp}{unitarity} from the lessons of quantum mechanical scattering.
Let us consider a scattering experiment in which a steady incident beam is maintained for an indefinitely long time, 
i.e. the incident flux, $F_{\rm in}$, is constant. Then, there will be a steady stream of scattered particles too.
In \fref{qm-scat}, the incident beam is parallel to the $z$-axis and is assumed to be much wider than the zone of influence of 
the potential, $V(\vec{r})$, centered at O. Far from this zone of influence a detector,
 D, measures the number, $dn$, of particles scattered per unit time into the solid angle $d\Omega$, 
centered around the direction defined by the polar angles $\theta$ and $\phi$. The number, $dn$, is proportional to $F_{\rm in}$ and to $d\Omega$; the constant of proportionality, $d\sigma/d\Omega$, is defined to be the differential scattering
 cross-section in the direction ($\theta$,$\phi$). Thus
 \begin{equation}
     dn=F_{\rm in}\cdot d\Omega \cdot\frac{d\sigma}{d\Omega} \,.
 \label{cross section}
 \end{equation}
 In the quantum theory of scattering, we imagine that an incident plane wave, $\psi_{\rm in}= Ae^{ikz}$, traveling along the
 $z$-axis, encounters a scattering potential producing an outgoing spherical wave. At large distances from the 
scattering center, the form of the wave function $\psi(\vec{r})$ must be such as to conform to the general pictures outlined
 below:
 \begin{enumerate}
 \item It must consist of a part of the incident wave corresponding to the parallel beam of incident particles which
  transmitted unmodified.
 \item Another part $\psi_{\rm sc}$ representing the scattered particles, having same energy as the incident particles (because of {\em elastic} scattering) and moving radially outwards from the center.
 \end{enumerate}
Thus we may write,
\begin{equation}
\psi(\vec{r})_{_{r\to\infty}} \approx e^{ikz}+\underbrace{f(\theta,\phi)\frac{e^{ikr}}{r}}_{\psi_{\rm sc}} \,.
\label{basic}
\end{equation}
In this expression, only the function $f(\theta,\phi$), which is called the scattering amplitude, depends on the potential 
$V(\vec{r})$. The $\phi$ dependence should be included in the general case, to account for the anisotropy of the potential. 
However, if the target is azimuthally symmetrical, the $\phi$ dependence would no longer be present. Note that, the 
spherical wave carries a factor of  $1/r$, because this portion of $|\psi|^2$ is spherically diverging and must go like  $1/r^2$  to conserve probability.

\subsection{Calculation of differential cross section}
We recall that the expression for the current density  $J(\vec{r})$ associated with a wave function $\psi(\vec{r})$ is:
\begin{equation}
\vec{J}(\vec{r})=\frac{\hbar}{2im}\left[\psi^*(\vec{\nabla}\psi)-(\vec{\nabla}\psi^*)\psi\right]=\frac{1}{m}{\rm Re}\left[\psi^*(\vec{r})\frac{\hbar}{i}\vec{\nabla}\psi(\vec{r})\right] \,.
\end{equation}
The incident and scattered fluxes are obviously proportional to the normal components of $\vec{J}_{\rm in}$ and $\vec{J}_{\rm sc}$ respectively. We will call the proportionality constant $C$.
Since $\psi_{\rm in} = e^{ikz}$, we obtain
\begin{eqnarray}
&& \left(J_{\rm in}\right)_z = \frac{1}{m}{\rm Re}\left[e^{-ikz}\frac{\hbar}{i}\frac{\partial}{\partial{z}}e^{ikz}\right]=\frac{\hbar k}{m}\,, \\
&\Rightarrow& F_{\rm in} = C \left(J_{\rm in}\right)_z \,.
\end{eqnarray}
For radially diverging scattered wave, the number of particles crossing an area $d\vec{s}=ds\hat{r}$, subtending solid angle $d\Omega$ at the origin is
\begin{eqnarray}
dn=\underbrace{C\vec{J}_{\rm sc}}_{\vec{F}_{\rm sc}}.(ds\hat{r})=C(J_{\rm sc})_rds \,.
\end{eqnarray}
Clearly, it is the $r$-th component of $\vec{J}_{sc}$ which receives our attention. Remembering $\psi_{\rm sc}=\frac{1}{r}f(\theta,\phi)e^{ikr}$ we may get,
\begin{eqnarray}
(J_{\rm sc})_r=\frac{1}{m} {\rm Re} \left[f^*(\theta,\phi)\frac{e^{-ikr}}{r}\frac{\hbar}{i}\frac{\partial}{\partial{r}}\{f(\theta,\phi)\frac{e^{ikr}}{r}\}\right]
=\frac{1}{m}|f(\theta,\phi)|^2\frac{\hbar{k}}{r^2}
\end{eqnarray}
Hence,
\begin{eqnarray}
dn=\frac{C}{m}|f(\theta,\phi)|^2\hbar{k}\frac{ds}{r^2}=F_{\rm in}|f(\theta,\phi)|^2d\Omega \,.
\end{eqnarray}
Comparing this with \Eqn{cross section} we obtain,
\begin{eqnarray}
\frac{d\sigma}{d\Omega}=|f(\theta,\phi)|^2 \,.
\end{eqnarray}
Thus problem of determining the scattering cross section reduces to finding the scattering amplitude, $f(\theta,\phi)$, in quantum mechanics. The quantity, $f(\theta,\phi)$, actually tells us about the `probability amplitude' for
scattering in a direction $(\theta,\phi)$, and hence is related to the differential cross-section which is the quantity of interest for the experimentalists. The scattering amplitude  is obtained by solving the Schr\"{o}dinger equation under the scattering potential. Depending on the mathematical form of the potential, there are several methods to find the scattering amplitude. The method of partial waves, in particular, comes in handy when the potential is central.

\subsection{Method of partial waves}
In the special case of a central potential $V(r)$, the orbital angular momentum $\vec{L}$ of the particle is a constant
 of motion. Therefore, there exists stationary states with well defined angular momentum, {\it i.e.}, eigenstates common to $H$, 
$L^2$ and $L_z$. We shall call such wave functions `partial waves' and denote them as
$\psi_{klm}(\vec{r})$. Their angular dependence is always given by the spherical harmonics $Y_l^m(\theta,\phi)$ --
the potential $V(r)$ influences their radial parts only.

We know that $e^{ikz}$ is a solution of the Schr\"{o}dinger equation with $V(r)=0$ in the $\{H,~p_x,~p_y,~p_z\}$ basis and may be denoted 
by $|0,0,k \rangle $ where z-axis is chosen as the direction of motion. Now if we wish, we may translate our wave function in 
terms of $\psi_{klm}(\vec{r}) \equiv$ $R_{kl}(r)Y_l^m(\theta,\phi)$ which are the eigenfunctions in the $\{H,~L^2,~L_z\}$ basis. For a free particle we know that $R_{kl}(r)$ is a linear combination of spherical Bessel and Neumann functions. 
But as Neumann function blows up at the origin it is dropped out. So we may write,
\begin{eqnarray}
\psi_{klm}^{(0)}(r,\theta,\phi)=j_l(kr)Y_l^m(\theta,\phi) \,,
\end{eqnarray}
where, the superscript `0' denotes that these are `free' (the potential is identically zero) spherical waves. Let us 
connect these two sets of bases as follows:
\begin{equation}
e^{ikz}=\sum_{l=0}^{\infty}\sum_{m=-l}^{+l}{\ma}_l^m(k)j_l(kr)Y_l^m(\theta,\phi) \,,
\end{equation}
where ${\ma}_l^m(k)$ are suitable expansion coefficients that can only be functions of the magnitude of the momentum. Since the LHS of above equation is independent of $\phi$, we require that RHS should also be independent of $\phi$, {\it i.e.} $m=0$. Thus, we are left with,
\begin{eqnarray}
e^{ikz}=\sum_{l=0}^{\infty}{\ma}_l^0(k)(k)j_l(kr)Y_l^0(\theta) \,,
\label{eikz}
\end{eqnarray}
where $Y_l^0(\theta)$ is given by
\begin{eqnarray}
Y_l^0(\theta)=\sqrt{\frac{(2l+1)}{4\pi}}P_l(\cos\theta) \,.
\end{eqnarray}
In view of this, we introduce the following shorthand:
\begin{eqnarray}
{\ma}_l=\sqrt{\frac{(2l+1)}{4\pi}}{\ma}_l^0 \,.
\end{eqnarray}
Using this, one may rewrite \Eqn{eikz} as
\begin{eqnarray}
e^{ikz}=\sum_{l=0}^{\infty}{\ma}_l(k)j_l(kr)P_l(\cos\theta) \,.
\label{eikz1}
\end{eqnarray}
To determine ${\ma}_l(k)$, we need to use the following integral representation for the Bessel function:
\begin{eqnarray}
j_l(kr)=\frac{1}{2i^l}\int_{-1}^{+1}P_l(\cos\theta)e^{ikr\cos\theta}d(\cos\theta) \,.
\end{eqnarray}
To illustrate the use of the above formula, let us multiply \Eqn{eikz1} by $P_{l'}(\cos\theta)d(\cos\theta)$ and integrate between $-1$ to $+1$ to obtain
\begin{eqnarray}
&& \int_{-1}^{+1}P_{l'}(\cos\theta)e^{ikr\cos\theta}d(\cos\theta) = \frac{2}{(2l'+1)}\sum\limits_{l=0}^{\infty}{\ma}_l(k) j_l(kr) \delta_{ll'} \,, \nonumber \\
&\Rightarrow& 2i^{l'}j_{l'}(kr) = \frac{2}{(2l'+1)} {\ma}_{l'}(k) j_{l'}(kr) \,, \nonumber \\
&\Rightarrow& {\ma}_{l'}(k) = i^{l'}(2l'+1) \,.
\end{eqnarray}
Plugging this into \Eqn{eikz1} we get the final expression as
\begin{eqnarray}
e^{ikz}=\sum_{l=0}^{\infty}i^l(2l+1)j_l(kr)P_l(\cos\theta) \,.
\label{eikz2}
\end{eqnarray}
In view of the asymptotic form of the Bessel function,
\begin{eqnarray}
j_l(kr)\xrightarrow{r\to\infty} \frac{\sin(kr-\frac{l\pi}{2})}{kr} \,,
\end{eqnarray}
we may rewrite \Eqn{eikz2} as
\begin{equation}
e^{ikz} \xrightarrow{r\to\infty} \frac{1}{2ikr}\sum_{l=0}^{\infty}(2l+1)\left[e^{ikr}-e^{-ikr}(-1)^l \right] P_l(\cos\theta) \,.
\label{asym1}
\end{equation}
Thus, at large distances, each $\psi_{klm}^{(0)}$ and so the `whole' $e^{ikz}$ results from the superposition of a converging spherical wave, $e^{-ikr}/r$, and a diverging spherical wave, $e^{ikr}/r$, whose amplitudes differ only by a phase. The fact that the squared amplitudes for both the incoming and outgoing spherical waves are same, simply reflects the conservation of probability. The presence of a scattering potential can only affect the amplitude of the outgoing spherical wave. Since probability conservation demands that the magnitude of the amplitude for the diverging wave should not change, it can only pick up additional phases arising due to the presence of a scattering potential. We will see the details in the following subsection.

\subsubsection*{Presence of a central potential -- asymptotic modification of the radial part}
The previous subsection was devoted for $V(r)=0$. Presence of a central potential simply modifies the wave function from the plane wave nature. But we know a special thing -- whatever be the form of $V(r)$, it dies out at a finite distance and in the asymptotic limit we should get the wave function in the form of Eq.~(\ref{basic}). 

In the presence of $V(r)$ the radial part of Schrodinger equation reads:
\begin{equation}
\left[\frac{d^2}{dr^2}+\frac{2}{r}\frac{d}{dr}+\left\{ k^2-\frac{2m}{\hbar^2}V(r)-\frac{l(l+1)}{r^2}\right\} \right] R_{kl}(r)=0 \,.
\label{tise}
\end{equation}
We assume that the potential is short ranged, {\it i.e.}, $V(r)\rightarrow{0}$ as $r\rightarrow\infty$. Then, at large distances, Eq.~(\ref{tise}) reduces to the free-particle equation. Therefore, the solution of Eq.~(\ref{tise}) should asymptotically approach the general solution for free particle:
\begin{equation}
R_{kl}(r)\xrightarrow{r\to\infty} A_lj_l(kr)+B_l\eta_l(kr)=\frac{C_l}{kr}\sin(kr-\frac{l\pi}{2}+\delta_l) \,,
\label{mod}
\end{equation}
where, the last step follows from the asymptotic forms of Bessel and Neumann functions:
\begin{eqnarray}
j_l(kr)\xrightarrow{r\to\infty}\frac{\sin(kr-\frac{l\pi}{2})}{kr} \,, ~~~~ \eta_l(kr)\xrightarrow{r\to\infty}-\frac{\cos(kr-\frac{l\pi}{2})}{kr} \,.
\end{eqnarray}
The quantities, $C_l$ and $\delta_l$ are related to $A_l$ and $B_l$  as follows:
\begin{eqnarray}
\tan\delta_l=-\frac{B_l}{A_l}\,, ~~~ {\rm and,}~~ C_l=\sqrt{A_l^2+B_l^2} \,.
\end{eqnarray}
Note that, unlike the free-particle case, here we did not demand $B_l=0$. This is due to the lack of information about 
the potential, we do not know the actual behavior of the wave function near the origin. \Eqn{mod} only represents the radial wave function at large distances where the potential is ineffective. Thus, the total wave function far away from the scatterer can be written as
\begin{eqnarray}
&& \psi(\vec{r})_{r\rightarrow\infty}=\sum_{l=0}^{\infty}R_{kl}(r \to \infty)P_l(\cos\theta) \,, \\
&\Rightarrow& \psi(\vec{r})_{r\rightarrow\infty}=\frac{1}{2ikr}\sum_{l=0}^{\infty}C_le^{-i\delta_l}e^{-i\frac{l\pi}{2}}
\left[e^{ikr}e^{2i\delta_l}-(-1)^le^{-ikr} \right] P_l(\cos\theta) \,.
\label{sc}
\end{eqnarray}
Now, this equation should be equivalent to Eq.~(\ref{basic}) with the expansion of $e^{ikz}$ in terms of partial waves given 
by Eq.~(\ref{asym1}). So, we can rewrite eqn~(\ref{basic}) as
\begin{equation}
\psi(\vec{r})_{r\rightarrow\infty}=\frac{1}{2ikr}\sum_{l=0}^{\infty}
(2l+1) \left[e^{ikr}-(-1)^le^{-ikr} \right] P_l(\cos\theta)+f_k(\theta)\frac{e^{ikr}}{r} \,.
\label{sc1}
\end{equation}
Comparing the co-efficients of ${e^{-ikr}}/{r}$ in Eqs.~(\ref{sc}) and (\ref{sc1}) one can easily get:
\begin{eqnarray}
C_l=i^l(2l+1)e^{i\delta_l} \,.
\end{eqnarray}
Once the value of $C_l$ is at hand, we can plug it into Eq.~(\ref{sc}), and then proceed to compare the co-efficients of ${e^{ikr}}/{r}$ to obtain the expression for $f_k(\theta)$. The final result is,
\begin{equation}
f_k(\theta)=\frac{1}{2ik}\sum_{l=0}^{\infty}(2l+1)(e^{2i\delta_l}-1)P_l(\cos\theta) \,.
\label{amp}
\end{equation}
Since $P_l(\cos\theta)$ serves as a complete set of basis vectors for any function of $\theta$, one can expand the scattering amplitude as follows:
\begin{equation}
f_k(\theta)=\frac{1}{k} \sum_{l=0}^{\infty}(2l+1)f_l(k)P_l(\cos\theta) \,.
\label{amp1}
\end{equation}
Comparing Eqs.~(\ref{amp}) and (\ref{amp1}) one can easily get
\begin{eqnarray}
&& f_l(k)=\frac{e^{2i\delta_l}-1}{2i} \,, \\
&\Rightarrow& f_l(k)=e^{i\delta_l} \sin\delta_l \,.
\label{flk}
\end{eqnarray}
\begin{wrapfigure}{r}{0.4\textwidth}
\centering
 \includegraphics[scale=0.5]{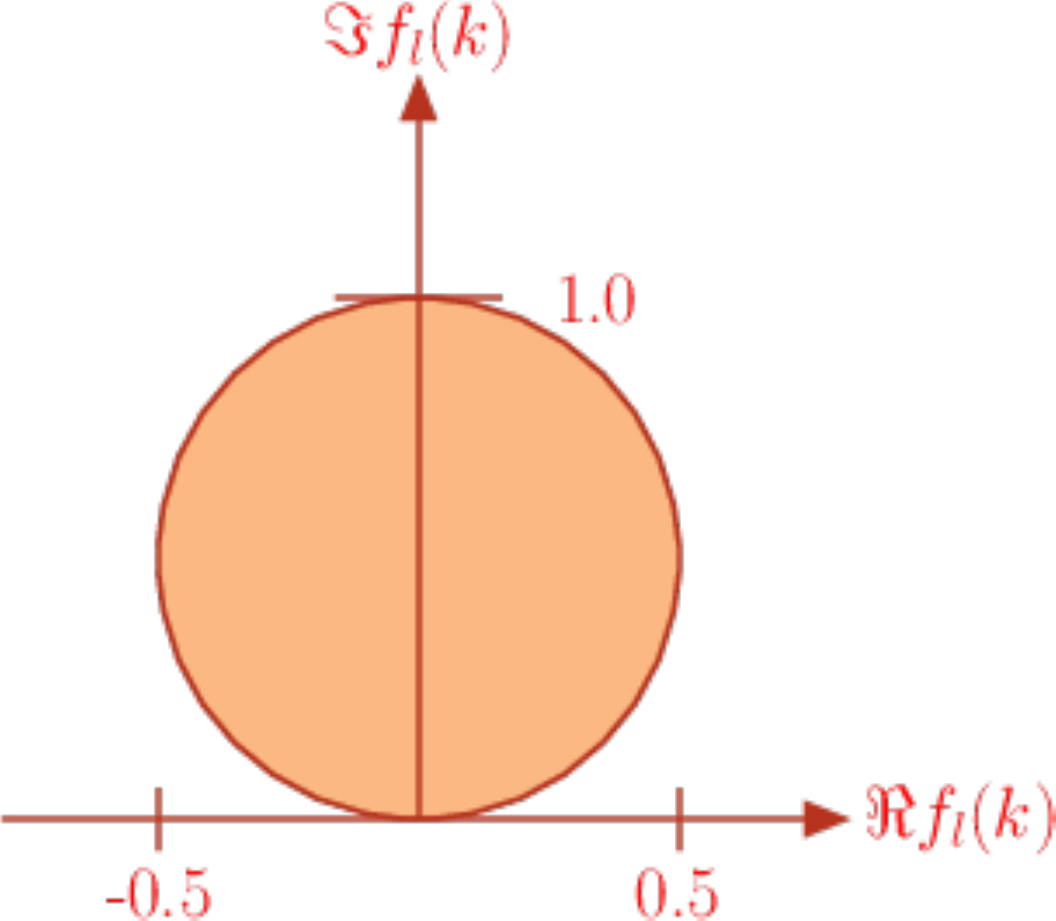}
 \caption{Unitarity circle.}
 \label{unicirc}
\end{wrapfigure}%
From \Eqn{flk} it follows that
\begin{eqnarray}
|f_l(k)| \le 1 ~~~ {\rm for~ all~ values~ of~} l \,,
\label{fun}
\end{eqnarray}
or, splitting $f_l(k)$ into its real and imaginary components, one can derive the equation of the {\em unitarity circle} (see Figure~\ref{unicirc}):
\begin{eqnarray}
\Big[{\Re} f_l(k)\Big]^2 +\left[{\Im} f_l(k)-\frac{1}{2}\right]^2 = \frac{1}{4} \,.
\end{eqnarray}

Now we have learned that the expansion coefficients, $f_l(k)$, of the quantum mechanical scattering amplitude obey the unitarity conditions of \Eqn{fun}. But, till now there is very little hint that these $f_l(k)$-s are the same as the $a_l$-s of \Eqn{uni_amp}. Some intuitive arguments to make the connections will be presented shortly. Before that, to make the discussion complete, we wish to present one important result that follows from \Eqn{amp1}.

Note that, using the value of $f_l(k)$ from \Eqn{flk}, one can rewrite \Eqn{amp1} as
\begin{eqnarray}
f_k(\theta)= \frac{1}{k}  \sum_{l=0}^{\infty}(2l+1)e^{i\delta_l} \sin\delta_lP_l(\cos\theta) \,.
\label{amp2}
\end{eqnarray}
We can now find the expression for the total scattering cross-section as
\begin{eqnarray}
\sigma = \int \frac{d\sigma}{d\Omega} d\Omega = \int |f_k(\theta)|^2 d\Omega \,.
\end{eqnarray}
Using the orthonormality of the Legendre polynomials, the final result becomes
\begin{eqnarray}
\sigma = \frac{4\pi}{k^2} \sum\limits_{l=0}^{\infty} (2l+1) \sin^2\delta_l \,.
\label{sigq}
\end{eqnarray}
Looking at \Eqs{amp2}{sigq} and keeping in mind tha $P_l(1)=1$ for any $l$, one can at once realize that
\begin{eqnarray}
\sigma= \frac{4\pi}{k} \Im \left\{f_k(\theta=0)\right\} \,,
\label{optical}
\end{eqnarray}
where, $\Im \left\{f_k(\theta=0)\right\}$ is the imaginary part of the forward scattering amplitude. \Eqn{optical} is known as the {\em optical theorem} in quantum mechanics.

\subsection{Connection with Quantum Field Theory}
We shall now give a hand waving argument on how the quantum mechanical scattering amplitude is related to the Feynman amplitude in Quantum Field Theory (QFT). We know the expression for differential scattering cross-section both in quantum mechanics and in QFT. This is given by
\begin{eqnarray}
\frac{d\sigma}{d\Omega}= \underbrace{\frac{1}{64\pi^2s}|{\cm}(\theta)|^2}_{\rm QFT} = \underbrace{|f_k(\theta)|^2}_{\rm QM} \,,
\label{qmqft}
\end{eqnarray}
where, ${\cm}(\theta)$ is the Feynman amplitude for the process and $s=4E^2$ is the CM energy squared. From \Eqn{qmqft} we can make a simple-minded connection:
\begin{eqnarray}
{\cm}(\theta) =16\pi{E}f_k(\theta) \,.
\label{Mf}
\end{eqnarray}
Now, plugging the expression of $f_k(\theta)$ from \Eqn{amp1} into \Eqn{Mf} and approximating $k\approx E$ at high energies, we may write
\begin{eqnarray}
{\cm}(\theta) = 16\pi \sum_{l=0}^{\infty}(2l+1)f_l(E)P_l(\cos\theta) \,.
\label{finale}
\end{eqnarray}
Thus, comparing \Eqs{uni_amp}{finale}, one can see that $a_l$s of \Eqn{uni_amp} are the same as $f_l(k)$s of \Eqn{amp1} and both of them must obey the unitarity condition of \Eqn{fun}. Extraction of each partial wave amplitude from the Feynman amplitude will now be a straightforward task:
\begin{eqnarray}
a_l =\frac{1}{32\pi}\int_{-1}^{+1}{\cm}(\theta)P_l(\cos\theta) d(\cos\theta) \,.
\label{partial_extraction}
\end{eqnarray}

\section{Fermi Theory}
\label{Fermi Theory}
The Fermi lagrangian (for the leptonic sector) is given by,
\begin{eqnarray}
{\mathscr L}_F = -\frac{G_F}{\sqrt{2}}J^{\mu}_{(\textrm{lep})} J^{\dagger}_{\mu(\textrm{lep})} \,.
\label{fermi lagrangian}
\end{eqnarray}
We know from the experiments that only the left-handed fields take part in $\beta$-decay. So the leptonic current should be written as
\begin{eqnarray}
 J^{\mu}_{(\textrm{lep})} = \bar{\nu}_e \gamma^{\mu}(1-\gamma^5)e \,.
\end{eqnarray}
The coupling constant $G_F$ has mass dimension $-2$. To explore the consequence of this negative mass dimension, let us consider the elastic scattering process $\nu_e e \to \nu_e e$. Neglecting the electron mass, the scattering amplitude will be non-zero for the following combination of helicities:
\begin{eqnarray}
h_1 = h_2 = h_3 = h_4 = -\frac{1}{2} \,. 
\end{eqnarray}
For these combination one has,
\begin{eqnarray}
\left|{\cal M}(\theta)\right|_{\nu_e e \to \nu_e e} = 4\sqrt{2}G_F s \,.
\end{eqnarray}
Thus, only the partial wave with $l=0$ contributes and that is given by,
\begin{eqnarray}
\left|a_0\right|_{\nu_e e \to \nu_e e} = \frac{1}{2\sqrt{2}\pi}G_F s \,.
\end{eqnarray}
The unitarity condition (\Eqn{unitarity}) then sets an upper bound on the CM energy, beyond which Fermi theory ceases to be valid. In this case the limit is,
\begin{eqnarray}
\left(E_{\rm CM}\right)_{\rm max} = \left(\frac{2\pi\sqrt{2}}{G_F}\right)^{\frac{1}{2}} = 870~{\rm GeV} \,.
\end{eqnarray}
However, one must remember that the exact value of this cut-off is process dependent.

\section{Intermediate Vector Bosons}
\label{IVB}
We have just learned that the direct four fermion interaction causes the corresponding coupling constant ($G_F$) to have negative mass dimension which causes the scattering amplitude to grow as $E^2$. A natural way out of this problem is to replace \Eqn{fermi lagrangian} with an interaction describe by `exchange' of another particle (which must be a boson) in analogy with photon in QED. This means, instead of \Eqn{fermi lagrangian} we write the following interaction:\footnote{At this point, one might wonder why we are starting with the possibility of a charged vector current. We should begin with charged scalar current which would be a more minimalistic choice. Admittedly, we are implicitly using some experimental inputs -- measurement of {\em Michel Parameters}\cite{Michel:1949qe} in polarized muon decay. These measurements confirm that the charged current processes are governed by $V-A$ type interactions and therefore, the possibility of a charged scalar as an intermediate boson is ruled out.}
\begin{eqnarray}
 {\mathscr L}_{\rm int}^{W} = \frac{g}{2\sqrt{2}}\left(J^{\mu}_{(\textrm{lep})}W_{\mu}^+ +J^{\mu\dagger}_{(\textrm{lep})}W_{\mu}^-\right) \,,
\label{W interaction}
 \end{eqnarray}
where $W_{\mu}^{\pm}$ is a vector field corresponding to a `mediating' particle (with spin-1) which is, therefore, usually called intermediate vector boson (IVB). The numerical factor of $(2\sqrt{2})^{-1}$ in front is purely conventional. One must remember that $W^{\pm}$ must be massive so that it can describe a short range force and we should write the propagator as:
\begin{eqnarray}
 D^{\mu\nu}(q) = \frac{-g^{\mu\nu} +\frac{q^\mu q^\nu}{M_W^2}}{q^2-M_W^2} \,. 
\label{W propagator}
\end{eqnarray}

The model of weak interactions defined by the Lagrangian of \Eqn{W interaction} must respect an experimentally established fact that the effective Fermi-type theory provides very good description of a considerable part of physical reality in the low energy region. Comparing low energy muon decay ($\mu^- \to e^- \nu_{\mu}\bar{\nu}_e$) amplitude for these two theories, we obtain,
\begin{eqnarray}
\frac{G_F}{\sqrt{2}} = \frac{g^2}{8M_W^2} \,.
\label{fermi constant}
\end{eqnarray}
Note that, in the derivation of \Eqn{fermi constant}, the negative sign in the Fermi Lagrangian plays an important role. It is just this convention which guarantees that $G_F>0$, if the Fermi theory is viewed as an effective low-energy approximation of the theory with IVB. Remember, as a consequence of IVB, there will be no energy growths in $\nu_e e \to \nu_e e$ and $\bar{\nu}_e e \to \bar{\nu}_e e$ scattering amplitudes. We should now investigate whether $W^\pm$ are sufficient for energy growth cancellation for all other possible processes.

\subsection{Electrodynamics of the $W$-bosons}
Since $W$-bosons carry electric charge, they must interact with the photon. The question is, how should we write the coupling. Let us motivate it in the following way:

The wave equation for a free particle with spin-1 and a non-zero mass,
\begin{eqnarray}
\partial_\mu B^{\mu\nu} + m^2 B^\nu = 0 \,, 
\end{eqnarray}
can be obtained from the Proca Lagrangian,
\begin{eqnarray}
{\mathscr L} &=& -\frac{1}{4}B^{\mu\nu}B_{\mu\nu} + \frac{1}{2}m^2 B^\mu B_\mu \nonumber \\
&=& -\frac{1}{4}(\partial_\mu B_\nu - \partial_\nu B_\mu)(\partial^\mu B^\nu - \partial^\nu B^\mu) + \frac{1}{2}m^2 B^\mu B_\mu \,.
\end{eqnarray}
For a complex (charged) spin-1 field, the generalization would be straightforward:
\begin{eqnarray}
{\mathscr L} &=& -\frac{1}{2}W^{\mu\nu}(W_{\mu\nu})^\dagger + M_W^2 W^{\mu+} W_\mu^- \nonumber \\
&=& -\frac{1}{2}(\partial_\mu W_\nu^- - \partial_\nu W_\mu^-)(\partial^\mu W^{\nu+} - \partial^\nu W^{\mu+}) + M_W^2 W^{\mu+} W_\mu^- \,.
\end{eqnarray}
We want the above Lagrangian to be invariant under the following $U(1)$ transformation:
\begin{subequations}
\begin{eqnarray}
 W_\mu^-(x) \to W_\mu^{-\prime}(x) &=& e^{-i\omega(x)}W_\mu^-(x) \,, \\
 W_\mu^+(x) \to W_\mu^{+\prime}(x) &=& e^{+i\omega(x)}W_\mu^+(x) \,.
\end{eqnarray}
\end{subequations}
It would be accomplished if we replace the ordinary derivatives with the covariant derivatives:
\begin{eqnarray}
D_\mu = \partial_\mu + ieA_\mu \,, 
\end{eqnarray}
with the $U(1)$ (electromagnetic) gauge field, $A_\mu (x)$, transforming as:
\begin{equation}
A_\mu^\prime (x) = A_\mu (x) + \frac{1}{e}\partial_\mu \omega(x) \,. 
\end{equation}
So our `minimal' Lagrangian will be
\begin{eqnarray}
{\mathscr L}^{\rm min} &=& -\frac{1}{2}(D_\mu W_\nu^- - D_\nu W_\mu^-)(D^{*\mu} W^{\nu+} - D^{*\nu} W^{\mu+}) + M_W^2 W^{\mu+} W_\mu^- \,.
\label{minimal lagrangian}
\end{eqnarray}
But there is more. One may add to the minimal Lagrangian of \Eqn{minimal lagrangian} another gauge invariant term:
\begin{eqnarray}
{\mathscr L}^\prime &=& -i\kappa e W_\mu^- W_\nu^+ F^{\mu\nu} \,, \\
{\rm where,} ~~~ F_{\mu\nu} &=& \partial_\mu A_\nu - \partial_\nu A_\mu \,.
\end{eqnarray}
So we write the total electromagnetic interaction as:
\begin{eqnarray}
{\mathscr L}_W^{\rm e.m.} = {\mathscr L}^{\rm min} + {\mathscr L}^\prime
\label{total lagrangian}
\end{eqnarray}
Till now there is no good reason why we should keep ${\mathscr L}^\prime$. The condition $\kappa=0$ will correspond to the minimal case. But let us keep our minds open and see, using unitarity arguments, what value of $\kappa$ should be chosen by nature. Assuming \Eqn{total lagrangian} to be our guiding Lagrangian, we may write the cubic and quartic couplings as:
\begin{eqnarray}
{\mathscr L}_{WW\gamma} &=& -ie \Big[ A^\mu \big\{W^{\nu-}(\partial_\mu W_\nu^+) -(\partial_\mu W_\nu^+)W^{\nu+}\big\} +W^{\mu-}\big\{\kappa W^{\nu+}(\partial_\mu A_\nu) \nonumber \\
 && -(\partial_\mu W_\nu^+)A^\nu\big\}+ W^{\mu+}\big\{A^\nu(\partial_\mu W_\nu^-)-\kappa(\partial_\mu A_\nu)W^{\nu-}\big\}\Big] 
 \label{WWg} \\
 {\mathscr L}_{WW\gamma\gamma} &=& -e^2(A_\mu A^\mu W_\nu^- W^{\nu+}-A^\mu A^\nu W_\mu^-W_\nu^-)
 \label{WWgg}
\end{eqnarray}
\paragraph*{Determination of $\kappa$:} Let us consider the following process:
$$W^-(p_1)+W^+(p_2) \to \gamma(k_1)+\gamma(k_2)$$
\begin{figure}
\includegraphics[scale=0.5]{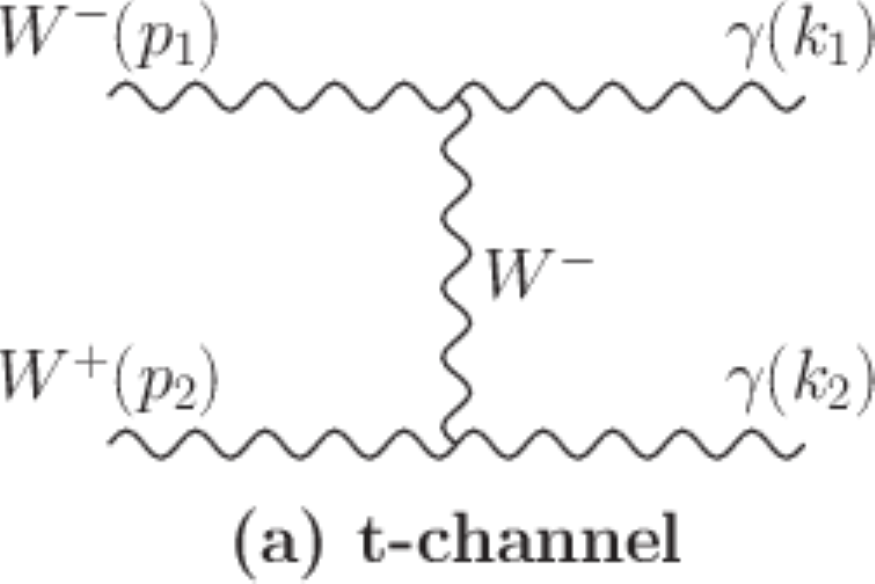}~~~
\includegraphics[scale=0.5]{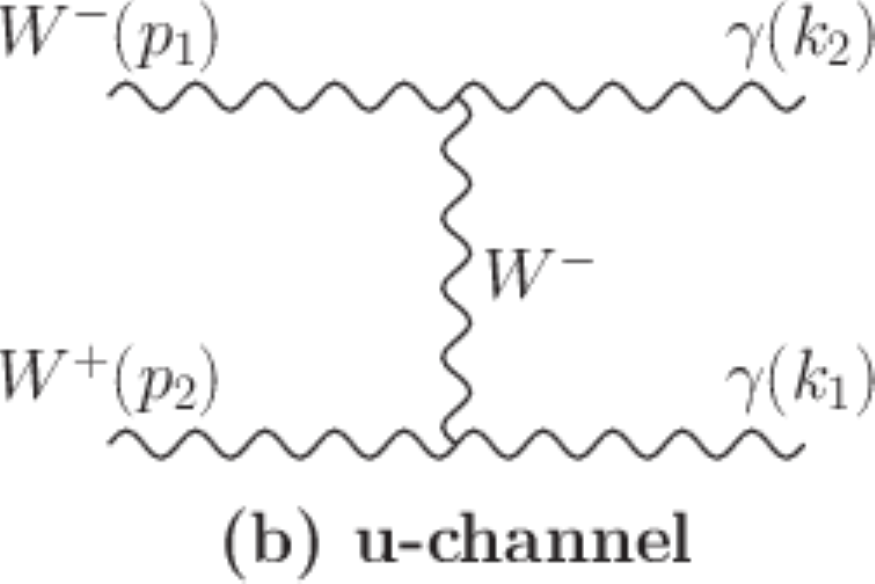}~~~
\includegraphics[scale=0.5]{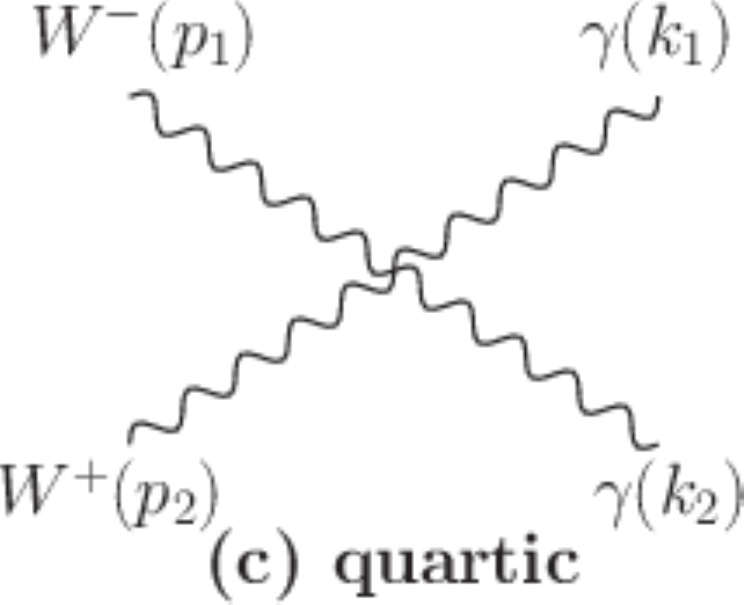}
\caption[Feynman diagrams for the process $W^-(p_1)+W^+(p_2) \to \gamma(k_1)+\gamma(k_2)$]{\em Feynman diagrams for the process $W^-(p_1)+W^+(p_2) \to \gamma(k_1)+\gamma(k_2)$.}
\label{fWWgg:subfigures}
\end{figure}
The Feynman diagrams for this process are shown in \fref{fWWgg:subfigures}. For diagrams (a) and (b) there will be a W propagator which contains a $g^{\mu\nu}$ term and a $q^\mu q^\nu$ term. From dimensional arguments one can understand that the leading order divergence comes from the $q^\mu q^\nu$ term. In view of this, we write the amplitude for diagram (a) as:
\begin{equation}
 {\cal M}_a = {\cal M}_a^{(1)} +{\cal M}_a^{(2)} \,.
\end{equation}
${\cal M}_a^{(1)}$ contains the contribution from the $g^{\mu\nu}$ term and ${\cal M}_a^{(2)}$ contains that from the $q^\mu q^\nu$ term. Calculating explicitly we have, at the leading order, 
\begin{eqnarray}
 {\cal M}_a^{(1)} &=& -\frac{e^2}{M_W^2}\frac{1}{t-M_W^2}\Big[(1-\kappa)^2\big\{(p_1.k_1)(\epsilon_1.\epsilon_1^\prime)-(p_1.\epsilon_1^\prime)(k_1.\epsilon_1)\big\} \nonumber \\
  && \times \big\{(p_2.k_2)(\epsilon_2.\epsilon_2^\prime)-(p_2.\epsilon_2^\prime)(k_2.\epsilon_2)\big\}+\dots\Big] 
\label{WWgg energy}
\end{eqnarray}
Diagram (b) will also have an analogous term, ${\cal M}_b^{(2)}$, with $t\leftrightarrow u$, $k_1 \leftrightarrow k_2$, $\epsilon_1^\prime \leftrightarrow \epsilon_2^\prime$ . Since ${\cal M}_b^{(2)}$ has $\frac{1}{u-M_W^2}$ as the prefactor, ${\cal M}_a^{(2)}$ and ${\cal M}_b^{(2)}$ will not cancel each other for any arbitrary set $\{p_1,p_2,k_1,k_2\}$ satisfying four-momentum conservation. So, ${\cal M}_a^{(2)}$ and ${\cal M}_b^{(2)}$ must vanish independently.

\paragraph*{Observation:} First of all, it should be noted that if at least one of the $W$ bosons has longitudinal polarization, the leading growth of \Eqn{WWgg energy} (quartic or cubic) vanishes for any arbitrary value of $\kappa$. This statement can immediately be verified if we replace, in such a case, the polarization vector $\epsilon_1 \equiv \epsilon_L(p_1)$ or $\epsilon_2 \equiv \epsilon_L(p_2)$ in \Eqn{WWgg energy} by the corresponding leading term $\frac{p_1}{M_W}$ or, $\frac{p_2}{M_W}$ according to the formula:
\begin{eqnarray}
\epsilon_L^\mu (p) = \frac{p^\mu}{M_W} + \order(\frac{M_W}{p^0}) \,.
\end{eqnarray}
The corresponding expression within the curly brackets will be zero causing the leading energy growth in \Eqn{WWgg energy} to vanish. But, if both the $W$ bosons have transverse polarizations, then the leading term in \Eqn{WWgg energy} will, in general,  grow quadratically unless $\kappa=1$.

Using this value of $\kappa$, \Eqn{WWg} can be rewritten as:
\begin{eqnarray}
{\mathscr L}_{WW\gamma} = -ie \big[A^\mu (W^{\nu-}\overleftrightarrow{\partial}_\mu W_\nu^+) +W^{\mu-}(W^{\nu+}\overleftrightarrow{\partial}_\mu A_\nu) + W^{\mu+}(A^\nu \overleftrightarrow{\partial}_\mu W_\nu^-)\big] \,.
\label{YM WWg}
\end{eqnarray}
The symbol $\overleftrightarrow{\partial}$ in \Eqn{YM WWg} is defined in the usual way as,
\begin{equation}
f\overleftrightarrow{\partial}_\mu g = f(\partial_\mu g) - (\partial_\mu f) g \,. 
\end{equation}
Therefore, we have arrived at the following important conclusion \cite{Horejsi:1993hz}:
\begin{quotation}
\emph{Leading power growth arising in the high energy limit in tree level diagrams involving both external and internal lines of vector bosons, $W^\pm$, are eliminated for an arbitrary combination of the $W^\pm$ polarizations if and only if the corresponding electromagnetic interaction is of the Yang-Mills type.} 
\end{quotation}
%
\subsection{Introducing the Z boson }
Now consider the process \cite{Cheng:1985bj}
$$e^-(p_1)+e^+(p_2) \to W_L^-(k_1)+W_L^+(k_2)$$
\begin{figure}
\centering
\includegraphics[scale=0.6]{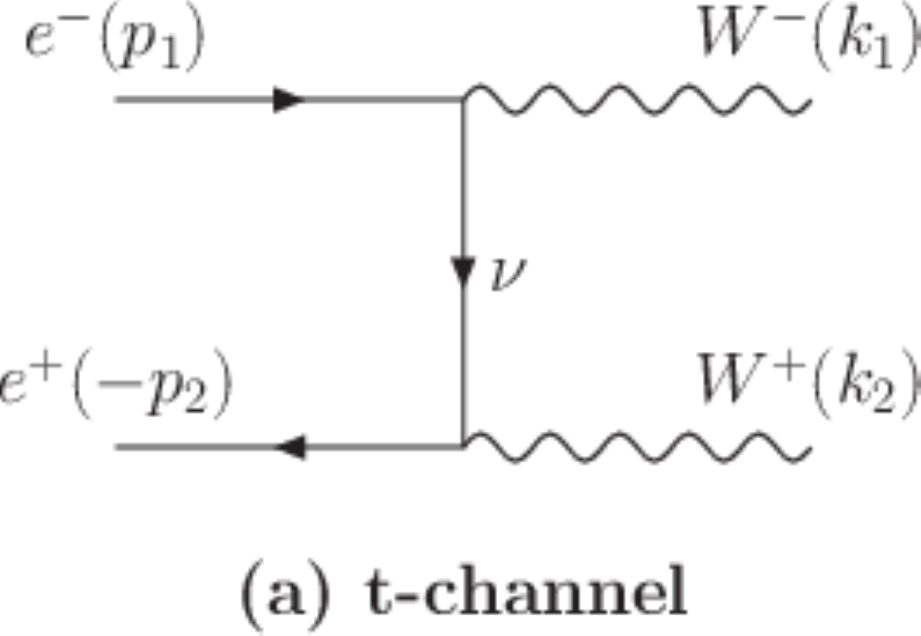}~~~~~~~~~
\includegraphics[scale=0.6]{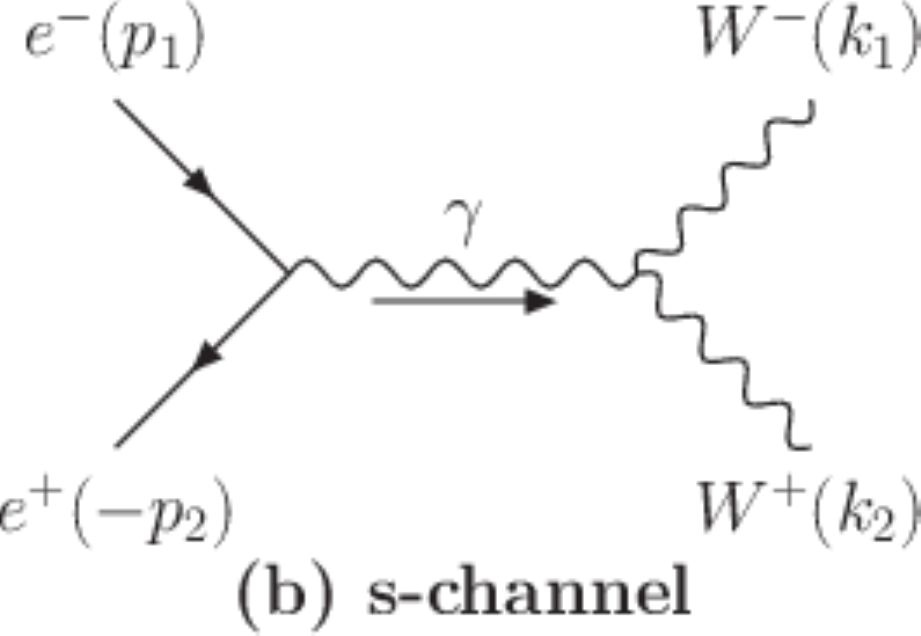}
\caption[Feynman diagrams for the process $e^-e^+ \to WW$]{\em Feynman diagrams for the process $e^-(p_1)+e^+(p_2) \to W_L^-(k_1)+W_L^+(k_2)$.}
\label{feeWW:subfigures}
\end{figure}
From what we have learned till now, there will be two possible diagrams for this process as shown in \fref{feeWW:subfigures}. The corresponding amplitudes are written below (see Appendix~\ref{AppendixB}):
\begin{eqnarray}
 {\cm}_a^\nu &=& -\frac{g^2}{4M_W^2}\underbrace{\bar{v}(p_2)\Slash{k_1}(1-\gamma_5)u(p_1)}_{\order(E^2)} + \order(E) \,, \label{EW1} \\
 {\cm}_b^\gamma &=& \frac{e^2}{M_W^2}\underbrace{\bar{v}(p_2)\Slash{k_1}u(p_1)}_{\order(E^2)} + \order(1) \,. \label{EW2}
\end{eqnarray}
Remember that the $u$ and $v$ spinors contain a factor of $\sqrt{E}$ in their normalization. So the leading terms grow as $E^2$. Let us collect the quadratic growths in a single equation
\begin{eqnarray}
 {\cm}^{\rm quadratic} &=& -\frac{g^2}{4M_W^2}\bar{v}(p_2)\slashed{k_1}(1-\gamma_5)u(p_1) + \frac{e^2}{M_W^2}\bar{v}(p_2)\Slash{k_1}u(p_1) \,.
 \label{growth in eeWW}
\end{eqnarray}
It is clear that one cannot arrange a mutual cancellation of quadratic energy growths between ${\cm}_a^\nu$ and ${\cm}_b^\gamma$ by tuning the relative magnitudes of the coupling constants $e$ and $g$. This is because the corresponding growth in \Eqn{EW1} contains a factor of $(1-\gamma_5)$ but in \Eqn{EW2} it does not. It is the consequence of the fact that the charged current interactions involve only the left handed (LH) fermions. 

Thus we have no other choices than to introduce new particles which can compensate the residual growth of \Eqn{growth in eeWW}. We shall restrict ourselves to particles with lowest possible spin ({\it i.e.} 0, $\frac{1}{2}$, 1) and allow only those interaction terms which satisfy the condition $[{\ml}_{\rm int}] \le 4$, so that renormalizability of the theory is not compromised.  
\subsubsection*{Choice 1 (Spin 0)}
Let us postulate the existence of spin 0 scalar ($h$) whose coupling with the fermions and $W$ bosons can be written as:
\begin{eqnarray}
 {\ml}_{WWh} &=& g_{WWh}~ W_\mu^- W^{\mu +}h \\ 
 {\ml}_{ffh} &=& g_{ffh} ~ \bar{f}\Gamma fh \,.
\end{eqnarray}
Where $\Gamma$ is, in general, a combination of the unit matrix and $\gamma_5$ and $g_{WWh}$, $g_{ffh}$ are corresponding coupling strengths. Note that $g_{ffh}$ is dimensionless but $g_{WWh}$ is not. In fact, without any loss of generality, we may express $g_{WWh}$ as
\begin{eqnarray}
 g_{WWh} = \alpha M_W \,, 
\end{eqnarray}
where $\alpha$ is a dimensionless constant. As a result of these new couplings there will be a $h$-mediated s-channel diagram similar to \fref{feeWW:subfigures}b. By simple dimension counting one can verify that this diagram can at best grow as $\order(E)$ for large CM energies when the external $W$ bosons are longitudinally polarized. An exchange of a spin 0 particle is therefore not suitable  for the desirable cancellation of the quadratic growths in \Eqn{growth in eeWW}. However, it is worth noting at this point that such a spin 0 particle can play a crucial role in suppressing linear energy growths in the $e^-e^+\to WW$ amplitude. This will be used later in this chapter.
\subsubsection*{Choice 2 (Spin $\frac{1}{2}$)}
In this case we shall assume the existence of a heavy neutrino-like fermion ($E^0$) which couples with the electron in the following way\cite{Georgi:1972cj}:
\begin{eqnarray}
 {\ml}_{\rm int}^{(E^0)} = \left( b_L\bar{E}_L^0\gamma^\mu e_L +b_R\bar{E}_R^0\gamma^\mu e_R \right) + {\rm h.c.} \,. 
\end{eqnarray}
We shall assume $b_L$ and $b_R$ to be real to respect CP invariance. This new interaction will lead to a new t-channel diagram similar to \fref{feeWW:subfigures}a mediated by $E^0$. Unlike the previous choice, the amplitude corresponding to this new diagram does contain terms which grow quadratically in the high-energy limit when the external $W$  bosons are longitudinally polarized. The requirement of a cancellation of quadratic growths in \Eqn{growth in eeWW} then yields the following conditions for the coupling constants $b_L,~b_R$:
\begin{eqnarray}
 b_L^2 &=& e^2 -\frac{g^2}{2} \,, \\
 b_R^2 &=& e^2 \,.
\end{eqnarray}
Since $b_L$ is real, the first relation leads to a constraint for relative strengths of weak and electromagnetic interactions, namely
\begin{eqnarray}
 g\le \sqrt{2}e \,. 
\end{eqnarray}
An interesting consequence of the above inequality and the general relation of \Eqn{fermi constant} is an upper bound for the $W^\pm$ mass:
\begin{eqnarray}
M_W \le \left(  \frac{\sqrt{2}\pi\alpha}{G_F}\right)^{\frac{1}{2}} = 53 ~{\rm GeV} \,. 
\end{eqnarray}
We now know that this is against the experimental facts. Therefore we shall not consider this scheme any further.
\subsubsection*{Choice 3 (Spin 1)}
Now we shall consider the next alternative, {\it i.e.}, the case where the ``compensation'' diagram for $e^-e^+\to W^-W^+$ corresponds to a s-channel exchange of a neutral spin 1 particle ($Z$) with non-zero mass. Note that a new massless neutral boson will imply the existence of a new kind of long range force (like electromagnetism) which is not observed in nature. But to calculate the amplitude we need to have some idea about the $WWZ$ coupling. Here also we will concentrate on interactions which satisfy $[{\ml}_{WWZ}] \le 4$. We proceed as follows \cite{LlewellynSmith:1973ey}:

\begin{wrapfigure}{r}{0.4\textwidth}
\centering
 \includegraphics[scale=0.5]{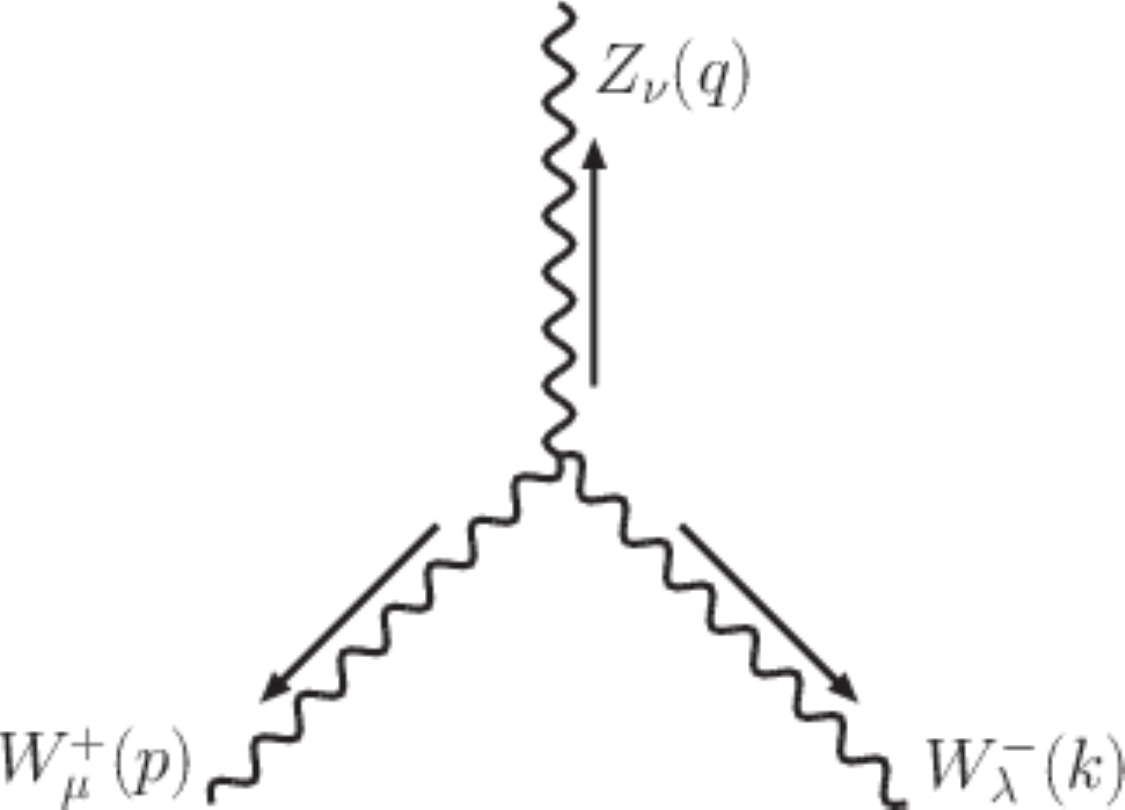}
\end{wrapfigure}%
It is obvious that a Lorentz invariant interaction involving three vector bosons and satisfying the above condition, must involve just one derivative of a vector boson field (the corresponding coupling constant is then of course dimensionless). In momentum space, this means that the interaction vertex shown in the adjacent picture can be represented by a linear polynomial in terms of the four momenta $k$, $p$ and $q$. Among these only two are independent because of the four momentum conservation $k+p+q = 0$. Choosing $k$ and $p$ to be independent variables, the most general linear polynomial representing the $W^-W^+Z$ interaction vertex  may be written as
\begin{eqnarray}
\label{WWZ structure}
V_{\lambda\mu\nu}(k,p,q) &=& (Ak_\lambda+Bp_\lambda)g_{\mu\nu} + (Ck_\mu+Dp_\mu)g_{\lambda\nu} + (Ek_\nu+Fp_\nu)g_{\lambda\mu} \nonumber \\
&& +G\epsilon_{\lambda\mu\nu\rho}k^\rho +H\epsilon_{\lambda\mu\nu\rho}p^\rho \,.
\end{eqnarray}
For comparison, let us write the Yang-Mills structure below 
\begin{eqnarray}
V_{\lambda\mu\nu}^{\rm (YM)}(k,p,q) = (k+2p)_\lambda g_{\mu\nu} + (-2k-p)_\mu g_{\lambda\nu} + (k-p)_\nu g_{\lambda\mu}
\end{eqnarray}
\begin{figure}
\centering
\includegraphics[scale=0.5]{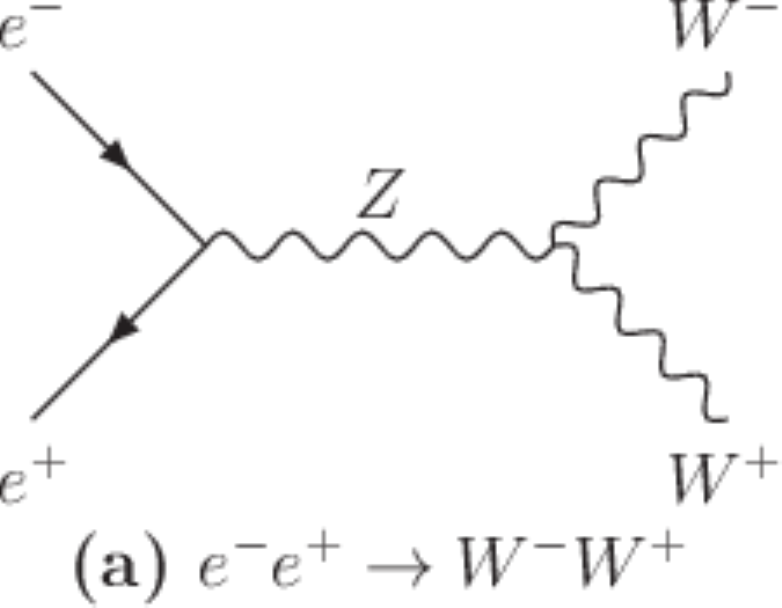}~~~
\includegraphics[scale=0.5]{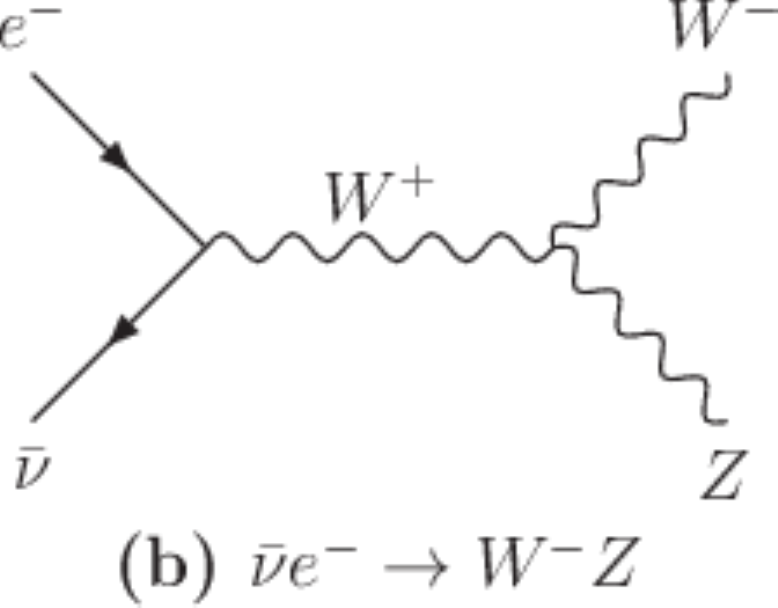}~~~
\includegraphics[scale=0.5]{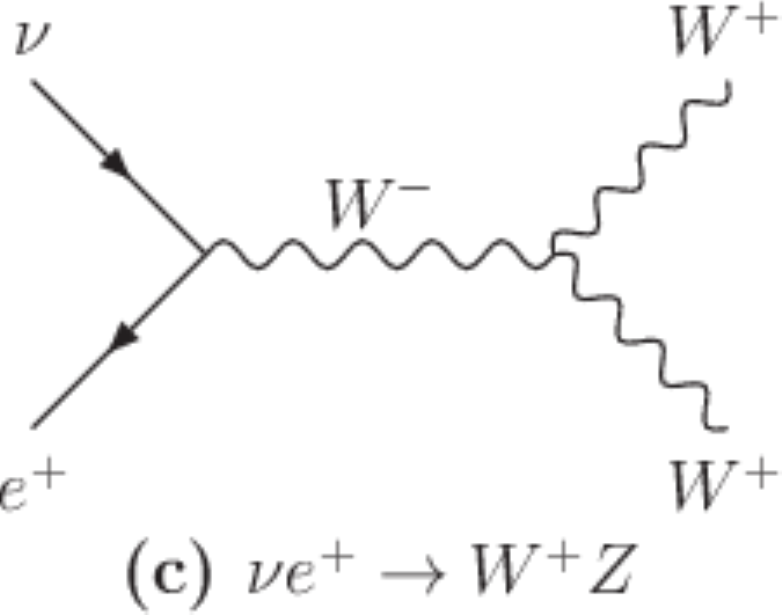}
\caption[Some Feynman diagrams to determine $WWZ$ vertex ]{\em Some $2\to 2$ scatterings to determine the $WWZ$ vertex.}
\label{WWZfigures}
\end{figure}
The leading growth in the three processes shown in \fref{WWZfigures} will come from the ${q^\mu q^\nu \over M_{W,Z}^2}$ term in the intermediate vector boson propagator. If we demand the vanishing of this leading growth then we get the following relations involving $A$, $B$, $\dots$, $G$ \cite{Horejsi:1993hz}:
\begin{subequations}
\begin{eqnarray}
B+C &=& 0 \,, \\
E+F &=& 0 \,, \\
B-E+F &=& 0 \,, \\
-C+2D &=& 0 \,, \\
C+E-F &=& 0 \,, \\
2A-B &=& 0 \,, \\
G = H &=& 0 \,.
\end{eqnarray}
\end{subequations}
Solving the above set of equations we may rewrite \Eqn{WWZ structure} as
\begin{eqnarray}
\label{YM WWZ}
V_{\lambda\mu\nu}(k,p,q) = g_{WWZ}\left[ (k+2p)_\lambda g_{\mu\nu} + (-2k-p)_\mu g_{\lambda\nu} + (k-p)_\nu g_{\lambda\mu}\right] \,,
\end{eqnarray}
where, we have relabeled $A$ as $g_{WWZ}$. Therefore, $V_{\lambda\mu\nu}$ has the exact Yang-Mills structure upto an overall multiplicative factor which, we hope, will be determined by the requirement of unitarity. 

Now, as a consequence of this new particle, the process $e^-e^+ \to W^-W^+$ will have an extra s-channel diagram similar to \fref{feeWW:subfigures}b mediated by the $Z$ boson. If we express the $eeZ$ coupling as
\begin{eqnarray}
\label{ffZ coupling}
{\ml}_{eeZ} = (g_L\bar{e}_L\gamma^\mu e_L + g_R\bar{e}_R\gamma^\mu e_R)Z_\mu \,,
\end{eqnarray}
Then the amplitude for the new diagram is found to be
\begin{eqnarray}
{\cm}_b^Z &=& -\frac{1}{2M_W^2}g_{WWZ}~ g_L \bar{v}(p_2)\slashed{k_1}(1-\gamma_5)u(p_1) \nonumber \\
                    &&   -\frac{1}{2M_W^2}g_{WWZ}~ g_R \bar{v}(p_2)\slashed{k_1}(1+\gamma_5)u(p_1) +\order(E)
\end{eqnarray}
Adding this with \Eqn{growth in eeWW} we immediately get the conditions for cancellation of leading energy growths in $e^-e^+\to WW$ at large values of CM energies:
\begin{eqnarray}
\label{SM1}
-\frac{g^2}{2} +e^2 -g_Lg_{WWZ} &=& 0 \,, \\
\label{SM2}
e^2-g_R g_{WWZ} &=& 0  \,.
\end{eqnarray}
%
\section{Retrieving vector and axial-vector couplings}
For this purpose we need to consider a few more processes. The first of these will be 
$$\nu(p_1)+\bar{\nu}(p_2) \to W_L^+(k_1) +W_L^-(k_2) \,.$$
Taking into account the experimental fact that only left-handed neutrinos are observed in nature, we can express the $\nu\nu Z$ coupling as follows:
\begin{eqnarray}
\label{nunuZ coupling}
{\ml}_{\nu\nu Z} = g_{\nu\nu Z} \bar{\nu}\gamma^\mu P_L \nu Z_\mu \,.
\end{eqnarray}
\begin{figure}
\centering
\includegraphics[scale=0.6]{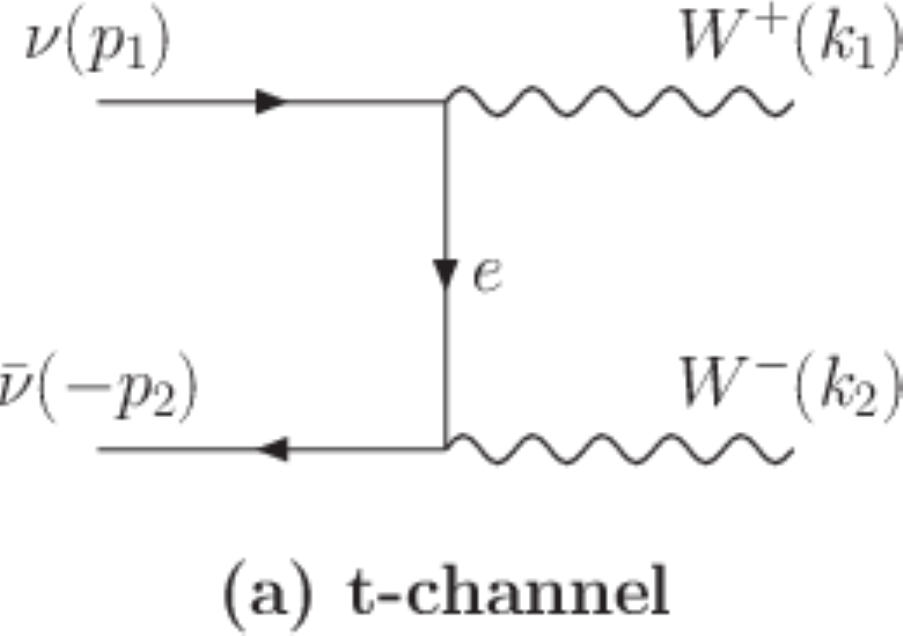}~~~~~~~~~
\includegraphics[scale=0.6]{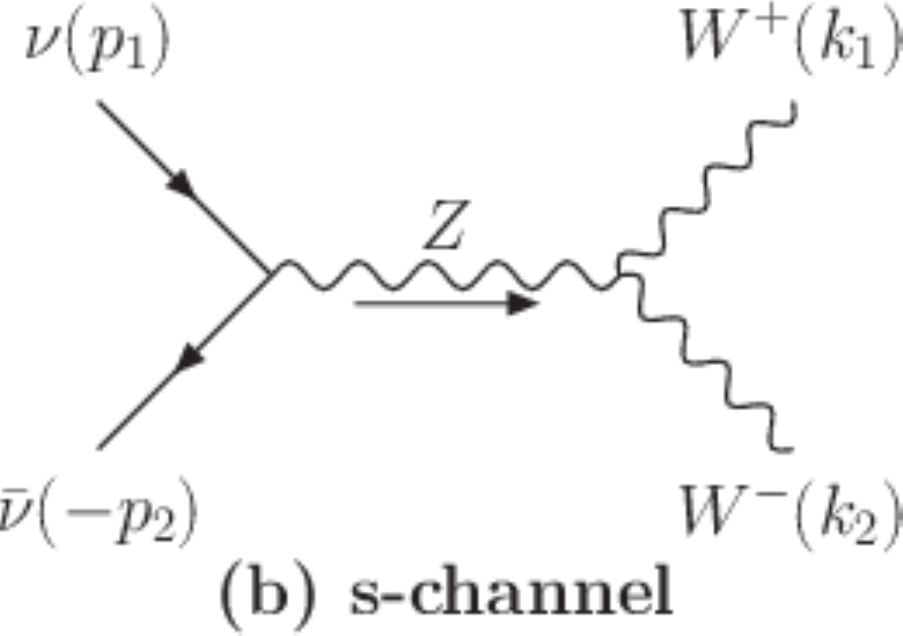}
\caption[Feynman diagrams for the process $\nu\bar{\nu} \to WW$]{\em Feynman diagrams for the process $\nu(p_1)+\bar{\nu}(p_2) \to W_L^+(k_1)+W_L^-(k_2)$.}
\label{fnunuWW:subfigures}
\end{figure}
There are two possible Feynman diagrams as shown in \fref{fnunuWW:subfigures}. The corresponding amplitudes are found to be
\begin{eqnarray}
{\cm}_a^e &=& -\frac{g^2}{4M_W^2}\underbrace{\bar{v}(p_2)\slashed{k_1}(1-\gamma_5)u(p_1)}_{\order(E^2)} + \order(1) \,, \\
{\cm}_b^Z &=& \frac{g_{\nu\nu Z} g_{WWZ}}{2M_W^2} \underbrace{\bar{v}(p_2)\slashed{k_1}(1-\gamma_5)u(p_1)}_{\order(E^2)} + \order(1) \,.
\end{eqnarray}
Note that the absence of $\order(E)$ terms in the above amplitudes is a manifest of the assumption that neutrinos are massless. As we will see later, this absence of linear growth will lead us to conclude that neutrinos do not {\emph need} to couple with the Higgs scalar. But one can see that there are quadratic growths present in the amplitude, cancellation of which would require 
\begin{eqnarray}
\label{SM3}
-\frac{g^2}{2}+g_{\nu\nu Z}g_{WWZ} = 0 \,.
\end{eqnarray}

The next process we consider is
$$e^-(p_1)+\bar{\nu}(p_2)\to Z_L(k_1)+W_L^-(k_2)$$
\begin{figure}
\centering
\includegraphics[scale=0.5]{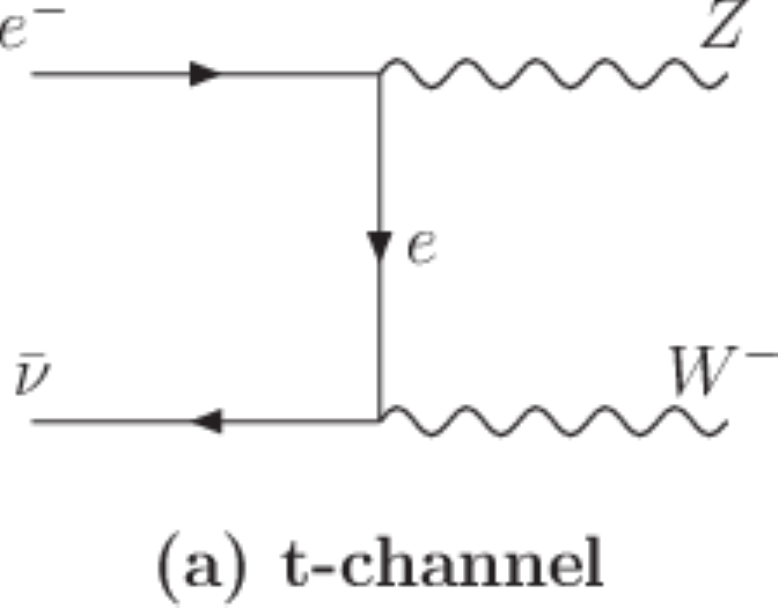}~~~
\includegraphics[scale=0.5]{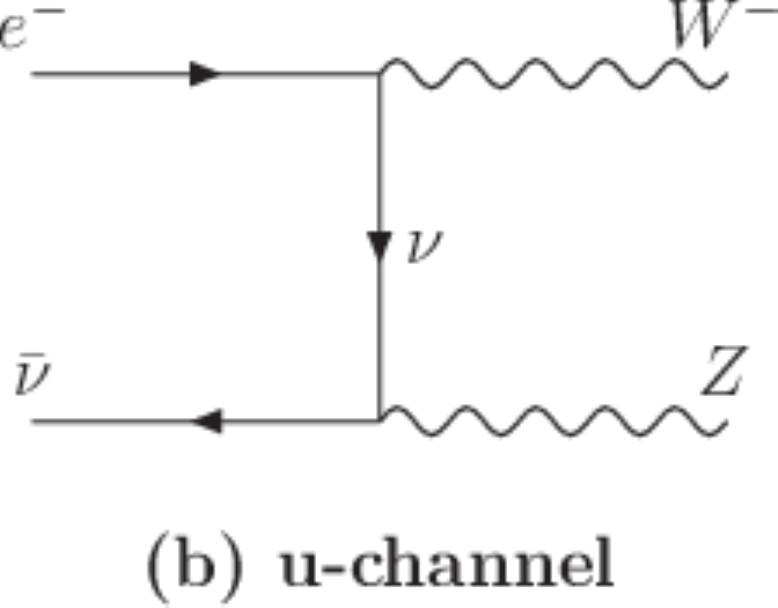}~~~
\includegraphics[scale=0.5]{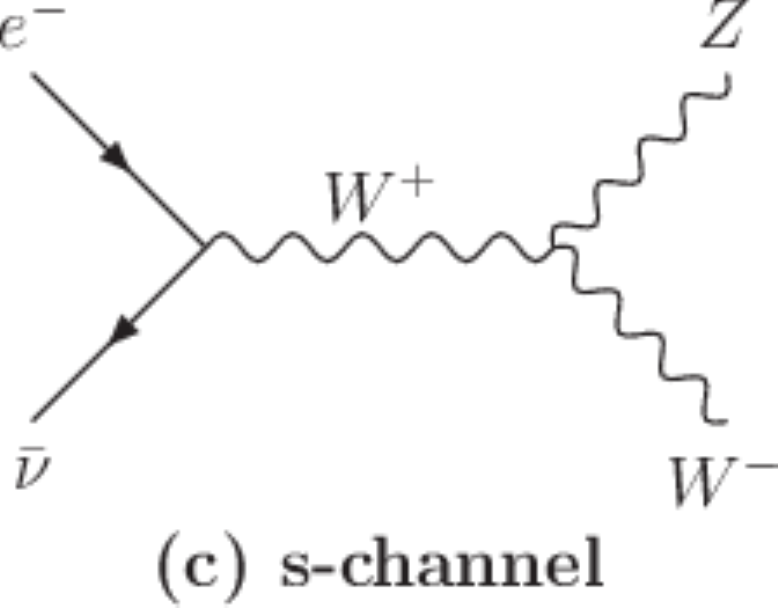}
\caption[Feynman diagrams for the process $e\bar{\nu}\to WZ$]{\em Feynman diagrams for $e^-(p_1)+\bar{\nu}(p_2)\to Z_L(k_1)+W_L^-(k_2)$ .}
\label{f:enuWZ}
\end{figure}
The amplitudes corresponding to the diagrams shown in \fref{f:enuWZ} are given below:
\begin{eqnarray}
\label{WZ mass ratio}
{\cm}_a &=& -\frac{gg_L}{2\sqrt{2}M_WM_Z}\underbrace{\bar{v}(p_2)\slashed{k_1}(1-\gamma_5)u(p_1)}_{\order(E^2)} +\order(E) \,, \\
{\cm}_b &=& -\frac{gg_{\nu\nu Z}}{2\sqrt{2}M_WM_Z}\underbrace{\bar{v}(p_2)\slashed{k_1}(1-\gamma_5)u(p_1)}_{\order(E^2)} +\order(E) \,, \\
{\cm}_c &=& -\frac{gg_{WWZ}}{2\sqrt{2}M_WM_Z}\underbrace{\bar{v}(p_2)\slashed{k_1}(1-\gamma_5)u(p_1)}_{\order(E^2)} +\order(E) \,. 
\end{eqnarray}
Clearly, the condition for cancellation of the quadratic growths reads:
\begin{eqnarray}
\label{SM4}
-g_L +g_{\nu\nu Z} -g_{WWZ} =0 \,.
\end{eqnarray}
Till now we have obtained four equations (Eqs.~(\ref{SM1}), (\ref{SM2}), (\ref{SM3}) and (\ref{SM4})) involving four unknowns $g_L$, $g_R$, $g_{\nu\nu Z}$, and $g_{WWZ}$. Let us rewrite them below:
\begin{subequations}
\begin{eqnarray}
-\frac{g^2}{2} +e^2 -g_Lg_{WWZ} &=& 0 \,, \\
e^2-g_R g_{WWZ} &=& 0  \,, \\
-\frac{g^2}{2}+g_{\nu\nu Z}g_{WWZ} &=& 0 \,, \\
-g_L +g_{\nu\nu Z} -g_{WWZ} &=& 0 \,.
\end{eqnarray}
\end{subequations}
The solution of the above set of equations is unique up to an overall sign. Choosing $g_{WWZ} = \sqrt{g^2-e^2}$ we have 
\begin{subequations}
\begin{eqnarray}
g_{WWZ} &=& \sqrt{g^2-e^2} \,, \\
g_{\nu \nu Z} &=& \frac{g^2}{2\sqrt{g^2-e^2}} \,, \\
g_L &=& \frac{-\frac{g^2}{2}+e^2}{\sqrt{g^2-e^2}} \,, \\
g_R &=& \frac{e^2}{\sqrt{g^2-e^2}} \,.
\end{eqnarray}
\end{subequations}
The above solutions clearly demands $e < g$. Therefore we can use the parametrization, 
\begin{eqnarray}
\sin\theta_w = \frac{e}{g}
\end{eqnarray}
to cast the solutions in their familiar form,
\begin{subequations}
\label{EW couplings}
\begin{eqnarray}
g_{WWZ} &=& g \cos\theta_w \,, \\
g_{\nu\nu Z} &=& \frac{g}{2\cos\theta_w} \,, \\
g_L &=& \frac{g}{2\cos\theta_w}(-1+2\sin^2\theta_w) \,, \\
g_R &=& \frac{g}{\cos\theta_w}\sin^2\theta_w \,.
\end{eqnarray}
\end{subequations}
There is more! One should remember that a residual linear growth (not shown in \Eqn{WZ mass ratio}) is still present in the amplitude  of the process $\bar{\nu}e^- \to W_L^- Z_L $. It can be easily shown that a cancellation of this growth can be arranged if we impose the following relation:
\begin{eqnarray}
g_R -g_{\nu\nu Z}+ g_{WWZ}\left( 1-\frac{M_Z^2}{2M_W^2}\right) = 0 \,.
\end{eqnarray}
Using the solutions of \Eqn{EW couplings} we can translate this into the following relation between the masses of the vector bosons:
\begin{eqnarray}
\label{rho parameter}
M_W = M_Z \cos\theta_w \,.
\end{eqnarray}
Using \Eqn{fermi constant} together with this, we obtain the standard formulae for $W$ and $Z$ masses:
\begin{subequations}
\label{SM mass prediction}
\begin{eqnarray}
M_W &=& \left( \frac{\pi\alpha}{G_F\sqrt{2}} \right)^{\frac{1}{2}} \frac{1}{\sin\theta_w} \,, \\
M_Z &=& \left( \frac{\pi\alpha}{G_F\sqrt{2}} \right)^{\frac{1}{2}} \frac{1}{\sin\theta_w\cos\theta_w} \,.
\end{eqnarray}
\end{subequations}
Now comes the experiment. If this theoretical model is true then the scattering $\nu_\mu e^- \to \nu_\mu e^-$ can only proceed via Z exchange. From these types of neutrino-fermion scatterings a preliminary value of $\sin^2\theta_w$ ($\approx 0.22$) was obtained \cite{LlewellynSmith:1983ie} and it was used in \Eqn{SM mass prediction} to predict $W$ and $Z$ boson masses as
\begin{eqnarray}
M_W \approx 79~{\rm GeV}\,, ~~~~ M_Z \approx 90~{\rm GeV} \,. 
\end{eqnarray}
Note that this is a robust prediction of the theory dictating not only what to look for but also where to look for! We now know that the experimental values are very close to these predictions vindicating the theory.
\section{Vector boson quartic self couplings}
Let us first investigate the process 
$$W_L^-(p_1) + W_L^-(p_2) \to W_L^-(k-1) + W_L^-(k_2) \,.$$
\begin{figure}
\centering
\includegraphics[scale=0.5]{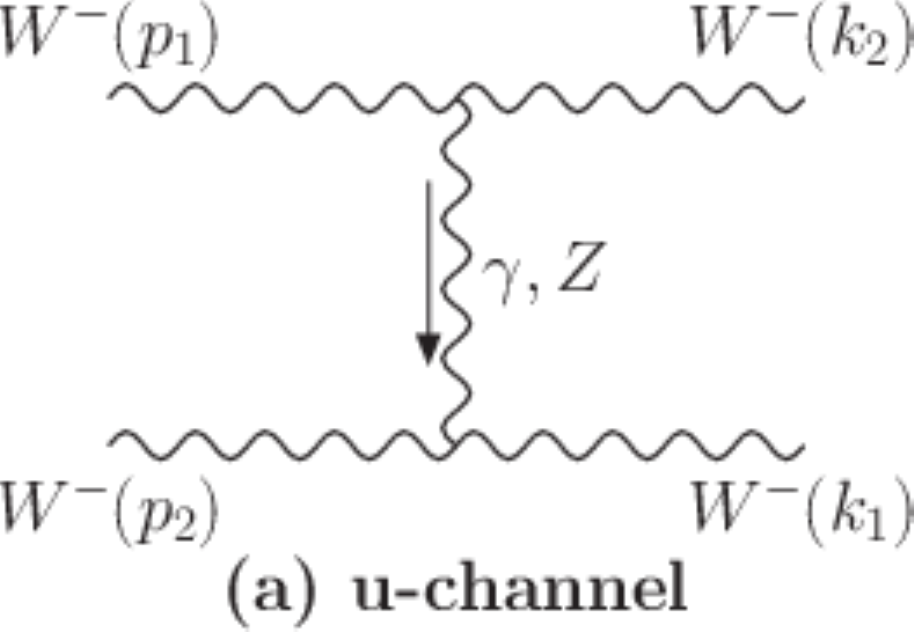}~~~
\includegraphics[scale=0.5]{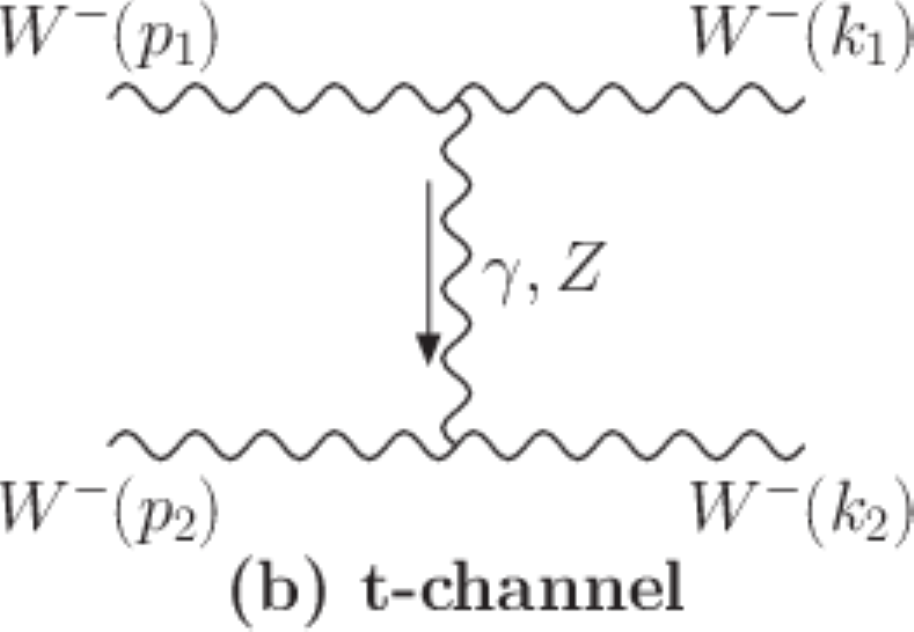}~~~
\includegraphics[scale=0.5]{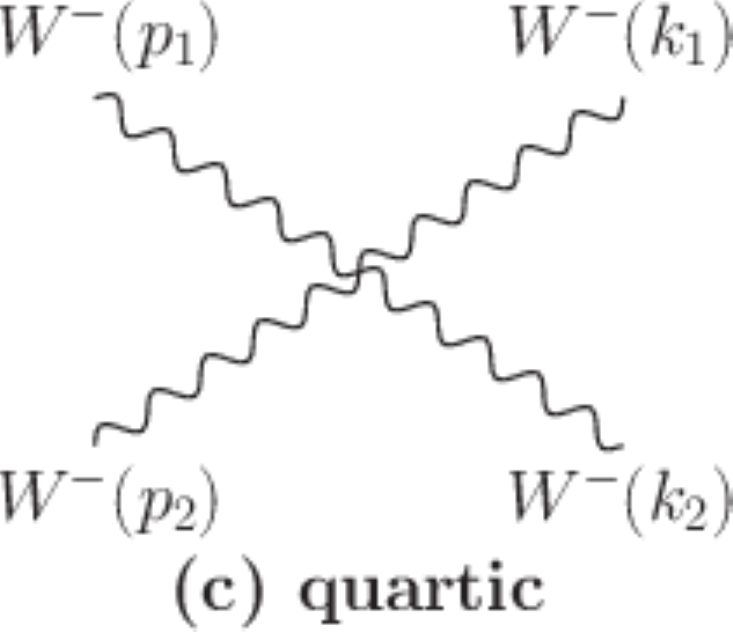}
\caption[$W_L^- W_L^-$ scattering]{\em Gauge diagrams for $W_L^- W_L^- \to W_L^- W_L^-$.}
%
\label{fWWWW-:subfigures}
\end{figure}
From what we have gathered till now, only (a) and (b) of \fref{fWWWW-:subfigures} exist. We can calculate their total amplitude as 
\begin{eqnarray}
\label{W-W-}
{\cm}_{a+b}^{\gamma +Z} = \frac{g^2}{4M_W^4}(t^2+u^2-2s^2) + \order(E^2) + \order(1)
\end{eqnarray}
Clearly, the quartic growth cannot be canceled by a scalar (spin 0) mediated diagram as it can give $\order(E^2)$ growth at best. The next possible choice would be to introduce another neutral vector boson ($Z^\prime$ say). But that will add same kind of quartic growth as already present in \Eqn{W-W-}. So these will not help.

It appears that the simplest possibility is to introduce a direct self interaction between the $W$ bosons. Since $[{\ml}_{\rm int}] \le 4$, it is clear that terms involving derivative of vector fields are not allowed. The most general interaction of this type can be written as 
\begin{eqnarray}
\label{WWWW quartic}
{\ml}_{WWWW} = a (W^-\cdot W^+)(W^-\cdot W^+) + b (W^-\cdot W^-)(W^+\cdot W^+) \,.
\end{eqnarray}
This new interaction will add one more diagram shown in \fref{fWWWW-:subfigures}c. The amplitude for this diagram can be found to be
\begin{eqnarray}
\label{W-W-2}
{\cm}_c = a \frac{1}{2M_W^4}(t^2+u^2) + b \frac{1}{M_W^4} s^2 + \order(E^2)  \,.
\end{eqnarray}
It is obvious that the leading growths of Eqs.~(\ref{W-W-}) and (\ref{W-W-2}) will cancel each other if
\begin{eqnarray}
&&  a = -b = -\frac{g^2}{2}\,, \\
&\Rightarrow& {\ml}_{WWWW} = \frac{g^2}{2} \left[(W^-\cdot W^-)(W^+\cdot W^+)-(W^-\cdot W^+)(W^-\cdot W^+)\right] \,.
\label{qua2}
\end{eqnarray}
The quadratic residual growth for $WW$ scattering is given by:
\begin{eqnarray}
\label{WWWW residual}
{\cm}^{\rm gauge} = -\frac{g^2}{4M_W^2}s + \order(1) \,.
\end{eqnarray}
Similarly, for the process 
$$W_L^-(p_1) + W_L^+(p_2) \to Z_L(k_1)+Z_L(k_2)\,,$$
\begin{figure}
\centering
\includegraphics[scale=0.5]{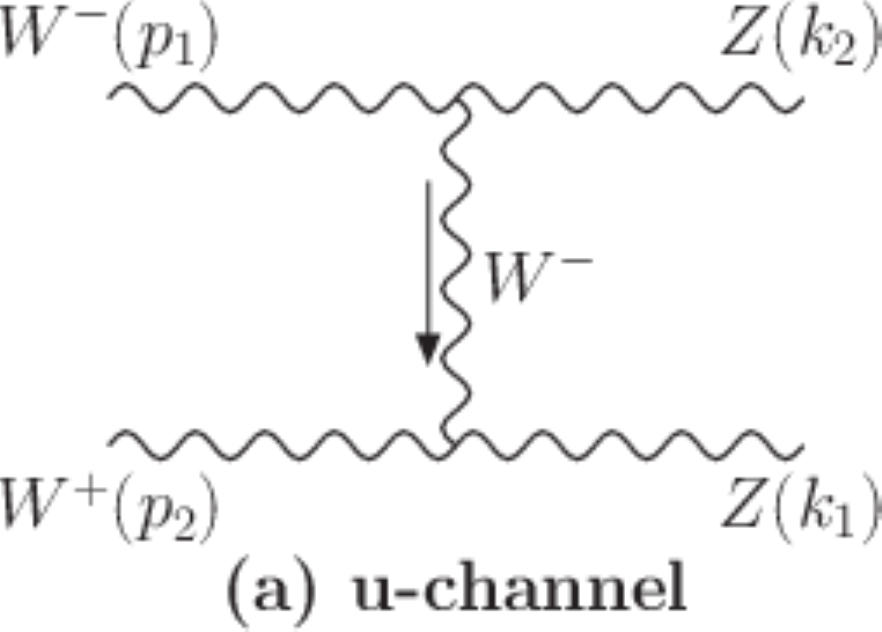}~~~
\includegraphics[scale=0.5]{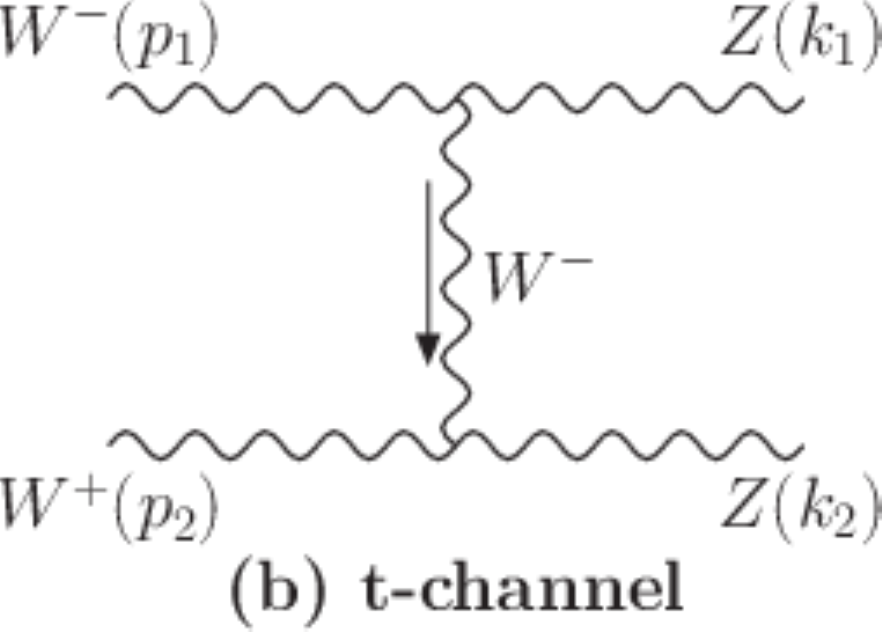}~~~
\includegraphics[scale=0.5]{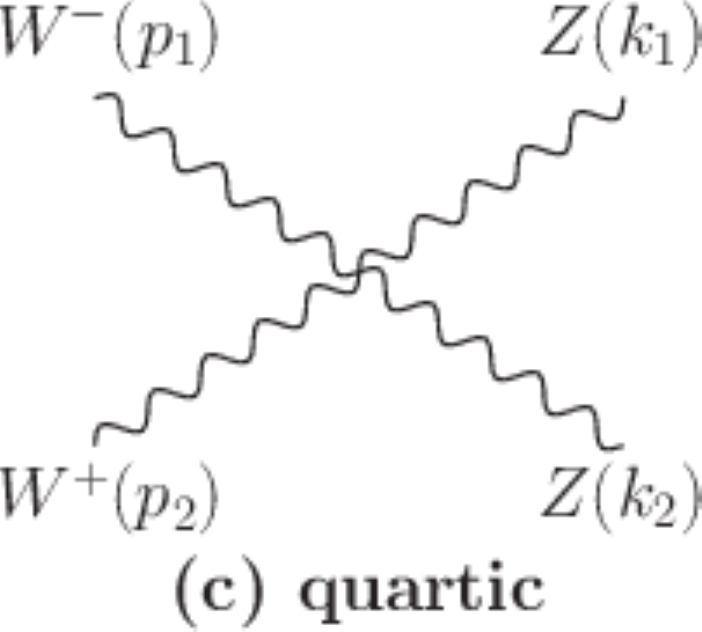}
\caption[$W_L^- W_L^+ \to Z_L Z_L$ scattering]{\em Gauge diagrams for $W_L^- W_L^+ \to Z_LZ_L$.}
%
\label{fWWZZ:subfigures}
\end{figure}
we have (see \fref{fWWZZ:subfigures}),
\begin{eqnarray}
{\cm}_{a+b} = -\frac{g_{WWZ}^2}{4M_W^2M_Z^2}(t^2+u^2-2s^2)+\order(E^2) + \order(1) \,,
\end{eqnarray}
which will invite us to introduce the following quartic interaction:
\begin{eqnarray}
{\ml}_{WWZZ} = c (W^-\cdot Z)(W^+\cdot Z) + d (W^-\cdot W^+)(Z\cdot Z) \,.
\end{eqnarray}
The amplitude of the corresponding diagram is found to be 
\begin{eqnarray}
{\cm}_c = \frac{1}{4M_W^2M_Z^2}\left[c(t^2+u^2)+2ds^2\right]+\order(E^2) + \order(1) \,.
\end{eqnarray}
Cancellation of the $\order(E^4)$ growth would require
\begin{eqnarray}
c = g_{WWZ}^2 \,; ~~~~~ d = -g_{WWZ}^2 \,,
\end{eqnarray}
which means
\begin{eqnarray}
\label{qua3}
{\ml}_{WWZZ} = g_{WWZ}^2\left[ (W^-\cdot Z)(W^+\cdot Z) -  (W^-\cdot W^+)(Z\cdot Z)\right] \,.
\end{eqnarray}

Similar considerations for $W_L^-W_L^+ \to Z_L\gamma$ leads to
\begin{eqnarray}
{\ml}_{WWZ\gamma} &=& g_{WWZ}g_{WW\gamma}\Big[ (W^-\cdot Z)(W^+\cdot A) +  (W^-\cdot A)(W^+\cdot Z) 
\nonumber \\
&& -  2(W^-\cdot W^+)(Z\cdot A)\Big] \,.
\label{qua4}
\end{eqnarray}
Thus collecting Eqs.~(\ref{WWgg}), (\ref{qua2}), (\ref{qua3}) and (\ref{qua4}) and defining 
\begin{eqnarray}
W^3_\mu = \cos\theta_w Z_\mu + \sin\theta_w A_\mu \,,
\end{eqnarray}
we can write the quartic vector boson self couplings in the following compact form:
\begin{eqnarray}
{\ml}_{VVVV} &=& {\ml}_{WWWW} + {\ml}_{WW\gamma\gamma} + {\ml}_{WWZZ} + {\ml}_{WWZ\gamma} \nonumber \\
&=& -g^2\Big\{\frac{1}{2}(W^-\cdot W^+)^2 -\frac{1}{2}(W^-)^2(W^+)^2 \nonumber \\
&& ~~~~~~~~ +(W^3)^2(W^-\cdot W^+) -(W^-\cdot W^3)(W^+\cdot W^3)\Big\} \,,
\label{SM quartic vector}
\end{eqnarray}
where, we have used the solution of \Eqn{EW couplings}. The same thing can be done with the cubic self couplings    
of Eqs.~(\ref{YM WWg}) and (\ref{YM WWZ})
\begin{eqnarray}
{\ml}_{VVV} &=& {\ml}_{WW\gamma} + {\ml}_{WWZ} \nonumber \\
&=& -ig\Big[W^{3\mu} (W^{\nu-}\overleftrightarrow{\partial}_\mu W_\nu^+) +W^{\mu-}(W^{\nu+}\overleftrightarrow{\partial}_\mu W^3_\nu) \nonumber \\
 && ~~~~~~~~~~~~~+ W^{\mu+}(W^{3\nu} \overleftrightarrow{\partial}_\mu W_\nu^-)\Big] \,.
\label{SM cubic vector}
 \end{eqnarray}
%
\section{Need for a neutral scalar}
As already mentioned in the preceding sections, some processes do contain some remnant energy growths even after the introduction of the quartic gauge self couplings. With no other couplings to tune freely, we must now extend the particle content of the theory to eliminate those residual energy growths. In doing this, we shall stick to the minimal choice, {\it i.e.}, we shall start from spin 0 particles and we shall not include any interaction with $[{\ml}]_{\rm int}>4$ for they will come with coupling constants having negative mass dimensions which is bad for the high energy behavior of the theory.

\subsection{Trilinear couplings with vector bosons and fermions} 
\begin{itemize}
\item
Let us now go back to $W_L^-W_L^- \to W_L^-W_L^-$ scattering again. For this process, there is a remaining quadratic growth even after the introduction of the quartic self couplings (see \Eqn{WWWW residual}).  We shall now try to eliminate this growth by introducing a new interaction of the $W$'s with a neutral scalar field which will be denoted by $h$. It is not difficult to realize that the only possible choice satisfying $[\ml_{\rm int}] \leq 4 $ is represented by the interaction Lagrangian 
\begin{eqnarray}
\ml_{WWh} = g_{WWh}W_\mu^-W^{\mu +}h
\label{WWh int}
\end{eqnarray}
\begin{figure}
\centering
\includegraphics[scale=0.6]{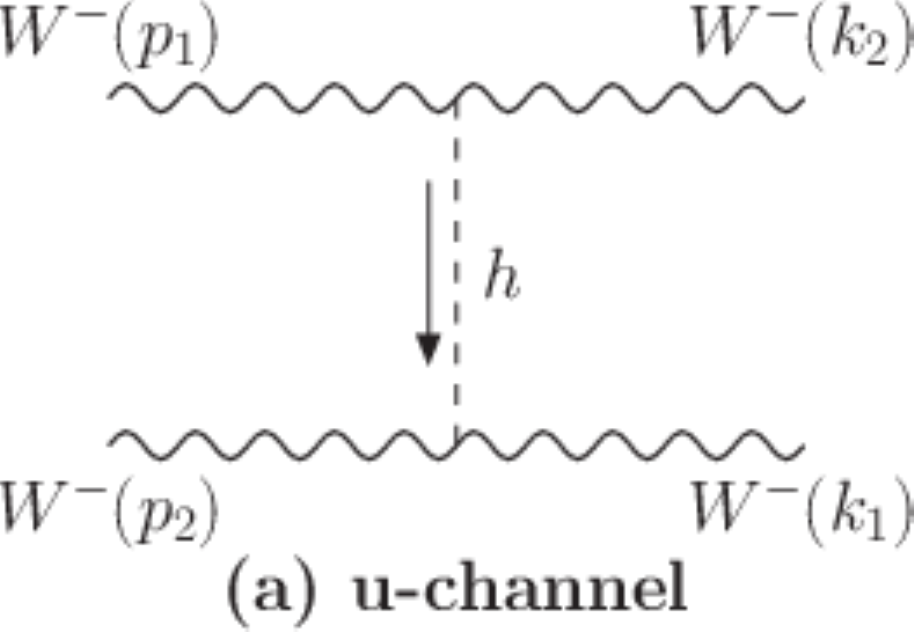}~~~~~~~~~
\includegraphics[scale=0.6]{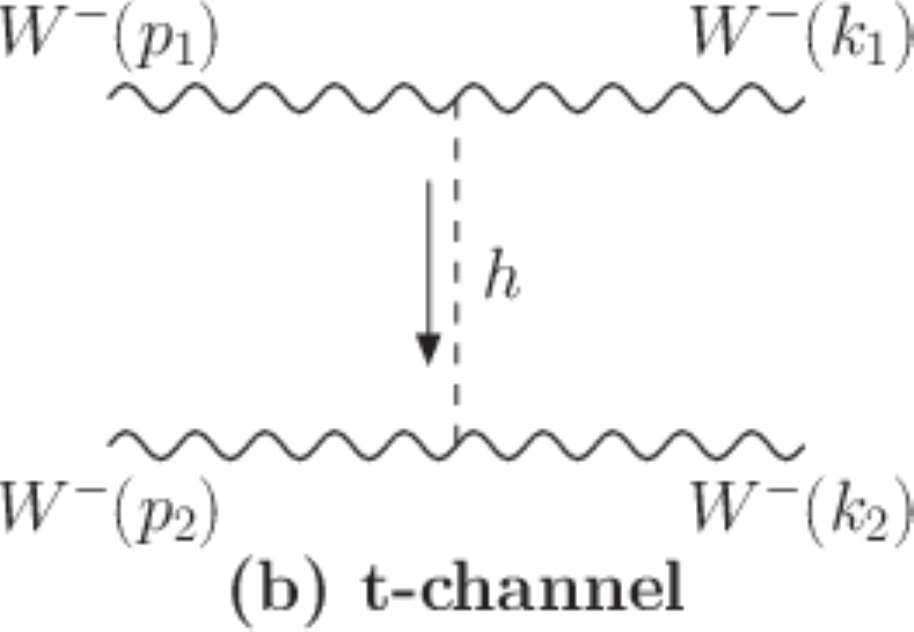}
\caption[$W_L^- W_L^-$ scattering (Higgs diagrams)]{\em Higgs diagrams for $W_L^- W_L^- \to W_L^- W_L^-$.}
\label{fWWWWh-:subfigures}
\end{figure}
Tree diagrams  for the process $W^-W^- \to W^- W^-$ corresponding to the interaction of \Eqn{WWh int}, have been shown in \fref{fWWWWh-:subfigures}. One can calculate:
\begin{eqnarray}
{\cm}^{(h)}={\cm}_a^{(h)}+{\cm}_a^{(h) }= g_{WWh}^2 \frac{s}{M_W^4}+\order(1) \,.
\end{eqnarray}
So, the desired cancellation would require:
\begin{eqnarray}
g_{WWh} = g M_W \,.
\label{gWWh}
\end{eqnarray}
\item
To have an idea of the coupling of this new scalar with massive fermions, let us consider the process 
$$e^-(p_1)+e^+(p_2)\to W_L^-(k_1)+ W_L^+(k_2)\,.$$
Without this new scalar, there will be three Feynman diagrams as shown in \fref{feeWW:subfigures} (don't forget to include the $Z$ mediated s-channel diagram similar to \fref{feeWW:subfigures}b). The total amplitude can be calculated as:
\begin{eqnarray}
{\cm}_{\rm without~ scalar}={\cm}_a^{\nu}+{\cm}_b^{\gamma,Z} = -\frac{g^2}{4M_W^2}m_e \underbrace{\bar{v}(p_2)u(p_1)}_{\order(E)}+\order(1) \,.
\label{residual eeWW}
\end{eqnarray}
To eliminate this linear energy growth, we introduce the following interaction:
\begin{eqnarray}
{\ml}_{eeh} = g_{eeh} \bar{e}eh \,.
\end{eqnarray}
Note that as there is no $\gamma_5$ in the residual growth of \Eqn{residual eeWW}, we do not introduce any pseudoscalar interaction with $h$. This new interaction will lead to a $h$ mediated s-channel diagram whose amplitude is found to be:
\begin{eqnarray}
{\cm}^h = -\frac{g_{eeh}g_{WWh}}{2M_W^2}\bar{v}(p_2)u(p_1)+ \order(1) \,,
\end{eqnarray}
where, $g_{WWh}$ is given by \Eqn{gWWh}. So the linear growth can be canceled by tuning $g_{eeh}$ as:
\begin{eqnarray}
g_{eeh}=-\frac{gm_e}{2M_W} \,.
\label{geeh}
\end{eqnarray}
\item
To get an idea of $ZZh$ coupling, we consider the process
$$e^-(p_1)+e^+(p_2)\to Z_L(k_1)+Z_L(k_2) \,. $$
\begin{figure}
\centering
\includegraphics[scale=0.5]{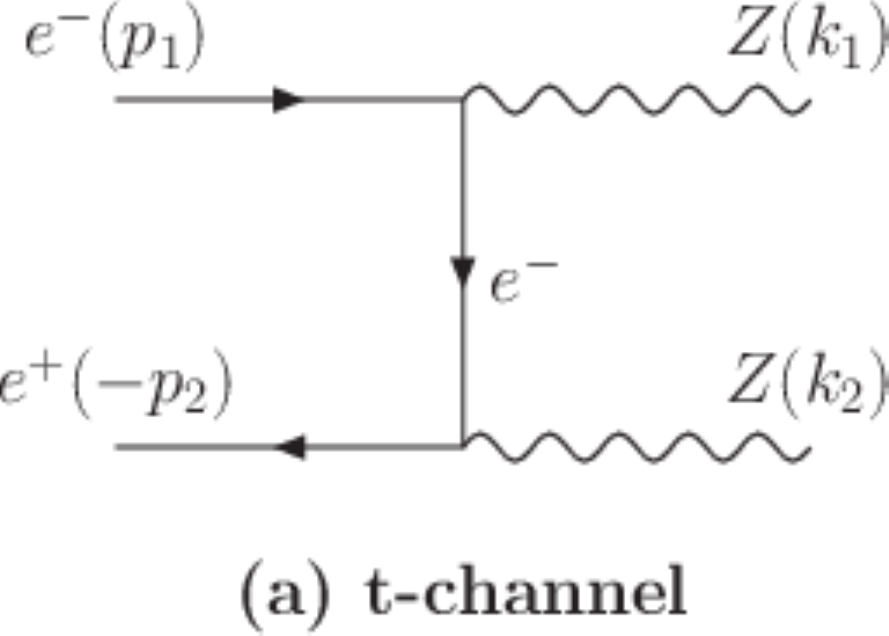}~~~
\includegraphics[scale=0.5]{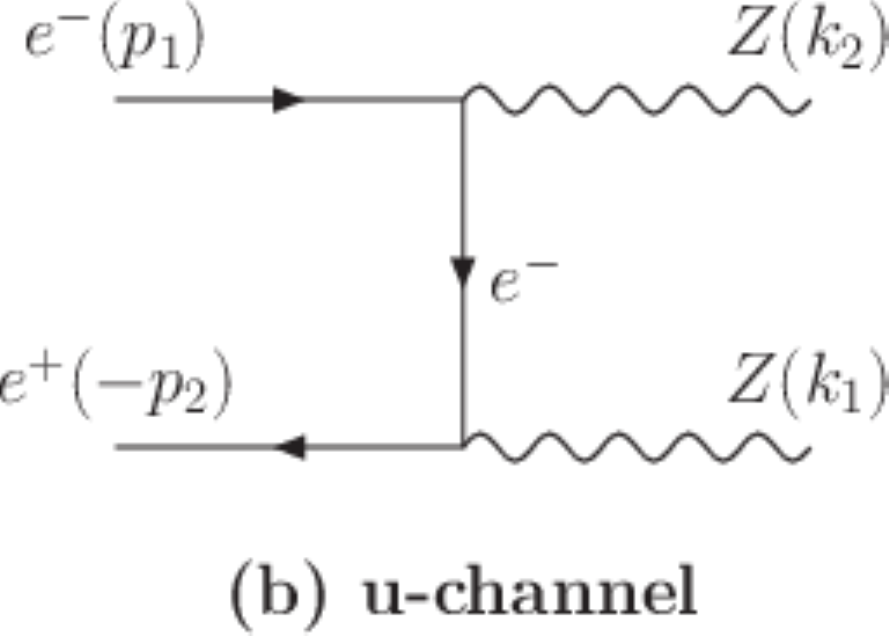}~~~
\includegraphics[scale=0.5]{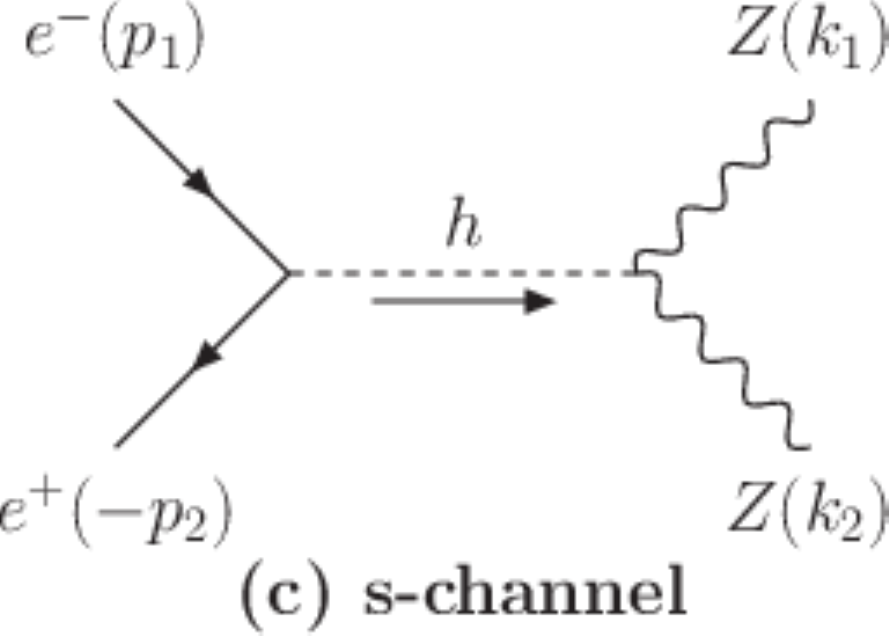}
\caption[$e^-e^+ \to Z_L Z_L$ scattering]{\em Feynman diagrams for $e^- e^+ \to Z_LZ_L$.}
%
\label{f:eeZZ}
\end{figure}
The Feynman diagrams for this process have been displayed in \fref{f:eeZZ}. Without the scalar mediated s-channel diagram, the total amplitude linearly grows with energy as follows:
\begin{eqnarray}
{\cm}_{\rm without~ scalar}={\cm}_a+{\cm}_b = -\frac{g^2m_e}{4M_Z^2\cos^2\theta_w} \underbrace{\bar{v}(p_2)u(p_1)}_{\order(E)}+\order(1) \,.
\label{residual eeZZ}
\end{eqnarray}
To eliminate this growth we introduce the $ZZh$ interaction
\begin{eqnarray}
{\ml}_{ZZh}=g_{ZZh} Z_\mu Z^\mu h
\end{eqnarray}
which brings in \dref{f:eeZZ}c with the following amplitude:
\begin{eqnarray}
{\cm}_c = -\frac{g_{eeh}g{ZZh}}{M_Z^2}\bar{v}(p_2)u(p_1) \,.
\end{eqnarray}
Using the value of $g_{eeh}$ from \Eqn{geeh}, one can find the condition for the cancellation of the $\order(E)$ growth in \Eqn{residual eeZZ} to be
\begin{eqnarray}
g_{ZZh}=\frac{gM_Z}{2\cos\theta_w}\,.
\label{gZZh}
\end{eqnarray}
\end{itemize}

From Eqs.~(\ref{residual eeWW}) and (\ref{residual eeZZ}) it is interesting to note that the amplitudes for the processes $\ell^+\ell^-\to V_LV_L$ ($\ell$ stands for leptons and $V=W,Z$) without the scalar contain energy growths which are proportional to the mass ($m_\ell$) of the lepton involved. Clearly, there will be no such energy growths even without this new scalar for $\nu_\ell \bar{\nu}_\ell \to VV$ as long as neutrinos are considered to be massless. Therefore, this new scalar need not couple to a pair of neutrinos. Thus, we have learned one remarkable feature of the trilinear interactions of this new scalar field, $h$, that a corresponding coupling constant is always proportional to the mass of the particle interacting with $h$.

\subsection{Quartic couplings with the vector bosons}
Since we already have concluded the existence of $WWh$ interaction the process $W_L^-W_L^+\to hh$ is possible.
\begin{figure}
\centering
\includegraphics[scale=0.5]{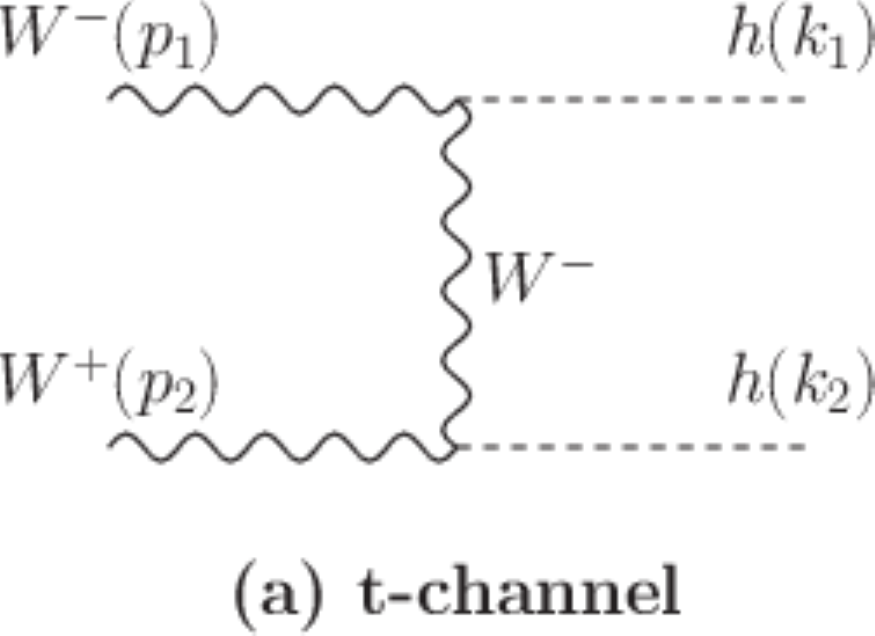}~~~
\includegraphics[scale=0.5]{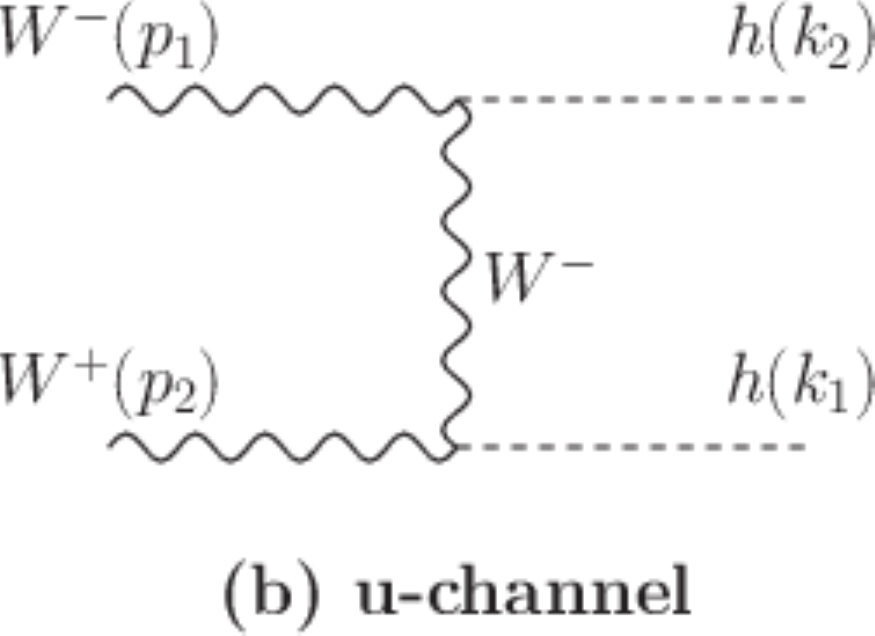}~~~
\includegraphics[scale=0.5]{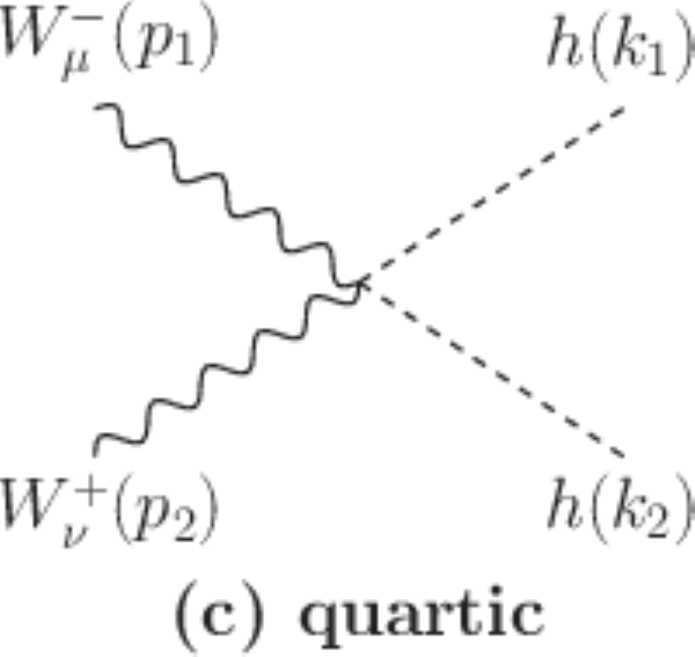}
\caption[$W^-W^+ \to hh$ scattering]{\em Feynman diagrams for $W_L^- W_L^+ \to hh$.}
%
\label{f:WWhh}
\end{figure}
From what we have gathered till now, only diagrams~\ref{f:WWhh}a and \ref{f:WWhh}b exist. Quadratic energy growths will appear from the $k^\mu k^\nu$ term in the intermediate $W$ propagator as follows:
\begin{eqnarray}
{\cm}_a+{\cm}_b = -\frac{g^2s}{4M_W^2} +\order(1) \,.
\label{residual WWhh}
\end{eqnarray}
To remove this growth we propose the existence of a quartic interaction of the form
\begin{eqnarray}
{\ml}_{WWhh} = g_{WWhh} W_\mu^-W^{+\mu} hh
\label{gWWhh}
\end{eqnarray}
which will introduce diagram~\ref{f:WWhh}c with the following amplitude
\begin{eqnarray}
{\cm}_c = g_{WWhh} \frac{s}{M_W^2} +\order(1) \,.
\end{eqnarray}
Clearly, the cancellation of the quadratic growth requires
\begin{eqnarray}
g_{WWhh}=\frac{g^2}{4} \,.
\label{gWWhh1}
\end{eqnarray}

Similar consideration of $Z_LZ_L\to hh$ scattering will lead to the existence of a $ZZhh$ quartic interaction of the form
\begin{eqnarray}
{\ml}_{ZZhh} = \frac{g^2}{8\cos^2\theta_w} Z_\mu Z^\mu hh \,.
\label{gZZhh}
\end{eqnarray}

\subsection{Scalar self couplings}
Scalar self couplings do not follow from the requirement of tree unitarity alone. For this, we need a stronger condition of renormalizability. For a renormalizable theory, it has been shown\cite{Cornwall:1974km} that the $k$-th loop amplitude of a scattering involving $n$ particles ($1+2 \to 3+4+\dots+n$) behaves at very high energies as follows:
\begin{eqnarray}
{\cm}^{(n)}|_{E\to \infty} =\order(E^{4-n}\ln^k E) \,.
\end{eqnarray}
Clearly, the tree level amplitude ($k=0$) of a process $1+2\to 3+4+5$ should go as $\order(1/E)$.
\begin{figure}
\centering
\includegraphics[scale=0.5]{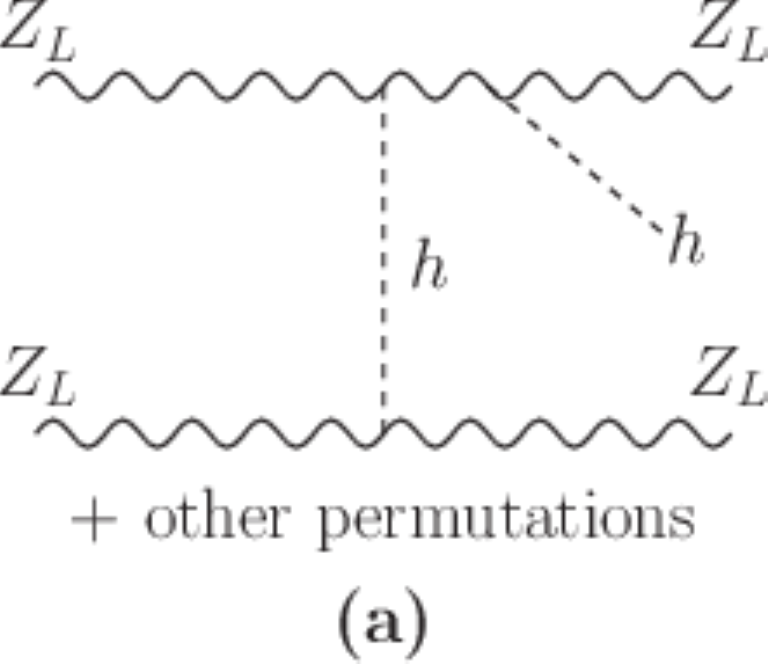}~~~
\includegraphics[scale=0.5]{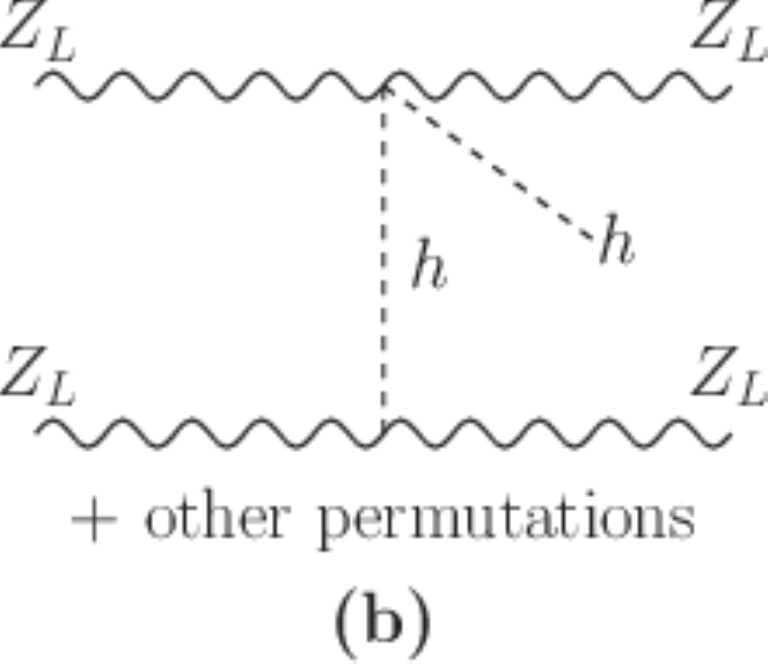}~~~
\includegraphics[scale=0.5]{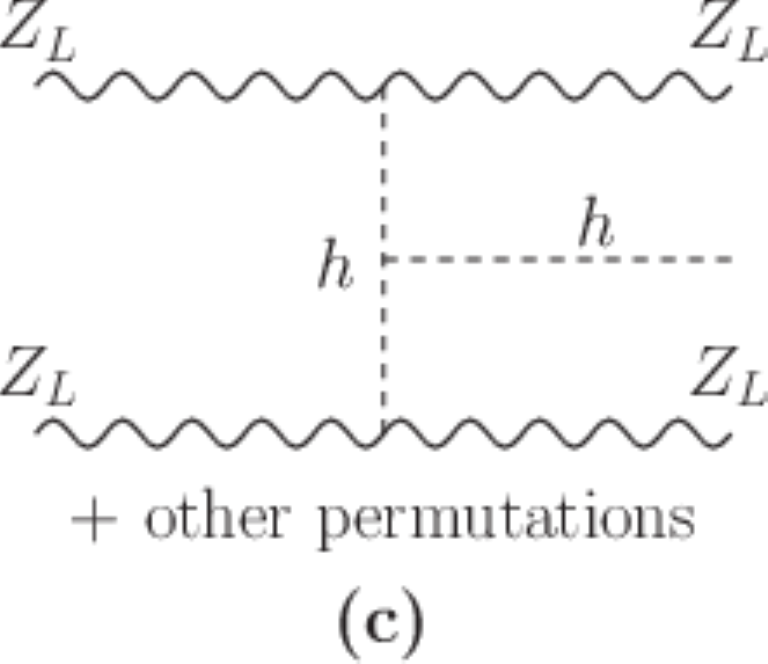}
\caption[$ZZ \to ZZh$ scattering]{\em Feynman diagrams for $Z_L Z_L \to Z_LZ_L h$.}
%
\label{f:ZZZZh}
\end{figure}
As an example, consider the process $Z_LZ_L\to Z_LZ_Lh$ which can proceed at the tree level due to the existence of interactions in the form of Eqs.~(\ref{gZZh}) and (\ref{gZZhh}). The Feynman diagrams stemming from these interactions have been displayed in Figures~\ref{f:ZZZZh}a and \ref{f:ZZZZh}b. But the total amplitude from these two types of diagrams behaves at $E\to \infty$ as a constant independent of the CM energy. An explicit calculation leads to the conclusion that the desired cancellation of the unwanted constant term occurs if we put in a self coupling of the form
\begin{eqnarray}
{\ml}_{hhh} =-\frac{gm_h^2}{4M_W}h^3
\label{ghhh}
\end{eqnarray}
leading to the diagram~\ref{f:ZZZZh}c. In this connection it is interesting to note that it is the first time when a non-zero mass ($m_h$) of the scalar $h$ became explicitly necessary.

\begin{figure}
\centering
\includegraphics[scale=0.5]{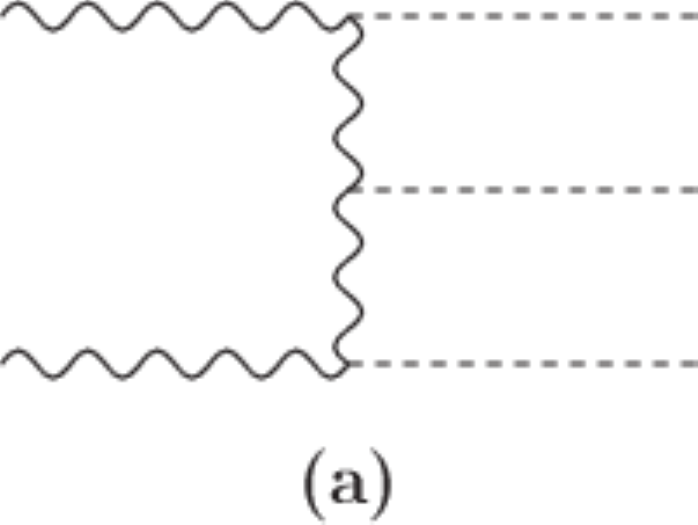}~~~
\includegraphics[scale=0.5]{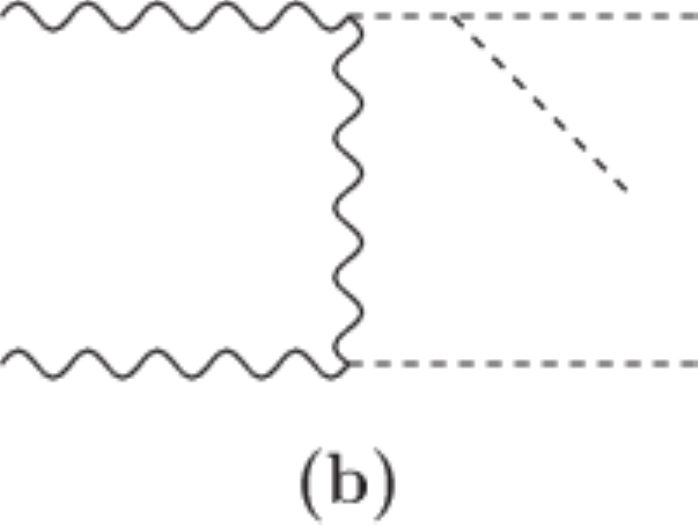}~~~
\includegraphics[scale=0.5]{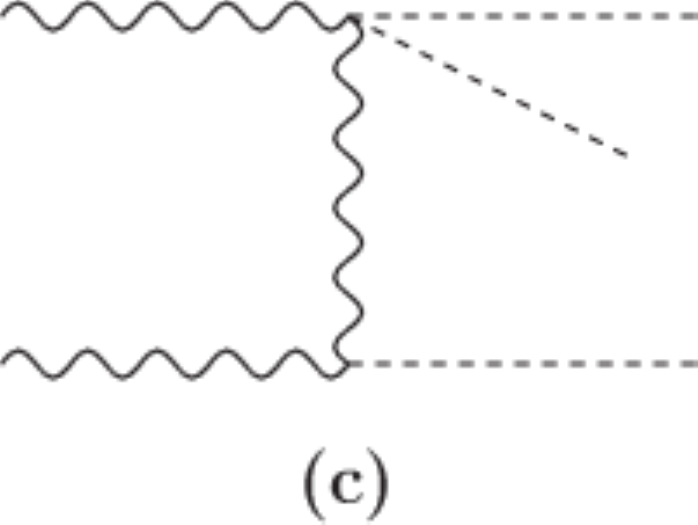} \\
\includegraphics[scale=0.5]{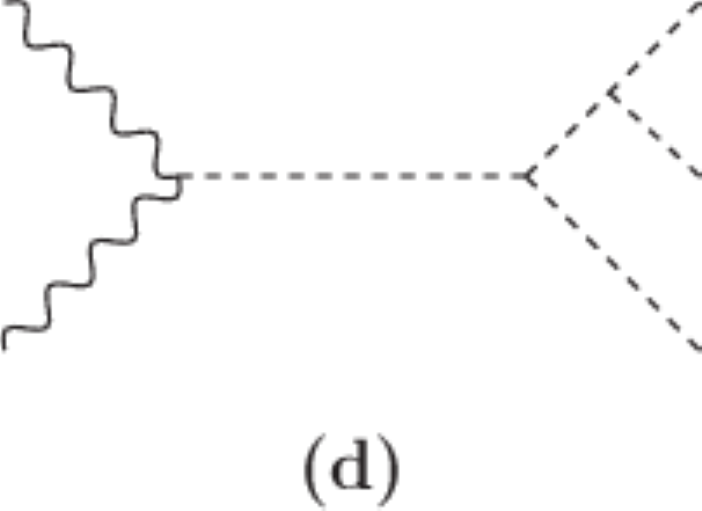}~~~
\includegraphics[scale=0.5]{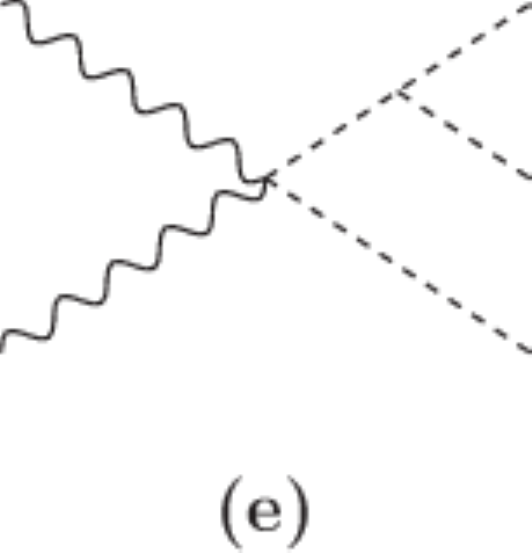}~~~
\includegraphics[scale=0.5]{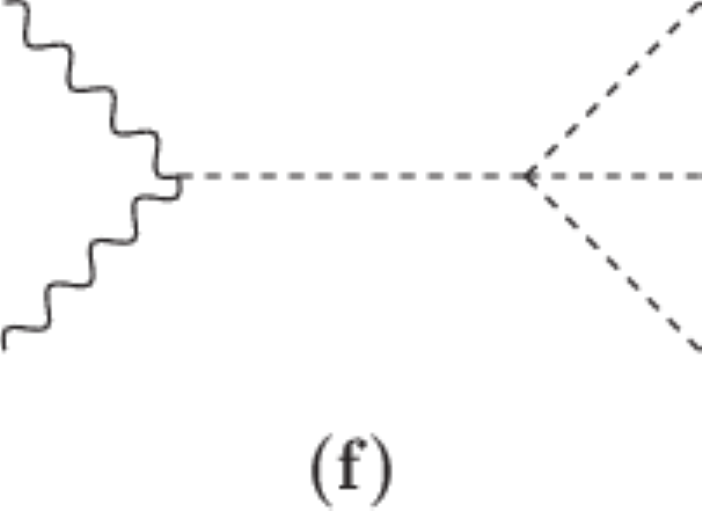}
\caption[$ZZ \to hhh$ scattering]{\em Feynman diagrams for $Z_L Z_L \to hhh$.}
%
\label{f:ZZhhh}
\end{figure}
Now let us turn our attention to another five particle scattering, $Z_LZ_L\to hhh$. Tree level Feynman graphs have been shown in \fref{f:ZZhhh}. Note that, without diagram \ref{f:ZZhhh}f the total amplitude would go as $\order(1)$ instead of $\order(1/E)$ at high energies. The requirement of cancellation of this constant term fixes the quartic coupling involved in diagram \ref{f:ZZhhh}f. The resulting interaction Lagrangian is 
\begin{eqnarray}
{\ml}_{hhhh} = -\frac{g^2m_h^2}{32M_W^2} h^4 \,.
\end{eqnarray}

This was the final piece. We now have the complete but minimal theory that is needed to describe all the electroweak phenomena involving electrons. We have obtained this by a systematic cancellation of energy growths appearing in different scattering amplitude and therefore this theory is ``ultraviolet safe'', {\it i.e.}, valid upto arbitrarily high energies. The different pieces of interactions now can be collected into a master Lagrangian,
\begin{eqnarray}
{\ml}_{\rm int} &=& {\ml}_{ee\gamma}+{\ml}_{\rm int}^W +\ml_{eeZ}+\ml_{\nu\nu Z} +\ml_{VVV}+\ml_{VVVV}+\ml_{WWh} \nonumber \\
&& +\ml_{ZZh}+\ml_{eeh}+\ml_{WWhh}+\ml_{ZZhh}+\ml_{hhh}+\ml_{hhhh} \,.
\label{master L}
\end{eqnarray}
Amazingly, \Eqn{master L} is what we get from the well known $SU(2)\times U(1)$ gauge theoretic construction of the Standard Model (SM) which has been tested experimentally with fantastic accuracy. Thus, the principle of tree unitarity which started off as an `educated guess' is now qualified enough to be declared as a rule which must be obeyed by all the future theories that promise to address the questions left unanswered by the SM. In the upcoming chapters, we will show how several new physics models can be constrained theoretically by employing the prescription of tree unitarity. In passing, we should keep in mind that as in the case of the SM, consideration of tree unitarity does not predict the number of lepton generations. But, given the number of lepton generations, one can show that the number of quark generations cannot be arbitrary. Historically it happened in the reverse order: from the observation of CP violation in kaon decay it was inferred that there must be at least three generations of quarks and three generations of quarks will require three generation of leptons. How this can be done using unitarity is the subject matter of the next section.

\section{ABJ anomaly and quarks}
\begin{figure}
\includegraphics[scale=0.6]{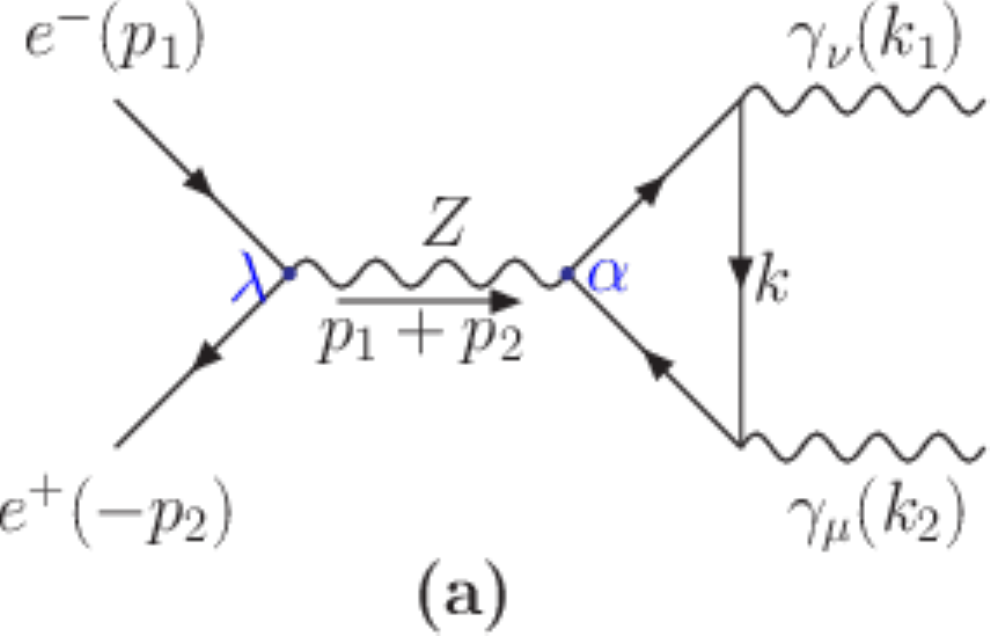}~~~~~~~~~
\includegraphics[scale=0.6]{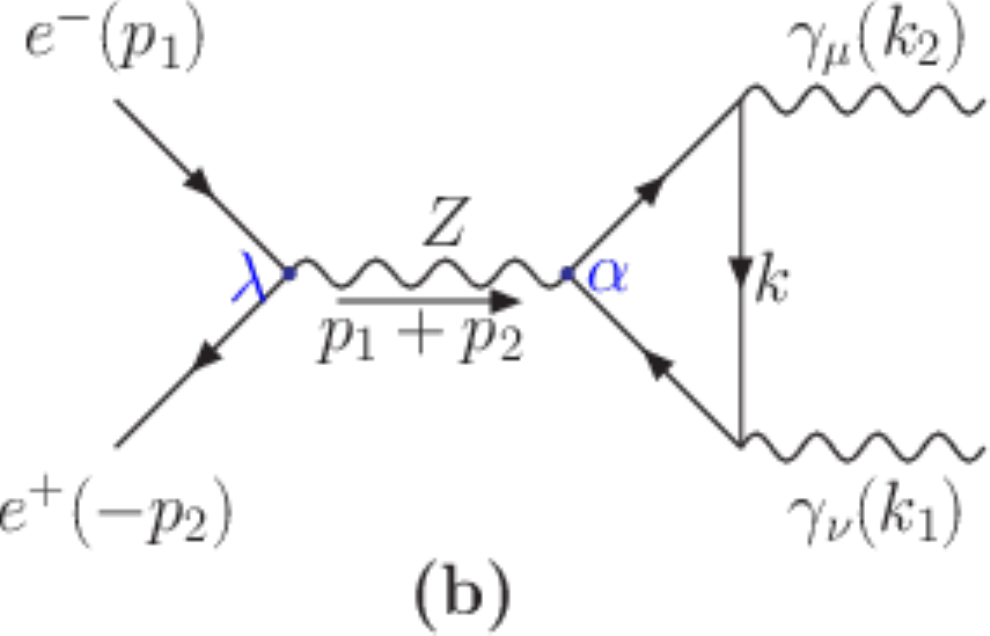}
\caption[One-loop diagrams for the process $e^-e^+ \to \gamma\gamma$]{\em One-loop diagrams for the process $e^-(p_1)+e^+(p_2) \to \gamma(k_1)+\gamma(k_2)$.}
\label{f:ABJ}
\end{figure}
Some processes are completely innocuous at the tree level but contain bad energy growths at the loop level. For example, consider the process
$$e^-(p_1)+e^+(p_2)\to \gamma(k_1)+\gamma(k_2) \,.$$
\fref{f:ABJ} shows the one-loop diagrams in which an effect of the Adler-Bell-Jackiw (ABJ) axial anomaly is manifested. Denoting $p_1+p_2$ by $p$, the total amplitude of these triangle graphs can be written as
\begin{eqnarray}
{\cm}_\Delta = {\cm}_a+{\cm}_b 
&=& i\frac{g^2 a_e Q_e^2e^2}{\cos^2\theta_w}\bar{v}(p_2)\gamma_\lambda (v_e-a_e\gamma_5)u(p_1) \nonumber \\
&& \times~\frac{-g^{\lambda\alpha}+\frac{p^\lambda p^\alpha}{M_Z^{2}}}{p^2-M_Z^2} T_{\alpha\mu\nu}(k_1,k_2) \epsilon^\nu(k_1)\epsilon^\mu(k_2) \,,
\label{ABJ1}
\end{eqnarray}
where,
\begin{eqnarray}
v_e &=& -\frac{1}{4}+\sin^2\theta_w \,, \\
a_e &=& -\frac{1}{4} \,, \\
T_{\alpha\mu\nu}(k_1,k_2) &=& \int \frac{d^4k}{(2\pi)^4} {\rm Tr}\left(\frac{1}{\Slash{k}-\Slash{k_2}-m_e} \gamma_\mu\frac{1}{\Slash{k}-m_e}\gamma_\nu\frac{1}{\Slash{k}+\Slash{k_1}-m_e }\gamma_\alpha\gamma_5  \right) \nonumber \\
&& + \left(k_1,\nu\right) \leftrightarrow \left(k_2,\mu\right) \,.
\label{TT}
\end{eqnarray}
A simple power counting reveals that $T_{\alpha\mu\nu}(k_1,k_2)\sim \order(E)$ as $E\to \infty$. Consequently ${\cm}_\Delta$ should go as $\order(E^2)$ due to the presence of the $p^\lambda p^\alpha$ term in the $Z$ propagator. But multiplying $p^\lambda$ with $\gamma_\lambda$ in the first neutral current vertex and using Dirac equation the electron mass ($m_e$) can be factored out. This compensates one factor of energy growth {\it i.e.}, ${\cm}_\Delta$ now grows linearly with energy.

Let us now investigate whether another factor of electron mass can be factorized or not. Rigorous calculation shows 
\begin{eqnarray}
p^\alpha T_{\alpha\mu\nu}(k_1,k_2) = 2m_e T_{\mu\nu}(k_1,k_2)+\frac{1}{2\pi^2} \epsilon_{\mu\nu\rho\sigma} k_2^\rho k_1^\sigma \,,
\label{Tmunu}
\end{eqnarray}
where $T_{\mu\nu}$ is given by \Eqn{TT} with $\gamma_\alpha\gamma_5$ replaced by $\gamma_5$. The second term in \Eqn{Tmunu} is just the celebrated ABJ axial anomaly. Since the fermion mass does not get factored out in this anomalous term, there remains an uncompensated factor of energy growth and therefore ${\cm}_\Delta$ continues to grow as $\order(E)$. It should be noted that the coefficient of the linearly growing term depends solely on the properties of the fermion which occurs in the loop. For a general fermion this coefficient can be calculated to be
\begin{eqnarray}
C_{\rm anomaly}^{(f)} = a_f Q_f^2 \,,
\label{anomaly}
\end{eqnarray}
where $a_f$ is the axial vector coupling and $Q_f$ is the electric charge of the fermion in units of the electronic charge. It is clear that adding more and more {\em electron-like} fermions will only worsen things because they will go on adding to the coefficient. A neutrino loop, of course, does not contribute. Therefore, we need some {\em other} kinds of fermions to cancel this.
But first we note that the contribution to $C_{\rm anomaly}$ due to a single lepton generation is
\begin{eqnarray}
C_{\rm anomaly}^{(\ell)} = -\frac{1}{4} \,.
\label{anomaly_lep}
\end{eqnarray}

Now, if we assume that the quarks, with similar axial vector couplings as the leptons, are the possible candidates to cancel the anomaly then we can determine the number of colors of the quarks. Suppose, a single generation of quarks contains an up-type and a down-type quark with $Q_u=+2/3$, $a_u=1/4$ and $Q_d=-1/3$, $a_d=-1/4$ respectively. If this generation has $N_c$ replicas, then
\begin{eqnarray}
C_{\rm anomaly}^{(q)} = \frac{N_c}{4}\left(Q_u^2-Q_d^2\right) \,.
\label{anomaly_q}
\end{eqnarray}
The condition for the anomaly cancellation then requires $N_c=3$, {\em i.e.}, each quark generation should come with three varieties which, with hindsight, we can connect with the {\em color} quantum number. Using similar unitarity arguments in association with simple phenomenology, it is also possible to obtain the couplings and spectrum for the quarks\cite{Horejsi:1993hz}.

\begin{savequote}[0.65\textwidth]  
It is a capital mistake to theorize before one has data. Insensibly one begins to twist facts to suit theories, instead of theories to suit facts.
\qauthor{Sherlock Holmes in ``A Scandal In Bohemia"}    
\end{savequote}
%


\chapter{Modified Higgs couplings and unitarity violation} 

\lhead{Chapter 2. \emph{Modified Higgs couplings and unitarity violation}}
\label{Chap2} 

One of the crucial arguments for the existence of the Higgs boson in
the Standard Model (SM) is that, without it, the longitudinal vector
boson ($V_L$, where $V=W,Z$) scattering amplitudes at the tree level
would uncontrollably grow with the center of mass energy ($E$).  This
will result in the violation of `unitarity', thus implying breakdown
of quantum mechanical sense of probability conservation in scattering
amplitudes.  In the SM, the Higgs boson possesses appropriate gauge
couplings to ensure exact cancellation of the residual $E^2$ growth in
the $V_L V_L \to V_L V_L$ scattering amplitude that survives after
adding the gauge boson contributions.  It has been explicitly shown in
\cite{Lee:1977eg} how, for $E \gg M_V$, the $E^2$ dependence is traded
in favor of the unknown $m_h^2$, where $m_h$ is the Higgs boson mass.
From this it was concluded that $m_h$ should be less than about a TeV
for unitarity not to be violated.  An intimate relationship between
unitarity and renormalizability adds a special relevance to this
issue.  For a renormalizable theory the tree level amplitude for $2
\to 2$ scattering should not contain any term which grows with energy
\cite{Cornwall:1974km}.  In perturbative expansion of scattering
amplitudes these energy growths must be canceled order by order
\cite{Chanowitz:1978mv}.  It has been shown that the energy dependent
terms in tree level amplitudes get exactly canceled if the couplings
satisfy certain sets of `unitarity sum rules' \cite{Gunion:1990kf}.
It has also been realized that the presence of the Higgs boson is not
the only option to satisfy these sum rules \cite{Csaki:2003dt,
  Lahiri:2011ic}.

Meanwhile, a Higgs-like particle has been observed with a mass of
around 125 GeV by the ATLAS and CMS collaborations of the LHC
\cite{Aad:2012tfa,Chatrchyan:2012ufa}.  This is much below the upper limit coming
from unitarity violation mentioned above.  If this particle indeed
turns out to be the SM Higgs, then the scattering amplitudes involving
not only the longitudinal vector bosons but any other SM particles as
external states would be well behaved for arbitrarily high energies.
However, the recent observation of some excess events in the $h \to
\gamma\gamma$ channel\footnote{This chapter is based on our paper \cite{Bhattacharyya:2012tj} which was written in view of the excess events observed at that time in the $h\to\gamma\gamma$ channel. Now, after an upgraded analysis of the data both by CMS and ATLAS, the excess seems to have gone away.}, as well as large errors associated with other
decay channels, has fuelled speculation that Higgs couplings to
fermions and/or gauge bosons might not be exactly as predicted by the
SM \cite{Carena:2012xa}.  There are more than one ways to modify the
Higgs couplings. One way is to hypothesize that the $WWh$ and the
$ZZh$ couplings are modified; more specifically, enhanced with respect
to their SM values.  This would result not only in an increase in the
Higgs production cross section via vector boson fusion and associated
production, but also in an enhancement of the $W$-loop contribution to
$h \to \gamma\gamma$ decay.  But this would at the same time lead to
excess events in the $h \to WW^*$ and $h \to ZZ^*$ channels, something
which is not obvious from data.  It would also result in the violation
of unitarity in longitudinal gauge boson scattering channels. This was
indeed explored long back \cite{Cheung:2008zh}, however, in the
absence of the LHC data there was no motivation to study the
correlation between unitarity violation and the Higgs decay branching
ratios at that time.  If we refrain from adding any extra particle to
the SM and yet attempt to account for the excess in the diphoton
channel, the next natural choice would be to modify the Yukawa
coupling of the top quark.  As is already known, if we put the sign of
the top Yukawa coupling opposite to what it is in the SM, the $h \to
\gamma\gamma$ rate gets enhanced due to a constructive interference
between the $W$-loop diagram and the top-loop diagram
\cite{Carena:2012xa}.  One of the fall-outs of this sign flip is that
$t\bar t \to V_L V_L$ scattering no longer remains unitary.  In fact,
as we shall show, any non-trivial admixture of CP-even and CP-odd
states in the composition of the scalar particle jeopardizes the good
high energy behavior of the $t\bar t \to V_L V_L$ amplitude even if we
keep the moduli of the top Yukawa coupling and the Higgs gauge
coupling to their SM values.  The purpose of this chapter is to
explicitly demonstrate how the scales of unitarity violation in $W_L
W_L \to W_L W_L$ and $t \bar t \to W_L W_L$ scattering processes
depend on the modification parameters of the gauge and the top Yukawa
couplings of the Higgs.  We demonstrate what an enhanced diphoton rate
may imply in this context.

\section{Modification of the Higgs couplings}
In our analysis, we modify only the top Yukawa coupling, since the
other Yukawa couplings are numerically much less relevant.  We take
\begin{subequations}
\label{xfdef}
\begin{eqnarray}
g_{tth} &=& 
(1-f) (\cos \delta - i \sin \delta \gamma_5) \, g_{tth}^{\rm SM} 
= (1-f) e^{- i\delta \gamma_5} g_{tth}^{\rm SM} \,.
\label{fdef}
\end{eqnarray}
The parameter $f$ is a measure of the overall coupling of the Higgs
boson to the top quark, whereas $\delta$ is a parameter that
quantifies the mixture of CP-even and CP-odd components in the Higgs
boson.  We also modify the gauge couplings of the Higgs boson as
\begin{eqnarray}
g_{VVh} &=& (1-x) \, g_{VVh}^{\rm SM} \,,
\label{xdef}
\end{eqnarray}
\end{subequations}
where $V$ can be $W$ or $Z$, as said before.  We maintain equality
between the $WWh$ and $ZZh$ couplings to respect custodial symmetry.
The parameters $x$, $f$ and $\delta$ are all real, and they all vanish
in the SM.

We now comment on the existing experimental constraints on these
modification parameters.  First, it has been shown in
\cite{Azatov:2012bz} that precision electroweak measurements imply
$-0.2 \le x \le 0.1$ at 95\% C.L.\ for $m_h = 125$ GeV and $m_t = 173$
GeV, while from the recent LHC Higgs data analysis the 95\%
C.L.\ range has been estimated to be $-0.4 \le x \le 0.4$
\cite{Plehn:2012iz, Giardino:2012dp}.  Second, the allowed range of
$f$ can be extracted from recent fits of modified Higgs couplings
against the LHC data.  For example, for $x=0$, the range is $-0.1 < f
< 0.6$ for values of $\delta$ fixed at $0$ and
$\pi$~\cite{Giardino:2012dp, Espinosa:2012im}. Note that similar
bounds have been obtained by the authors of
Ref.~\cite{Banerjee:2012xc}, who considered a phase in the effective
coupling due to an absorptive part in the amplitude.  In this chapter,
we take a more conservative approach and consider a hermitian Yukawa
Lagrangian.

\section{Impact on high energy unitarity}
\begin{figure}
\includegraphics[width = 0.49\textwidth]{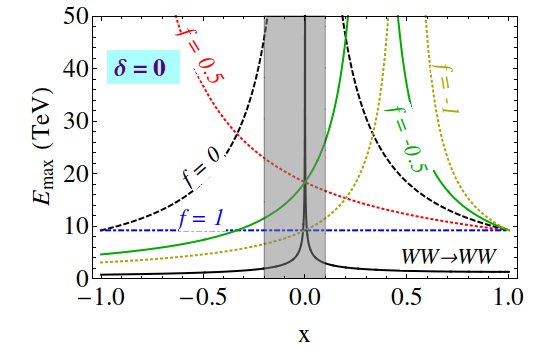} 
\includegraphics[width = 0.49\textwidth]{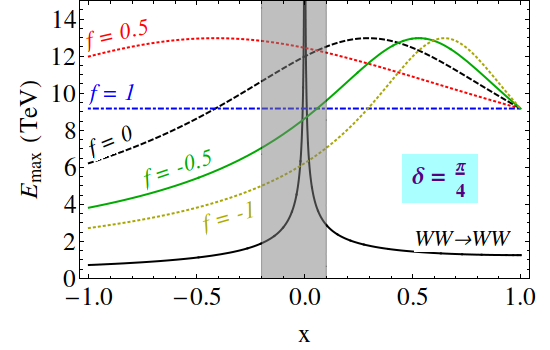} 
\\
 \includegraphics[width = 0.49\textwidth]{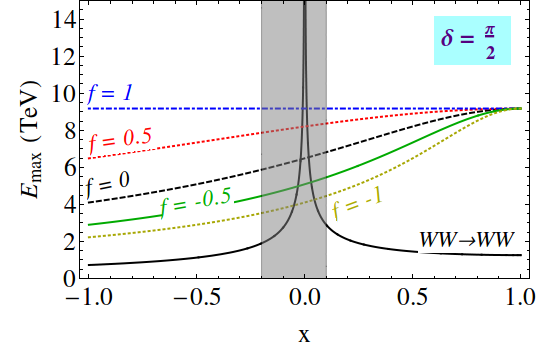} 
 \includegraphics[width = 0.49\textwidth]{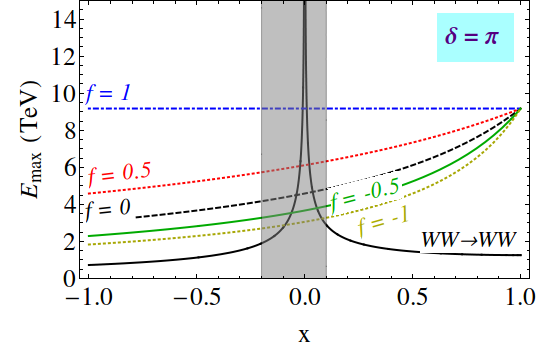}
 \caption[Unitarity violation scale as a function of $x$]{\em Unitarity violation scale as a function of $x$, for
   specific values of $f$ and $\delta$.  For each panel, the scale
   coming from the elastic $WW\to WW$ scattering has been marked.  The
   other lines come from $t\bar t \to WW$ scattering for various
   values of $f$.  The vertical shaded region represents the range of
   $x$ consistent with electroweak precision data.  Note the different
   scale on the vertical axis for the plot with $\delta=0$.}
 \label{xEplot}
\end{figure}
With the modifications prescribed in \Eqn{xfdef}, one should examine
unitarity constraints on scattering processes involving the top quark
and the $W$-boson.  Note that we will talk about the longitudinally
polarized component of the $W$-boson only, dropping the polarization
subscript $L$ which is implicitly assumed.  We have looked at the
energy dependence of the elastic scattering $WW \to WW$ and the
inelastic scattering $t\bar t \to WW$.  The scattering amplitudes that
we find are as follows (see Appendices~\ref{AppendixA} and \ref{AppendixB} for details):
\begin{subequations}
\label{amplitudes}
\begin{eqnarray}
{\cal A}^{WW\to WW} &=& 2\sqrt{2}G_F E^2 (2x-x^2)(1+\cos\theta) +
\cdots \,, 
\label{ampWWWW}  \\ 
 {\cal A}^{t\bar t \to WW} &=& 
2\sqrt{2} G_F E m_t Y(x,f,\delta) + \cdots \,, 
\label{ampWWtt}
\end{eqnarray}
\end{subequations}
where the dots indicate sub-leading terms in energy which do not
concern us, $\theta$ is the scattering angle, and 
\begin{eqnarray}
Y(x,f,\delta) = \mp \Big[ 1 - (1-x)(1-f)e^{\mp i\delta} \Big] \,,
\label{Y}
\end{eqnarray}
where different signs correspond to different combinations of
helicities \cite{Whisnant:1994fh}.  The scattering amplitude can be
expanded in terms of partial waves \cite{Lee:1977eg}:
\begin{eqnarray}
 {\cal A}(\theta)=16\pi\sum_{l=0}^{\infty}(2l+1)a_lP_l(\cos\theta) \, .  
\end{eqnarray}
The unitarity condition $|a_0|\le 1$ puts upper limits on the center
of mass energy in each of these processes.  These limits are as follows:
\begin{subequations}
\label{u12}
\begin{equationarray}{rcll}
E \leq E_{\rm max}^{WW} & = & \left(\frac{4\sqrt{2}\pi}{G_F} \;
\frac{1}{|2x-x^2|}\right)^\frac{1}{2} & \text{[from $WW \to
     WW$]} \, ;\label{u1} \\ 
E \leq E_{\rm max}^{tt} & = & \frac{4\sqrt{2}\pi}{G_Fm_t} \;
\frac{1}{|Y(x,f,\delta)|} \hspace{20mm} &  \text{[from $t\bar
    t \to WW$]} \, .\label{u2} 
\end{equationarray}
\end{subequations}
Because only $\cos\delta$ appears in $|Y|$, we can take $\delta$ in
the range $[0,\pi]$.  Without any loss of generality, we can take
$1-f\geq 0$ to cover the entire parameter space.  In passing, let us
add that the constraints from $t \bar t \to ZZ$ is the same in the
leading order in $E$ as that given in \Eqn{u2}.

We now discuss the numerical dependence of the unitarity violation
scale on the nonstandard parameters expressed through our master
equations given in \Eqn{u12}.  Our results are displayed in
Fig.~\ref{xEplot}.  The different panels correspond to different
choices of $\delta$, as indicated in the figure.  For the $WW \to WW$
scattering amplitude which grows as $E^2$, there is contribution
coming from Higgs mediated diagram and therefore it depends on $x$,
but there is no dependence on $f$ and $\delta$ since the top-Higgs
coupling is not involved.  The latter coupling is of course relevant
for the $t\bar t \to WW$ scattering, and the Higgs mediated graph is
sensitive to all the three nonstandard parameters, i.e.  $x$, $f$ and
$\delta$.  In all the panels the lines titled $WW \to WW$, obtained by
plotting \Eqn{u1}, show the scale of unitarity violation as the $WWh$
coupling departs from its SM value.  The other lines mark the
unitarity violation scale arising from $t \bar t \to WW$, and are
obtained from Eq.~(\ref{u2}).  In the limit $x=1$, i.e.  when the
Higgs either does not exist or does not couple to $W$, unitarity is
\begin{wrapfigure}{r}{0.43\textwidth}
\includegraphics[scale=0.38]{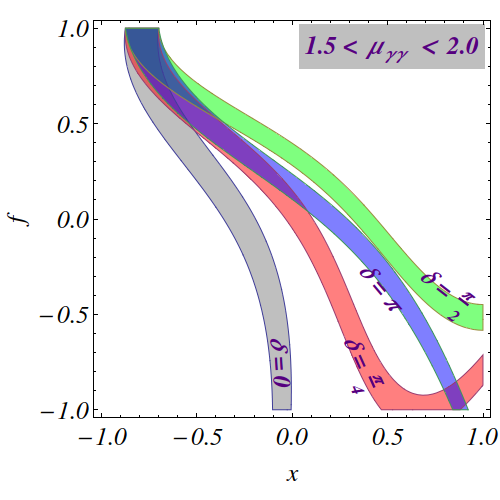}
 \caption[Allowed regions in the $x$-$f$ plane]{\em Allowed regions in the $x$-$f$ plane that correspond to 
   the diphoton enhancement ratio $\mu_{\gamma\gamma}$ lying between 1.5 and 2, for
different values of $\delta$.}
 \label{xf}
\end{wrapfigure}
violated at a pretty low scale, $E_{\rm max}^{WW} \approx 1.3$ TeV.
As $x$ approaches zero, $E_{\rm max}^{WW}$ goes up.  On the other
hand, the limit $f=1$ implies that the Higgs does not couple to the
top quark, so in this limit the Higgs mediated graph for $t\bar t \to
WW$ would not exist, and hence, the unitarity violation scale arising
from the above scattering would be independent of $x$ and $\delta$.
Similar things happen in the limit $x=1$, causing the unitarity
violation scale from $t\bar t \to WW$ to be independent of $f$ and
$\delta$.  This is precisely the reason as to why the horizontal $f=1$
line in all the panels meet the curvy lines for other values of $f$ at
one single point which is at $x=1$ corresponding to $E_{\rm max}^{tt}
\approx 9$~TeV.

An important observation at this stage is the following: for
$\delta\neq0$ and $\delta\neq\pi$, the process $t\bar t \to WW$ is not
unitary regardless of the choice of $x$ and $f$.  The vertical shades
in the four panels restrict the values of $x$ within the zone allowed
by precision tests.  One thing is quite clear that if $x$ happens to
take a value near the edge of the shade in any panel, the unitarity
violation would set in for $WW \to WW$ at a scale much lower than
where it would happen for $t\bar t \to WW$, which is easily understood
from the $E^2$ versus $E$ growth in the two amplitudes.  But if $x$
settles at a much smaller value, as one can see from the different
panels, the unitarity violation scales from these two amplitudes get
closer and at some point the hierarchy mentioned earlier is reversed.

\section{Impact on diphoton signal strength}
We now consider the decay of the 125\,GeV particle into two photons.
Two-photon final states have a definite CP property, more
specifically, a definite parity.  As a result, if the initial
spin-zero state is not an eigenstate of parity, the parity-even and
parity-odd components will contribute incoherently, {\it i.e.}, their loop contributions
can be added together separately in the amplitude squared level. 

The decay $h\to \gamma\gamma$ proceeds dominantly through a $W$ boson
loop and a top loop diagram.  For a  CP-mixed $h$ whose couplings are given by \Eqs{xdef}{fdef}, the decay width is
given by \cite{Djouadi:2005gi}:
\begin{eqnarray}
 \Gamma(h\to \gamma\gamma) =
\frac{\alpha^2g^2}{2^{10}\pi^3} \frac{m_h^3}{M_W^2}\left[ \left|(1-x)F_W+\frac{4}{3}(1-f)\cos\delta ~ F_t\right|^2
+\left|\frac{4}{3}(1-f)\sin\delta ~ P_t\right|^2\right] \,,
\label{diphoton_1}
\end{eqnarray}
where, the values of $F_W$, $F_t$ and $P_t$ are given by
 \begin{subequations}
 \begin{eqnarray}
  F_W &=& 2+3\tau_W+3\tau_W(2-\tau_W)f(\tau_W) \,,  \\
  F_t &=& -2\tau_t \left[1+(1-\tau_t)f(\tau_t)\right] \,,  \\
  P_t &=& 2\tau_t ~f(\tau_t)  \,, \\
{\rm with}~~~ \tau_x &\equiv& (2m_x/m_h)^2 \,.
 \end{eqnarray}
 \end{subequations}
In the above equations, $F_t$ and $P_t$ represent the top-loop contributions
from the scalar and pseudoscalar parts respectively. For $m_h\approx125 ~\textrm{GeV}$, $\tau_x >1$ for both $x=W,t$.  In
this situation, 
\begin{eqnarray}
f(\tau) =
\left[\sin^{-1}\left(\sqrt{1/\tau}\right)\right]^2 \,.
\end{eqnarray}
The SM expression for the Higgs to diphoton decay width is obtained by putting $x,~f,~\delta =0$ in \Eqn{diphoton_1}.
In view of this, the modification factor for the partial decay width can be expressed as:
\begin{eqnarray}
 R_{\gamma\gamma} = \frac{ \Gamma(h\to \gamma\gamma)}{ \Gamma^{\rm SM}(h\to \gamma\gamma)}
 = \frac{|(1-x)F_W+\frac{4}{3}(1-f)\cos\delta ~F_t|^2+|\frac{4}{3}(1-f)\sin\delta ~P_t|^2}{|F_W+\frac{4}{3}F_t|^2} \,.
\end{eqnarray}
It should also be noted that due to the modification of the $tth$ Yukawa coupling, the $ggh$ effective vertex will also be modified. Denoting the modification factor for $h\to gg$ decay width by $\cal{G}$, one can easily find
\begin{eqnarray}
{\cal{G}} = \frac{ \Gamma^{\rm SM}(h\to gg)}{ \Gamma(h\to gg)} 
 = (1-f)^2\frac{|\cos\delta ~F_t|^2+|\sin\delta ~P_t|^2}{|F_t|^2} \,.
\end{eqnarray}
We should also remember that the same factor, $\cal{G}$, also controls the modification of the production cross-section through the $gg\to h$ channel.

We now estimate how the Higgs production cross section would be
modified.  For 7(8)-TeV LHC, the top loop driven gluon-gluon fusion
channel contributes around 85\% of the total cross section, while the
associated production and the vector boson fusion together almost
account for the remaining 15\% \cite{Djouadi:2005gi}.  The production
cross section would then be modified roughly by the factor
\begin{eqnarray}
 \frac{\sigma(pp\to h)}{\sigma^{\rm SM}(pp\to h)} 
= \frac{{\cal{G}}\cdot \sigma_{\rm G}+(1-x)^2\sigma_{\rm V}}{\sigma_{\rm
    G}+\sigma_{\rm V}} \approx {\cal{G}}\cdot 85\% + (1-x)^2 15\% \,.
\end{eqnarray}
As far as the different decay channels of the Higgs are concerned, for
$m_h \approx 125$ GeV, branching ratios of the SM Higgs boson are
roughly as follows: $58\%$ to $b\bar{b}$, $7\%$ to $\tau^+\tau^-$,
$3\%$ to $c\bar{c}$, $24\%$ to $VV^{*}$ and $8\%$ to $gg$
\cite{Djouadi:2005gi}.  We then express the modification of the total
decay width by the ratio:
\begin{eqnarray}
 \frac{\Gamma_h}{\Gamma^{\rm SM}_h} = (58\%+7\%+3\%) +
 (1-x)^2 24\% + {\cal{G}}\cdot 8\% \, .
\end{eqnarray}
The above expressions lead us to define
\begin{eqnarray}
 \mu_{\gamma\gamma} &=& \frac{\sigma(pp\to h)}{\sigma^{\rm SM}(pp\to h)} \times
 \frac{ \Gamma (h \to \gamma\gamma)}{ \Gamma^{\rm SM}(h
 \to \gamma\gamma)} \times \frac{\Gamma^{\rm SM}_h}{\Gamma_h}  \,.  
\label{mu-gam}
\end{eqnarray}

\begin{figure}
\centering
\includegraphics[scale=0.38]{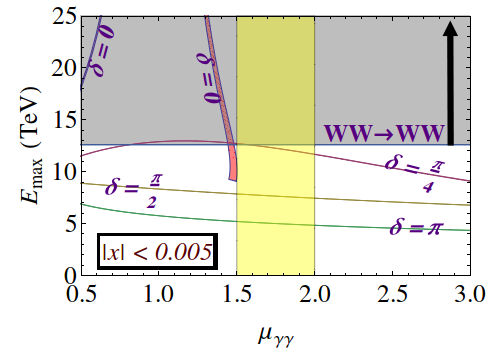}~~
\includegraphics[scale=0.38]{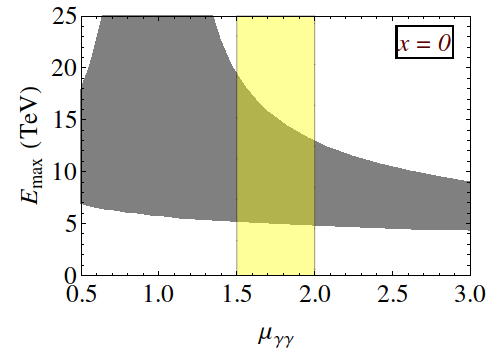}
 \caption[Correlation between $E_{\rm max}$ and $\mu_{\gamma\gamma}$]{\em Unitarity violation scale plotted against diphoton enhancement.  The vertical shaded region (in yellow) corresponds to $1.5 < \mu_{\gamma\gamma} < 2.0$. In the left panel we vary $x$ in a very narrow range: [-0.005, +0.005]. In the right panel, we set $x=0$ so that $WW$ scattering remain always unitary and vary $f$ and $\delta$ within [-1, +1] and [0, $\pi$] respectively to see the correlation between $E_{\rm max}$ for $t\bar{t}\to WW$ and $\mu_{\gamma\gamma}$.}
\label{emaxmu}
\end{figure}
In Fig.~\ref{xf} we have shaded different regions in the $x$-$f$
plane, for different choices of $\delta$, which can account for
the apparent excess of the diphoton events.  Motivated by the recent
LHC data, we choose $\mu_{\gamma\gamma}$ in the range $1.5$ to $2$ for the sake of
illustration.  For $x \approx0$ and $\delta=\pi$, we observe that 
\begin{eqnarray}
0.1 < f < 0.25 
\label{flimit}
\end{eqnarray}
which is roughly consistent with the limit quoted earlier in
connection with global fits.  Thus a {\em top-phobic Higgs}, which
corresponds to $f \to 1$, is highly unlikely.  We must admit though
that this comparison is not entirely fair as we have modified only the
top Yukawa coupling, while in the global fits all the Yukawa couplings
were modified.  We also admit that for the simplicity of illustration
we have not taken into account the efficiency factors in the
estimation of $\mu_{\gamma\gamma}$.

\section{Correlation between \texorpdfstring{$E_{\rm max}$}{TEXT} and \texorpdfstring{$\mu_{\gamma\gamma}$}{TEXT}}
In the left panel of Fig.~\ref{emaxmu}, we have exhibited the correlation between the
unitarity violation scale and the diphoton enhancement ratio $\mu_{\gamma\gamma}$.
For drawing this plot, we have varied $f$ between $-1$ and $+1$.
Keeping in mind the relative sensitivity of the two scattering
processes, we restrict $x$ in a rather narrow range: $-0.005 < x <
0.005$ in the left panel.  The lower horizontal boundary of the (gray) shaded region around $E_{\rm max} = 13$~TeV, appropriately labeled, corresponds to
the unitarity violation scale in $WW\to WW$ scattering with
$|x|=0.005$.  For smaller values of $x$, this line will appear at
higher energy.  The other curvy lines come from $t \bar t \to WW$ and
they correspond to four different choices of $\delta$, viz., $0$, $\pi/4$, $\pi/2$ and
$\pi$.  The thickness of these lines for different values of $\delta$
come from the range of $x$ just mentioned.  For $\delta=0$, it is hard
to achieve a value of $\mu_{\gamma\gamma}$ as large as 1.5.  For $\delta=\pi$, it is
possible to obtain a value of $\mu_{\gamma\gamma}$ in the range 1.5 to 2, as can be
seen by the corresponding line going through the vertical shade.  The
corresponding range of $f$, which can be read from Fig.~\ref{xf}, has
been mentioned in \Eqn{flimit}. It is worth noting from this figure
that for $\delta=\pi$, which facilitates diphoton rate enhancement,
the unitarity violation scale comes down to around 5\,TeV.  This is
true even when $x=0$, i.e., when the gauge coupling of the Higgs boson
matches the SM value and therefore the $WW\to WW$ scattering is
perfectly unitary. For visual clarification, we have set $x=0$ in the right panel
of Fig.~\ref{emaxmu} so that $WW$-scattering remains unitary up to arbitrary 
high energies. Here we can see (from the overlap between vertical yellow band
and gray shaded region) that if the value of $\mu_{\gamma\gamma}$ eventually
settles somewhere within $1.5$ to $2$, then high scale unitarity of the process
$t\bar t \to WW$ is bound to be violated somewhere between $5$ and $19$~TeV.

\section{Conclusions}
To summarize, even though the existence of a Higgs-like particle has
been announced, precise measurements of its couplings to gauge bosons
and fermions would take quite a while.  If the measured
couplings eventually match their SM values, the theory is unitary,
i.e.  well-behaved up to arbitrarily high energies.  Otherwise, the
extent of departure of the measured values of the couplings from their
SM predictions would mark the scale where unknown dynamics would set
in (see e.g.~\cite{Bellazzini:2012tv}).  We have carried out a
quantitative study of this scale as a function of the deviation of the
Higgs couplings from their SM values through studies of the $WW \to
WW$ and $t\bar t \to WW$ scattering processes.  We have specifically
focused on nonstandard effects on the gauge coupling of the Higgs and
the top Yukawa coupling, as these two couplings play a crucial r\^ole
in the stability of the electroweak vacuum and the perturbative
unitarity of the theory.  If future measurements favor Higgs couplings
closer to its SM values, the expected scale of unitarity saturation
would go up.
 

\begin{savequote}[0.85\textwidth]  
`That's a rather broad idea', I remarked. \\
`One's ideas must be as broad as Nature if they are to interpret Nature', he answered.
 \qauthor{Watson and Holmes in ``A study in scarlet''}    
\end{savequote}
%

\chapter{Post Higgs overview of two Higgs-doublet models} 

\lhead{Chapter 3. \emph{Post Higgs overview of two Higgs-doublet models}}
\label{Chap3} 

The discovery of a new boson in July of 2012 at the LHC\cite{Aad:2012tfa,Chatrchyan:2012ufa} is undoubtedly the
greatest achievement of this decade in the field of Particle Physics. This might be the final missing piece of the SM. But at the same time, SM does not account for observations like neutrino oscillations, dark matter. Phenomena like these constitute the primary motivation to look for other avenues beyond the SM (BSM). The SM relies on the minimal choice of a single $SU(2)$ scalar doublet giving masses to all the massive particles contained in the SM. Extension of the SM scalar sector is a common practice in constructing BSM models. While extending the scalar sector, one runs into the risk of altering the tree level value of the electroweak $\rho$-parameter. If we construct an $SU(2)\times U(1)$ gauge theory with $n$ scalar multiplets, then the general expression for the tree level $\rho$-parameter reads
\begin{eqnarray}
\rho^{\rm tree} = \frac{\sum\limits_{i=1}^{n}\left\{T_i(T_i+1)-\frac{Y_i^2}{4} \right\}v_i }{\frac{1}{2} \sum\limits_{i=1}^{n}Y_i^2v_i } \,,
\end{eqnarray}
where, $T_i$ and $Y_i$ denote the weak isospin and hypercharge of the $i$-th scalar multiplet respectively and $v_i$ refers to the vacuum expectation value (vev) picked up by the neutral component of the $i$-th multiplet. One can easily verify that if the scalar sector contains only $SU(2)$ singlets ($T_i=0$) and doublets ($T_i=1/2$) with hypercharges $0$ and $\pm 1$ respectively, the $\rho^{\rm tree}=1$ is automatically recovered without requiring any fine tuning among the vevs. In this article we restrict ourselves to the doublet extensions only. This simplest extension of this type is the case of two Higgs-doublet model (2HDM)\cite{Branco:2011iw} which receives a lot of attention because minimal supersymmetry relies on it. In a general 2HDM both the doublets can couple to each type of fermions. Consequently, there will be two Yukawa matrices which, in general, are not diagonalizable simultaneously. This will introduce new flavor changing neutral current (FCNC) couplings mediated by neutral Higgses. It was shown by Glashow and Weinberg\cite{Glashow:1976nt} and independently by Pascos\cite{Paschos:1976ay} that Higgs mediated FCNC can be avoided altogether if fermions of a particular charge get their masses from a single scalar doublet. This prescription was realized by employing a $Z_2$ symmetry under which one of the doublet is odd. Then there are four different possibilities for assigning $Z_2$ parities to the fermions so that Glashow-Weinberg-Pascos theorem is satisfied. This leads to the following four types of 2HDMs:
\begin{itemize}
\item Type I: all quarks and leptons couple to only one scalar doublet $\Phi_2$ ;
\item Type II: $\Phi_2$ couples to up-type quarks, while $\Phi_1$ couples to down-type quarks and charged leptons (minimal supersymmetry conforms to this category);
\item Type X or lepton specific: $\Phi_2$ couples to all quarks, while $\Phi_1$ couples to all leptons;
\item Type Y or flipped: $\Phi_2$ couples to up-type quarks and leptons, while $\Phi_1$ couples to down-type quarks.
\end{itemize}
There is also the option for preventing tree level FCNC by assuming the two Yukawa matrices proportional to each other. This gives rise to the so called aligned 2HDM. However Branco, Grimus and Lavoura (BGL) employed a global $U(1)$ symmetry which textures both Yukawa matrices in a certain way\cite{Branco:1996bq}. As a result of this the tree level Higgs FCNC couplings get related to the off diagonal elements of the CKM matrix and thereby are naturally suppressed. In this chapter, I intend to highlight some major phenomenological aspects of these different types of 2HDMs with special emphasis on the BGL models. We will also discuss how the recent LHC Higgs data constrain these models.

\section{The scalar potential}
There are two equivalent notations that are used in the literature to write the 2HDM scalar potential invariant under a $Z_2$ symmetry ($\Phi_2\to -\Phi_2$)~:
\paragraph*{$\blacksquare$ Notation 1}
\begin{eqnarray}
V(\Phi_1,\Phi_2) &=& m_{11}^2 \Phi_1^\dagger\Phi_1  + m_{22}^2\Phi_2^\dagger\Phi_2 -\left(m_{12}^2 \Phi_1^\dagger\Phi_2 +{\rm h.c.} \right) +\frac{\beta_1}{2} \left(\Phi_1^\dagger\Phi_1 \right)^2  +\frac{\beta_2}{2} \left(\Phi_2^\dagger\Phi_2 \right)^2 \nonumber \\
&& +\beta_3 \left(\Phi_1^\dagger\Phi_1 \right) \left(\Phi_2^\dagger\Phi_2 \right) +\beta_4 \left(\Phi_1^\dagger\Phi_2 \right) \left(\Phi_2^\dagger\Phi_1 \right) +\left\{\frac{\beta_5}{2} \left(\Phi_1^\dagger\Phi_2 \right)^2 +{\rm h.c.} \right\}
\label{notation1}
\end{eqnarray}
\paragraph*{$\blacksquare$ Notation 2}
\begin{eqnarray}
 V &=& 
 \lambda_1 \left( \Phi_1^\dagger\Phi_1 - \frac{v_1^2}{2} \right)^2 
+\lambda_2 \left( \Phi_2^\dagger\Phi_2 - \frac{v_2^2}{2} \right)^2 
 +\lambda_3 \left( \Phi_1^\dagger\Phi_1 + \Phi_2^{\dagger}\Phi_2 
- \frac{v_1^2+v_2^2}{2} \right)^2
\nonumber \\
&& +\lambda_4 \left(
(\Phi_1^{\dagger}\Phi_1) (\Phi_2^{\dagger}\Phi_2) -
(\Phi_1^{\dagger}\Phi_2) (\Phi_2^{\dagger}\Phi_1)
\right)  + \lambda_5 \left( {\rm Re}~  \Phi_1^\dagger\Phi_2  - \frac{v_1v_2}{2}  \right)^2 
+ \lambda_6 \left({\rm Im}~ \Phi_1^\dagger\Phi_2  \right)^2 \,.
\label{notation2}
\end{eqnarray}
The bilinear terms proportional to $m_{12}^2$ in \Eqn{notation1} and $\lambda_5$ in \Eqn{notation2} break the $Z_2$ symmetry softly. The significance of these types of soft breaking term will be discussed later. We assume all the potential parameters to be real so that CP symmetry is conserved in the scalar sector. Note that, when we minimize the potential of \Eqn{notation1}, the two minimization conditions can be used to trade $m_{11}^2$ and $m_{22}^2$ for $v_1$ and $v_2$ and the potential can be cast in the form of \Eqn{notation2}. Note that, unlike \Eqn{notation1}, \Eqn{notation2} implicitly assumes that the $Z_2$ symmetry is also broken spontaneously, {\it i.e.}, both the doublets receive vevs. In this chapter, we shall only consider 2HDMs where the value of $\tan\beta$ ($\equiv v_2/v_1$) is nonzero and finite. The connections between the parameters of \Eqn{notation1} and \Eqn{notation2} are given below:
\begin{eqnarray}
&&  m_{11}^2 =-(\lambda_1v_1^2+\lambda_3v^2)~;~ m_{22}^2= -(\lambda_2v_2^2+\lambda_3v^2)~;~ m_{12}^2=\frac{\lambda_5}{2}v_1v_2~;~ \beta_1=2(\lambda_1+\lambda_3)~;  \nonumber \\
&& \beta_2=2(\lambda_2+\lambda_3)~;~ \beta_3=2\lambda_3+\lambda_4~;~ \beta_4=\frac{\lambda_5+\lambda_6}{2}-\lambda_4~;~ \beta_5=\frac{\lambda_5-\lambda_6}{2}~.
\label{connections}
\end{eqnarray}
In \Eqn{connections} $v=\sqrt{v_1^2+v_2^2} = 246$ GeV, where $v_1$ and $v_2$ are the vevs of the two doublets $\Phi_1$ and $\Phi_2$ respectively. We also remember that it is the combination $m_{12}^2/(\sin\beta\cos\beta)$, not $m_{12}^2$ itself, which controls the nonstandard masses\cite{Gunion:2002zf}. In view of these facts, $\l_5$, rather than $m_{12}^2$, constitutes a convenient parameter that can track down the effect of soft breaking. Therefore, for most part of this chapter, we choose to work with the notation of \Eqn{notation2}.

Before we move on, it should be reemphasized that the parametrization of Eq.~(\ref{notation2}) is less
general than that of Eq.~(\ref{notation1}). Any connection between the
two sets of parameters can be established only when both the scalars
receive vevs.  The inert doublet scenario can be very easily realized
in the parametrization of Eq.~(\ref{notation1}), while just setting
$v_2 = 0$ in the parametrization of Eq.~(\ref{notation2}) does
not lead us to the same limit. To appreciate this salient aspect, we
consider a simpler scenario when we have only one Higgs doublet. Then
the potential can be written in two equivalent ways: $V \sim \mu^2
|\phi|^2 + \lambda |\phi|^4$, and $V' \sim \lambda \left(|\phi|^2 -
v^2/2\right)^2$. They become truly equivalent when $\mu^2 < 0$, and
consequently, the scalar receives a vev. But when $\mu^2 > 0$, the
scalar remains inert. In that case, putting $v = 0$ in $V'$ does not
take us to the physical situation given by $V$, as the latter still
contains, in addition to $\lambda$, an independent dimensionful
parameter $\mu^2$. Our Eqs.~(\ref{notation2}) and
(\ref{notation1}) are 2HDM generalizations of $V'$ and $V$,
respectively.

\subsection{Physical eigenstates}
Expressing the scalar doublets as
\begin{eqnarray}
\Phi_i =\frac{1}{\sqrt{2}} \begin{pmatrix} \sqrt{2} w_i^+ \\ (h_i+v_i) +iz_i \end{pmatrix} \,,
\label{phi}
\end{eqnarray}
we will be able to construct the mass matrices using \Eqn{notation2}. Since we have assumed all the potential parameters to be real, there will be no bilinear mixing term of the form $h_iz_j$. As a result, the neutral mass eigenstates will also be the eigenstates of CP. For the charged sector we get the following mass matrix~:
\begin{eqnarray}
V_{\rm mass}^{\rm charged} = \begin{pmatrix} w_1^+ & w_2^+ \end{pmatrix} M_C^2 \begin{pmatrix} w_1^- \\ w_2^- \end{pmatrix}~~ {\rm with,}~ M_C^2 = \frac{\lambda_4}{2} \begin{pmatrix}
v_2^2 & -v_1v_2 \\ -v_1v_2 & v_1^2 \end{pmatrix} \,.
\end{eqnarray}
$M_C^2$ can be diagonalized to obtain a physical charged Higgs pair ($H_1^\pm$) and a pair of charged Goldstones as follows~:
\begin{eqnarray}
\begin{pmatrix} \omega^\pm \\ H_1^\pm \end{pmatrix} = \begin{pmatrix} \cos\beta & \sin\beta \\ -\sin\beta & \cos\beta \end{pmatrix} \begin{pmatrix} w_1^\pm \\ w_2^\pm \end{pmatrix} \,,
\end{eqnarray}
where, the rotation angle, $\beta$, is defined through the relation $\tan\beta=v_2/v_1$. The mass of the charged Higgs pair ($H_1^\pm$) is found to be
\begin{eqnarray}
m_{1+}^2 = \frac{\lambda_4}{2}v^2 \,.
\label{m1+}
\end{eqnarray}
Similarly for the pseudoscalar part one can easily find
\begin{eqnarray}
V_{\rm mass}^{\rm CP~odd} = \begin{pmatrix} z_1 & z_2 \end{pmatrix}\frac{1}{2} M_P^2 \begin{pmatrix} z_1 \\ z_2 \end{pmatrix}~~ {\rm with,}~ M_P^2 = \frac{\lambda_6}{2} \begin{pmatrix}
v_2^2 & -v_1v_2 \\ -v_1v_2 & v_1^2 \end{pmatrix} \,.
\end{eqnarray}
The diagonalization is similar to that in the charged sector. Here we shall get a physical pseudoscalar ($A$) and a neutral Goldstone ($\zeta$) as follows~:
\begin{eqnarray}
\begin{pmatrix} \zeta \\ A \end{pmatrix} = \begin{pmatrix} \cos\beta & \sin\beta \\ -\sin\beta & \cos\beta \end{pmatrix} \begin{pmatrix} z_1 \\ z_2 \end{pmatrix} \,.
\label{pseudoscalar}
\end{eqnarray}
The mass of the pseudoscalar is given by
\begin{eqnarray}
m_{A}^2 = \frac{\lambda_6}{2}v^2 \,.
\label{mA}
\end{eqnarray}
For the scalar part we find
\begin{subequations}
\begin{eqnarray}
&& V_{\rm mass}^{\rm CP~even} = \begin{pmatrix} h_1 & h_2 \end{pmatrix}\frac{1}{2} M_S^2 \begin{pmatrix} h_1 \\ h_2 \end{pmatrix}~~ {\rm with,}~ M_S^2 =  \begin{pmatrix} A_S & B_S \\ B_S & C_S \end{pmatrix} \,, \\
{\rm where,} && A_S = 2(\lambda_1+\lambda_3)v_1^2 +\frac{\lambda_5}{2}v_2^2 \,, \\
&& B_S =2(\lambda_3+\frac{\lambda_5}{4})v_1v_2 \,, \\
&& C_S = 2(\lambda_2+\lambda_3)v_2^2 +\frac{\lambda_5}{2}v_1^2 \,.
\end{eqnarray}
\end{subequations}
The masses of the physical eigenstates, $H$ and $h$, can be readily found as
\begin{subequations}
\begin{eqnarray}
m_{H}^2 &=& \frac{1}{2}\left[(A_S+C_S)+\sqrt{(A_S-C_S)^2+B_S^2} \right] \,, \\
m_{h}^2 &=& \frac{1}{2}\left[(A_S+C_S)-\sqrt{(A_S-C_S)^2+B_S^2} \right] \,. 
\end{eqnarray}
\end{subequations}
The physical scalars are obtained by rotating the original basis by an angle $\alpha$~:
\begin{eqnarray}
\begin{pmatrix} H \\ h \end{pmatrix} = \begin{pmatrix} \cos\alpha & \sin\alpha \\ -\sin\alpha & \cos\alpha \end{pmatrix} \begin{pmatrix} h_1 \\ h_2 \end{pmatrix} \,.
\end{eqnarray}
This rotation angle is defined through the following relation
\begin{eqnarray}
\tan 2\alpha =\frac{2B_S}{A_S-C_S} = \frac{2\left(\lambda_3+\frac{\lambda_5}{4} \right)v_1v_2}{\lambda_1v_1^2- \lambda_2v_2^2+\left(\lambda_3+\frac{\lambda_5}{4}\right)(v_1^2-v_2^2)} \,.
\end{eqnarray}
Note that there were eight parameters to start with: $v_1,~v_2$ and 6 lambdas. We trade $v_1$ and $v_2$ for $v$ and $\tan\beta$. All the lambdas except $\lambda_5$ can be traded for 4 physical Higgs masses and $\alpha$. The relations between these two equivalent sets of parameters are given below~:
\begin{subequations}
\begin{eqnarray}
\lambda_1 &=& \frac{1}{2v^2\cos^2\beta}\left[m_H^2\cos^2\alpha  +m_h^2\sin^2\alpha  -\frac{\sin\alpha\cos\alpha}{\tan\beta}\left(m_H^2-m_h^2\right)\right] \nonumber \\
&& -\frac{\lambda_5}{4}\left(\tan^2\beta-1\right) \,, \\
\lambda_2 &=& \frac{1}{2v^2\sin^2\beta}\left[m_h^2\cos^2\alpha  +m_H^2\sin^2\alpha  -\sin\alpha\cos\alpha\tan\beta\left(m_H^2-m_h^2\right) \right] \nonumber \\
&& -\frac{\lambda_5}{4}\left(\cot^2\beta-1\right) \,, \\
\lambda_3 &=& \frac{1}{2v^2} \frac{\sin\alpha\cos\alpha}{\sin\beta\cos\beta} \left(m_H^2-m_h^2\right) -\frac{\lambda_5}{4} \,, \\
\lambda_4 &=& \frac{2}{v^2} m_{1+}^2 \,, \\
\lambda_6 &=& \frac{2}{v^2} m_A^2 \,.
\end{eqnarray}
\label{inv2HDM}
\end{subequations}
Among these, $v$ is already known and if we assume that the lightest CP-even Higgs is what has been observed at the LHC, then $m_h$ is also known. The rest of the parameters need to be constrained from theoretical as well as experimental considerations.

\subsection{The alignment limit}
The {\em alignment limit} addresses the possibility of recovering a CP-even mass eigenstate with exactly same couplings as the SM Higgs with the SM particles. To start with, it is instructive to look at the trilinear gauge-Higgs couplings which stem from the Higgs kinetic terms. As an example, consider the case of a 2HDM~:
\begin{eqnarray}
{\mathscr L}_{\rm kin}^{\rm scalar} = |D_\mu\Phi_1|^2+|D_\mu\Phi_2|^2 \ni \frac{g^2}{2}W_\mu^+W^{\mu -}(v_1h_1+v_2h_2) \,.
\end{eqnarray}
Clearly the combination
\begin{eqnarray}
H^0=\frac{1}{v}(v_1h_1+v_2h_2)
\end{eqnarray}
will carry away the exact SM-like gauge couplings and its orthogonal combination ($R$ say) will not have any trilinear couplings with the gauge bosons. Obviously, for $n$ Higgs-doublet case, the definition of $H^0$ will be
\begin{eqnarray}
H^0=\frac{1}{v}(v_1h_1+v_2h_2+\dots +v_nh_n) \,.
\label{H0gen}
\end{eqnarray}
As we will illustrate later, this $H^0$ will mimic the SM Higgs in its Yukawa couplings also. For the case of 2HDM this combination can be obtained by applying the same rotation as in the charged and pseudoscalar cases:
\begin{eqnarray}\begin{pmatrix} H^0 \\ R \end{pmatrix} = \begin{pmatrix} \cos\beta & \sin\beta \\ -\sin\beta & \cos\beta \end{pmatrix} \begin{pmatrix} h_1 \\ h_2 \end{pmatrix} \,.
\label{H0R-h1h2}
\end{eqnarray}
Now, this SM-like state, $H^0$, is not guaranteed to be a mass eigenstate in general. The alignment limit specifically points towards the condition under which $H^0$ coincides with one of the CP-even physical eigenstates. For the 2HDM case, the relationships are~:
 \begin{subequations}
 \label{align_def}
 \begin{eqnarray}
 H &=& \cos(\beta-\alpha) H^0-\sin(\beta-\alpha) R \,, \\
 h &=& \sin(\beta-\alpha) H^0+\cos(\beta-\alpha)R \,.
 \end{eqnarray}
 \end{subequations}
Clearly, if we want the lightest CP-even scalar, $h$, to posses SM-like couplings, we must set 
\begin{eqnarray}
\sin(\beta-\alpha)=1
\label{alignment}
\end{eqnarray}
 which is the definition of the alignment limit in the 2HDM context. So, going to the alignment limit reduces one more parameter from the theory.

\begin{figure}
\includegraphics[scale=0.27]{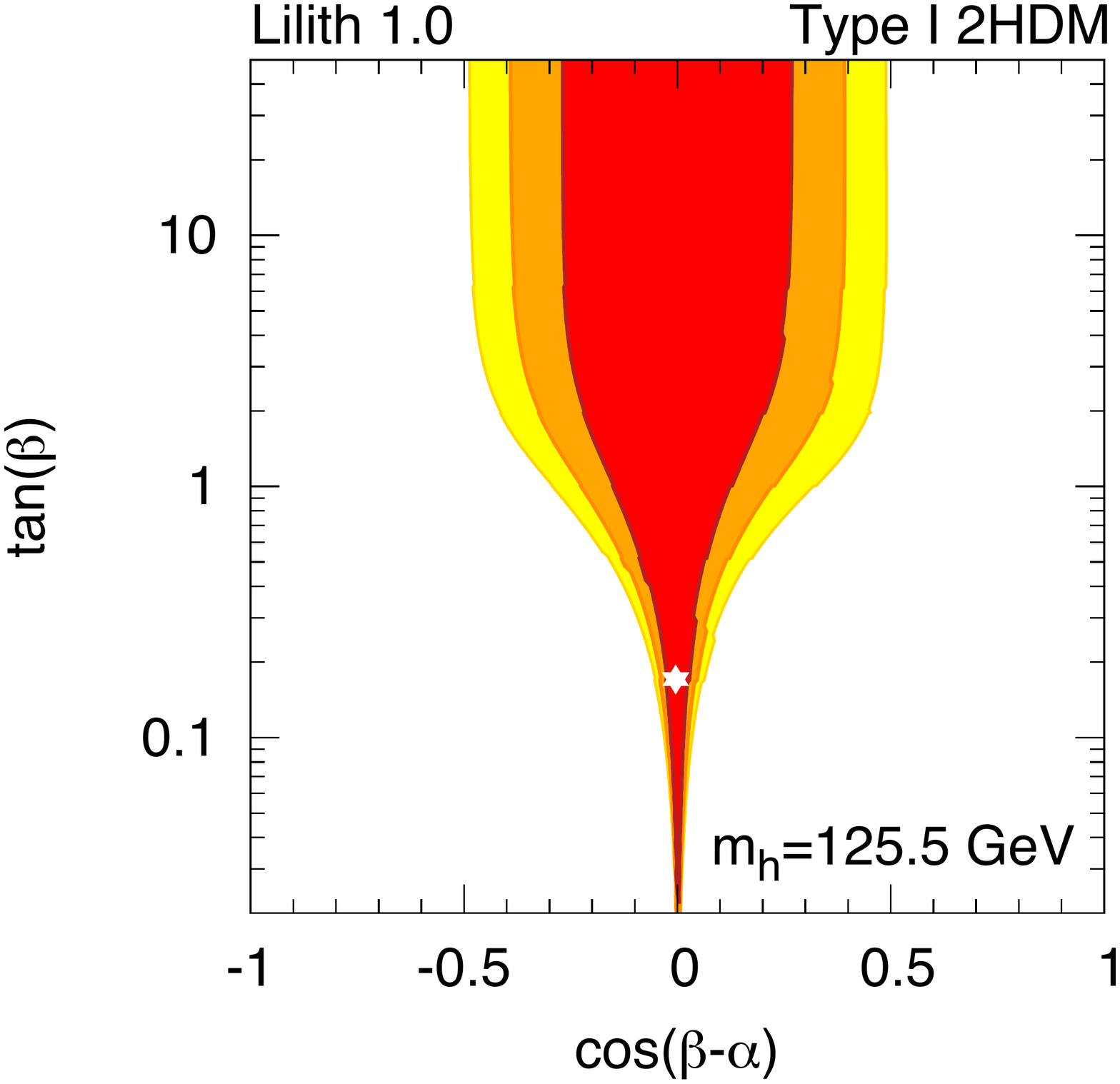} 
\includegraphics[scale=0.27]{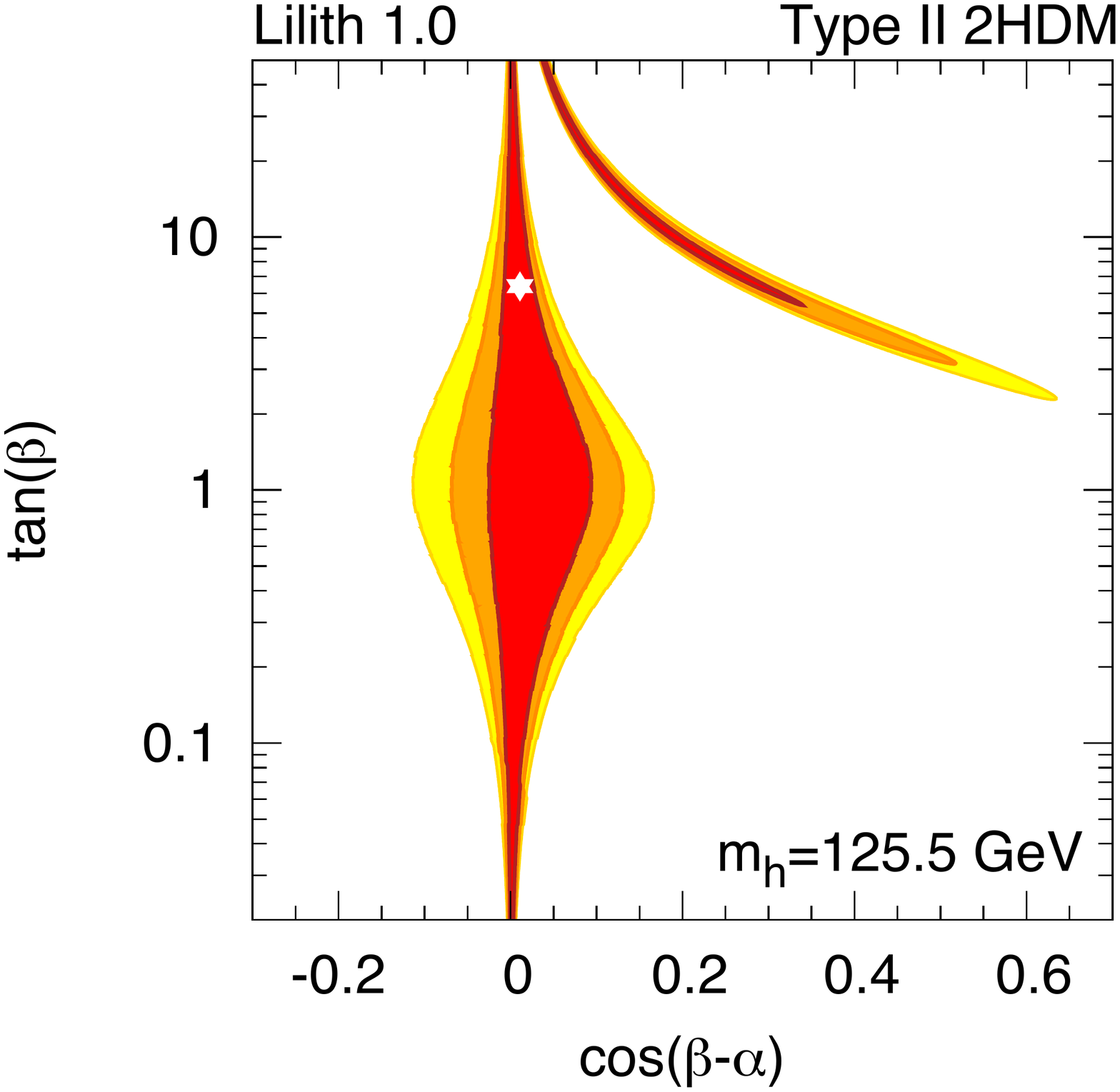}
\caption[Current fit for 2HDMs using the LHC data]{\em The red, orange and yellow regions represent the 68\%, 95\% and 99\% CL allowed regions respectively coming from the Higgs signal strength measurement at the LHC. The left panel shows the situation for Type I model, whereas, the right panel shows that of Type II model. The figures have been taken from \cite{Bernon:2014vta}.}
\label{current_fit}
\end{figure}

\begin{figure}
\centering
\includegraphics[scale=0.3]{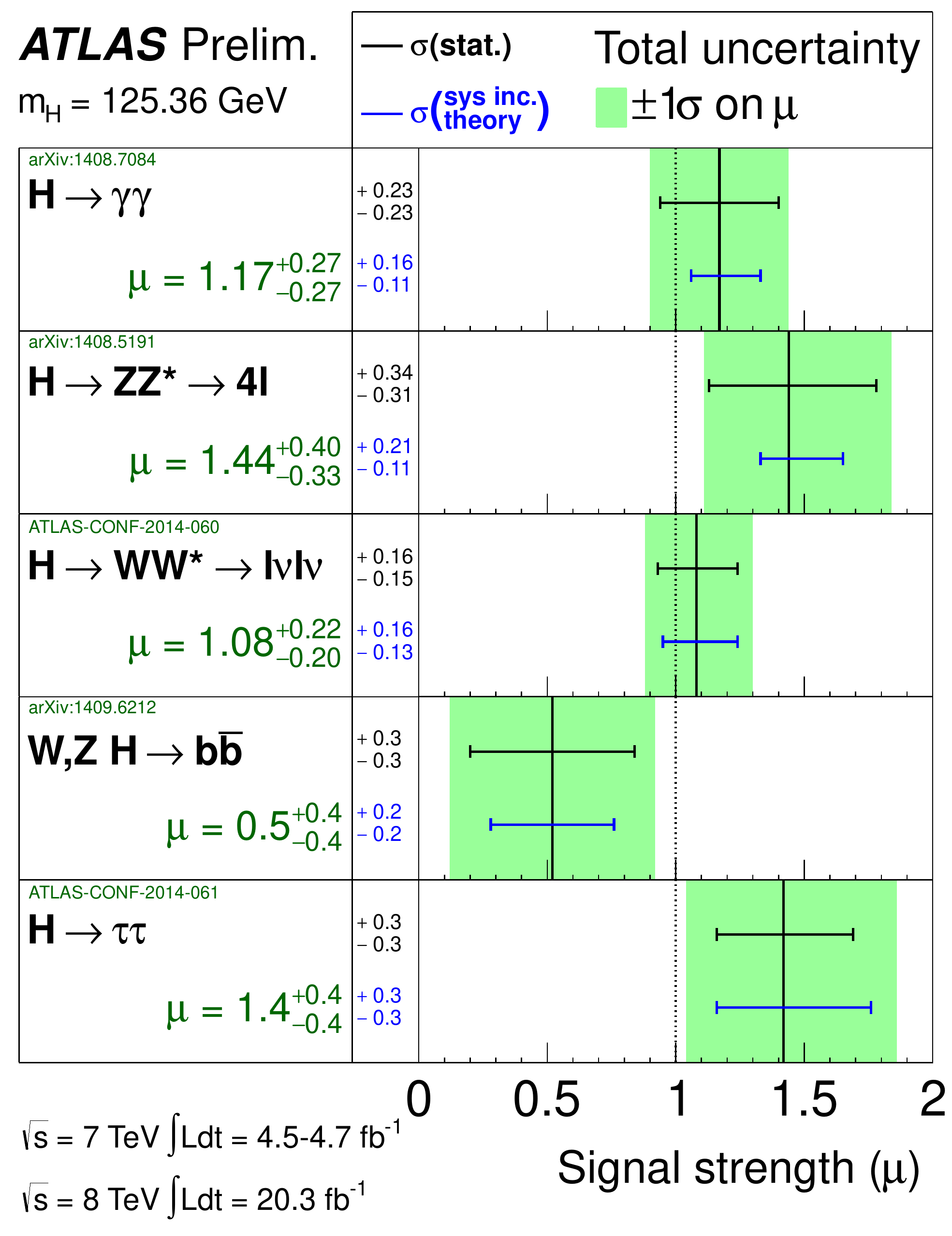} ~
\includegraphics[scale=0.4]{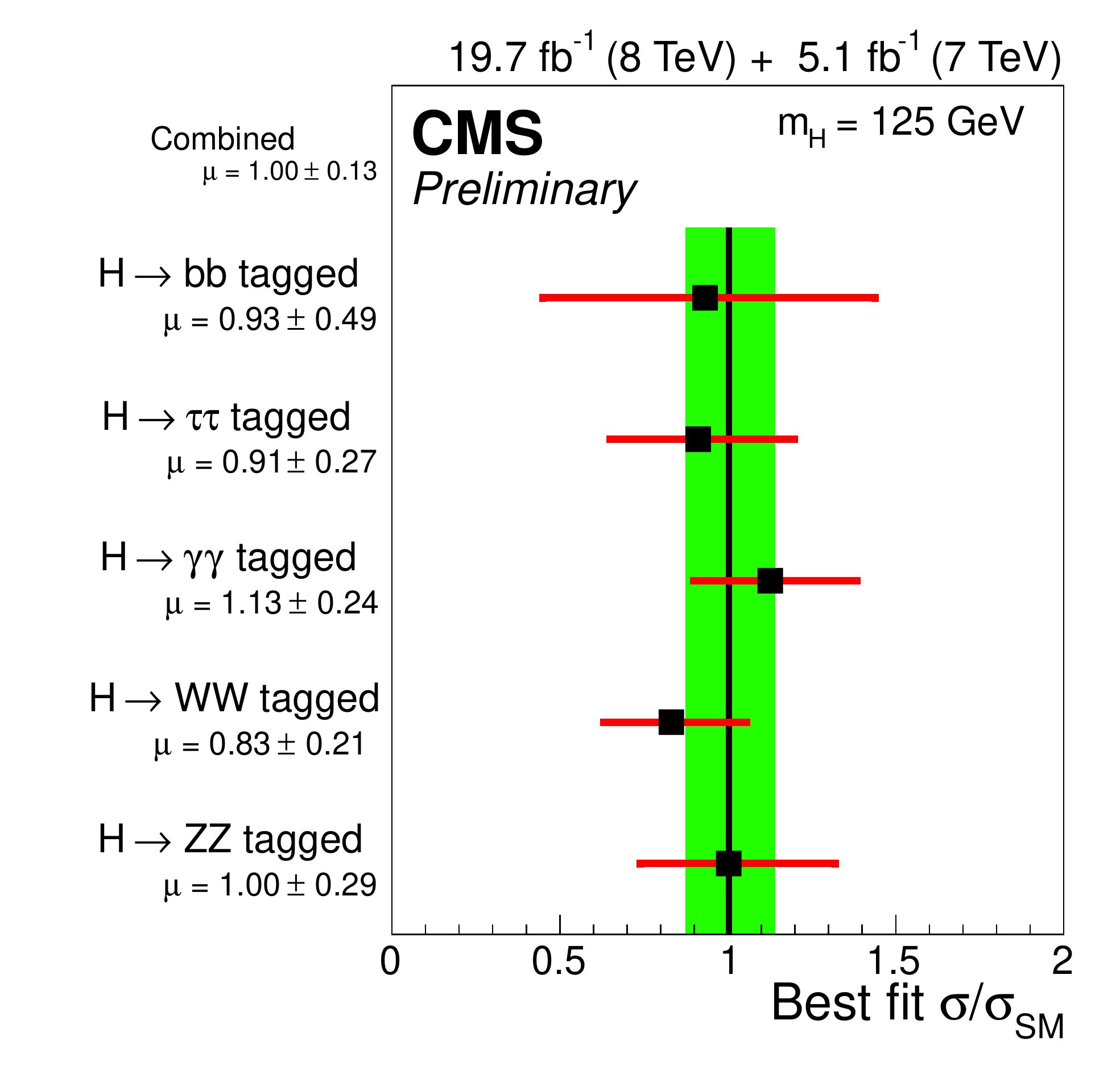}
\caption[Current values of the Higgs signal strengths]{\em Current measurements of the Higgs signal strengths into different channels by ATLAS \cite{Atlas:signal} and CMS \cite{cms:signal}.}
\label{signal_strengths}
\end{figure}

Now we come to the important question of how crucial this limit is in the context of current LHC Higgs data. Many global fit results in view of the recent data can be found in the literature\cite{Eberhardt:2013uba,Coleppa:2013dya,Chen:2013rba,Craig:2013hca,Dumont:2014wha,Bernon:2014vta}. In Fig.~\ref{current_fit} we choose to display the result of a recent analysis\cite{Bernon:2014vta}. The orange regions represent the 95\% CL allowed region from measurements of the Higgs signal strengths in various channels (See Fig.~\ref{signal_strengths}). Since the data is compatible with the SM prediction, one can easily see that the alignment limit is preferred. The horizontal widths of the allowed  regions reflect the present accuracy of measurements. In ref \cite{Dumont:2014wha}, it has also been projected how this region will shrink if future measurement continues to agree with the SM predictions with greater amount of accuracy. We can easily guess that if this is the case, we will continuously be pushed closer and closer to the alignment limit. Thus finding an alignment limit might be crucial for survival of the different BSM scenarios.

\subsection{Bounded from below constraints}
For this particular part, it might be convenient to work with the notation of \Eqn{notation1}. Here we try to derive the constraints on parameters $\beta_i$  such that the scalar potential, $V$, is bounded from below in any direction in the field space. It is sufficient to examine the quartic terms of the scalar potential (which we denote by $V_4$) because only this part of the potential will be dominant for large values of the field components of $\Phi_1$ and $\Phi_2$. We define $a=\Phi_1^\dagger\Phi_1$, $b=\Phi_2^\dagger\Phi_2$, $c= {\rm Re}~\Phi_1^\dagger\Phi_2$, $d={\rm Im}~\Phi_1^\dagger\Phi_2$ and note that
\begin{eqnarray}
ab \ge c^2+d^2 \,.
\end{eqnarray}
Using these definitions we can rewrite the quartic part of the scalar potential as follows~\cite{Gunion:2002zf}:
\begin{eqnarray}
V_4 &=& \frac{1}{2}\left(\sqrt{\beta_1}a-\sqrt{\beta_2}b \right)^2+ \left(\beta_3+\sqrt{\beta_1\beta_2}\right) \left(ab-c^2-d^2\right) +2\left(\beta_3+\beta_4+\sqrt{\beta_1\beta_2}\right)c^2 \nonumber \\
&& +\left({\rm Re}~\beta_5-\beta_3-\beta_4-\sqrt{\beta_1\beta_2} \right)\left(c^2-d^2\right) -2cd~ {\rm Im}~\beta_5 \,.
\label{quartic_stab}
\end{eqnarray}
Although we shall assume all the potential parameters to be real for our phenomenological studies, here we wish to keep things general because our arguments in this subsection do not depend on the reality of $\beta_5$. We have to ensure that $V_4$ never becomes infinitely negative in any direction of the field space, {\it i.e.}, for any choice of  8 independent field parameters (4 of $\Phi_1$ and 4 of $\Phi_2$). Note that, since $\Phi_1$ and $\Phi_2$ are two component column matrices, it is possible to choose arbitrary nonzero values for $a$ and $b$ even when we make $c=d=0$. But if $a$ and/or $b$ becomes zero, then $c=d=0$ for sure. Keeping these facts in mind, we now proceed to find the constraints for the potential to be bounded from below.
\begin{itemize}
\item Consider the field direction $b=0$ (and therefore $c=d=0$) and $a\to\infty$; then $V_4=\beta_1/2a^2$. So, $V_4$ is not largely negative requires
\begin{equation}
\beta_1 \ge 0\,. \label{stab1}
\end{equation}
\item Consider the field direction $a=0$ (and therefore $c=d=0$) and $b\to\infty$; then $V_4=\beta_2/2b^2$. So, $V_4$ is not largely negative requires
\begin{equation}
\beta_2 \ge 0 \,. \label{stab2}
\end{equation}
\item Consider the field direction along which $a=\sqrt{\beta_2/\beta_1}b$ (so that the first term in \Eqn{quartic_stab} vanishes) and $c=d=0$. In addition to this we go to large field values in that direction, {\it i.e.}, $a,~b\to\infty$. Then, $V_4=(\beta_3+\sqrt{\beta_1\beta_2})ab$. Now, as $a,~b >0$ by definition, the condition for the potential not to hit $-\infty$ becomes
\begin{equation}
\beta_3+\sqrt{\beta_1\beta_2} \ge 0 \,. \label{stab3}
\end{equation}
\item Again consider the field direction in which $a=\sqrt{\beta_2/\beta_1}b$ along with $ab=c^2+d^2$. Along this direction, $V_4$ is of the form
 \begin{subequations}
 \begin{eqnarray}
 V_4 &=& Pc^2+2Qcd+Rd^2 \,, \label{v4_4th} \\
 {\rm where,~~} P &=& {\rm Re}~\beta_5+\Lambda \,, \\
 Q &=& -{\rm Im}~\beta_5 \,, \\
 R &=& -{\rm Re}~\beta_5+\Lambda \,, \\
 {\rm with,}~~ \Lambda &=& \beta_3+\beta_4+\sqrt{\beta_1\beta_2} \,.
 \end{eqnarray}
 \end{subequations}
 Since $c$ and $d$ are still arbitrary, by choosing $d=0,~c\to\infty$ and $c=0,~d\to\infty$ successively, we require
  \begin{subequations}
  \label{Lamb_pre}
  \begin{eqnarray}
  P &=& {\rm Re}~\beta_5+\Lambda \ge 0 \,, \label{P_vac} \\
  R &=& -{\rm Re}~\beta_5+\Lambda \ge 0 \,, \\
  {\rm and~ hence,}~~ && \Lambda \ge 0 \,.
  \label{stab_aux}
  \end{eqnarray}
  \end{subequations}
\end{itemize}
To have another condition, let us recast \Eqn{v4_4th} into the following form~:
\begin{eqnarray}
V_4 &=& P\left(c+\frac{Q}{P}d\right)^2 + \left(R-\frac{Q^2}{P}\right)d^2 \,.
\label{v4recast}
\end{eqnarray}
We can now choose a direction along which $c=-Q/P~d$ with $d\to\infty$ so that we have the following condition~:
\begin{eqnarray}
R-\frac{Q^2}{P} > 0 ~~\Rightarrow PR>Q^2 \,.
\label{vac_final}
\end{eqnarray}
For the last step, remember that $P>0$ (\Eqn{P_vac}) so that we can multiply both sides by $P$ without flipping the inequality sign. After substituting for $P$, $Q$ and $R$ we get from \Eqn{vac_final}~:
 \begin{subequations}
 \begin{eqnarray}
&& \Lambda^2-\left({\rm Re}~\beta_5\right)^2 > \left({\rm Im}~\beta_5\right)^2  ~~\Rightarrow \Lambda^2 >|\beta_5|^2 \,, \\
&& \Lambda > |\beta_5| \,,
 \label{Lamb}
 \end{eqnarray}
 \end{subequations}
 where, in the last step we have used the fact that $\Lambda>0$ (\Eqn{stab_aux}). Since $|\beta_5|>\pm\beta_5,~0$, \Eqn{Lamb} puts  a stronger constrain on $\Lambda$ than \Eqn{Lamb_pre}. Therefore, substituting for $\Lambda$, \Eqn{Lamb} becomes
 \begin{eqnarray}
 \beta_3+\beta_4+\sqrt{\beta_1\beta_2} > |\beta_5| \,.
 \label{stab4}
 \end{eqnarray}
 We now collect Eqs.~(\ref{stab1}), (\ref{stab2}), (\ref{stab3}) and (\ref{stab4}) together and, using \Eqn{connections}, express them in terms of lambdas for later use~:
  \begin{subequations}
  \label{stability}
  \begin{eqnarray}
&&  \lambda_1+\lambda_3 >0 \,, \\
&&  \lambda_2+\lambda_3 >0 \,, \\
&& (2\lambda_3+\lambda_4) +2\sqrt{(\lambda_1+\lambda_3)(\lambda_2+\lambda_3)} >0 \,, \\
&& 2\lambda_3+\frac{\lambda_5+\lambda_6}{2} -\frac{|\lambda_5-\lambda_6|}{2}+2\sqrt{(\lambda_1+\lambda_3)(\lambda_2+\lambda_3)} >0 \,.
  \end{eqnarray}
  \end{subequations}
The way we have presented the derivation, it appears that these are only the {\em necessary} conditions for the potential to be bounded from below. But rigorous analysis\cite{Klimenko:1984qx, Maniatis:2006fs}  shows that these are indeed the sufficient conditions also.
\subsection{Constraints from unitarity}
As mentioned in Chapter~\ref{Chap1}, any scattering amplitude can be expanded in terms of the partial waves as follows:
\begin{equation}
\mathcal{M}(\theta) = 16 \pi\sum\limits_{\ell=0}^{\infty} a_\ell (2\ell+1)P_\ell (\cos\theta) \,,
\label{def:feynman}
\end{equation}
where, $\theta$ is the scattering angle and $P_\ell (x)$ is the Legendre polynomial of order $\ell$. The prescription is simple~: once we calculate the Feynman amplitude of a certain $2\to 2$ scattering process, each of the partial wave amplitude ($a_\ell$), in \Eqn{def:feynman}, can be extracted by using the orthonormality of the Legendre polynomials. 
In the context of SM, the pioneering work has been done by Lee, Quigg and Thacker (LQT) \cite{Lee:1977eg}. They have analyzed several two body scatterings involving longitudinal gauge bosons and physical Higgs in the SM. All such scattering amplitudes are proportional to Higgs quartic coupling in the high energy limit. The $\ell=0$ partial wave amplitude
$(a_0)$ is then extracted from these amplitudes and cast in the form of an S-matrix having different two-body states
as rows and columns. The largest eigenvalue of this matrix is bounded by the unitarity constraint, $|a_0 | < 1$.
This restricts the quartic Higgs self coupling and therefore the Higgs mass to a maximum value.

The procedure has been extended to the case of a 2HDM scalar potential \cite{Maalampi:1991fb,Kanemura:1993hm,Akeroyd:2000wc,Horejsi:2005da}. 
Here also same types of two body scattering channels are considered. Thanks to the equivalence theorem \cite{Pal:1994jk,Horejsi:1995jj}, we can use unphysical Higgses instead of actual longitudinal components of the gauge bosons when considering the
high energy limit. So, we can use the Goldstone-Higgs potential of \Eqn{notation2} for this analysis. Still it will be a
much involved calculation. But we notice that the diagrams containing trilinear vertices will be suppressed by
a factor of $E^2$ coming from the intermediate propagator. Thus they do not contribute at high energies, -- only
the quartic couplings contribute. Clearly the physical Higgs masses that could come from the propagators, do
not enter this analysis. Since we are interested only in the eigenvalues of the S-matrix, this allows us to work
with the original fields of \Eqn{notation2} instead of the physical mass eigenstates.

As already argued, only the dimensionless quartic couplings will contribute to the amplitudes under consideration at high energies. As a result, only $\ell=0$ partial amplitude ($a_0$) will receive nonzero contribution from the leading order term in the scattering amplitude. It is our purpose, then, to find the expressions of $a_0$ for every possible $2\to 2$ scattering process and cast them in the form of an S-matrix which is constructed by taking the different two-body channels as rows and columns. Unitarity will restrict the magnitude of each of the eigenvalues of this S-matrix to lie below unity. 

First important part of the calculation is to identify all the possible two-particle channels. These two-particle states are made of the fields $w_{k}^{\pm},~h_{k}$ and $z_{k}$ corresponding to the parametrization of \Eqn{phi}. For our calculation, we consider neutral two-particle states (e.g., $w_{i}^{+}w_{j}^{-},$ $h_i h_j,~z_i z_j,~h_i z_j$) and singly charged two-particle states (e.g., $w_{i}^{+}h_j,~w_{i}^{+}z_j$). In general, if we have $n$-number of doublets $\phi_{k}~(k=1,\ldots,n)$ there will be $(3n^2+n)$-number of neutral  and $2n^2$-number of charged two-particle states. Clearly, the dimensions of S-matries formed out of these two-particle states will be a $(3n^2+n)\times(3n^2+n)$  and $2n^2\times 2n^2$ for the neutral and charged cases respectively. The eigenvalues of these matrices should be bounded by the unitarity constraint.

We will exemplify these by considering the case of a 2HDM. In this case, the neutral channel S-matrix will be a $14\times 14$ matrix with the following two-particle states as rows and columns~:
\begin{eqnarray}
&& w_1^+w_1^-,~ w_2^+w_2^-,~w_1^+w_2^-,~w_2^+w_1^-,~\frac{h_1h_1}{\sqrt{2}},~\frac{z_1z_1}{\sqrt{2}}, \nonumber \\
&& \frac{h_2h_2}{\sqrt{2}},~\frac{z_2z_2}{\sqrt{2}},~ h_1z_2,~h_2z_1,~z_1z_2,~h_1h_2,~h_1z_1,~h_2z_2\,. \nonumber
\end{eqnarray}
The factor of $1/\sqrt{2}$ associated with the identical particle states arises due to Bose symmetry. In the most general case, finding the eigenvalues of the $14\times 14$ matrix would be a tedious job. But the potential of \Eqn{notation2} contains some obvious symmetries in its quartic terms. These symmetries will allow us to decompose the full matrix in smaller blocks. One must note that the quartic part of the potential always contain even number of indices, 1 or 2. Consequently a state $x_1y_1$ or $x_2y_2$ will always go into $x_1y_1$ or $x_2y_2$ but not into $x_1y_2$ or $x_2y_1$ and vice versa. Furthermore, CP symmetry is conserved. This implies, a neutral state with combination $h_ih_j$ or $z_iz_j$ will never go into $h_iz_j$. Keeping these facts in mind we can now decompose the S-matrix in the neutral sector into smaller blocks as follows~:
\begin{equation}
\mathcal{M}_{N}= \begin{pmatrix}
(\mathcal{M}_N^{11})_{6\times 6} & 0 & 0 \\ 0 & (\mathcal{M}_N^{11})_{2\times 2} & 0 \\ 0 & 0 & (\mathcal{M}_N^{12})_{6\times 6}
\end{pmatrix} \,.
\label{mat_neutral}
\end{equation} 
The submatrices are given below~:
%
\begin{subequations}
\begin{eqnarray}
(\mathcal{M}_N^{11})_{6\times 6}= \begin{pmatrix}
(\mathcal{A}_N^{11})_{3\times 3} & (\mathcal{B}_N^{11})_{3\times 3} \\ (\mathcal{B}_N^{11})^\dagger_{3\times 3} & (\mathcal{C}_N^{11})_{3\times 3} &
\end{pmatrix} \,,
\end{eqnarray}
where,
\begin{eqnarray}
(\mathcal{A}_N^{11})_{3\times 3} &=& 
\bordermatrix{
  &\m w_1^+ w_1^-&\m  w_2^+ w_2^-&\m \frac{z_1z_1}{\sqrt{2}}  \cr\vbox{\hrule}
\m w_1^+ w_1^- & 4(\l_1+\l_3) & 2\l_3+\frac{\l_5+\l_6}{2}  & \sqrt{2}(\l_1+\l_3) 
   \cr
\m w_2^+ w_2^-& 2\l_3+\frac{\l_5+\l_6}{2} & 4(\l_2+\l_3) & \sqrt{2}\left(\l_3+\frac{\l_4}{2}\right) 
   \cr
\m \frac{z_1z_1}{\sqrt{2}} &  \sqrt{2}(\l_1+\l_3) & \sqrt{2}\left(\l_3+\frac{\l_4}{2}\right) & 3(\l_1+\l_3) 
   \cr
   }\,,  \\
(\mathcal{B}_N^{11})_{3\times 3} &=& 
  \bordermatrix{
 &\m \frac{h_1h_1}{\sqrt{2}}&\m \frac{z_2z_2}{\sqrt{2}}&\m \frac{h_2h_2}{\sqrt{2}} \cr\vbox{\hrule}
\m w_1^+ w_1^-  & \sqrt{2}(\l_1+\l_3) & \sqrt{2}\left(\l_3+\frac{\l_4}{2}\right)  & \sqrt{2}\left(\l_3+\frac{\l_4}{2}\right)
   \cr
\m w_2^+ w_2^- & \sqrt{2}\left(\l_3+\frac{\l_4}{2}\right) &  \sqrt{2}(\l_2+\l_3) & \sqrt{2}(\l_2+\l_3)
   \cr
\m \frac{z_1z_1}{\sqrt{2}}  & (\l_1+\l_3) & \l_3+\frac{\l_5}{2} & \l_3+\frac{\l_6}{2}
   \cr
   }\,, \\
(\mathcal{C}_N^{11})_{3\times 3} &=& 
  \bordermatrix{
 &\m \frac{h_1h_1}{\sqrt{2}}&\m \frac{z_2z_2}{\sqrt{2}}&\m \frac{h_2h_2}{\sqrt{2}} \cr\vbox{\hrule}
\m \frac{h_1h_1}{\sqrt{2}} & 3(\l_1+\l_3) & \l_3+\frac{\l_6}{2} & \l_3+\frac{\l_5}{2} 
   \cr 
\m \frac{z_2z_2}{\sqrt{2}} & \l_3+\frac{\l_6}{2} & 3(\l_2+\l_3)  & (\l_2+\l_3)  
   \cr
\m \frac{h_2h_2}{\sqrt{2}} & \l_3+\frac{\l_5}{2} & (\l_2+\l_3)  & 3(\l_2+\l_3)  \cr
   }\,.
\end{eqnarray}
\begin{eqnarray}
 (\mathcal{M}_N^{11})_{2\times 2} =
  \bordermatrix{
  &\m h_1z_1 &\m  h_2z_2  \cr\vbox{\hrule}
\m h_1z_1 & 2(\l_1+\l_3) & \frac{\l_5-\l_6}{2}
   \cr
\m h_2z_2 & \frac{\l_5-\l_6}{2} & 2(\l_2+\l_3)
   \cr
   }\,, 
\end{eqnarray}
\begin{eqnarray}
(\mathcal{M}_N^{12})_{6\times 6}= \begin{pmatrix}
(\mathcal{A}_N^{12})_{3\times 3} & (\mathcal{B}_N^{12})_{3\times 3} \\ (\mathcal{B}_N^{12})^\dagger_{3\times 3} & (\mathcal{C}_N^{12})_{3\times 3} &
\end{pmatrix} \,,
\end{eqnarray}
where,
\begin{eqnarray}
(\mathcal{A}_N^{12})_{3\times 3} &=& 
\bordermatrix{
  &\m w_1^+ w_2^-&\m  w_2^+ w_1^-&\m h_1z_2  \cr\vbox{\hrule}
\m w_1^+ w_2^- & 2\l_3+\frac{\l_5+\l_6}{2} & \l_5-\l_6  & -\frac{i}{2}(\l_4-\l_6) 
   \cr
\m w_2^+ w_1^-& \l_5-\l_6 & 2\l_3+\frac{\l_5+\l_6}{2} & \frac{i}{2}(\l_4-\l_6) 
   \cr
\m h_1z_2 & \frac{i}{2}(\l_4-\l_6)  & -\frac{i}{2}(\l_4-\l_6)  & 2\l_3+\l_6 
   \cr
 }\,,  \\
(\mathcal{B}_N^{12})_{3\times 3} &=& 
\bordermatrix{
  &\m h_2z_1 &\m z_1z_2 &\m h_1h_2 \cr\vbox{\hrule}
\m w_1^+ w_2^-  & \frac{i}{2}(\l_4-\l_6) & \frac{\l_5-\l_4}{2} & \frac{\l_5-\l_4}{2}
   \cr
\m w_2^+ w_1^- &-\frac{i}{2}(\l_4-\l_6)  & \frac{\l_5-\l_4}{2} &  \frac{\l_5-\l_4}{2}
   \cr
\m h_1z_2  & \frac{\l_5-\l_6}{2} & 0 & 0
   \cr
 }\,, \\
(\mathcal{C}_N^{12})_{3\times 3} &=& 
\bordermatrix{
 &\m h_2z_1 &\m z_1z_2 &\m h_1h_2 \cr\vbox{\hrule}
\m h_2z_1 & 2\l_3+\l_6  & 0 & 0
   \cr 
\m z_1z_2  & 0 & 2\l_3+\l_5  & \frac{\l_5-\l_6}{2}  
   \cr
\m h_1h_2 & 0 & \frac{\l_5-\l_6}{2}  & 2\l_3+\l_5  \cr
   }\,.
\end{eqnarray}
\end{subequations}
\normalsize
The same exercise can be repeated for the charged two-particle states. For the singly charged sector, it will be a $8\times 8$ matrix which will take the following block diagonal form~:
\begin{equation}
\mathcal{M}_{C}= \begin{pmatrix}
(\mathcal{M}_C^{11})_{4\times 4} & 0 \\ 0 & (\mathcal{M}_C^{12})_{4\times 4} 
\end{pmatrix} \,.
\label{mat_charged}
\end{equation} 
The submatrices are given below~:
 \begin{subequations}
 \begin{eqnarray}
  (\mathcal{M}_C^{11})_{4\times 4} &=&
   \bordermatrix{
   &\m h_1w_1^+ &\m  h_2w_2^+ &\m z_1w_1^+ &\m z_2w_2^+  \cr\vbox{\hrule}
 \m h_1w_1^+ & 2(\l_1+\l_3) & \frac{\l_5-\l_4}{2} & 0 & -\frac{i}{2}(\l_4-\l_6)
    \cr
 \m h_2w_2^+ & \frac{\l_5-\l_4}{2} & 2(\l_2+\l_3) & -\frac{i}{2}(\l_4-\l_6) & 0
    \cr
 \m z_1w_1^+ & 0 & \frac{i}{2}(\l_4-\l_6) & 2(\l_1+\l_3) & \frac{\l_5-\l_4}{2} 
  \cr 
 \m z_2w_2^+  & \frac{i}{2}(\l_4-\l_6)  & 0 & \frac{\l_5-\l_4}{2} & 2(\l_2+\l_3) \cr 
    }\,,  \\
  (\mathcal{M}_C^{12})_{4\times 4} &=&
   \bordermatrix{
   &\m h_1w_2^+ &\m  h_2w_1^+ &\m z_1w_2^+ &\m z_2w_1^+  \cr\vbox{\hrule}
 \m h_1w_2^+ & 2\l_3+\l_4 & \frac{\l_5-\l_4}{2} & 0 & \frac{i}{2}(\l_4-\l_6)
    \cr
 \m h_2w_1^+ & \frac{\l_5-\l_4}{2} & 2\l_3+\l_4 & \frac{i}{2}(\l_4-\l_6) & 0
    \cr
 \m z_1w_2^+ & 0 & -\frac{i}{2}(\l_4-\l_6) & 2\l_3+\l_4 & \frac{\l_5-\l_4}{2} 
  \cr 
 \m z_2w_1^+  & -\frac{i}{2}(\l_4-\l_6)  & 0 & \frac{\l_5-\l_4}{2} & 2\l_3+\l_4 \cr 
    }\,.
 \end{eqnarray}
 \end{subequations}
The eigenvalues for these matrices are listed below~:
\begin{itemize}
\item $({\mathcal{M}_N^{11}})_{6\times 6}$~: $a_1^\pm,~a_2^\pm,~a_3^\pm$.
\item $({\mathcal{M}_N^{11}})_{2\times 2}$~: $a_3^\pm$.
\item $({\mathcal{M}_N^{12}})_{6\times 6}$~: $b_1,~b_2,~b_3,~b_4,~b_5$ with $b_5$ twofold degenerate.
\item $({\mathcal{M}_C^{11}})_{4\times 4}$~: $a_2^\pm,~a_3^\pm$.
\item $({\mathcal{M}_C^{12}})_{4\times 4}$~: $b_2,~b_4,~b_5,~b_6$.
\end{itemize}
We also enlist below the explicit expressions for these eigenvalues:
 \begin{subequations}
 \begin{eqnarray}
 a_1^\pm &=& 3(\l_1+\l_2+2\l_3) \pm \sqrt{9(\l_1-\l_2)^2+ \left(4\l_3+\l_4+\frac{\l_5+\l_6}{2} \right)^2} \,, \\
 a_2^\pm &=& (\l_1+\l_2+2\l_3) \pm \sqrt{(\l_1-\l_2)^2+\frac{1}{4} \left(2\l_4-\l_5-\l_6 \right)^2} \,, \\
 a_3^\pm &=& (\l_1+\l_2+2\l_3) \pm \sqrt{(\l_1-\l_2)^2+\frac{1}{4} \left(\l_5-\l_6 \right)^2} \,, \\
 b_1 &=& 2\l_3-\l_4-\frac{1}{2}\l_5+\frac{5}{2}\l_6 \,, \\
 b_2 &=& 2\l_3+\l_4-\frac{1}{2}\l_5+\frac{1}{2}\l_6 \,, \\
 b_3 &=& 2\l_3-\l_4+\frac{5}{2}\l_5-\frac{1}{2}\l_6 \,, \\
 b_4 &=& 2\l_3+\l_4+\frac{1}{2}\l_5-\frac{1}{2}\l_6 \,, \\
 b_5 &=& 2\l_3+\frac{1}{2}\l_5+\frac{1}{2}\l_6 \,, \\
 b_6 &=& 2(\l_3+\l_4)-\frac{1}{2}\l_5-\frac{1}{2}\l_6 \,.
 \end{eqnarray}
 \label{unieigenvalues}
 \end{subequations}
Each of the above eigenvalues should be bounded from the unitarity constraint as 
\begin{eqnarray}
|a_i^\pm|,~|b_i| \le 16\pi \,.
\end{eqnarray}

\subsection{Numerical constraints on the scalar masses in the alignment limit}
Next important thing is to investigate the implications of these conditions on the physical scalar masses especially the nonstandard ones.
\begin{figure}
\includegraphics[scale=1]{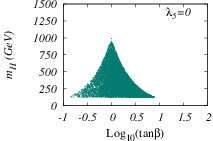}~~
\includegraphics[scale=1]{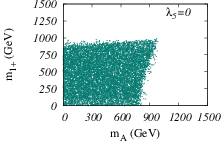}
\caption[Unitarity and stability constraints for 2HDMs with exact $Z_2$ symmetry]{\em Allowed region from unitarity and stability for exact $Z_2$ symmetry. The figures have been taken from \cite{Das:2015qva}.}
\label{f:uniZ2}
\end{figure}
Fig.~\ref{f:uniZ2} shows the region allowed by the combined constraints coming from unitarity and boundedness of the potential for the case $\lambda_5=0$, {\it i.e.}, exact $Z_2$ symmetry. Noteworthy features are listed below~:
\begin{itemize}
\item From the left panel, we can read the limit on $\tan\beta$, $1/8<\tan\beta<8$.
\item Limits on the masses are, $m_H,~m_A,~m_{1+}<1$ TeV.
\end{itemize}
The reason for the above mentioned bounds can be traced back to the eigenvalues of \Eqn{unieigenvalues}. First two constraints for boundedness in \Eqn{stability} can be combined into
\begin{eqnarray}
\l_1+\l_2+2\l_3 > 0 \,.
\end{eqnarray}
This, then together with the condition $|a_1^\pm| < 16\pi$, implies
\begin{eqnarray}
&& 0< \l_1+\l_2+2\l_3 < \frac{16\pi}{3} \,, \\
\Rightarrow && 0 < \left( m_H^2 -\frac{1}{2}\l_5v^2 \right)(\tan^2\beta+\cot^2\beta) +2m_h^2 <\frac{32\pi v^2}{3} \,,
\label{unitb}
\end{eqnarray}
where the last expression is obtained from the previous one by using \Eqn{inv2HDM} in the alignment limit. Since $m_H>125$ GeV, this will put a limit on $\tan\beta$ (as well as $\cot\beta$) when $\lambda_5=0$. Since the minimum value of ($\tan^2\beta+\cot^2\beta$) is 2 when $\tan\beta=1$, the maximum possible value of $m_H$ occurs at $\tan\beta=1$. In summary, \Eqn{unitb} explains the $\tan\beta$ dependent bound on $m_H$ as depicted in the left panel of Fig.~\ref{f:uniZ2}.

\Eqn{unitb} also implies that the restriction on $\tan\beta$ will be lifted for $1/2\l_5 v^2 > (m_H^2)_{\rm min} = (125~{\rm GeV})^2$. Once this condition is satisfied, $m_H^2$ will have the chance to saturate to $1/2\l_5 v^2$ making the difference between them to vanish in \Eqn{unitb}. In fact, to a very good approximation, one can use 
\begin{eqnarray}
m_H^2 \approx 1/2\l_5v^2
\label{corr}
\end{eqnarray}
 for $\tan\beta > 5$.

To understand the restrictions on $m_A$ and $m_{1+}$, we use the triangle inequality to note the following:
 \begin{subequations}
 \begin{eqnarray}
 |b_1-b_3| \equiv 3|\l_6-\l_5| < 32\pi \,, && \Rightarrow~ |m_A^2-\frac{1}{2}\l_5v^2| < \frac{16\pi v^2}{3} \,, \label{limma} \\
 |b_6-b_3| \equiv 3|\l_4-\l_5| < 32\pi \,, && \Rightarrow~ |m_{1+}^2-\frac{1}{2}\l_5v^2| < \frac{16\pi v^2}{3} \,.
 \label{limm1+}
 \end{eqnarray}
 \end{subequations}
Because of Eqs.~(\ref{limma}) and (\ref{limm1+}) we expect to put limits on $m_A$ and $m_{1+}$ respectively, when $\l_5=0$. Additionally, note that due to the inequality 
\begin{eqnarray}
|b_1-b_6| \equiv 3|\l_6-\l_4| < 32\pi \,, && \Rightarrow~ |m_A^2-m_{1+}^2| < \frac{16\pi v^2}{3} \,,
\label{diff}
\end{eqnarray}
we expect the splitting between $m_A$ and $m_{1+}$ to be {\em always} restricted in a 2HDM. It is also interesting to note that the conclusions obtained from Eqs.~(\ref{limma}), (\ref{limm1+}) and (\ref{diff}) {\em do not} depend on the imposition of the alignment condition.

\begin{figure}
\includegraphics[scale=1]{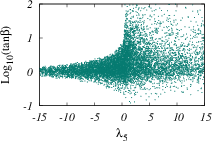}~~
\includegraphics[scale=1]{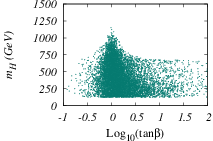}
\caption[Relaxation of the constraints on $\tan\beta$]{\em  Relaxation of the unitarity and stability constraints on $\tan\beta$ in the presence of soft breaking. In the right panel, the vertical width of the of the tail in the region where $\tb$ is much away from unity is caused by the variation of $\l_5$ in the range $[-15,15]$. The figures have been taken from \cite{Das:2015qva}.}
\label{f:l5tb}
\end{figure}

\begin{figure}
\centering
\includegraphics[scale=0.2]{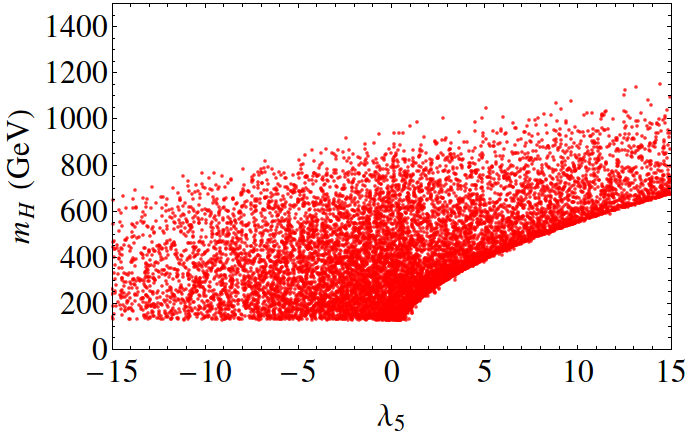}~~
\includegraphics[scale=0.2]{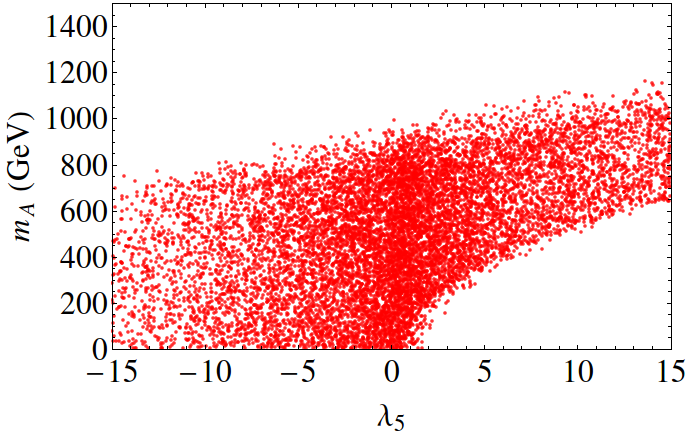}~~
\includegraphics[scale=0.2]{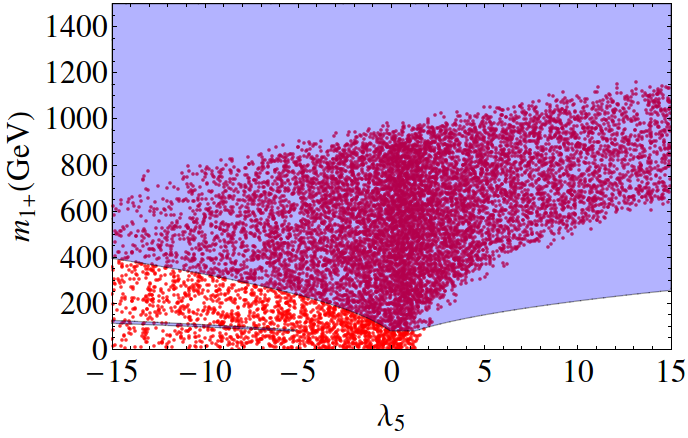}
\caption[Effect of soft breaking on the constraints on the nonstandard masses]{\em Effect of soft breaking on the constraints on the nonstandard masses. The (light blue) shaded region in the rightmost panel represents the combined allowed region from direct search and the diphoton signal strength at 95\% C.L. The figures have been taken from \cite{Das:2015qva}.}
\label{f:l5mass}
\end{figure}

Next we shall investigate the implications of the soft breaking parameter on these constraints. We have varied $\l_5$ in the range $[-15,15]$ for this purpose. From \Eqn{unitb} one can observe that the space for $\tan\beta$ is squeezed further if $\l_5 <0$ but the bound is relaxed if $\l_5>0$. This feature emerges from the left panel of Fig.~\ref{f:l5tb}. One can also see from \Eqn{unitb} that $m_H^2$ must follow $1/2\l_5v^2$ if $\tan\beta$ moderately deviates from unity. This feature is reflected by the horizontal tail in the right panel of Fig.~\ref{f:l5tb} on both sides of the peak. The vertical width of the tail is caused by the variation of $\l_5$ in the range $[-15, 15]$. On the other hand, from Eqs.~(\ref{limma}), (\ref{limm1+}) and (\ref{unitb}), it should be noted that the upper bounds on the nonstandard scalar masses will be relaxed for $\l_5>0$ but will get tighter for $\l_5<0$. Fig.~\ref{f:l5mass} reflects these features where one can see that this dependence is rather weak.

It is also important to note that the production as well as the tree-level decay widths of $h$ remain unaltered from the corresponding SM expectations due to the imposition of alignment limit of \Eqn{alignment}. But the loop induced decay modes of $h$, such as $h\to\gamma\gamma$ and $h\to Z\gamma$, will pick up additional contributions due the presence of the charged scalar in loops. For example, the diphoton signal strength ($\mu_{\gamma\gamma}$), in general, depends on both $\l_5$ and $m_{1+}$\cite{Bhattacharyya:2014oka}. The current measurement by CMS gives $\mu_{\gamma\gamma}=1.14^{+0.26}_{-0.23}$\cite{Khachatryan:2014ira}, whereas ATLAS measures $\mu_{\gamma\gamma}$ to be $1.17\pm 0.27$\cite{Aad:2014eha}. In addition to this, the direct search limit of $m_{1+}>80$~GeV\cite{Searches:2001ac} should also be taken into account. Considering all of these experimental constraints, the allowed region at 95\% C.L. has been shaded (in light blue) in the rightmost panel of Fig.~\ref{f:l5mass}. Only those points that lie within the shaded region survive both the theoretical and experimental constraints.

\begin{figure}
\rotatebox{90}{\quad\quad\quad\quad$m_A$ (GeV)}
\includegraphics[scale=0.6]{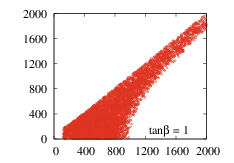} 
\includegraphics[scale=0.6]{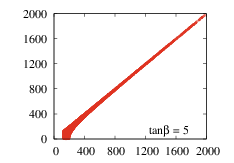}
\includegraphics[scale=0.6]{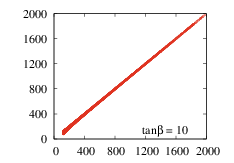}
\centerline{ \null \hfill $m_H$ (GeV) \quad}

\rotatebox{90}{\quad\quad\quad\quad$m_{1+}$ (GeV)}
\includegraphics[scale=0.6]{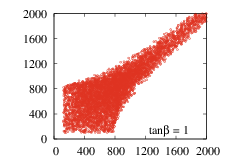} 
\includegraphics[scale=0.6]{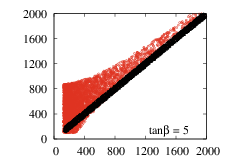}
\includegraphics[scale=0.6]{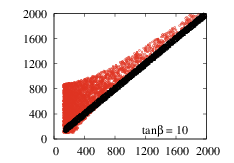}
\centerline{ \null \hfill $m_H$ (GeV) \quad}

\rotatebox{90}{\quad\quad\quad\quad$m_{1+}$ (GeV)}
\includegraphics[scale=0.6]{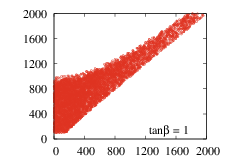} 
\includegraphics[scale=0.6]{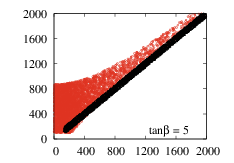}
\includegraphics[scale=0.6]{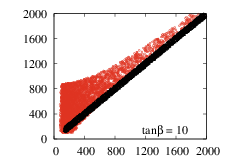}
\centerline{ \null \hfill $m_A$ (GeV) \quad}

\caption[Unitarity and stability constraints for 2HDMs with U(1) symmetry]{\em 2HDM potential with softly broken U(1) symmetry: regions allowed in $m_H$-$m_A$, $m_H$-$m_{1+}$ and
  $m_A$-$m_{1+}$ planes from unitarity and stability (red points), and
  from $T$-parameter (black points), for three choices of
  $\tan\beta$.  The plots have been taken from \cite{Bhattacharyya:2013rya} where $m_{1+} > 100$\,GeV was assumed to respect LEP direct search bound \cite{Searches:2001ac}.}
\label{f:uniu1}
\end{figure}

An interesting alternative arises if, instead of $Z_2$, one imposes an U(1) symmetry under which $\Phi_2\to e^{i\alpha}\Phi_2$. This U(1) symmetry needs to be broken softly to forbid the appearance of a massless pseudoscalar. This symmetry will imply $\beta_6=0$ in \Eqn{notation1} or $\l_5=\l_6$ in \Eqn{notation2}. Thus,  the soft breaking parameter now gets related to the pseudoscalar mass as $m_A^2=1/2\l_5v^2$. Consequently, the correlation between $m_H$ and $\l_5$ in the leftmost panel of Fig.~\ref{f:l5mass} transforms into the degeneracy between $m_H$ and $m_A$ 
\cite{Bhattacharyya:2013rya}. The constraints on the scalar masses imposed by the stability and unitarity conditions in \Eqs{stability}{unieigenvalues} have been plotted in Fig.~\ref{f:uniu1} for $\tan\beta=1$, 5 and 10 by performing random scan over all non-standard scalar masses.The following salient features emerge from the
plots.

\begin{itemize}
\item There is a correlation between $m_A$ and $m_H$ which gets
  stronger for larger values of $\tan\beta$, to the extent that they
  become nearly degenerate once  $\tan\beta$ crosses 10.  To
  understand this, we note that for $\l_5=\l_6$, \Eqn{unitb} reduces
  to the following form
\begin{eqnarray}
0 \leq (m_H^2-m_A^2)(\tan^2\beta+\cot^2\beta)+2m_h^2 
\leq {\frac{32\pi v^2}{3}} \,.
\label{uniM1}
\end{eqnarray}
Clearly, for $\tan\beta$ away from unity, $H$ and $A$ are almost
degenerate.

\item There is a similar correlation between $m_H$ and $m_{1+}$, but
  this time without any dependence on $\tan\beta$.  This can again be
  seen from the inequalities of \Eqs{limma}{limm1+} keeping in mind that now $m_A^2=1/2\l_5v^2$.

\item As regards the non-standard scalars, the unitarity conditions
  essentially apply on the difference of their squared masses.  Thus,
  any individual mass can be arbitrarily large without affecting the
  unitarity conditions.  This conclusion crucially depends on the
  existence of a U(1) symmetry of the potential.  When the symmetry of
  the potential is only a discrete $Z_2$, considerations of unitarity
  do restrict the individual non-standard masses as has already been demonstrated.

\item We note at this point that the splitting between the heavy scalar
masses is also constrained by the oblique electroweak $T$-parameter.
In the present case, the expression of the $T$-parameter in the
alignment limit is given by~\cite{He:2001tp, Grimus:2007if}
\begin{eqnarray}
T = {1 \over 16\pi \sin^2 \theta_w M_W^2} \Big[ F(m_{1+}^2,m_H^2) +
  F(m_{1+}^2, m_A^2) - F(m_H^2,m_A^2) \Big] \,,
\label{T}
\end{eqnarray}
with
\begin{eqnarray}
F(x,y) = {x+y \over 2} - {xy \over x-y} \; \ln (x/y) \,.
\end{eqnarray}
The new physics contribution to the $T$-parameter can be found to be~\cite{Baak:2013ppa} 
\begin{eqnarray}
T = 0.05 \pm 0.12 \,.
\end{eqnarray}

To provide an intuitive feel on the constraints from the
  $T$-parameter, we assume $m_H=m_A$, which is anyway dictated by the
  unitarity constraints for $\tan\beta$ somewhat away from unity.  It
  then follows from \Eqn{T} that the splitting between $m_{1+}$ and
  $m_H$ is approximately 50\,GeV, for $|m_{1+}-m_H| \ll m_{1+},m_H$.  It
  turns out from Fig.~\ref{f:uniu1} that the constraints from the
  $T$-parameter are stronger than that from unitarity and stability.

  For $\tan\beta=1$, unitarity and stability do not compel $m_H$ and
  $m_A$ to be very close.  In this case, the $T$-parameter cannot give
  any definitive constraints in the planes of the heavy scalar masses,
  unlike the unitarity and stability constraints.  For this reason, we
  have shown only the latter constraints in Fig.~\ref{f:uniu1} for
  $\tan\beta=1$.

\item Thus, for moderate or large $\tan\beta$, the unitarity and
  stability constraints, together with the constraints coming from the
  $T$-parameter, imply that all three heavy scalar states are nearly
  degenerate in the alignment limit.

\end{itemize} 
\section{Yukawa part}
We shall start by proving the assertion made below \Eqn{H0gen} that $H^0$ carries SM-like Yukawa couplings too. The most general Yukawa interaction for $n$ Higgs-doublet model can be written as:
\begin{eqnarray}
\ml_{Y} = -\sum\limits_{j=1}^{n} \left[\bar{Q}_L\Gamma_j\Phi_jn_R +\bar{Q}_L\Delta_j\tilde{\Phi}_jn_R  \right] + {\rm h.c.}
\label{yukawa-n}
\end{eqnarray}
Here $Q_L=(p_L~n_L)^T$ is the left-handed quark doublet and, $p_R$ and $n_R$ are up- and down-type singlets respectively. In writing the Yukawa Lagrangian, we have suppressed the flavor indices. $\Gamma_j$ and $\Delta_j$ are actually $3\times 3$ Yukawa matrices in the down- and up-sectors respectively. After spontaneous symmetry breaking
\begin{eqnarray}
\Phi_j =\frac{1}{\sqrt{2}} \begin{pmatrix} \sqrt{2} w_j^+ \\ (h_j+v_j) +iz_j \end{pmatrix} \,,~~
\tilde{\Phi}_j =i\sigma_2\Phi_2^* =\frac{1}{\sqrt{2}} \begin{pmatrix} (h_j+v_j) -iz_j \\ -\sqrt{2} w_j^- \end{pmatrix} \,.
\label{phitil}
\end{eqnarray}
Hence the mass-matrices take the following form in the gauge basis:
 \begin{subequations}
 \label{massmat}
 \begin{eqnarray}
 M_n &=& \frac{1}{\sqrt{2}}\sum\limits_{j=1}^{n} v_j\Gamma_j \,, \\
 M_p &=& \frac{1}{\sqrt{2}}\sum\limits_{j=1}^{n} v_j\Delta_j \,.
 \end{eqnarray}
 \end{subequations}
The diagonal mass matrices can be obtained via the following biunitary transformations:
 \begin{subequations}
 \label{biunitary}
 \begin{eqnarray}
&& D_d = U_L^\dagger\cdot M_n\cdot U_R = {\rm diag}(m_d,~m_s, ~m_b) \,, \\
&& D_u = V_L^\dagger\cdot M_p\cdot V_R = {\rm diag}(m_u,~m_c, ~m_t) \,.
 \end{eqnarray}
 \end{subequations}
The matrices, $U$ and $V$ relates the quark fields in the gauge basis to those in the mass basis:
 \begin{subequations}
 \label{bibasis}
 \begin{eqnarray}
n_L = U_Ld_L\,,~~ && n_R=U_Rd_R\,; \\
p_L = V_Lu_L\,,~~ && p_R=V_Ru_R\,.
 \end{eqnarray}
 \end{subequations}
Clearly, the CP-even Yukawa interaction arising from \Eqn{yukawa-n} becomes
\begin{eqnarray}
\ml_Y^{\rm CP even} = -\frac{1}{\sqrt{2}}\bar{n}_L\left(\sum\limits_{j=1}^{n}\Gamma_jh_j \right)n_R
-\frac{1}{\sqrt{2}}\bar{p}_L\left(\sum\limits_{j=1}^{n}\Delta_jh_j \right)p_R + {\rm h.c.}
\label{genY}
\end{eqnarray}
Let us now rotate the $\{h_1,~h_2,\dots h_n \}$ basis, via an orthogonal transformation, to a new basis containing the state $H^0$ which has been defined in \Eqn{H0gen}. The transformation will look like
\begin{eqnarray}
&& \begin{pmatrix} H^0 \\ \checked \\ \checked \\ \vdots \end{pmatrix} = \frac{1}{v}
\begin{pmatrix} v_1 & v_2 & \dots & v_n \\ \checked & \checked & \dots & \checked \\ \checked & \checked & \dots & \checked \\ \vdots & \vdots & \vdots & \vdots \end{pmatrix}
\begin{pmatrix} h_1 \\ h_2 \\ \vdots \\ h_n \end{pmatrix} \,, \\
&\Rightarrow & \begin{pmatrix} h_1 \\ h_2 \\ \vdots \\ h_n \end{pmatrix}= \frac{1}{v}
\begin{pmatrix} v_1 & \checked & \checked & \dots \\ v_2 & \checked & \checked & \dots \\ \vdots & \vdots & \vdots & \vdots \\ v_n & \checked & \checked & \dots \end{pmatrix}  \begin{pmatrix} H^0 \\ \checked \\ \checked \\ \vdots \end{pmatrix}
\label{invH0}
\end{eqnarray}
Using \Eqn{invH0}, we can extract the Yukawa coupling of $H^0$ from \Eqn{genY} as follows~:
\begin{eqnarray}
\ml_Y^{H^0} &=& \frac{H^0}{v}\left[-\frac{1}{\sqrt{2}}\bar{n}_L\left(\sum\limits_{j=1}^{n}\Gamma_jv_j \right)n_R
-\frac{1}{\sqrt{2}}\bar{p}_L\left(\sum\limits_{j=1}^{n}\Delta_jv_j \right)p_R\right] + {\rm h.c.} \\
&=& -\frac{H^0}{v}\left[\bar{n}_LM_nn_R +\bar{p}_LM_pp_R\right] + {\rm h.c.}
\label{gbasis}
\end{eqnarray}
Using \Eqs{biunitary}{bibasis} we can rewrite the coupling of \Eqn{gbasis} in the mass basis as
\begin{eqnarray}
\ml_Y^{H^0} =-\frac{H^0}{v}\left[\bar{d}_LD_dd_R +\bar{u}_LD_uu_R\right] + {\rm h.c.} \equiv -\frac{H^0}{v}\left[\bar{d}D_dd +\bar{u}D_uu\right] \,,
\label{SM-like}
\end{eqnarray}
where, the last step follows from the fact that $D_{u,d}$ are diagonal. \Eqn{SM-like} shows that $H^0$, by construction, possesses SM-like Yukawa couplings with the SM fermions.

\subsection{The problem of FCNC }
The general Yukawa Lagrangian for the 2HDM case is given by \Eqn{yukawa-n} with $n=2$. Following the definitions of \Eqs{H0R-h1h2}{phitil} we can take out the Yukawa coupling for the CP even part~:
\begin{eqnarray}
\ml_{\rm 2HDM}^{\rm CP~even} &=& -\frac{H^0}{v}\left\{\bar{n}_L\left[\frac{1}{\sqrt{2}}\left(\Gamma_1v_1+\Gamma_2v_2 \right) \right]n_R + \bar{p}_L\left[\frac{1}{\sqrt{2}}\left(\Delta_1v_1+\Delta_2v_2 \right) \right]p_R \right\} \nonumber \\
&&+ \frac{R}{v}\left\{\bar{n}_L\left[\frac{1}{\sqrt{2}}\left(\Gamma_1v_2-\Gamma_2v_1 \right) \right]n_R + \bar{p}_L\left[\frac{1}{\sqrt{2}}\left(\Delta_1v_2-\Delta_2v_1 \right) \right]p_R \right\} \\
&=& -\frac{H^0}{v}\left(\bar{d}D_dd+\bar{u}D_uu \right) \nonumber \\
&& +\frac{R}{v} \left\{\bar{d}\left(N_dP_R+N_d^\dagger P_L \right)d +\bar{u}\left(N_uP_R+N_u^\dagger P_L \right)u \right\} \,.
\label{CP even}
\end{eqnarray}
In \Eqn{CP even}, $N_{u,d}$ are defined as follows~:
 \begin{subequations}
 \label{Nudgen}
 \begin{eqnarray}
 N_d = \frac{1}{\sqrt{2}} U_L^\dagger (\Gamma_1v_2-\Gamma_2v_1) U_R &=& \frac{v_2}{v_1}D_d-\frac{v_2}{\sqrt{2}} \left(\frac{v_2}{v_1}+\frac{v_1}{v_2} \right) U_L^\dagger \Gamma_2 U_R \,, \\
 \label{Ndgen}
 N_u = \frac{1}{\sqrt{2}} V_L^\dagger (\Delta_1v_2-\Delta_2v_1) V_R &=& \frac{v_2}{v_1}D_u-\frac{v_2}{\sqrt{2}} \left(\frac{v_2}{v_1}+\frac{v_1}{v_2} \right) V_L^\dagger \Delta_2 V_R \,.
 \label{Nugen}
 \end{eqnarray}
 \end{subequations}
In writing \Eqn{Nudgen} we have made use of \Eqs{massmat}{biunitary} with $n=2$. Clearly, $N_{u,d}$ are non-diagonal in general and consequently, the state $R$ in \Eqn{CP even} carries tree-level flavor changing couplings which are very much constrained from the experiments. Following the definition of \Eqn{pseudoscalar}, it can be shown that the pseudoscalar ($A$) also mediates tree-level flavor changing processes~:
\begin{eqnarray}
\ml_{\rm 2HDM}^{\rm CP~odd} = -\frac{iA}{v} \left\{\bar{u}\left(N_uP_R-N_u^\dagger P_L \right)u -\bar{d}\left(N_dP_R-N_d^\dagger P_L \right)d \right\} \,.
\label{Yuk-n}
\end{eqnarray}
For completeness, we record the general charged Higgs Yukawa interaction as follows~:
\begin{eqnarray}
\ml_{\rm 2HDM}^{\rm charged} = \frac{\sqrt{2} H_1^+}{v} \bar{u} \left[VN_dP_R-N_u^\dagger VP_L \right]d + {\rm h.c.} \,,
\label{Yuk-c}
\end{eqnarray}
where, $V=V_L^\dagger U_L$ is the CKM matrix.

\subsection{Natural flavor conservation (NFC)}
As prescribed by Glashow, Weinberg and Pascos, the tree-level Higgs mediated FCNC can be avoided altogether if all the fermions of a particular charge get their masses from a single scalar doublet. As the easiest example, if all the fermions couple to only $\Phi_1$ (say) then $\Gamma_2=\Delta_2=0$. Consequently, from \Eqn{Nudgen} one can easily see that $N_{u,d}$ are diagonal and so there will be no Higgs mediated FCNC. These arrangements are generally made by employing a $Z_2$ symmetry under which $\Phi_2 \to -\Phi_2$. Proper assignments of $Z_2$ charges to the different fermions then dictate which fermion couples to which doublet. To fix our convention, we shall always call $\Phi_2$ the doublet which couples to the up-type quarks. Then, there are only two possibilities for the quark sector -- either $\Phi_1$ or $\Phi_2$ gives masses to the down-type quarks. For each possibility in the quark sector, either $\Phi_1$ or $\Phi_2$ can give masses to the charged leptons making a total of four variants of 2HDMs as mentioned in Sec.~1. However, in this review, we shall primarily concentrate only on the phenomenology of the quark sector. In the following, we spell out the relevant Yukawa interaction in the {\em alignment limit} for the two possibilities in the quark sector.

\begin{itemize}
\item {\bf Type-I and Lepton specific or X:} In these cases $\Gamma_1=\Delta_1=0$. Consequently, the Yukawa Lagrangian becomes
\begin{eqnarray}
\ml_Y^{\rm I} &=& -\frac{h}{v}\left(\bar{u}D_uu+\bar{d}D_dd \right)  
+\frac{H}{v}\cot\beta\left(\bar{u}D_uu+\bar{d}D_dd \right) \nonumber \\
&& +\frac{iA}{v}\cot\beta \left(\bar{u}D_u\gamma_5 u-\bar{d}D_d \gamma_5d \right) \nonumber \\
&& +\left[\frac{\sqrt{2}H_1^+}{v}\cot\beta  \left\{\bar{u}_R\left(D_uV\right)d_L-\bar{u}_L\left(VD_d\right)d_R \right\} +{\rm h.c.} \right]
\label{1y}
\end{eqnarray}
\item {\bf Type-II and Flipped or Y:}  In these cases $\Gamma_2=\Delta_1=0$. Consequently, the Yukawa Lagrangian becomes
\begin{eqnarray}
\ml_Y^{\rm II} &=& -\frac{h}{v}\left(\bar{u}D_uu+\bar{d}D_dd \right)  
+\frac{H}{v}\left(\cot\beta\bar{u}D_uu-\tan\beta\bar{d}D_dd \right)  \nonumber \\
&& +\frac{iA}{v} \left(\cot\beta\bar{u}D_u\gamma_5 u+ \tan\beta \bar{d}D_d \gamma_5d \right) \nonumber \\
&& +\left[\frac{\sqrt{2}H_1^+}{v} \left\{\cot\beta \bar{u}_R\left(D_uV\right)d_L +\tan\beta\bar{u}_L\left(VD_d\right)d_R \right\} +{\rm h.c.} \right]
\label{2y}
\end{eqnarray}
\end{itemize}

\paragraph{Flavor constraints on the NFC variants:}
\begin{figure}
\begin{center}
\includegraphics[scale=0.39]{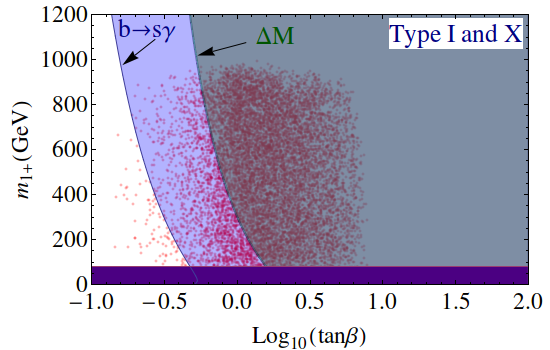} ~
\includegraphics[scale=0.39]{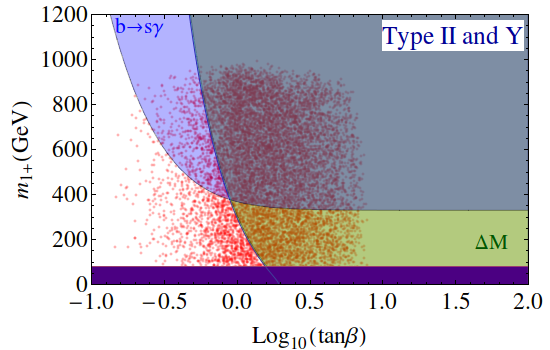}
\end{center} 
\caption[Flavor constraints on Type-I and II 2HDMs]{\em Constraints on $\tan\beta$ and the charged Higgs mass
  from unitarity and flavor physics. The left panel corresponds to
  Type~I and X and the right panel to Type~II and Y scenarios. The lower
  horizontal dark (purple) strip in both the panels corresponds to the direct search
  limit of 80~GeV\cite{Searches:2001ac}. The ligther shades represent allowed regions from individual flavor observables. The scattered points are allowed from unitarity and stability for 2HDMs with exact $Z_2$ symmetry. The figures have been taken from \cite{Das:2015qva}.}
\label{flavor}
\end{figure}
Now we shall concentrate on the constraints on the charged scalar mass ($m_{1+}$), imposed by the measured values of $b\to s\gamma$ branching ratio\cite{Amhis:2012bh} and neutral meson mass differences ($\Delta M$)\cite{Agashe:2014kda}. Since we are concerned with the quark sector only, the constraints will be the same for Type~I and Type~X models. The same is true for Type~II and Type Y models.

For the process $b\to s\gamma$, the major new physics contributions come from charged scalar loops. We have added the new physics contribution to the SM one at the amplitude level and therefore have taken the interference into account. The branching ratio is then compared with the experimental value, $(3.55\pm 0.26)\times 10^{-4}$\cite{Amhis:2012bh}, to obtain the allowed region at 95\% C.L. in Fig.~\ref{flavor}.
As can be seen from \Eqn{2y}, for Type~II and Y models, in the charged Higgs Yukawa interaction, the up-type Yukawa coupling is multiplied by $\cot\beta$ while the down-type Yukawa is multiplied by $\tan\beta$. Their product is responsible for setting $\tan\beta$-independent limit $m_{1+}>320$ GeV for $\tan\beta >1$\cite{Mahmoudi:2009zx,Deschamps:2009rh}. This feature has been depicted in the right panel of Fig.~\ref{flavor}. In Type~I and X models, on the other hand, each of these couplings picks up a $\cot\beta$ factor. This is why there is essentially no bound on $m_{1+}$ for $\tan\beta >1$ in these models\cite{Mahmoudi:2009zx}. This character of Type~I and X models emerges from the left panel of Fig.~\ref{flavor}.

The dominant new physics contributions to neutral meson mass differences come from the charged scalar box diagrams. In Fig.~\ref{flavor}, allowed regions have been shaded assuming that the new physics contributions saturate the experimental values of $\Delta M$\cite{Agashe:2014kda}. Since the amplitudes for the new box diagrams receive prevailing contributions from the up-type quark masses which, for all four variants of 2HDMs, comes with a $\cot\beta$ prefactor, the overall charged scalar contribution to the amplitude goes as $\cot^4\beta$ due to the presence of four charged scalar vertices in the box diagram. Not surprisingly, $\Delta M$ offers a stronger constraint than $b\to s\gamma$ for $\tan\beta <1$ because, in this region, the new physics amplitude for the latter goes as $\cot^2\beta$.

Things become more interesting when the above flavor constraints are superimposed on top of the constraints from unitarity and stability. Most stringent constraints are obtained when $Z_2$ symmetry is exact in the scalar potential, {\it i.e.}, $\l_5=0$. In Fig.~\ref{flavor}, the scattered points span the region allowed by the combined constraints of unitarity and stability for the case of exact $Z_2$ symmetry. Only those points which lie within the common shaded region survive when all the constraints are imposed. For 2HDMs of all four types, one can read the bound on $\tb$ as
\begin{eqnarray}
0.5 < \tb <8 \,.
\end{eqnarray}
However, in order to allow a lighter charged scalar in the ballpark of 400~GeV or below, one must require $1<\tb<8$. It should be remembered that, the lower bound on $\tb$ mainly comes from the flavor data, whereas the upper limit, for the case of exact $Z_2$ symmetry, is dictated by unitarity and stability. In the presence of a soft breaking parameter, the upper bound will be lifted allowing $\tb$ to take much larger values at the expense of a strong correlation between the soft breaking parameter and $m_H$ as depicted by \Eqn{corr}.
 
\subsection{An alternative to NFC -- BGL models}
In this subsection, we shall study a special category of 2HDM formulated by
Branco, Grimus and Lavoura \cite{Branco:1996bq}, where tree level FCNC exists with appropriate
suppression arising from the elements of the Cabibbo-Kobayashi-Maskawa
(CKM) matrix. Unlike the general 2HDM with tree level FCNC\cite{Crivellin:2013wna,Atwood:1996vj}, 
the BGL models introduce no
new parameters in the Yukawa sector, and therefore, are more
predictive.  In this scenario, instead of the discrete $Z_2$ symmetry
a global $U(1)$ symmetry acts on a particular generation $i$ at a time,
as follows:
\begin{eqnarray}
{\cal S}:~~~ Q_{Li}\to e^{i\theta}Q_{Li}\,,~~p_{Ri}\to
e^{2i\theta} p_{Ri}\,,~~\Phi_2\to e^{i\theta}\Phi_2 \,.
\label{BGL symmetry}
\end{eqnarray}
Here $Q_{Li}=(p_{Li}\,,n_{Li})^T$ is the left-handed quark doublet
for the $i$-th generation ($i=1,2,3$), while $p_R$ denotes the
up-type right-handed quark singlets, all in the weak basis.  The
scalar doublet $\Phi_1$ and the other quark fields remain unaffected
by this transformation.  For this particular choice of the symmetry,
there will be no FCNC in the up sector and the FCNC in the down sector
will be controlled by the $i$-th row of the CKM matrix. This will lead
to three variants which will be called u-, c- and t-type models
according to $i=$ 1, 2, and 3 respectively. The other three
variants can be obtained by replacing $p_R$ in Eq.\ (\ref{BGL
symmetry}) with $n_R$ (down-type singlet), as a result of which
there will be no FCNC in the down sector and the FCNC in the up
sector will be controlled by the $i$-th column of the CKM matrix. We will
not consider the later scenario here primarily because the FCNC in the up
sector is less restrictive.

To understand the details of the BGL models, let us consider, as an example, the case
when $i=3$ in \Eqn{BGL symmetry}. As has been already asserted in the previous
paragraph, the FCNC couplings, in this particular example, will be controlled by the third row of 
the CKM matrix. To begin with, one should realize that certain terms in the Yukawa Lagrangian of \Eqn{yukawa-n}
(with $n=2$) will be forbidden when the BGL symmetry is imposed. In our case with $i=3$, the following terms 
will not remain invariant under the symmetry ${\cal S}$:
\begin{eqnarray}
\overline{Q}_{L3}(\Gamma_1)_{3A}\Phi_1(n_R)_A \,, ~~ \overline{Q}_{L1}(\Gamma_2)_{1A}\Phi_2(n_R)_A \,, ~~
\overline{Q}_{L2}(\Gamma_2)_{2A}\Phi_2(n_R)_A \,,
\end{eqnarray}
where, `$A$' represents the generation index and can take values $1,2,3$. Consequently, the third row of $\Gamma_1$ and the first and second rows of $\Gamma_2$ will be zero:
\begin{eqnarray}
\Gamma_1 &=& \begin{pmatrix}\checkmark & \checkmark & \checkmark \\ \checkmark & \checkmark & \checkmark \\ 0 & 0 & 0 \end{pmatrix} \,, ~~~~
 \Gamma_2 = \begin{pmatrix}0 & 0 & 0 \\ 0& 0& 0 \\ \checkmark & \checkmark & \checkmark  \end{pmatrix} \,.
\label{Gamma12}
\end{eqnarray}
In a similar way, one can easily verify that the matrices $\Delta_1$ and $\Delta_2$ will assume the following textures:
\begin{eqnarray}
 \Delta_1 = \begin{pmatrix} \checkmark & \checkmark & 0 \\ \checkmark & \checkmark & 0\\0& 0& 0  \end{pmatrix} \,, ~~~~
 \Delta_2 = \begin{pmatrix} 0& 0& 0\\ 0& 0& 0 \\ 0& 0& \checkmark \end{pmatrix} \,.
\label{Delta12}
\end{eqnarray}
Due to \Eqn{Delta12}, the mass matrix, $M_p$ (see \Eqn{massmat}), in the up sector will be block diagonal as follows:
\begin{eqnarray}
M_p = \begin{pmatrix} \checkmark & \checkmark & 0 \\ \checkmark & \checkmark & 0\\0& 0& \checkmark  \end{pmatrix} \,.
\end{eqnarray}
Consequently, the matrices which bi-diagonalize $M_p$ will have the same block diagonal structure as $M_p$:
\begin{eqnarray}
V_L = \begin{pmatrix} \checkmark & \checkmark & 0 \\ \checkmark & \checkmark & 0\\0& 0& \checkmark  \end{pmatrix} \,, 
~~~~
V_R = \begin{pmatrix} \checkmark & \checkmark & 0 \\ \checkmark & \checkmark & 0\\0& 0& \checkmark  \end{pmatrix} \,.
\end{eqnarray}
At this point, it is important to realize that we are at liberty to choose
\begin{eqnarray}
V_L = \begin{pmatrix} \checkmark & \checkmark & 0 \\ \checkmark & \checkmark & 0\\0& 0& 1  \end{pmatrix} \,, 
\label{VL}
\end{eqnarray}
with the understanding that the phase of $(M_p)_{33}$ can always be dumped into $(V_R)_{33}$. Denoting the bi-diagonalizing matrices in the down sector by $U_{L,R}$, we can see that the particular form of $V_L$, given by \Eqn{VL}, has the following consequence on the CKM matrix:
 \begin{subequations}
 \begin{eqnarray}
V &=& V_L^\dagger U_L = \begin{pmatrix} \checkmark & \checkmark & 0 \\ \checkmark & \checkmark & 0\\0& 0& 1  \end{pmatrix} \begin{pmatrix} \checkmark & \checkmark & \checkmark \\ \checkmark & \checkmark & \checkmark \\ (U_L)_{31} & (U_L)_{32} & (U_L)_{33}  \end{pmatrix}  \\
\Rightarrow &&  \begin{pmatrix} \checkmark & \checkmark & \checkmark \\ \checkmark & \checkmark & \checkmark \\ V_{td} & V_{ts} & V_{tb}  \end{pmatrix} ~=~  \begin{pmatrix} \checkmark & \checkmark & \checkmark \\ \checkmark & \checkmark & \checkmark \\ (U_L)_{31} & (U_L)_{32} & (U_L)_{33}  \end{pmatrix} \,.
 \end{eqnarray}
 \end{subequations}
 This implies that the third row of the CKM matrix is identical to the third row of $U_L$:
  \begin{subequations}
 \label{U-t-model}
  \begin{eqnarray}
(U_L)_{3A} &=& V_{tA} \equiv V_{3A} \,,\\
\Rightarrow ~~~ (U_L^\dagger)_{A3} &=& (U_L)_{3A}^* = V_{3A}^* \,.
  \end{eqnarray}
  \end{subequations}
Note that, \Eqn{U-t-model} is a direct consequence of our choice $i=3$ in \Eqn{BGL symmetry}. In general, for FCNC in the down sector, the $i$-th row of $U_L$ should be identical with the $i$-th row of the CKM matrix. To proceed further with our choice of $i=3$, it is useful to define the following projection matrix:
\begin{eqnarray}
P = \begin{pmatrix} 0 & 0 & 0 \\ 0 & 0 & 0\\ 0 & 0 & 1  \end{pmatrix} \,.
\label{projection}
\end{eqnarray}
Using \Eqn{massmat} in conjunction with \Eqn{Gamma12}, we see that it is only $\Gamma_2$ that contributes to the third row of $M_n$, {\em i.e.},
\begin{eqnarray}
\frac{v_2}{\sqrt{2}} \Gamma_2 = P\cdot M_n \,.
\label{PMn}
\end{eqnarray}
We also note the following:
\begin{eqnarray}
P\cdot \Gamma_1 =0 \,, ~~ P\cdot \Gamma_2 =\Gamma_2 \,, ~~ P\cdot\Delta_1=0 \,, ~~ P\cdot\Delta_2=\Delta_2 \,.
\end{eqnarray}
From \Eqn{Nudgen} we recall that the matrices $N_{u,d}$ actually control the FCNC couplings. The non-diagonal part of $N_d$ is given by 
\begin{subequations}
\begin{eqnarray}
X &=& U_L^\dagger \cdot \frac{v_2}{\sqrt{2}} \Gamma_2 \cdot U_R  \\
&=& U_L^\dagger \cdot (P\cdot M_n) \cdot U_R  ~~~~~~ {\rm [Using~\Eqn{PMn}]} \\
&=& U_L^\dagger P U_L\cdot U_L^\dagger M_n U_R \\
&=& U_L^\dagger P U_L\cdot D_d \,.
\end{eqnarray}
\end{subequations}
Thus,
 \begin{subequations}
 \begin{eqnarray}
X_{AB} &=& \sum_C \left(U_L^\dagger P U_L \right)_{AC} (D_d)_{CB} \\
&=& \sum_C \left(U_L^\dagger P U_L \right)_{AC}m_B^d~ \delta_{CB} ~=~ m_B^d \left(U_L^\dagger P U_L \right)_{AB} \,.
 \end{eqnarray}
 \end{subequations}
But note that
\begin{eqnarray}
U_L^\dagger =  \begin{pmatrix} \checkmark & \checkmark & (U_L)_{31}^* \\ \checkmark & \checkmark & (U_L)_{32}^* \\ \checkmark  & \checkmark & (U_L)_{33}^*  \end{pmatrix} \,, ~~~ {\rm and,}~~~
PU_L = \begin{pmatrix} 0 & 0 & 0 \\ 0 & 0 & 0 \\ (U_L)_{31}  & (U_L)_{32} & (U_L)_{33}  \end{pmatrix} \,.
\end{eqnarray}
Thus,
\begin{eqnarray}
 \left(U_L^\dagger P U_L \right)_{AB} = (U_L)^*_{3A} (U_L)_{3B} \equiv V_{3A}^*V_{3B} \,,
\end{eqnarray}
where, in the last step, we have used \Eqn{U-t-model}. Hence, we obtain
\begin{eqnarray}
X_{AB} = m^d_B V^*_{3A} V_{3B} \,.
\end{eqnarray}
Substituting this into \Eqn{Ndgen} we get
\begin{eqnarray}
(N_d)^{\bf t}_{AB} &=& \tb (D_d)_{AB} -(\tb+\cot\beta) X_{AB} \nonumber \\
&=& \tb~m_A^d~\delta_{AB} -(\tb +\cot\beta) V_{3A}^*V_{3B}~ m_B^d \,.
\label{Nd-t}
\end{eqnarray}

To obtain a simplified form for $N_u^{\bf t}$, we note, similar to \Eqn{PMn}, that
\begin{eqnarray}
\frac{v_2}{\sqrt{2}} \Delta_2 = P\cdot M_p \,.
\label{PMp}
\end{eqnarray}
In a similar way, we can write the following for the non-diagonal part of $N_u^{\bf t}$:
\begin{eqnarray}
Y ~=~ V_L^\dagger \cdot \frac{v_2}{\sqrt{2}} \Delta_2 \cdot V_R ~=~ V_L^\dagger P V_L\cdot D_u \,.
\label{Y-t}
\end{eqnarray}
Because of the special form of $V_L$ in \Eqn{VL}, we have
\begin{eqnarray}
PV_L = \begin{pmatrix} 0 & 0 & 0 \\ 0 & 0 & 0 \\ 0  & 0 & 1  \end{pmatrix} ~~~ 
\Rightarrow ~~~  V_L^\dagger P V_L =  \begin{pmatrix} 0 & 0 & 0 \\ 0 & 0 & 0 \\ 0  & 0 & 1  \end{pmatrix} \,.
\end{eqnarray}
Substituting this into \Eqn{Y-t}, one can easily verify
\begin{eqnarray}
Y = {\rm diag} \{0,~0,~m_t \} \,.
\end{eqnarray}
Plugging this into \Eqn{Nugen} we get the final expression for $N_u^{\bf t}$:
\begin{eqnarray}
N_u^{\bf t} &=& \tb~D_u -(\tb +\cot\beta)Y \nonumber \\
&=& \tb ~{\rm diag}\{m_u,~m_c,~0 \} -\cot\beta~{\rm diag}\{0,~0,~m_t\} \,.
\label{Nu-t}
\end{eqnarray}
Since $N_u^{\bf t}$ is diagonal, there will be no Higgs mediated tree-level FCNC in the up sector and from \Eqn{Nd-t} one can see that the tree-level FCNC in the down sector is controlled by the off-diagonal elements of the third row of the CKM matrix. Had we considered u- or c-type models, the tree-level FCNC in the down sector would be controlled by the first or second row of the CKM matrix respectively.

\paragraph{Constraints from flavor data:}
Here we will only consider the u-, c- and t-type BGL scenarios where the tree-level FCNC exists in the down sector. To begin with, it might be useful to rewrite the explicit expressions for the matrices that control the tree-level FCNC couplings for these models.
 The matrices $N_u$ and $N_d$, for the u-, c- and
t-type models, have the following form (the $(i,j)$ indices in $N_d$
refer to $(d,s,b)$ quarks and the superscripts in bold font refer to
the model type):
\begin{subequations}
\begin{eqnarray}
N_u^{\bf u} &=& {\rm diag}\{-m_u \cot\beta\,, m_c\tan\beta\,,
m_t\tan\beta\}\,, \\
(N_d)_{ij}^{\bf u} &=& \tan\beta~m_i \delta_{ij}
-(\tan\beta+\cot\beta)V_{ui}^* V_{uj} m_j\, , \\
N_u^{\bf c} &=& {\rm diag}\{m_u \tan\beta\,, -m_c\cot\beta\,,
m_t\tan\beta \}\,, \\
(N_d)_{ij}^{\bf c} &=& \tan\beta~m_i \delta_{ij}
-(\tan\beta+\cot\beta)V_{ci}^* V_{cj} m_j\, , \\
N_u^{\bf t} &=& {\rm diag}\{m_u \tan\beta\,, m_c\tan\beta\,,
-m_t\cot\beta \}\,, \\
(N_d)_{ij}^{\bf t} &=& \tan\beta~m_i \delta_{ij}
-(\tan\beta+\cot\beta)V_{ti}^* V_{tj} m_j\,.
\end{eqnarray}
\label{nund}
\end{subequations}
In the leptonic sector (with only left-handed neutrinos), the Yukawa
couplings of \Eqs{Yuk-n}{Yuk-c} should be read with the replacement
$(N_u, D_u)\to 0$, $V=1$, and $N_d(D_d)\to N_e(D_e)$, with $N_e$
resembling the diagonal $N_u$ matrices in Eq.~(\ref{nund}) with
appropriate replacement of quark masses by the charged lepton masses.
This means that there is no FCNC in the leptonic sector when the
neutrinos are considered to be massless. But if we assume the neutrinos to
be massive, then the tree-level FCNC in the leptonic sector will be controlled
by the rows and columns of the PMNS matrix.

The CP-odd scalar mass eigenstate $A$ would be massless if the
symmetry of \Eqn{BGL symmetry} is exact in the Higgs potential. Thus,
in the 't~Hooft sense, a light pseudoscalar will be natural in these
models.  While there are five free parameters in any BGL model,
namely, $\alpha$, $\beta$, $m_{1+}$, $m_H$, and $m_A$, we can make some
reasonable simplifications.  Considerations of perturbativity and
stability of scalar potential ensure that $m_A \sim m_H$ if $\tan\beta
\geq 10$ \cite{Bhattacharyya:2013rya}. If $m_A$ and $m_H$ are large,
we can even bring down the $\tan\beta$ limit further, say up to
$\tan\beta = 5$. However, for the sake of simplicity and economy of
parameters, we will assume $m_H = m_A$ for the remainder of this chapter
unless explicitly mentioned otherwise.  Thus, in the alignment limit,
i.e. $\sin(\beta - \alpha) = 1$, we are left with only three unknown
parameters: $\tan\beta$, $m_{1+}$ and $m_{H/A}$. It should be noted
though that consistency with the oblique $T$-parameter requires $m_{1+}
\sim m_H$ once we assume $m_H = m_A$ \cite{Bhattacharyya:2013rya}.

\paragraph{Neutral meson mixing:}
Neutral meson mass
differences offer important constraints. 
The tree-level scalar exchange contribution to the
off-diagonal element of the $2\times 2$ Hamiltonian matrix is given by
\cite{Branco:1996bq}
\begin{equation}
(M_{12}^K)^{\rm BGL} \approx 
\frac{5}{24}\frac{f_K^2 m_K^3}{v^2} (V_{id}^* V_{is})^2 \frac{1}{{\cal
    A}^2} \,,
\label{m12}
\end{equation}
where $m_K$ is the neutral kaon mass and $f_K$ is the decay constant.
Similar expressions exist for $B_d$ and $B_s$ systems.
 The mass difference is given by $\Delta M_K \approx 2 \vert M_{12}^K\vert$. 
The contributions of three neutral scalars are contained in
\begin{eqnarray}
\frac{1}{{\cal A}^2}& =& (\tan\beta +\cot\beta)^2\left( \frac{\cos^2(\beta-\alpha)}{m_h^2}
+\frac{\sin^2(\beta-\alpha)}{m_H^2} -\frac{1}{m_A^2} \right) \nonumber \\
&=& (\tan\beta +\cot\beta)^2  \left(\frac{1}{m_H^2} -
\frac{1}{m_A^2}\right)\,.
\label{onebya2}
\end{eqnarray}
The last equality in Eq.~(\ref{onebya2}) holds in the alignment
limit.  The size of the prefactors in Eq.~(\ref{m12}) tells us that
$m_A=m_H$ is very well motivated from the neutral kaon mass difference
for the u- and c-type models.  For the t-type model, however, this
degeneracy is more of an assumption than a requirement especially for
$\tan\beta \sim 1$.

With the assumption $m_H = m_A$, the dominant contributions to neutral
meson mass differences come from the charged Higgs box diagrams. The
expressions for the loop-induced amplitudes are given explicitly in
Appendix~\ref{AppendixC}. In Fig.~\ref{tmodel1}, constraints have been placed
assuming that the new physics contributions saturate the experimental
values of $\Delta M$ \cite{Beringer:1900zz}.  For $\tb > 1$, $\Delta
M_d$ and $\Delta M_s$ severely restrict the u- and c-type models,
whereas the t-type model can admit a light charged Higgs, at least for
$m_H = m_A$. For large $\tb$, $\Delta M_K$ offers a stronger
constraint than $b \to s \gamma$ (discussed later) in the t-type model
due to the dominance of the charm-induced box graph.

\begin{figure}
\includegraphics[scale=0.4]{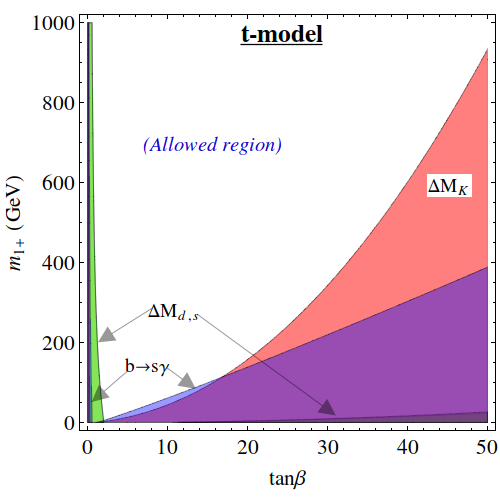}
~~~~~~~~~ 
\includegraphics[scale=0.4]{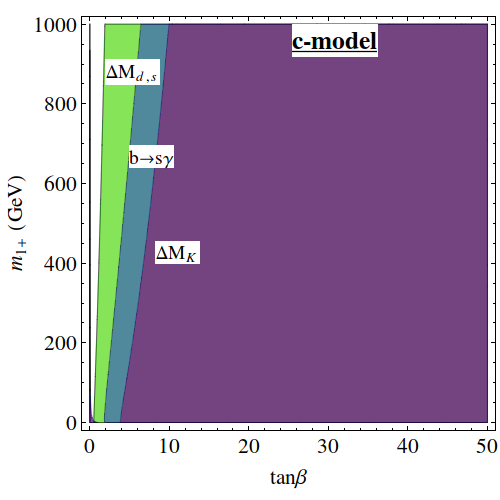}
\caption[Flavor constraints on $m_{1+}$ in BGL models]{\em Constraints from various observables for t- and
  c-(u-) type BGL models. In the left panel (t-type), for large
  $\tan\beta$, $\Delta M_K$ offers a stronger constraint than $b \to
  s \gamma$. The vertical spiked shaded region in the extreme
  left also correspond to the entire disallowed region in Type I and X
  models. In the right panel (c- or u-types), $\Delta M_d$ and $\Delta M_s$
  provide the most stringent constraints. Note that an assumption $m_H
= m_A$ has been made to switch off the tree level contribution to
the neutral meson mass differences. The figures have been taken from \cite{Bhattacharyya:2014nja}.}
\label{tmodel1}
\end{figure}
\paragraph{\bf \texorpdfstring{$\mathbf{b\to s\gamma}$}{TEXT}:}
The process $b\to s\gamma$ offers severe constraint on the charged
Higgs mass \cite{Grinstein:1987pu,Borzumati:1998tg}.  For Type II and
Y models, in the charged Higgs Yukawa interaction, the up-type Yukawa
coupling is multiplied by $\cot\beta$ while the down-type Yukawa is
multiplied by $\tan\beta$.  Their product is responsible for setting
$\tb$-independent limit $m_{1+} > 300$ GeV for $\tan\beta>1$
\cite{Deschamps:2009rh,Mahmoudi:2009zx,Cheng:2014ova}.  In Type I and
X models, each of these couplings picks up a $\cot\beta$ factor, which
is why there is essentially no bound on charged Higgs mass for
$\tan\beta>1$ in these models \cite{Mahmoudi:2009zx}.

\begin{figure}
\begin{center}
\includegraphics[scale=0.36]{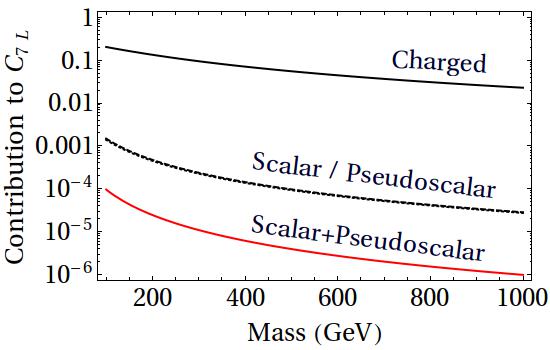}~~~
\includegraphics[scale=0.36]{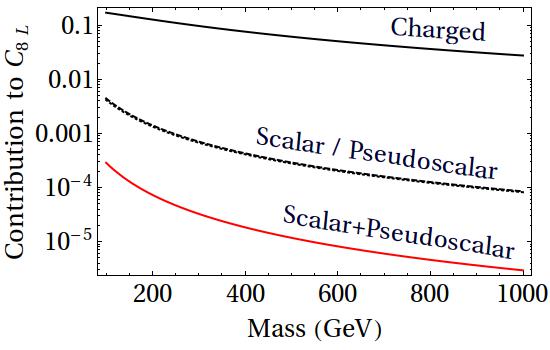}
\caption[Neutral Higgs contributions to the $b\to s\gamma$ amplitude]{\em Magnitude of the contributions to the effective Wilson
  coefficients $C_{7L}$ and $C_{8L}$ for $b\to s\gamma$, coming from
  $H_1^+$, $H$, and $A$, plotted against the corresponding masses. The
  middle curve in each panel shows the magnitude of the individual
  scalar and pseudoscalar contributions; they are too close to be
  differentiated in the shown scale. The lowest curve in each panel
  shows the sum of $H$ and $A$ contributions for the case $m_H = m_A$,
  which shows that the scalar and pseudoscalar contributions interfere
  destructively. $C_{7R}$ and $C_{8R}$ are suppressed by $m_s/m_b$ and
  are not shown here. The figures have been taken from \cite{Bhattacharyya:2014nja}.}
\label{fig:c78}
  \end{center}
\end{figure}

In the BGL class of models, the constraint on $m_{1+}$ is different from
that in Type I or Type X 2HDM (detailed expressions are displayed in
Appendix~\ref{AppendixC}). This is because the BGL symmetry of \Eqn{BGL symmetry}
does not respect family universality.  For the $i$-type BGL model, the
relevant Yukawa couplings contain an overall factor of $(-\cot\beta)$
for vertices involving the $i$-th generation up-type fermion and a
factor of $\tan\beta$ for the others. Consequently, the top loop
contribution to the $b\to s\gamma$ amplitude will grow as
$\tan^2\beta$ for u- and c-type models resulting in very tight
constraints on $m_{1+}$ for $\tan\beta>1$. On the contrary, for t-type
models, the top-loop contribution will decrease with increasing
$\tan\beta$ and will hardly leave any effect for $\tan\beta>1$,
similar to what happens in the Type I and X models.  But unlike in the
latter scenarios, the charm loop amplitude in t-type BGL grows as
$\tan^2\beta$.  It becomes numerically important for large $\tb$ and
does not allow $H_1^+$ to be very light.

Taking the branching ratio ${\rm Br}(b \to s \gamma) _{\rm SM}
=(3.15\pm 0.23)\times 10^{-4}$
\cite{Gambino:2001ew,Misiak:2006zs,Misiak:2006ab} and ${\rm Br}(b \to
s \gamma)_{\rm exp} = (3.55\pm 0.26)\times 10^{-4}$
\cite{Amhis:2012bh}, these features of the BGL models have been
displayed in Fig.~\ref{tmodel1}. The regions excluded at 95\% CL from
$b\to s\gamma$ have been shaded and appropriately marked.  Note that
we have considered not only the contributions from $(H_1^+, u_i)$
loops, but also from $(H/A, d_i)$ loops (due to tree level FCNC
couplings of $H$ and $A$). The numerical effects of the latter are
found to be small; we refer the reader to Fig.\ \ref{fig:c78}, where
separate contributions from the charged and the neutral scalars to
$C_{7L}$ and $C_{8L}$ are shown. The behavior can also be intuitively
understood from the following comparison of the dominant contributions
from the charged and neutral scalar induced loops to the $b \to s
\gamma$ amplitude.  The ratio of $H_1^+$ and $(H/A)$-induced loop
contributions roughly goes like $(m_c^2 \tan^2\beta / m_b m_s)$ for
large $\tan\beta$, and $(m_t^2 \cot^2 \beta / m_b m_s)$ for
$\tan\beta$ of the order of one.  This justifies that the constraint
from $b \to s \gamma$ essentially applies on the charged Higgs mass.
In other words, that $H_1^+$ can be really light does not crucially
depend on the values of $m_H$ and $m_A$.  From now on, we stick only
to the t-type model to promote light charged Higgs phenomenology.

\begin{wrapfigure}{r}{0.44\textwidth}
\includegraphics[scale=0.35]{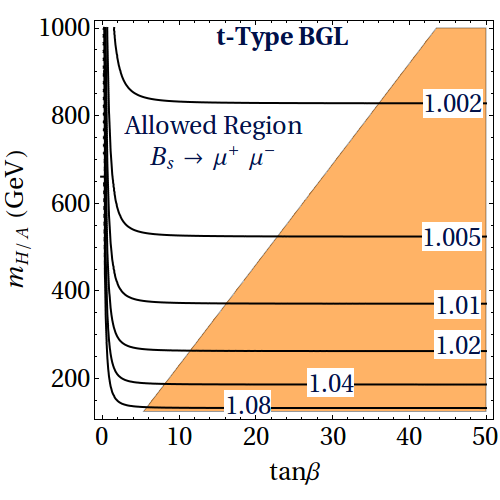}
\caption[Constraints from $B_s \to \mu^+ \mu^-$]{\em The shaded region is disallowed by $B_s \to \mu^+ \mu^-$
  at 95\% CL. Contours of enhancements in $B_s\to\tau^+\tau^-$ over
  the SM estimate are also shown.}
  \label{fig:dilepton}
\end{wrapfigure}
\paragraph{Other constraints:}
For t-type model, the branching ratios ${\rm Br}(B \to D^{(*)}
\tau\nu)$ and ${\rm Br}(B^+\to\tau\nu)$ do not receive any appreciable
contributions unless the charged Higgs mass is unnaturally small
defying the LEP2 direct search limit of 80 GeV
\cite{Searches:2001ac}. The process $B_s\to \ell^+\ell^-$ proceeds at
the tree level mediated by $H/A$ providing important constraints. The
amplitudes are proportional to $(\tan^2\beta+1)/m_{H/A}^2$ for $\ell =
e,\mu$, and $(\cot^2\beta+1)/m_{H/A}^2$ for $\ell=\tau$.  In
Fig.~\ref{fig:dilepton} we have shaded the region excluded at 95\% CL,
obtained by comparing the SM expectation of ${\rm Br}
(B_s\to\mu^+\mu^-) = (3.65\pm 0.23) \times 10^{-9}$
\cite{Bobeth:2013uxa} with its experimental value $(3.2\pm 1.0)\times
10^{-9}$ \cite{HFAGbs}. The details are provided in Appendix~\ref{AppendixC}. In the
same plot we display different contours for ${\rm
  Br}(B_s\to\tau^+\tau^-)/{\rm Br}(B_s\to\tau^+\tau^-)_{\rm SM}$,
where we observe slight enhancement over the SM expectation.
 
\section{Decay of the nonstandard Higgs bosons}
\begin{figure}
\begin{center}
\includegraphics[height=5.5cm,width=7.5cm]{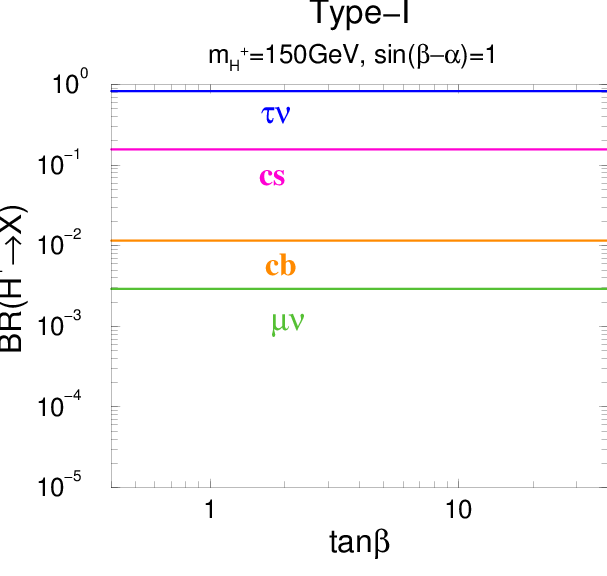}~~~
\includegraphics[height=5.5cm,width=6.5cm]{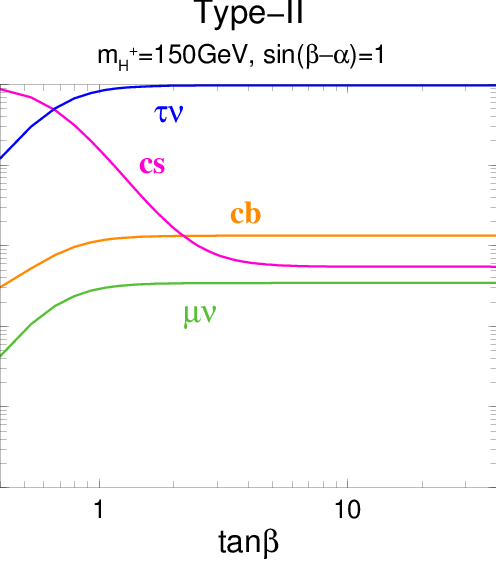} \\
\includegraphics[height=5.5cm,width=7.5cm]{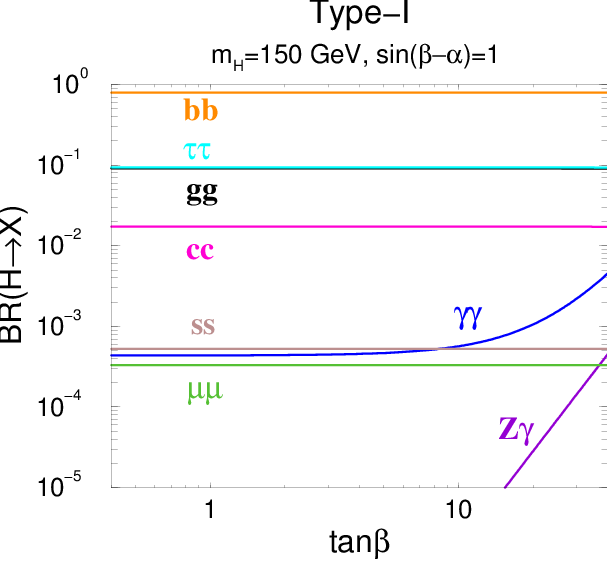}~~~
\includegraphics[height=5.5cm,width=6.5cm]{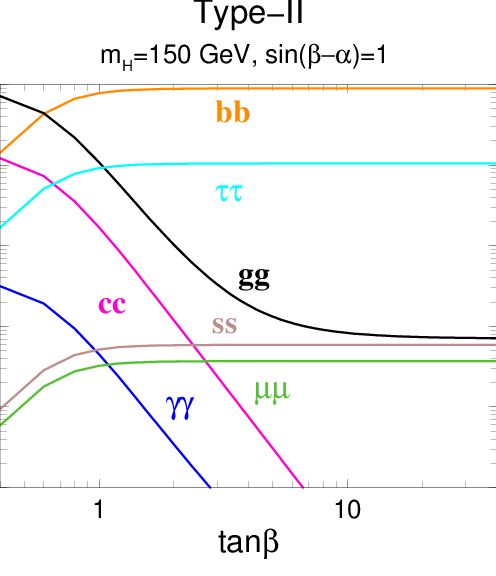}
\caption[Nonstandard Higgs decays in conventional 2HDMs]{\em Branching ratios of
$H$ and $H_1^+$ in Type I and Type II models as a function of $\tb$ for $m_H=m_{1+}=150$~GeV in 
the alignment limit. The plots have been taken from \cite{Aoki:2009ha}.}
\label{fig:decay_conv}
  \end{center}
\end{figure}
\begin{figure}
\centering
\includegraphics[scale=0.38]{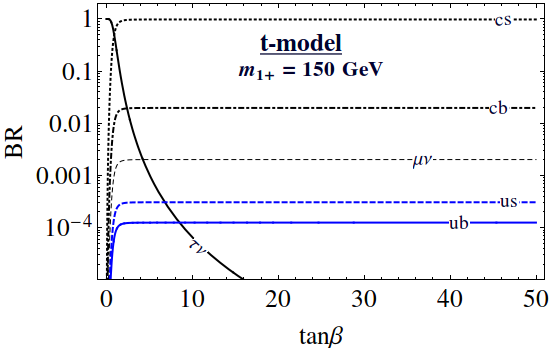}
~~~~ 
\includegraphics[scale=0.38]{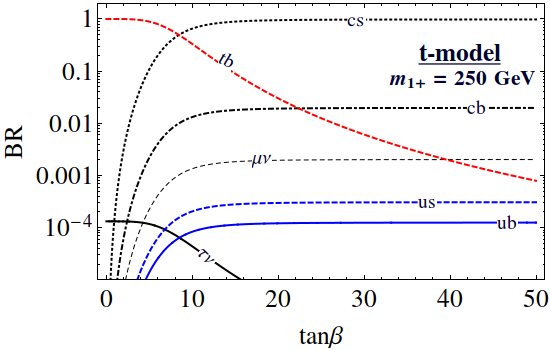}
\caption[Charged Higgs branching ratios in t-Type BGL]{\em The charged Higgs branching ratios to two-body final
  states for two benchmark choices of $m_{1+}$. The figures have been taken from \cite{Bhattacharyya:2014nja}.}
\label{charged-br}
\end{figure}

In Type I model, the light charged Higgs goes to $\tau\nu$ and $cs$
(below the $tb$ threshold), and the branching ratios are independent
of $\tan\beta$ (see Fig.~\ref{fig:decay_conv}), because both the leptonic and the quark couplings have
the same $\cot\beta$ prefactor \cite{Aoki:2009ha,Aoki:2011wd}. In Type
X model, the leptonic part has an overall $\tan\beta$ multiplicative
factor, so the charged Higgs preferentially decays into third
generation leptonic channels for large $\tan\beta$ (e.g. almost
entirely so for $\tb \geq 2.5$).  In the t-type BGL scenario, the
charged Higgs branching ratios into two-body fermionic final states
have been plotted in Fig.~\ref{charged-br}.  We have considered two
benchmark values for $m_{1+}$, one below the $tb$ threshold and the
other well above it. To a good approximation it is enough to consider
fermionic final states, because in the alignment limit the $W^\pm
hH_1^\mp$ coupling vanishes and if we consider near degeneracy of
$m_{1+}$ and $m_{H/A}$ to satisfy the $T$-parameter constraint, then
$H_1^+$ cannot decay into $W^+ S^0$ ($S^0=H,A$) channel. Two
noteworthy features which distinguish the t-type BGL model from others
are: ($i$) the $\mu\nu$ final state dominates over $\tau\nu$ for
$\tan\beta > 5$, which is a distinctive characteristic of t-type BGL
model unlike any of the Type I, II, X or Y models (due to family
nonuniversal BGL Yukawa couplings); ($ii$) for $\tb > 10$, the
branching ratio into $cs$ significantly dominates over other channels
including $tb$, again a unique feature of t-type BGL.  The reason for
the latter can be traced to the relative size of the top and charm
quark masses {\em vis-\`{a}-vis} the $\tan\beta$ or $\cot\beta$
prefactor.  This will result in a dijet final state at the LHC,
without any $b$-jet, and hence the signal will be extremely difficult
to be deciphered over the standard QCD background.

We now discuss the decay branching ratios of the neutral scalar
$H$. In the alignment limit $HVV$ ($V=W,Z$) coupling vanishes. Hence
we discuss flavor diagonal $ff$ final states (flavor violating modes
are CKM suppressed), together with $\gamma\gamma$ and $gg$ final
states.  In conventional types of 2HDM, the $bb$ and $\tau\tau$ final states
dominate over $cc$ and $\mu\mu$ channels, respectively
\cite{Aoki:2009ha}. Here, the hierarchy is reversed, which transpires
from the expressions of $N_d$ and $N_u$ in \Eqn{nund}.  To provide an
intuitive estimate of the signal strength, we define the following
variable:
\begin{eqnarray}
R_X = \frac{\sigma(pp\to H \to X)}{\sigma(pp\to h \to
  \gamma\gamma)} \, , 
\label{rx}
\end{eqnarray}
\begin{figure}[!htbp]
\centering
\includegraphics[width=5cm,height=4cm]{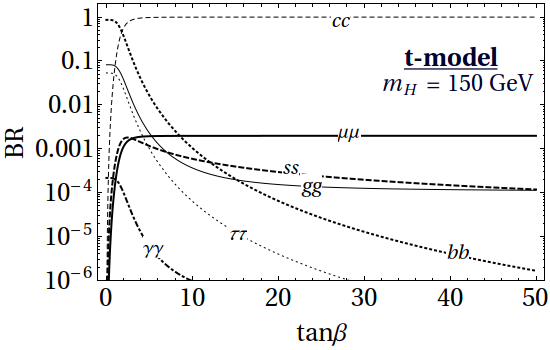}
\includegraphics[width=5cm,height=4cm]{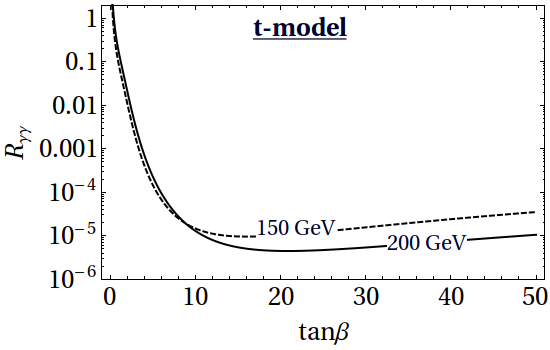}
\includegraphics[width=5cm,height=4cm]{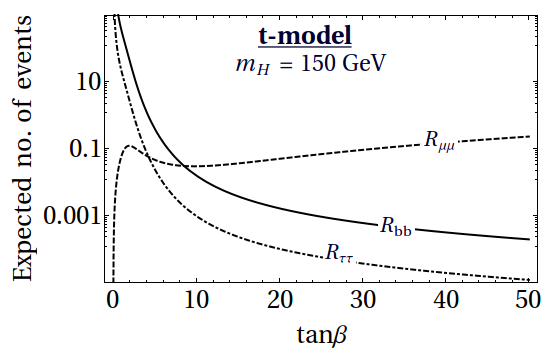}
\caption[Neutral Higgs branching ratios in t-Type BGL]{\em For various two-body final states, the $R$ values, and the
  branching ratios of $H$. The figures have been taken from \cite{Bhattacharyya:2014nja}.}
\label{neutral-br}
\end{figure}
where the normalization has been done with respect to the SM Higgs
production and its diphoton decay branching ratio.  We recall that the
loop contributions of charged scalars to $h\to \gamma\gamma$ is tiny
as long as $m_{H/A} \simeq m_{1+}$ \cite{Bhattacharyya:2013rya}.  The
relative merits of various channels have been plotted in
Fig.~\ref{neutral-br}. The crucial thing to observe is that although
for $\tan\beta > 5$, $H$ decays entirely into dijet ($cc$), the
$\mu^+\mu^-$ mode may serve as a viable detection channel for $H$ in
future. With 20 fb$^{-1}$ luminosity at LHC8, the expected number of
diphoton events from the SM Higgs decay is about
400. Fig.~\ref{neutral-br} shows that $R_{\mu\mu} \sim 0.1$,
i.e. about 40 dimuon events from $H$ decay should have been
observed. However, they are going to be swamped by huge background
(mainly Drell-Yan, also QCD jets faking dimuon) \cite{cms-bkg,atlas-bkg}. At
LHC14 with an integrated luminosity of 300 fb$^{-1}$, we expect about
39000 $h\to\gamma\gamma$ events \cite{cerntwiki}, which means about
3900 $H\to\mu\mu$ events for $m_H=150$ GeV. Dimuon background studies
at 14 TeV are not yet publicly available. A rough conservative
extrapolation of the existing 7 and 8 TeV studies of the dimuon
background \cite{cms-bkg,atlas-bkg} gives us hope that the signal can be
deciphered over the background.  Note that these are all crude
estimates, made mainly to get our experimental colleagues interested
in probing such exotic decay modes.  A more careful study including,
e.g.  detection efficiencies and detailed background estimates, is
beyond the scope of this chapter.  We emphasize that our scenario does
not say that $H, A$ or $H_1^+$ have to be necessarily light. If they
are heavy as they are forced to be in many other 2HDMs ($\sim 500$ GeV
or more), their direct detection in early LHC14 would be that much
difficult. The feature that makes our scenario unique is the {\em
  possibility} of their relative lightness as well as unconventional
decay signatures.

\section{Loop induced decays of the SM-like Higgs} \label{s:decay}
Since we are working in the alignment limit, the couplings of $h$
with the fermions and gauge bosons will be exactly like in the SM.
The production cross section of $h$ will therefore be as expected in
the SM.  All the tree level decay widths of $h$ will also have the SM
values for the same reason.  Loop induced decays like $h\to
\gamma\gamma$ and $h\to Z\gamma$ will however have additional
contributions from virtual charged scalars ($H_1^\pm$).  Since the
branching fractions of such decays are tiny, the total decay width is
hardly modified.

The contribution of the $W$-boson loop and the top loop diagrams to
$h\to \gamma\gamma$ and $h\to Z\gamma$ are same as in the SM.  As
regards the charged scalar induced loop, we first parametrize the
cubic coupling $g_{hH_1^+H_1^-}$ in the following
way: 
\begin{eqnarray}
g_{hH_1^+H_1^-} =  \kappa_1 \; \frac{g m_{1+}^2}{M_W} \,,
\label{defkappa}
\end{eqnarray}
where $\kappa_1$ is dimensionless.  The diphoton decay width is then
given by \cite{Djouadi:2005gi, Djouadi:2005gj}:
\begin{eqnarray}
 \Gamma (h\to \gamma\gamma) = \frac{\alpha^2g^2}{2^{10}\pi^3}
 \frac{m_h^3}{M_W^2} \Big|F_W + \frac{4}{3}F_t  + \kappa_1 F_{1+} \Big|^2
 \,, 
\label{h2gg}
\end{eqnarray}
where, introducing the notation
\begin{eqnarray}
\tau_x \equiv (2m_x/m_h)^2 \,,
\end{eqnarray}
the values of $F_W$, $F_t$ and $F_{i+}$ are given by
\begin{subequations}
\begin{eqnarray}
 F_W &=& 2+3\tau_W+3\tau_W(2-\tau_W)f(\tau_W) \,,  \\
 F_t &=& -2\tau_t \big[1+(1-\tau_t)f(\tau_t)\big] \,,  \\
 F_{i+} &=& -\tau_{i+} \big[ 1-\tau_{i+}f(\tau_{i+}) \big] \,.
\end{eqnarray}
\label{Fs}
\end{subequations}
If we assume $m_{1+}>80$ GeV to respect the direct search bound \cite{Searches:2001ac}, then $\tau_x >1$
for $x=W$, $t$, $H_1^\pm$.   In this limit
\begin{eqnarray}
f(\tau) =
\left[\sin^{-1}\left(\sqrt{1/\tau}\right)\right]^2 \,.
\label{f}
\end{eqnarray}
The decay width for $h\to Z\gamma$ can analogously be written as:
\begin{eqnarray}
 \Gamma (h\to Z\gamma) = \frac{\alpha^2g^2}{2^{9}\pi^3}
 \frac{m_h^3}{M_W^2} \Big|A_W + A_t  + \kappa_1 A_{1+}\Big|^2
 \left(1-\frac{M_Z^2}{m_h^2}\right)^3 \,,
\label{h2Zg}
\end{eqnarray}
where, introducing
\begin{eqnarray}
\eta_x = (2m_x/M_Z)^2 \,,
\end{eqnarray}
the values of $A_W$, $A_t$ and $A_{i+}$ are given by~\cite{Gunion:1989we}
\begin{subequations}
\begin{eqnarray}
 A_W &=& \cot \theta_w\bigg[ 4(\tan^2\theta_w - 3)I_2(\tau_W,\eta_W)
 \nonumber \\*
&& \null +\bigg\{ \left(5+\frac{2}{\tau_W}\right) -
 \left(1+\frac{2}{\tau_W}\right)\tan^2\theta_w \bigg\}
 I_1(\tau_W,\eta_W)\bigg] \,, 
 \\ 
 A_t &=&
 \frac{4\Big(\frac{1}{2}-\frac{4}{3}\sin^2\theta_w\Big)}{\sin\theta_w
   \cos\theta_w} \; \Big[I_2(\tau_t,\eta_t)-I_1(\tau_t,\eta_t) \Big] \,, \\
 A_{i+} &=& \frac{(2\sin^2\theta_w-1)}{\sin\theta_w \cos\theta_w} \;
 I_1(\tau_{i+},\eta_{i+}) \,.
\end{eqnarray}
\label{As}
\end{subequations}
The functions $I_1$ and $I_2$ are given by
\begin{subequations}
\begin{eqnarray}
 I_1(\tau,\eta) &=& \frac{\tau \eta}{2(\tau -\eta)} +
 \frac{\tau ^2\eta^2}{2(\tau -\eta)^2}\Big[f(\tau )-f(\eta)\Big]
 +\frac{\tau ^2\eta}{(\tau -\eta)^2}\Big[g(\tau )-g(\eta)\Big]  \,, \\
 I_2(\tau ,\eta) &=& -\frac{\tau \eta}{2(\tau -\eta)}\Big[f(\tau
   )-f(\eta)\Big] \,,  
\end{eqnarray}
\end{subequations}
where the function $f$ has been defined in \Eqn{f}.  Since 
$\tau_x,\eta_x > 1$ for $x=W,t,H_1^\pm$, the function $g$ assumes the
following form:
\begin{eqnarray}
 g(a) = \sqrt{a-1}\sin^{-1}\left(\sqrt{1/a}\right) \,.
\end{eqnarray}
In the alignment limit with softly broken $U(1)$ symmetry, the parameter $\kappa_1$ which appears in
Eqs.\ (\ref{defkappa}), (\ref{h2gg}) and (\ref{h2Zg}) is given by
\begin{eqnarray}
\kappa_1 = \frac{1} {m_{1+}^2} (m_A^2 - m_{1+}^2 - \frac12 m_h^2) \,.
\label{BGLkappa}
\end{eqnarray}
\begin{figure}
\centering
\includegraphics[scale=0.36]{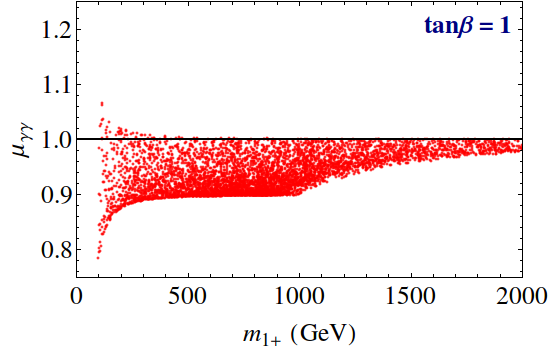} 
\includegraphics[scale=0.36]{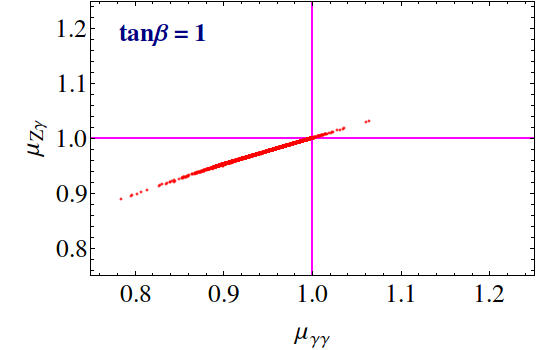} 

\includegraphics[scale=0.36]{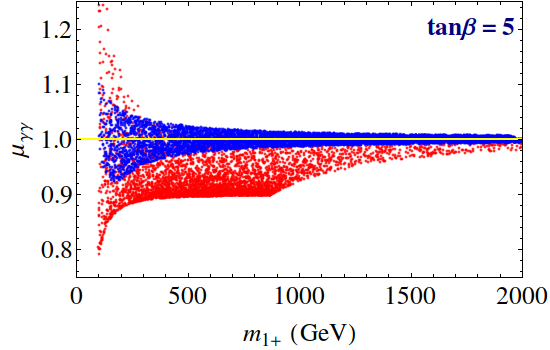}
\includegraphics[scale=0.36]{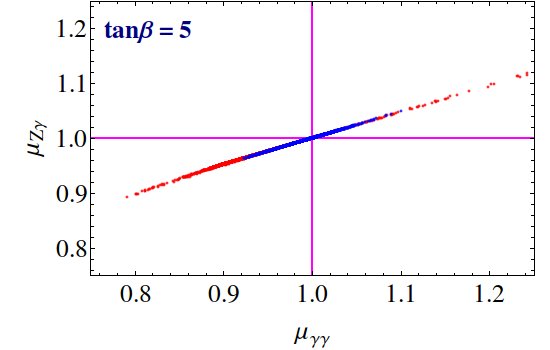}

\includegraphics[scale=0.36]{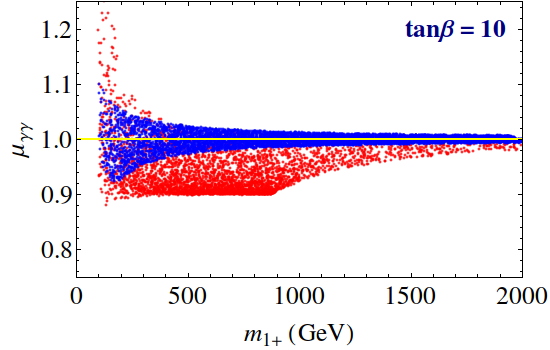}
\includegraphics[scale=0.36]{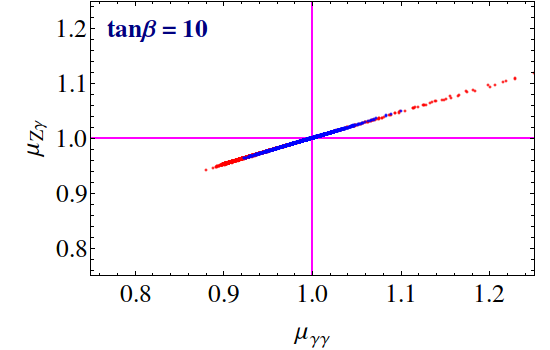}

\caption[Impact on loop induced Higgs decays]{\em (2HDMs with softly broken $U(1)$ symmetry in the scalar potential.)
 The left panels show allowed regions in
  $m_{1+}$-$\mu_{\gamma\gamma}$ plane for three values of $\tan\beta$,
  while the right panels show correlation between $\mu_{\gamma\gamma}$
  and $\mu_{Z\gamma}$ for the same choices of $\tan\beta$.  The
  regions shown in red  are allowed by unitarity and stability, while
  the blue regions additionally pass the $T$-parameter test.  We put
  $m_{1+} > 100$\,GeV and $m_H > m_h$. These figures have been taken from \cite{Bhattacharyya:2013rya}.} 
\label{f:mu}
\end{figure}
The appearance of $m_A$ in \Eqn{BGLkappa} is merely an artefact of the
$U(1)$ symmetry in the scalar potential which enforces $\l_5=\l_6$.  In
the more general potential of \Eqn{notation2}, the expression for
$\kappa_1$ involves $\lambda_5$, which has nothing to do with $m_A$.
The decoupling behavior of $\kappa_1$ for large $m_{1+}$ is not then
guaranteed, as noted in Refs.~\cite{Djouadi:1996yq,Bhattacharyya:2014nja}.
Later we will see that such nondecoupling of charged scalars are not unique to the
2HDMs. In fact, whenever we have a scalar potential invariant under an exact discrete symmetry 
with multiple Higgs doublets, this kind of nondecoupling phenomena are expected.
However, in the present scenario, unitarity conditions bound the splitting between heavy scalar masses \cite{Bhattacharyya:2013rya},
ensuring smooth decoupling of $\kappa_1$ with increasing $m_{1+}$.  As we
have noticed, this splitting is also controlled by the $T$-parameter.

In our case, the quantities $\mu_{\gamma\gamma}$ and $\mu_{Z\gamma}$,
defined through the equations
\begin{eqnarray}
 \mu_{\gamma\gamma} &=& {\sigma(pp\to h) \over \sigma^{\rm
     SM}(pp\to h)} \cdot {\mbox{BR} (h \to 
   \gamma\gamma) \over \mbox{BR}^{\rm SM} (h \to \gamma\gamma)} \,,
 \\ 
 \mu_{Z\gamma} &=& {\sigma(pp\to h) \over \sigma^{\rm
     SM}(pp\to h)} \cdot {\mbox{BR} (h \to 
   Z\gamma) \over \mbox{BR}^{\rm SM} (h \to Z\gamma)} \,,
\label{mu}
\end{eqnarray}
assume the following forms:
\begin{eqnarray}
 \mu_{\gamma\gamma} &=&  \frac{\Gamma(h \to \gamma\gamma)}{\Gamma^{\rm
     SM}(h\to \gamma\gamma)} 
 = \frac{\Big|F_W + \frac{4}{3}F_t  +
   \kappa_1 F_{1+}\Big|^2}{\Big|F_W + \frac{4}{3}F_t \Big|^2} \,,
 \\ 
 \mu_{Z\gamma} &=&  \frac{\Gamma(h \to
  Z\gamma)}{\Gamma^{\rm SM}(h\to Z\gamma)} = \frac{\Big|A_W + A_t  +
  \kappa_1 A_{1+} \Big|^2}{ \Big|A_W + A_t \Big|^2} \,.
\end{eqnarray}

In Fig.~\ref{f:mu}, we show the variation of $\mu_{\gamma\gamma}$
against $m_{1+}$ and the correlation between $\mu_{\gamma\gamma}$ and
$\mu_{Z\gamma}$ for $\tan\beta=1$, 5, 10.  When we show the variation
of $\mu_{\gamma\gamma}$ with $m_{1+}$, we take into consideration all
values of $m_H$ and $m_A$ which are allowed in Fig.~\ref{f:uniu1}.
In case of Fig.~\ref{f:mu}, the red points are those which are allowed
by unitarity and stability constraints, while the superimposed blue
points are allowed by the $T$-parameter.  The points allowed by
unitarity prefer suppression in $\mu_{\gamma\gamma}$ compared to the
SM expectation.  The correlation between $\mu_{\gamma\gamma}$ and
$\mu_{Z\gamma}$ can in principle be used for discriminating new
physics models with increased sensitivity in the future course  of the
LHC run.

\section{Conclusions}
In the first part of this chapter, we have revisited the constraints from tree-unitarity and stability in the context of 2HDMs. The observed scalar at LHC has been identified with the lightest CP-even scalar of the model. The {\em alignment limit} has been imposed in view of the conformity of the LHC Higgs data with the SM predictions. These are the {\em new} informations that became available only after the Higgs discovery. If the $Z_2$ symmetry is exact in the potential, it is found that all the nonstandard masses are restricted below 1~TeV from unitarity with the upper limit on $m_H$ being highly correlated to $\tb$. The value of $\tb$ is also confined within the range $1/8<\tb<8$ from unitarity and stability. The constraints from flavor data severely restrict the region with $\tb<1$. Therefore, for an exact $Z_2$ symmetry, $\tb$ is bounded within a very narrow range of $1<\tb<8$ when a light charged scalar with mass around 400~GeV is looked for.

In the presence of an appropriate soft breaking parameter the bounds on $\tb$ will be diluted. However, for large values of $\tb$, the unitarity and stability conditions will render a strong correlation between the soft breaking parameter and $m_H$ as appears in \Eqn{corr}. It is also worth noting that the value of $\mu_{\gamma\gamma}$ can play a crucial role in the presence of soft breaking. For example, if $\mu_{\gamma\gamma}$ is measured to be consistent with the SM expectation with 5\% accuracy then one can conclude $m_{H^+}^2 \approx 1/2\l_5v^2$\cite{Bhattacharyya:2014oka} for any value of $\tb$. Thus, for large values of $\tb$ one may expect $m_H^2\approx m_{H^+}^2\approx 1/2\l_5v^2$. We have also studied quantitative correlation among the non-standard scalar masses in a class
of 2HDM with a global $U(1)$ symmetry in the potential. In this context, we have observed that for values of $\tan\beta\sim5$ or 
larger, all the three non-standard scalar masses are roughly degenerate.  More
specifically, in this limit unitarity dictates $m_H$ and $m_A$ to be
almost equal and $|m_{1+}^2-m_H^2|$ to be small, while the
$T$-parameter restricts $|m_{1+}-m_H|$ to be very small. It is also important to note that, due
to the global $U(1)$ symmetry, unitarity restricts mass-squared
differences and not the individual masses of the non-standard scalars.

Later in this chapter, we have shown that BGL class of two-Higgs-doublet model
admits charged and additional neutral scalars which can be as light as
$\sim$ 150 GeV. They successfully negotiate the stringent constraints
from radiative $b$-decay, neutral meson mass differences, and dimuon
decays of $B$ mesons. Special features of Yukawa couplings in this
model lead to characteristic decay signatures of the nonstandard
scalars, which are different from the signatures of similar scalars in
other 2HDM variants.  Preferential decays of both the charged and
additional neutral scalars into {\em second}, rather than the {\em
  third}, generation fermions for $\tb > 5$ constitute the trademark
distinguishing feature of this scenario, which can be tested in the
high luminosity option of the LHC or at the ILC.
 

\begin{savequote}[0.65\textwidth]  
Symmetry is what we see at a glance; based on the fact that there is no reason for any difference \dots
\qauthor{Blaise Pascal}    
\end{savequote}
%

\chapter[Analysis of an extended scalar sector with \texorpdfstring{$S_3$}{TEXT} symmetry]{Analysis of an extended scalar sector \\ with \texorpdfstring{$S_3$}{TEXT} symmetry} 

\lhead{Chapter 4. \emph{Analysis of an extended scalar sector with $S_3$ symmetry}}
\label{Chap4} 

The newly observed boson at the Large Hadron Collider (LHC)  \cite{Aad:2012tfa,Chatrchyan:2012ufa} fits very well to the description of the Higgs scalar in the Standard Model (SM). The SM relies on the minimal choice that a single Higgs doublet provides masses to all particles. But unexplained phenomena like neutrino masses and existence dark matter motivate us to contemplate other avenues beyond the SM (BSM). Majority of these BSM scenarios extend the SM Higgs sector predicting a richer scalar spectrum. One of them $-$ the $S_3$ flavor model $-$ stems from an effort to answer the aesthetic question as to why there are precisely three fermion generations \cite{Kubo:2003pd}. Keeping the fermions in appropriate $S_3$-multiplets, it is possible to reproduce all the measured parameters of the CKM and PMNS matrices as well as make testable predictions for the unknown parameters of the PMNS matrix \cite{ Koide:1999mx,Harrison:2003aw,Kubo:2003iw,Teshima:2005bk,Koide:2006vs,Chen:2007zj,Mondragon:2007af,Jora:2009gz,Xing:2010iu,Kaneko:2010rx,Zhou:2011nu,Teshima:2011wg,Dev:2011qy,Dev:2012ns,Meloni:2012ci, Dias:2012bh,Siyeon:2012zu,Canales:2012dr, Canales:2013cga,Benaoum:2013ji,Hernandez:2014vta}. But one needs at least three Higgs doublets to achieve this goal \cite{Kubo:2003iw}. However, the $S_3$ invariant scalar potential contains some new parameters which are difficult to constrain phenomenologically. Although some lower bounds on the additional scalar masses can be placed from the Higgs mediated flavor changing neutral current (FCNC) processes \cite{Ma:2013zca}, these bounds rely heavily on the Yukawa structure of the model. In this article we will present some new  bounds on the physical scalar masses which do not depend on the parameters of the Yukawa sector. To achieve this, we will employ the prescription of tree unitarity which is known to be able to set upper limits on different scalar masses \cite{Lee:1977eg}. Although various aspects of the $S_3$ scalar potential have been discussed in the literature \cite{Pakvasa:1977in,Kubo:2004ps}, to the best of our knowledge, this is the first attempt to derive the exact unitarity constraints on the quartic couplings in the $S_3$ invariant three-Higgs-doublet model (S3HDM) scalar potential. We also identify an \emph{ alignment limit} in the context of S3HDM where a CP-even Higgs with SM-like properties can be obtained. Since the recent LHC Higgs data seem to increasingly leaned towards the SM expectations, our numerical analysis will be restricted to this limit.  
 
The chapter is organized as follows: in Section~\ref{s:potential} we discuss the scalar potential and derive necessary conditions for the potential to be bounded from below. In Section~\ref{s:physeig} we minimize the potential and calculate the physical scalar masses. In this section we also figure out an  alignment limit in which one neutral CP-even physical scalar behaves exactly like the SM Higgs. In Section~\ref{s:unitarity} we derive the exact constraints arising from the considerations of tree level unitarity and use them to constrain the nonstandard scalar masses. In Section~\ref{s:decay-s3} we quantitatively investigate the effect of the charged scalar induced loops on $h\to \gamma \gamma$ and $h\to Z \gamma$ signal strengths. Finally, we summarize our findings in Section~\ref{s:concl}.

\section{The scalar potential} \label{s:potential}
%
$S_3$ is the permutation group involving three objects, $\{\Phi_a, \Phi_b, \Phi_c\}$. The three dimensional representation of $S_3$ is not an irreducible one simply because we can easily construct a linear combination of the elements, $\Phi_a+\Phi_b+\Phi_c$, which remains unaltered under the permutation of the indices. We choose to decompose the three dimensional representation into a singlet and doublet as follows:
\begin{subequations}
\begin{eqnarray}
{\bf 1}: && ~~ \Phi_3=\frac{1}{\sqrt{3}}(\Phi_a+\Phi_b+\Phi_c) \,, \\
{\bf 2}: && ~ \begin{pmatrix}\Phi_1 \\ \Phi_2 \end{pmatrix} = 
 \begin{pmatrix} \frac{1}{\sqrt{2}}(\Phi_a-\Phi_b) \\ \frac{1}{\sqrt{6}}(\Phi_a+\Phi_b-2\Phi_c) \end{pmatrix}\,.
\end{eqnarray}
\end{subequations}
The elements of $S_3$ for this particular doublet representation are given by:
\begin{eqnarray}
 \begin{pmatrix}\cos\theta & \sin\theta \\ -\sin\theta & \cos\theta \end{pmatrix} \,,~~
 \begin{pmatrix}\cos\theta & \sin\theta \\ \sin\theta & -\cos\theta \end{pmatrix}\,, ~~
 {\rm for}~~ \left(\theta=0,\pm \frac{2\pi}{3}\right) \,.
\end{eqnarray}
The most general renormalizable potential invariant under $S_3$ can be written in terms of $\Phi_3$, $\Phi_1$ and $\Phi_2$ as follows~\cite{Pakvasa:1977in,Kubo:2004ps,Koide:2005ep,Teshima:2012cg,Machado:2012ed}:
\begin{subequations}
\begin{eqnarray}
V(\Phi)&=& V_2(\Phi) +V_4(\Phi) \,, \\
{\rm where,}~~ V_2(\Phi) &=& \mu_1^2(\Phi_1^\dagger\Phi_1+\Phi_2^\dagger\Phi_2)+ \mu_3^2\Phi_3^\dagger\Phi_3 \,, \\
V_4(\Phi)&=& \lambda_1 (\Phi_1^\dagger\Phi_1+\Phi_2^\dagger\Phi_2)^2 +\lambda_2 (\Phi_1^\dagger\Phi_2 -\Phi_2^\dagger\Phi_1)^2 \nonumber \\
&& +\lambda_3 \left\{(\Phi_1^\dagger\Phi_2+\Phi_2^\dagger\Phi_1)^2 +(\Phi_1^\dagger\Phi_1-\Phi_2^\dagger\Phi_2) ^2\right\} \nonumber \\
&& +\lambda_4 \left\{(\Phi_3^\dagger\Phi_1)(\Phi_1^\dagger\Phi_2+\Phi_2^\dagger\Phi_1) +(\Phi_3^\dagger\Phi_2)(\Phi_1^\dagger\Phi_1-\Phi_2^\dagger\Phi_2) + {\rm h. c.}\right\} \nonumber \\
&& +\lambda_5(\Phi_3^\dagger\Phi_3)(\Phi_1^\dagger\Phi_1+\Phi_2^\dagger\Phi_2) + \lambda_6 \left\{(\Phi_3^\dagger\Phi_1)(\Phi_1^\dagger\Phi_3)+(\Phi_3^\dagger\Phi_2)(\Phi_2^\dagger\Phi_3)\right\} \nonumber \\
&& +\lambda_7 \left\{(\Phi_3^\dagger\Phi_1)(\Phi_3^\dagger\Phi_1) + (\Phi_3^\dagger\Phi_2)(\Phi_3^\dagger\Phi_2) +{\rm h. c.}\right\} +\lambda_8(\Phi_3^\dagger\Phi_3)^2 \,.
\label{quartic}
\end{eqnarray}
\label{potential}
\end{subequations}
In general $\lambda_4$ and $\lambda_7$ can be complex, but we assume them to be real so that CP symmetry is not broken explicitly. For the stability of the vacuum in the asymptotic limit we impose the requirement that there should be no direction in the field space along which the potential becomes infinitely negative. The necessary and sufficient conditions for this is well known in the context of two Higgs-doublet models (2HDMs) \cite{Gunion:2002zf}. For the potential of \Eqn{potential}, a 2HDM equivalent situation arise if one of the doublets is made identically zero. Then it is quite straightforward to find the following {\em necessary} conditions for the global stability in the asymptotic limit:
\begin{subequations}
\begin{eqnarray}
\lambda_1 &>& 0 \,, \\
\lambda_8 &>& 0 \,, \\
\lambda_1+\lambda_3 &>& 0 \,, \\
2\lambda_1 +(\lambda_3-\lambda_2) &>& |\lambda_2+\lambda_3| \,, \\
\lambda_5 +2\sqrt{\lambda_8(\lambda_1+\lambda_3)} &>& 0 \,, \\
\lambda_5+\lambda_6+ 2\sqrt{\lambda_8(\lambda_1+\lambda_3)} &>& 2|\lambda_7| \,, \\
\lambda_1+\lambda_3+\lambda_5+\lambda_6+2\lambda_7+\lambda_8 &>& 2|\lambda_4| \,.
\end{eqnarray}
\label{stability-s3}
\end{subequations}
To avoid confusion, we wish to mention that an equivalent doublet representation,
\begin{eqnarray}
\begin{pmatrix}\chi_1 \\ \chi_2 \end{pmatrix} =  \frac{1}{\sqrt{2}}\begin{pmatrix}i & 1 \\ -i & 1 \end{pmatrix}
 \begin{pmatrix} \Phi_1 \\ \Phi_2 \end{pmatrix}\,,
\end{eqnarray}
has also been used in the literature. In terms of this new doublet, the quartic part of the scalar potential is written as~\cite{Bhattacharyya:2010hp,Bhattacharyya:2012ze,Chen:2004rr}:
\begin{eqnarray}
V_4 &=& \frac{\beta_1}{2}\left(\chi_1^\dagger\chi_1+\chi_2^\dagger\chi_2 \right)^2 +\frac{\beta_2}{2}\left(\chi_1^\dagger\chi_1- \chi_2^\dagger\chi_2 \right)^2 +\beta_3(\chi_1^\dagger\chi_2)(\chi_2^\dagger\chi_1) +\frac{\beta_4}{2}(\Phi_3^\dagger\Phi_3)^2 \nonumber \\
&& +\beta_5(\Phi_3^\dagger\Phi_3)(\chi_1^\dagger\chi_1+\chi_2^\dagger\chi_2 ) +\beta_6\Phi_3^\dagger (\chi_1\chi_1^\dagger +\chi_2\chi_2^\dagger )\Phi_3+ \beta_7\left\{(\Phi_3^\dagger\chi_1)(\Phi_3^\dagger\chi_2)+{\rm h.c.}\right\} \nonumber \\
&& +\beta_8\left\{\Phi_3^\dagger(\chi_1\chi_2^\dagger\chi_1+\chi_2\chi_1^\dagger\chi_2) + {\rm h.c.}\right\} \,.
\label{alternative}
\end{eqnarray}
It is easy to verify that the parameters of \Eqn{alternative} are related to the parameters of \Eqn{quartic} in the following way:
\begin{eqnarray}
&& \beta_1 = 2\lambda_1~; ~~\beta_2=-2\lambda_2~; ~~\beta_3=4\lambda_3~; ~~ \beta_4=2\lambda_8~; \nonumber \\ && \beta_5=\lambda_5~; ~~ \beta_6=\lambda_6~;~~ \beta_7=2\lambda_7~;~~ \beta_8=-\sqrt{2}\lambda_4 \,.
\end{eqnarray}
This mapping can be used to translate the constraints on $\lambda$s into constraints on $\beta$s.
In this chapter we choose to work with the parametrization of \Eqn{potential}.

\section{Physical eigenstates} \label{s:physeig}
We represent the scalar doublets in the following way:
\begin{eqnarray}
\Phi_k= \begin{pmatrix} w_k^+ \\ \frac{1}{\sqrt{2}}(v_k+h_k+iz_k) \end{pmatrix} ~~~~{\rm for}~k=1,~2,~3\,.
\label{e:doub}
\end{eqnarray}
We shall assume that CP symmetry is not spontaneously broken and so the vacuum expectation values (vevs) are taken to be real. They also satisfy the usual vev relation: $v=\sqrt{v_1^2+v_2^2+v_3^2}=$ 246 GeV. The minimization conditions for the scalar potential of \Eqn{potential} reads:
\begin{subequations}
\begin{eqnarray}
2\mu_1^2 &=& -2\lambda_1(v_1^2+v_2^2)-2\lambda_3(v_1^2+v_2^2)-v_3\{6\lambda_4v_2 +(\lambda_5+\lambda_6+2\lambda_7)v_3\} \,, \label{mu11}\\
2\mu_1^2 &=& -2\lambda_1(v_1^2+v_2^2)-2\lambda_3(v_1^2+v_2^2) -\frac{3v_3}{v_2}\lambda_4(v_1^2-v_2^2) - (\lambda_5+\lambda_6+2\lambda_7)v_3^2 \,, \label{mu22}\\
2\mu_3^2 &=& \lambda_4\frac{v_2}{v_3}(v_2^2-3v_1^2) -(\lambda_5+\lambda_6+2\lambda_7)(v_1^2+v_2^2)-2\lambda_8 v_3^2 \,.
\end{eqnarray}
\label{minimization}
\end{subequations}
For the self-consistency of Eqs.~(\ref{mu11}) and (\ref{mu22}), two possible scenarios arise\footnote{Another possibility, $v_3=0$, while mathematically consistent, is unattractive. This is because, in some $S_3$ structure of the Yukawa sector, the $S_3$-singlet fermion generation will the remain massless.}:
\begin{subequations}
\begin{eqnarray}
\lambda_4 &=& 0 \,, \\
{\rm or,}~~v_1 &=& \sqrt{3}v_2 \,.
\end{eqnarray}
\label{consistency}
\end{subequations}
In the following subsections we shall discuss each of the above scenarios separately.

\subsection{Case-I ($\lambda_4=0$)}
Since CP symmetry is assumed to be exact in the scalar potential, the neutral physical states will be eigenstates of CP too. We find that the mass-squared matrices in the scalar($M_S^2$), pseudoscalar($M_P^2$) and charged($M_C^2$) sectors are  simultaneously block diagonalizable by the following matrix:
\begin{eqnarray}
X= \begin{pmatrix}
\cos\gamma & -\sin\gamma & 0 \\ \sin\gamma & \cos\gamma & 0 \\ 0 & 0 & 1
\end{pmatrix} ~~~{\rm with} ~~\tan\gamma=\frac{v_1}{v_2} \,.
\end{eqnarray}
For the charged mass matrix, we obtain:
\begin{eqnarray}
XM_C^2X^T= \begin{pmatrix}
m_{1+}^2 & 0 & 0 \\ 0 & -\frac{1}{2}v_3^2(\lambda_6+2\lambda_7) & \frac{1}{2}v_3\sqrt{v_1^2+v_2^2} (\lambda_6+2\lambda_7) \\ 0& \frac{1}{2}v_3\sqrt{v_1^2+v_2^2}(\lambda_6+2\lambda_7) & -\frac{1}{2}(v_1^2+v_2^2) (\lambda_6+2\lambda_7)
\end{pmatrix} \,,
\end{eqnarray}
where, one of the charged Higgs ($H_1^+$) with mass $m_{1+}$ is defined as:
\begin{subequations}
\begin{eqnarray}
H_1^+ &=& \cos\gamma ~w_1^+ -\sin\gamma ~w_2^+ \,, \\
m^2_{1+} &=& -\left\{2\lambda_3\sin^2\beta +\frac{1}{2}(\lambda_6+2\lambda_7)\cos^2\beta \right\}v^2 \,, \\
{\rm with,} ~~\tan\beta &=& \frac{\sqrt{v_1^2+v_2^2}}{v_3} \,.
\label{tanb}
\end{eqnarray}
\end{subequations}
The second charged Higgs ($H_2^+$) along with the massless Goldstone ($\omega^+$), which will appear as the longitudinal component of the $W$-boson, can be obtained by diagonalizing the remaining $2\times 2$ block:
\begin{eqnarray}
\begin{pmatrix}
H_2^+ \\ \omega^+
\end{pmatrix}= \begin{pmatrix}
\cos\beta & -\sin\beta \\ \sin\beta & \cos\beta
\end{pmatrix} \begin{pmatrix}
w_2'^+ \\ w_3^+
\end{pmatrix} ~~~{\rm with,} ~~ w_2'^+ = \sin\gamma~w_1^+ +\cos\gamma~w_2^+ \,.
\end{eqnarray}
The mass of the second charged Higgs is given by:
\begin{eqnarray}
m_{2+}^2 = -\frac{1}{2}(\lambda_6+2\lambda_7) v^2 \,.
\end{eqnarray}
Similar considerations for the pseudoscalar part gives:
\begin{eqnarray}
XM_P^2X^T= \begin{pmatrix}
\frac{1}{2} m_{A1}^2 & 0 & 0 \\ 0 & -v_3^2\lambda_7 & v_3\sqrt{v_1^2+v_2^2}\lambda_7 \\ 0& v_3\sqrt{v_1^2+v_2^2}\lambda_7 & -(v_1^2+v_2^2)\lambda_7
\end{pmatrix} \,,
\end{eqnarray}
where, the pseudoscalar state ($A_1$) with mass eigenvalue $m_{A1}$ is defined as:
\begin{subequations}
\begin{eqnarray}
A_1 &=& \cos\gamma ~z_1 -\sin\gamma ~z_2 \,, \\
m^2_{A1} &=& -2\left\{(\lambda_2+\lambda_3)\sin^2\beta +\lambda_7\cos^2\beta \right\}v^2 \,, 
\end{eqnarray}
\end{subequations}
where, $\tan\beta$ has already been defined in \Eqn{tanb}. Similar to the charged part, here also the second pseudoscalar ($A_2$) along with the massless Goldstone ($\zeta$) can be obtained as follows:
\begin{subequations}
\begin{eqnarray}
\begin{pmatrix}
A_2 \\ \zeta
\end{pmatrix}&=& \begin{pmatrix}
\cos\beta & -\sin\beta \\ \sin\beta & \cos\beta
\end{pmatrix} \begin{pmatrix}
z_2' \\ z_3
\end{pmatrix} ~~~{\rm with,} ~~ z_2' = \sin\gamma~z_1 +\cos\gamma~z_2 \,, \\
{\rm and,} ~~~ m_{A2}^2 &=& -2\lambda_7 v^2 \,.
\end{eqnarray}
\end{subequations}
Finally, for the CP-even part we have:
\begin{subequations}
\begin{eqnarray}
XM_S^2X^T &=& \begin{pmatrix}
0 & 0 & 0 \\ 0 & A'_S & -B'_S \\ 0& -B'_S &  C'_S
\end{pmatrix} \,, \\
{\rm where,} ~~~ A'_S &=& (\lambda_1+\lambda_3)(v_1^2+v_2^2) \,, \\
B'_S &=& -\frac{1}{2}v_3\sqrt{v_1^2+v_2^2}(\lambda_5+\lambda_6+2\lambda_7) \,, \\
C'_S &=& \lambda_8 v_3^2 \,.
\end{eqnarray}
\end{subequations}
The massless state ($h^0$), as also noted in \cite{Beltran:2009zz}, is given by:
\begin{eqnarray}
h^0 &=& \cos\gamma ~h_1 -\sin\gamma ~h_2 \,.
\label{h0}
\end{eqnarray}
But we wish to add here that the appearance of a massless scalar is not surprising. One can easily verify that the potential of \Eqn{potential} has the following $SO(2)$ symmetry for $\lambda_4=0$:
\begin{eqnarray}
\begin{pmatrix}
\Phi_1' \\ \Phi_2'
\end{pmatrix}&=& \begin{pmatrix}
\cos\theta & -\sin\theta \\ \sin\theta & \cos\theta
\end{pmatrix} \begin{pmatrix}
\Phi_1 \\ \Phi_2
\end{pmatrix}
\end{eqnarray}
Since $SO(2)$ is a continuous symmetry isomorphic to $U(1)$, a massless physical state is expected. Other two physical scalars are obtained as follows:
\begin{subequations}
\begin{eqnarray}
\begin{pmatrix}h \\ H \end{pmatrix} &=& \begin{pmatrix}
\cos\alpha & -\sin\alpha \\ \sin\alpha & \cos\alpha
\end{pmatrix} \begin{pmatrix}
h_2' \\ h_3
\end{pmatrix} ~~~{\rm with,} ~~ h_2' = \sin\gamma~h_1 +\cos\gamma~h_2 \,, \\
{\rm and,}~~~ \tan2\alpha &=&\frac{2B'_S}{A'_S-C'_S} \,.
\end{eqnarray}
\end{subequations}
We assume $H$ and $h$ to be the heavier and lighter CP-even mass eigenstates respectively, with the following eigenvalues:
\begin{subequations}
\begin{eqnarray}
m_H^2 &=& (A'_S+C'_S)+\sqrt{(A'_S-C'_S)^2+4B_S^{'2}} \,, \\
m_h^2 &=& (A'_S+C'_S)-\sqrt{(A'_S-C'_S)^2+4B_S^{'2}} \,.
\end{eqnarray}
\end{subequations}
At this stage, it is worth noting that we can define two intermediate scalar states, $H^0$ and $R$, as
\begin{eqnarray}
\begin{pmatrix}R \\ H^0 \end{pmatrix} &=& \begin{pmatrix}
\cos\beta & -\sin\beta \\ \sin\beta & \cos\beta
\end{pmatrix} \begin{pmatrix}
h_2' \\ h_3
\end{pmatrix} \,,
\label{H0R}
\end{eqnarray}
with the property that $H^0$ has the exact SM couplings with the vector boson pairs and fermions. $H^0$ does not take part in the flavor changing processes as well. Of course, $H^0$ and $R$ are not the physical eigenstates in general but are related to them in the following way:
\begin{subequations}
\begin{eqnarray}
h &=& \cos(\beta-\alpha)R +\sin(\beta-\alpha) H^0 \,, \\
H &=& -\sin(\beta-\alpha) R + \cos(\beta-\alpha) H^0 \,.
\end{eqnarray}
\end{subequations}
In view of the fact that a $125$ GeV scalar with SM-like properties has already been observed at the LHC, we wish the lighter CP-even mass eigenstate ($h$) to coincide with $H^0$. Then we must require:
\begin{eqnarray}
\cos(\beta-\alpha) \approx 0 \,.
\end{eqnarray}
In analogy with the 2HDM case \cite{Gunion:2002zf}, this limit can be taken as the {\em  alignment limit} in the context of a 3HDM with an $S_3$ symmetry. We must emphasize though, the term ` alignment limit' does not necessarily imply the heaviness of the additional scalars. Considering Eqs.~(\ref{h0}) and (\ref{H0R}), it is also interesting to note that the state $h^0$, being orthogonal to $H^0$, does not have any trilinear $h^0VV$ ($V=$ $W$,$Z$) coupling. But, in general, it will have flavor changing coupling in the Yukawa sector. This type of neutral massless state with flavor changing fermionic coupling will be ruled out from the well measured values of neutral meson mass differences. This means that the choice $\lambda_4=0$ is phenomenologically unacceptable and we shall not pursue this scenario any further.

\subsection{Case-II ($v_1=\sqrt{3}v_2$)}
This situation has recently been analyzed in \cite{Barradas-Guevara:2014yoa}. We, however, use a convenient parametrization
 that can provide intuitive insight into the scenario and additionally, we also discuss the possibility of a {\em  alignment limit} in the same way as done in the previous subsection.

The definitions for the angles, $\gamma$ and $\beta$, and the digonalizing  matrix, $X$, remain the same as before. Only difference is that, due to the vev alignment ($v_1=\sqrt{3}v_2$), $\tan\gamma$ ($=\sqrt{3}$) and hence $X$ is determined completely. Now only two of the vevs, $v_2$ and $v_3$ (say), can be considered independent and $\tan\beta$ is given in terms of them as follows~:
\begin{eqnarray}
\tan\beta = \frac{2v_2}{v_3} \,.
\end{eqnarray}
The charged and pseudoscalar mass eigenstates have the same form as before; only the mass eigenvalues get modified due to the presence of $\lambda_4$~:
\begin{subequations}
\begin{eqnarray}
m^2_{1+} &=& -\left\{2\lambda_3\sin^2\beta+ \frac{5}{2}\lambda_4\sin\beta\cos\beta +\frac{1}{2}(\lambda_6+2\lambda_7)\cos^2\beta \right\}v^2 \,, \\
m^2_{2+} &=& -\frac{1}{2}\left\{\lambda_4\tan\beta+(\lambda_6+2\lambda_7) \right\}v^2 \,, \\
m^2_{A1} &=& -\left\{2(\lambda_2+\lambda_3)\sin^2\beta +\frac{5}{2}\lambda_4\sin\beta\cos\beta +2\lambda_7\cos^2\beta \right\}v^2 \,, \\
m^2_{A2} &=& -\left(\frac{1}{2}\lambda_4\tan\beta+2\lambda_7 \right) v^2 \,.
\end{eqnarray}
\end{subequations}
In the presence of $\lambda_4$, analysis of the scalar part will be slightly different~:
\begin{subequations}
\begin{eqnarray}
XM_S^2X^T &=& \begin{pmatrix}
\frac{1}{2}m_{h0}^2 & 0 & 0 \\ 0 & A_S & -B_S \\ 0& -B_S &  C_S
\end{pmatrix} \,, \\
{\rm where,} ~~~ A_S &=& (\lambda_1+\lambda_3)v^2\sin^2\beta +\frac{3}{4}\lambda_4v^2 \sin\beta\cos\beta \,, \\
B_S &=& -\frac{1}{2} \left\{\frac{3}{2}\lambda_4\sin^2\beta + (\lambda_5+\lambda_6+2\lambda_7)\sin\beta\cos\beta \right\}v^2 \,, \\
C_S &=& -\frac{\lambda_4}{4}v^2\sin^2\beta\tan\beta + \lambda_8 v^2\cos^2\beta \,.
\end{eqnarray}
\end{subequations}
The state, $h^0$, will no longer be massless, in fact,
\begin{eqnarray}
m_{h0}^2 = -\frac{9}{2}\lambda_4 v^2\sin\beta\cos\beta \,.
\end{eqnarray}
The angle $\alpha$, which was used to rotate from $(h'_2,~h_3)$ basis to the physical $(H,~h)$ basis, should be redefined as~:
\begin{eqnarray}
 \tan2\alpha &=&\frac{2B_S}{A_S-C_S} \,,
\end{eqnarray}
and corresponding mass eigenvalues should have the following expressions~:
\begin{subequations}
\begin{eqnarray}
m_H^2 &=& (A_S+C_S)+\sqrt{(A_S-C_S)^2+4B_S^{2}} \,, \\
m_h^2 &=& (A_S+C_S)-\sqrt{(A_S-C_S)^2+4B_S^{2}} \,.
\end{eqnarray}
\end{subequations}

The conclusion of the previous subsection that in the  alignment limit, $\cos(\beta-\alpha)=0$, $h$  possesses SM-like gauge and  Yukawa couplings, still holds. It should be emphasized that the Yukawa couplings of $h$ in this limit, resembles that of the SM, do not depend on the transformation properties of the fermions under $S_3$. Also, the self couplings of $h$ coincides with the corresponding SM expressions in the  alignment limit~:
\begin{eqnarray}
\mathscr{L}^{\rm self}_h = -\frac{m_h^2}{2v} h^3 -\frac{m_h^2}{8v^2} h^4 \,.
\end{eqnarray}
 Similar to the case described in the previous subsection, $h^0$ will not have any $h^0VV$ ($V=W$, $Z$) couplings, but in the present scenario, we may identify a symmetry which forbids such couplings. Note that, when the specified relation between $v_1$ and $v_2$ is taken, there exists a two dimensional representation of $Z_2$~:
\begin{eqnarray}
 \begin{pmatrix}1 & 0 \\ 0 & 1 \end{pmatrix} \,,~~
 \frac{1}{2}\begin{pmatrix}1 & \sqrt{3} \\ \sqrt{3} & -1 \end{pmatrix}\,, 
\end{eqnarray}
which was initially a subgroup of the original $S_3$ symmetry, remains intact even after the spontaneous symmetry breaking, {\it i.e.}, the vacuum is invariant under this $Z_2$ symmetry. This allows us to assign a $Z_2$ parity for different physical states and this should be conserved in the theory. The state $h^0$ is odd under this $Z_2$ and this is what forbids it to couple with the $VV$ pair. In fact, using the assignments of Table~\ref{table1}, together with CP symmetry, many of the scalar self couplings can be inferred to be zero.

\begin{table}
\begin{center}
\begin{tabular}{|c|c|}
\hline
Physical States &  Transformation under $Z_2$ \\
\hline\hline
$h^{0}$, $H_{1}^{\pm}$, $A_{1}$  & Odd \\
\hline
$H^{0}$, $R$, $H_{2}^{\pm}$, $A_{2}$ & Even \\
\hline
\end{tabular}
\end{center}
\caption[$Z_2$ parity assignments to the physical mass eigenstates]{\em $Z_2$ parity assignments to the physical mass eigenstates.}
\label{table1}
\end{table}

In connection with the number of independent parameters in the Higgs potential, we note that there were ten to start with ($\mu_{1,3}$ and $\lambda_{1,2, \dots, 8}$). $\mu_1$ and $\mu_3$ can be traded for $v_2$ and $v_3$ or, equivalently for $v$ and $\tan\beta$. The remaining eight $\lambda$s can be traded for seven physical Higgs masses and $\alpha$. The connections are given below~:
\begin{subequations}
\begin{eqnarray}
\lambda_1 &=& \frac{1}{2v^2\sin^2\beta} \left\{\left(m_h^2\cos^2\alpha+m_H^2\sin^2\alpha\right)+ \left(m_{1+}^2 -m_{2+}^2\cos^2\beta -\frac{1}{9} m_{h0}^2 \right) \right\} \,, \\
\lambda_2 &=& \frac{1}{2v^2\sin^2\beta} \left\{(m_{1+}^2-m_{A1}^2)- (m_{2+}^2-m_{A2}^2)\cos^2\beta \right\} \,, \\
\lambda_3 &=& \frac{1}{2v^2\sin^2\beta} \left(\frac{4}{9} m_{h0}^2 +m_{2+}^2\cos^2\beta - m_{1+}^2\right) \,, \\
\lambda_4 &=& -\frac{2}{9} \frac{m_{h0}^2}{v^2}\frac{1}{\sin\beta\cos\beta} \,, \\
\lambda_5 &=& \frac{1}{v^2} \left\{\frac{\sin\alpha\cos\alpha}{\sin\beta\cos\beta}\left(m_H^2-m_h^2 \right) +2 m_{2+}^2 +\frac{1}{9}\frac{m_{h0}^2}{\cos^2\beta} \right\} \,, \\
\lambda_6 &=& \frac{1}{v^2}\left(\frac{1}{9}\frac{m_{h0}^2}{\cos^2\beta}+m_{A2}^2-2m_{2+}^2 \right) \,, \\
\lambda_7 &=& \frac{1}{2v^2}\left(\frac{1}{9}\frac{m_{h0}^2}{\cos^2\beta}-m_{A2}^2 \right) \,, \\
\lambda_8 &=& \frac{1}{2v^2\cos^2\beta}\left\{\left(m_h^2\sin^2\alpha+m_H^2\cos^2\alpha \right) -\frac{1}{9} m_{h0}^2\tan^2\beta \right\} \,.
\end{eqnarray}
\end{subequations}
In passing, we wish to state that for the analysis purpose we will always be working in the  alignment limit with $v_1=\sqrt{3}v_2$.

\begin{figure} 
%
\includegraphics[scale=0.75]{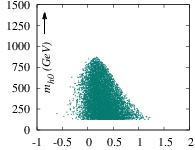} 
\includegraphics[scale=0.75]{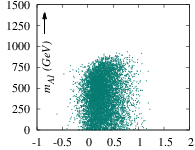}
\includegraphics[scale=0.75]{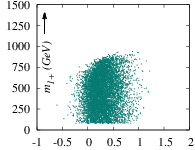}
\centerline{ \null \hfill $\log_{10}(\tan \beta) ~~ \rightarrow$ \quad}

\vspace{4mm}
%
\includegraphics[scale=0.75]{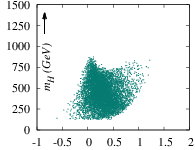} 
\includegraphics[scale=0.75]{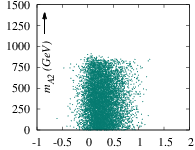}
\includegraphics[scale=0.75]{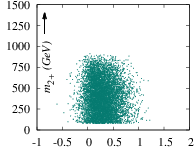}
\centerline{ \null \hfill $\log_{10}(\tan \beta) ~~ \rightarrow$ \quad}

\caption[Unitarity constrains on the nonstandard masses in S3HDM]{\em (Case-II) Regions allowed from unitarity and stability. We have fixed $m_h$ at 125 GeV and taken $m_{1+},~m_{2+} > 80$\,GeV and $m_H, m_{h0} > m_h$.}
\label{f:planes}
\end{figure}

\section{Constraints from unitarity} \label{s:unitarity}
In this context, the pioneering work has been done by Lee, Quigg and Thacker (LQT)  \cite{Lee:1977eg}. They have analyzed
several two body scatterings involving longitudinal gauge bosons and physical Higgs in the SM. All such scattering amplitudes are proportional to Higgs quartic coupling in the high energy limit. The $\ell=0$ partial wave amplitude
$(a_0)$ is then extracted from these amplitudes and cast in the form of an S-matrix having different two-body states
as rows and columns. The largest eigenvalue of this matrix is bounded by the unitarity constraint, $|a_0 | < 1$.
This restricts the quartic Higgs self coupling and therefore the Higgs mass to a maximum value.

The procedure has been extended to the case of a 2HDM scalar potential \cite{Maalampi:1991fb,Kanemura:1993hm,Akeroyd:2000wc,Horejsi:2005da}. We take it one step further and apply it in the context of 3HDMs.
Here also same types of two body scattering channels are considered. Thanks to the equivalence theorem \cite{Pal:1994jk,Horejsi:1995jj}, we can use unphysical Higgses instead of actual longitudinal components of the gauge bosons when considering the
high energy limit. So, we can use the Goldstone-Higgs potential of \Eqn{potential} for this analysis. Still it will be a
much involved calculation. But we notice that the diagrams containing trilinear vertices will be suppressed by
a factor of $E^2$ coming from the intermediate propagator. Thus they do not contribute at high energies, -- only
the quartic couplings contribute. Clearly the physical Higgs masses that could come from the propagators, do
not enter this analysis. Since we are interested only in the eigenvalues of the S-matrix, this allows us to work
with the original fields of \Eqn{quartic} instead of the physical mass eigenstates. After an inspection of all the neutral and charged two-body channels, we find the following eigenvalues to be bounded from unitarity:
\begin{eqnarray}
|a_i^\pm|,~|b_i| \le 16\pi, ~\mbox{for}~i=1,2,\ldots,6\,.
\end{eqnarray}
The expressions for the individual eigenvalues in terms of $\lambda$s are given below:
\begin{subequations}
\footnotesize
\begin{eqnarray}
a_1^\pm &=& \left(\lambda_1-\lambda_2+ \frac{\lambda_5+\lambda_6}{2}\right) \nonumber \\
&& \pm \sqrt{\left(\lambda_1-\lambda_2+ \frac{\lambda_5+\lambda_6}{2}\right)^2 -4\left\{(\lambda_1-\lambda_2)\left(\frac{\lambda_5+\lambda_6}{2}\right) -\lambda_4^2 \right\} } \,, \\
a_2^\pm &=& \left(\lambda_1+\lambda_2+2\lambda_3+\lambda_8\right) \pm \sqrt{\left(\lambda_1+\lambda_2+2\lambda_3+\lambda_8\right)^2 -4\left\{\lambda_8(\lambda_1+\lambda_2+2\lambda_3) -2\lambda_7^2 \right\} } \,, \\
a_3^\pm &=& \left(\lambda_1-\lambda_2+2\lambda_3+\lambda_8\right) \pm \sqrt{\left(\lambda_1-\lambda_2+2\lambda_3+\lambda_8\right)^2 -4\left\{\lambda_8(\lambda_1-\lambda_2+2\lambda_3) -\frac{\lambda_6^2}{2} \right\} } \,, \\
a_4^\pm &=& \left(\lambda_1+\lambda_2+\frac{\lambda_5}{2}+\lambda_7\right) \nonumber \\
&& \pm  \sqrt{\left(\lambda_1+\lambda_2+\frac{\lambda_5}{2}+\lambda_7\right)^2 -4\left\{(\lambda_1+\lambda_2)\left(\frac{\lambda_5}{2}+\lambda_7\right) -\lambda_4^2 \right\} } \,, \\
a_5^\pm &=& \left(5\lambda_1-\lambda_2+2\lambda_3+3\lambda_8\right) \nonumber \\
&& \pm \sqrt{\left(5\lambda_1-\lambda_2+2\lambda_3+3\lambda_8\right)^2 -4\left\{3\lambda_8(5\lambda_1-\lambda_2+2\lambda_3) -\frac{1}{2}(2\lambda_5+\lambda_6)^2 \right\} } \,, \\
a_6^\pm &=& \left(\lambda_1+\lambda_2+4\lambda_3+\frac{\lambda_5}{2}+\lambda_6+3\lambda_7\right) \nonumber \\
&& \pm \sqrt{\left(\lambda_1+\lambda_2+4\lambda_3+\frac{\lambda_5}{2}+\lambda_6+3\lambda_7\right)^2 -4\left\{(\lambda_1+\lambda_2+4\lambda_3)\left(\frac{\lambda_5}{2}+\lambda_6+3\lambda_7\right) -9\lambda_4^2 \right\} } \,, \\
b_1 &=& \lambda_5+2\lambda_6-6\lambda_7 \,, \\
b_2 &=& \lambda_5 -2\lambda_7 \,, \\
b_3 &=& 2(\lambda_1-5\lambda_2-2\lambda_3) \,, \\
b_4 &=& 2(\lambda_1-\lambda_2-2\lambda_3) \,, \\
b_5 &=& 2(\l_1+\l_2-2\l_3) \,, \\
b_6 &=& \lambda_5-\lambda_6\,.
\end{eqnarray}
\label{unitarity eq}
\end{subequations}
In passing, we remark that the perturbativity criteria, $|\lambda_i| < 4\pi$, coming from the requirement that the leading order contribution to the physical amplitude must have higher magnitude than the subleading order, may have some ambiguity in this context. This is due to the fact the individual $\lambda$s do not appear in the quartic couplings involving the physical scalars. Hence the combination of $\lambda$s, that constitute the physical couplings,  should be used for this purpose and it does not necessarily imply that the individual $\lambda$s should be bounded. We have presented here the exact constraints on $\lambda$s which should be satisfied for unitarity not to be violated.

Eqs.~(\ref{stability-s3}) and (\ref{unitarity eq}) can be used to put limits on the physical Higgs masses. For this purpose, we work in the alignment limit taking the lightest scalar ($h$) to be the SM-like Higgs that has been found at the LHC and we set its mass at 125 GeV. We also assume the charged scalars ($m_{1+}$ and $m_{2+}$) to be heavier than 80 GeV to respect the direct search bound from LEP2 \cite{Searches:2001ac}. To collect sufficient number of data points we have generated fifty million random sets of \{$\tan\beta,~m_{h0},~m_{H},~m_{A1},~m_{A2},~m_{1+},~m_{2+}$\} by varying $\tan\beta$ from 0.1 to 100 and filter them through the combined constraints from unitarity and stability. The sets that survive the filtering are plotted in Figure~\ref{f:planes}. The bounds that follow from these figures are listed below:
\begin{itemize}
\item $\tan\beta~\in$  [0.3, 17],
\item $m_{h0}<$ 870 GeV, $m_{H}<$ 880 GeV, $m_{A1}<$ 940 GeV, $m_{A2}<$ 910 GeV, $m_{1+}<$ 940, $m_{2+}<$ 910 GeV.
\end{itemize}
It is interesting to note that if the observed scalar at the LHC has its root in the S3HDM, then there must be several other nonstandard scalars with masses below 1 TeV. 

\section{Impact on loop induced Higgs decays} \label{s:decay-s3}
\begin{figure}
\includegraphics[scale=0.4]{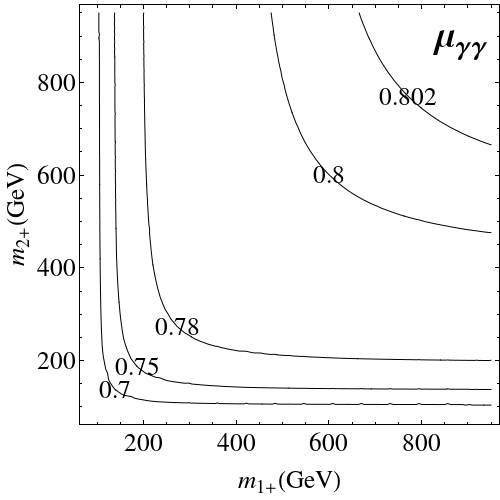}~~~~~~~
\includegraphics[scale=0.4]{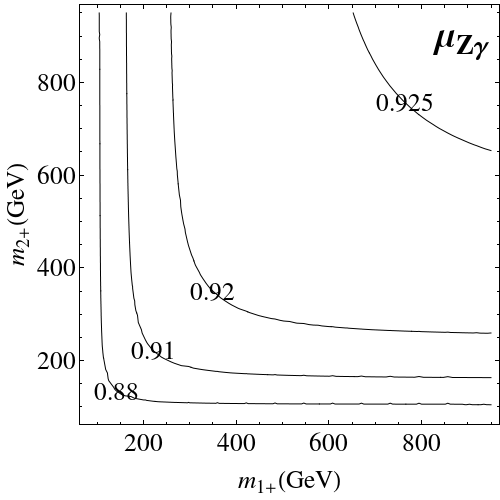}
\caption[Impact on diphoton signal strength]{\em Signal strengths for diphoton and Z-photon decay modes within the allowed range for charged Higgs masses.}
\label{f:decay}
\end{figure}
Similar to the 2HDM case, to display the contribution of the charged scalar loops to the decay amplitudes in a convenient form, we define dimensionless parameters, $\kappa_{i}~(i=1,2)$, in the following way:
\begin{eqnarray}
g_{hH_{i}^{+}H_{i}^{-}}=\kappa_{i}\frac{gm_{i+}^{2}}{M_{W}}\,.
\label{defkappa-s3}
\end{eqnarray}
Following this definition in analogy with the 2HDM case, we can write the diphoton and $Z$-photon signal strengths in the alignment limit as follows:
\begin{eqnarray}
 \mu_{\gamma\gamma} &=&  \frac{\Gamma(h \to \gamma\gamma)}{\Gamma^{\rm
     SM}(h\to \gamma\gamma)} 
 = \frac{\Big|F_W + \frac{4}{3}F_t  +
    \sum_{i=1}^{2}\kappa_{i} F_{i+}\Big|^2}{\Big|F_W + \frac{4}{3}F_t \Big|^2} \,,
 \\ 
 \mu_{Z\gamma} &=&  \frac{\Gamma(h \to
  Z\gamma)}{\Gamma^{\rm SM}(h\to Z\gamma)} = \frac{\Big|A_W + A_t  +
   \sum_{i=1}^{2}\kappa_{i} A_{i+} \Big|^2}{ \Big|A_W + A_t \Big|^2} \,,
\end{eqnarray}
where the expressions for the quantities $F_x$ and $A_x$ already appear in \Eqs{Fs}{As}.
In the alignment limit, the parameters $\kappa_{i}~(i=1,2)$, which appear in
Eq.\ (\ref{defkappa-s3}) are given by,
\begin{eqnarray}
\kappa_{i} = -\left(1 + \frac{m_h^2}{2m_{i+}^2}\right)\,.
\label{kappa}
\end{eqnarray}
As the charged Higgs becomes heavy, the quantity $F_{i+}$, for example, saturates to $\frac{1}{3}$.  So the
 decoupling of charged Higgs from loop induced Higgs decay depends on how $\kappa_{i}$ behaves with increasing $m_{i+}$. 
 It follows from \Eqn{kappa} that $\kappa_{i}\to -1$ if $m_{i+}\gg m_{h}$. Consequently, the charged Higgs never decouples 
 from the diphoton or $Z$-photon decay amplitudes. In fact, it reduces the decay widths from their corresponding SM 
 expectations. These features have been displayed in Figure~\ref{f:decay} where we have made a contour plot by varying the 
 charged Higgs masses within the allowed ranges coming from unitarity and vacuum stability. We find that $\mu_{\gamma \gamma}$ and $\mu_{Z \gamma}$ should lie within  [0.42, 0.80] and [0.73, 0.93] for $m_{1+}\in [80, 950]$ and $m_{2+}\in [80, 950]$. We must admit though, this nondecoupling of charged scalar is not a unique feature of a S3HDM as it is also known 
 to be present in the context of a 2HDMs \cite{Djouadi:1996yq, Arhrib:2003ph, Bhattacharyya:2013rya, Ferreira:2014naa}. 
 Currently both the ATLAS and CMS data shows agreement with the SM expectations \cite{Atlas:signal,cms:signal}. Thus a precise measurement of the diphoton and $Z$-photon signal strengths can pin down the difference between the 
 SM Higgs and a SM-like Higgs arising from an extended scalar sector. 

\section{Conclusions} \label{s:concl}
In this chapter the scalar sector of an S3HDM has been analyzed in detail. The major findings are listed below:
\begin{itemize}
\item The minimization of the scalar potential leads to a specific relation between the vevs of the first two doublets, $v_{1}=\sqrt{3} v_{2}$ in particular.

\item  In this limit we find a $Z_{2}$ subgroup of $S_{3}$ that remains unbroken even after the spontaneous symmetry breaking. The different scalar mass eigenstates can then be assigned with appropriate $Z_{2}$ parity which can help us understand why certain couplings do not appear in the theory.

\item Additionally, we have identified a alignment limit for this model where the lightest CP-even scalar has the exact same coupling as the SM Higgs with the other SM particles.

\item We have also derived the exact tree unitarity constraints and exploited them, in the decoupling limit, to put new bounds on the physical nonstandard Higgs masses, which we consider to be an important development in the multi-Higgs context.

\item From unitarity and stability $\tan \beta$ is likely to be in the range [0.3,17] and all the nonstandard Higgs masses lie below 1~TeV.

\item Regarding the decay of the SM-like $S_{3}$ Higgs, we have observed that the charged Higgs never decouples from the diphoton or $Z$-photon decay modes. The additional contributions from the charged Higgs loops to the decay amplitudes actually reduces the signal strengths of these modes. Although this depletion may not be a unique property of this scenario, but any statistically significant enhancement in $h\to \gamma \gamma$ and/or $h\to Z \gamma$ modes will certainly disfavor the possibility of an SM-like Higgs arising from an S3HDM.
\end{itemize}


\begin{savequote}[0.4\textwidth]  
I dwell in possibility \dots
\qauthor{Emily Dickinson}    
\end{savequote}
%

\chapter[The Higgs or a Higgs?]{The Higgs or a Higgs? \\ Will the LHC be able to decipher?} 

\lhead{Chapter 5. \emph{The Higgs or a Higgs?}}
\label{Chap5} 

The behavior of the scalar boson observed at the CERN LHC is tantalizingly close to that of the 
SM Higgs boson. A very timely and relevant question is whether this
scalar is the only one of its type as predicted by the SM, or it is
the first to have been discovered in a family of more such species
arising from an underlying extended scalar sector.  A natural
extension of the SM scalar structure is realized by adding more $SU(2)$  scalar doublets,
which we consider in this chapter. There are two advantages for choosing
doublets.  First, the $\rho$-parameter remains unity at tree
level. Second, it is straightforward to find a combination, namely,
\begin{eqnarray}
h=\frac{1}{v}\sum\limits_{i=1}^{n} v_ih_i \,, ~~~~{\rm with}~~
v^2=\sum\limits_{i=1}^{n}v_i^2=(246~{\rm GeV})^2\,,
\label{SMh}
\end{eqnarray}
($v_i$ is the vev of the $i$-th doublet and
$h_i$ is the corresponding real scalar field), which has SM-like
couplings with fermions and gauge bosons. This is not in general a
mass eigenstate.  But when we demand that this is indeed the physical
state observed at the LHC with a mass $m_h \approx$ 125 GeV, we are
automatically led to the so called {\em alignment limit}. This limit
is motivated by the LHC data on the Higgs boson signal strengths in
different channels which are showing increasing affinity towards the
SM predictions.  In this chapter we pay specific attention to the $h \to
\gamma\gamma$ process. Though this process is loop driven and has a
small branching ratio, it played an important role in the Higgs
discovery. Importantly, this branching ratio is expected to be
measured in LHC-14 with much greater accuracy.  Now, additional $SU(2)$ 
scalar doublets would bring in additional states, both charged and
neutral, in the spectrum. Here our primary concern is how those
charged scalars couple to $h$ and how much they contribute to the $h
\to \gamma \gamma$ rate as virtual states in loops. This leads to the
observation that even when the masses of the charged scalars floating
in the loop are taken to very large values, they do not {\em
  necessarily} decouple from this process. Deciphering the underlying
reasons behind this constitutes the motive of this chapter. Although
this has been noted in the past in the context of 2HDMs, only some cursory remarks were made on it without
exploring its full implications~\cite{Djouadi:1996yq, Arhrib:2003ph,
  Chang:2012ta, Bhattacharyya:2013rya,
  Ferreira:2014naa,Fontes:2014tga}.  We investigate the r\^ole of
symmetries which are imposed on the scalar potential in figuring out
under what conditions the decoupling of heavy charged scalars in the
$h \to \gamma \gamma$ loop takes place. The upshot is that if the
potential has an exact $Z_2$ symmetry {\em and} both the scalars
receive vevs, which is the case for a large class of 2HDM
scenarios~\cite{Branco:2011iw}, the contribution of the charged scalar
does not decouple. If $Z_2$ is softly broken by a term in the
potential then decoupling can be achieved at the expense of tuning of
parameters.  On the other hand, a global $U(1)$ symmetry followed by its
soft breaking can ensure decoupling.  For simplicity, we first
demonstrate this behavior in the context of 2HDM.  We then address the
same question, for the first time, in the context of
3HDMs. It is not difficult to foresee what
happens if we add more doublets, which leads us to draw an important
conclusion: unless decoupling is ensured, e.g. as we did by imposing a
global $U(1)$ symmetry in the 2HDM potential, precision measurements of
$h \to \gamma \gamma$ branching ratio can put constraints on the
number of additional non-inert scalar doublets regardless of how heavy
the charged scalars are.  It should be recalled that only lower bounds on charged
scalar masses have been placed from processes like $b \to s \gamma$,
as the effects decouple when their masses are heavy for all such
flavor observables.  Thus, precision measurements of $h \to
\gamma\gamma$ would provide complementary information.  Incidentally,
whatever we comment on $h \to \gamma \gamma$ applies for $h \to Z
\gamma$ as well at least on a qualitative level.

It has already been emphasized in the preceding chapters that in multi doublet scalar models, the production
cross section as well as the tree-level decay widths of the Higgs
boson remain unaltered from their respective SM expectations in the
alignment limit. Only the loop induced decay modes like
$h\to\gamma\gamma$ and $h\to Z\gamma$ will pick up additional
contributions induced by virtual charged scalars. However, the
branching ratios into these channels are too tiny compared to other
dominant modes. As a result, the total Higgs decay width will be
hardly modified. Considering all these, the expression for the
diphoton signal strength is simplified to
\begin{eqnarray}
 \mu_{\gamma\gamma} &\equiv & {\sigma(pp\to h) \over \sigma^{\rm SM}(pp\to
   h)} \cdot {\mbox{BR} (h \to \gamma\gamma) \over \mbox{BR}^{\rm SM}
   (h \to \gamma\gamma)}~=~\frac{\Gamma(h \to
   \gamma\gamma)}{\Gamma^{\rm SM}(h\to \gamma\gamma)} \, .
\label{mugg}
\end{eqnarray}
For convenience, we parametrize again the coupling of $h$ to the charged
scalars in the following generic way:
\begin{eqnarray}
g_{hH_{i}^{+}H_{i}^{-}}=\kappa_{i}\frac{gm_{i+}^{2}}{M_{W}} \,,
\label{defkappa-1}
\end{eqnarray}
where $m_{i+}$ is the mass of the $i$-th charged scalar
($H_i^\pm$). As we will see later, the decoupling or nondecoupling
behavior of the $i$-th charged scalar from $\mu_{\gamma\gamma}$ is
encoded in $\kappa_i$. The expression of the Higgs to diphoton signal strength will be given by
\begin{eqnarray}
 \mu_{\gamma\gamma} &=& \frac{\Big|F_W + \frac{4}{3}F_t  +
    \sum_{i}\kappa_{i} F_{i+}\Big|^2}{\Big|F_W +
   \frac{4}{3}F_t \Big|^2} \,,
\end{eqnarray}
where the expressions for $F_x$ appear in \Eqn{Fs}. In the limit the charged scalar is very heavy, the quantity
$F_{i+}$ saturates to $1/3$.  If $\kappa_i$ also saturates
to some finite value in that limit then the charged scalar would not
decouple from the $h \to \gamma\gamma$ loop. Then no matter how heavy
the charged scalar is, $\mu_{\gamma\gamma}$ will differ from its SM
value.  If the experimental value of $\mu_{\gamma\gamma}$ eventually
settles on very close to the SM prediction then such nondecoupling
scenarios will be disfavored. The decoupling would happen only if
$\kappa_i$ falls with increasing charged scalar mass.  In what
follows, we will illustrate these features by considering some popular
doublet extensions of the SM scalar sector.

\section{Two Higgs-doublet models} 
\begin{figure}
\begin{center}
\includegraphics[scale=0.35]{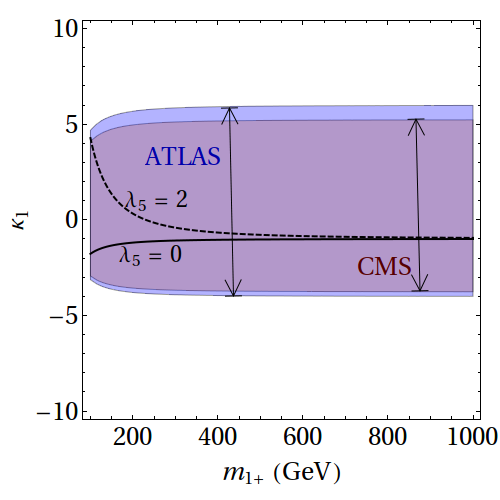} ~~~~~
\includegraphics[scale=0.35]{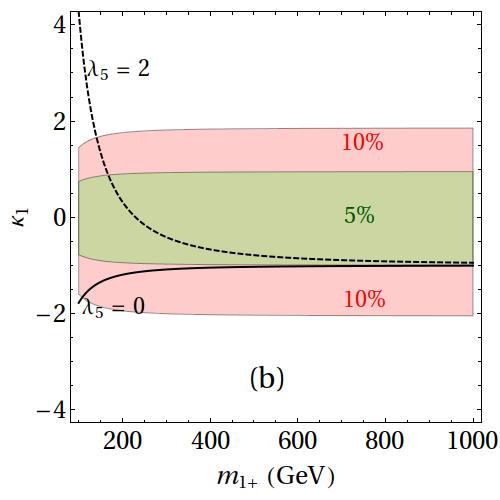}
\end{center} 
\caption[Allowed region for $\kappa_1$ in 2HDMs]{\em In the left panel (a) we display the constraints on
  $\kappa_1$ in 2HDM coming from the measured values of
  $\mu_{\gamma\gamma}$ at 95\% C.L. by the CMS ($1.14
  ^{+0.26}_{-0.23}$ \cite{Khachatryan:2014ira}) and ATLAS ($1.17
  \pm 0.27$ \cite{Aad:2014eha}) Collaborations. In the right panel
  (b) we show what would be the 95\% C.L. allowed range of $\kappa_1$
  if $\mu_{\gamma\gamma}$ is hypothetically measured to be as $1 \pm
  0.1 (0.05)$ in future colliders.  In both panels we have plotted
  Eq.~(\ref{kappaz2}) for two different values of $\lambda_5$. The figures have been taken from \cite{Bhattacharyya:2014oka}.}
\label{fig1}
\end{figure}

The 2HDM potential has been discussed in detail in Chapter~\ref{Chap3}. Here we will briefly highlight
the main conclusions in the context of $h\to\gamma\gamma$. First, it is important to note that the charged scalar contribution to $\mu_{\gamma\gamma}$ is controlled
by (putting $i=1$ in $\kappa_i$)~\cite{Djouadi:1996yq, Arhrib:2003ph,
  Bhattacharyya:2013rya, Ferreira:2014naa, Swiezewska:2012eh}
\begin{eqnarray}
\kappa_1 =
-\frac{1}{m_{1+}^2}\left(m_{1+}^2+\frac{m_h^2}{2}-\frac{\lambda_5
  v^2}{2} \right) \, .
\label{kappaz2}
\end{eqnarray}
Clearly, $\kappa_1$ saturates to $-1$ as the charged scalar becomes
excessively heavy. Decoupling can be achieved by tuning $m_{1+}^2
\simeq \lambda_5 v^2/2$ \cite{Gunion:2002zf}. Recalling our counting
of independent parameters, any adjustment between the charged scalar
mass and $\lambda_5$ is nothing short of fine-tuning.  On the other
hand, if the $Z_2$ symmetry in the scalar potential is exact, {\it
  i.e.}  $\lambda_5=0$, then the charged scalar will never decouple
and will cause $\mu_{\gamma\gamma}$ to settle below its SM prediction.
In Fig.~\ref{fig1} we have plotted the allowed range of $\kappa_1$ in
2HDM from the present LHC data as well as from an anticipation of
future sensitivity.

An interesting possibility arises when we employ a $U(1)$ symmetry,
instead of the usual $Z_2$ symmetry, in the potential. The choice
$\lambda_5=\lambda_6$ will ensure $U(1)$ symmetry in the quartic
terms. The bilinear term involving $\lambda_5$ still breaks the $U(1)$
symmetry softly.  Then the mass of the pseudoscalar gets related to
the soft breaking parameter $\lambda_5$ as $m_A^2 = \lambda_5 v^2/2 $.
In this case, the expression for $\kappa_1$ reads \cite{Bhattacharyya:2013rya}:
\begin{eqnarray}
\kappa_1 = -\frac{1}{m_{1+}^2}\left(m_{1+}^2-m_A^2+\frac{m_h^2}{2} \right) \,.
\label{u1kappa}
\end{eqnarray}
In the paper~\cite{Bhattacharyya:2013rya}, a
detailed analysis on the unitarity and stability constraints on
various combination of $\lambda_i$ couplings when the 2HDM scalar
potential has a softly broken $U(1)$ symmetry has been provided. We cite some of them here
to demonstrate `decoupling' for large individual quartic couplings, as
what is constrained from unitarity is only their differences in
certain combinations. For example, $(2 \lambda_3 + \lambda_4) \leq
16\pi$ implies $(2 m_{1+}^2 - m_H^2 - m_A^2 +m_h^2) \leq 16\pi
v^2$. Also, $(\lambda_1 + \lambda_2 + 2 \lambda_3) \leq 16\pi/3$
implies $(m_H^2 - m_A^2)(\tan^2\beta + \cot^2\beta) + 2m_h^2 \leq
32\pi v^2/3$. These relations, together with $|m_{1+} - m_H| \ll
(m_{1+}, m_H)$ arising from the electroweak $T$ parameter, restrict
the splitting between the charged scalar and the pseudoscalar mass
($|m_{1+}^2-m_A^2|$). As displayed through more such relations among
quartic couplings and the associated plots in the plane of non-SM
scalar masses in \cite{Bhattacharyya:2013rya}, the individual scalar
masses can become very large without violating unitarity as long as
their mass-square differences are within certain limits.
Consequently, the numerator in \Eqn{u1kappa} cannot grow indefinitely
with increasing $m_{1+}$.  Thus $\kappa_1$ becomes very small in that
limit and $\mu_{\gamma\gamma}$ reaches the SM predicted value.  The
key issue is that the $Z_2$ symmetry breaking $\lambda_5$ term was not
related to the mass of any particle in the spectrum, and hence its
adjustment {\em vis-\`a-vis} the charged scalar mass was nothing short
of fine-tuning.  Now, the global $U(1)$ breaking $\lambda_5$ is related
to the pseudoscalar mass whose splitting with the charged scalar mass
is restricted from unitarity.

\subsection{Underlying dynamics behind decoupling}
We now discuss the underlying reason behind decoupling or
nondecoupling of nonstandard scalars from physical processes in the
2HDM context. The conclusion is equally applicable for $n$HDM where $n
> 2$. We first recall the 2HDM potential in \Eqs{notation1}{notation2}.
The parametrization of \Eqn{notation1} does not {\em a priori} assume, unlike the one
 in Eq.~(\ref{notation2}), that $\Phi_1$ 
or $\Phi_2$ necessarily acquires any vev. In this parametrization, in
the limit when the dimensionless couplings $\beta_2 = \beta_3 =
\beta_4 = \beta_5 = 0$, the mass mixing parameter $m_{12}^2 = 0$ the second Higgs doublet $\Phi_2$ does not acquire any
vev, and the SM scalar potential is recovered with the relation $v^2 =
v_1^2 = - m_{11}^2/\beta_1$.  This is  one special case of the more general {\em inert doublet} scenario
with a perfectly $Z_2$ symmetric potential, in which all the
nonstandard scalars decouple from physical processes when the
parameter $m_{22}^2$ controlling their masses is taken to infinitely
large value.  Note that $m_{22}^2$, in this case, does not have its
origin in spontaneous symmetry breaking (SSB), and this is why its
large value could ensure decoupling. But when both the doublets receive vevs, one can trade the two
parameters $m_{11}^2$ and $m_{22}^2$ in favor of $v_1$ and $v_2$. Then
the magnitude of the third parameter $m_{12}^2$, or equivalently
$\lambda_5$, has nothing to do with SSB, and this parameter provides
the regulator whose large value ensures decoupling of all nonstandard
scalars from physical processes. However, while employing $m_{12}^2$
(or equivalently $\lambda_5$ in our parametrization) for decoupling,
one cannot escape from some tuning of parameters for softly broken
$Z_2$ as explained around Eq.~(\ref{kappaz2}), but no such tuning is
required for softly broken $U(1)$ (discussed before).  Nondecoupling
would result when the symmetry of the potential is exact ($m_{12}^2 =
0$), {\em and at the same time}, both the scalars receive vevs (which
implies $\lambda_5 = 0$).  In this case all the non-SM physical scalar
masses would be proportional to the electroweak vev, and there is no
independent mass-dimensional parameter which has non-SSB origin.  As
illustrated in the inert doublet case, even with exactly symmetric
potential, decoupling is achieved in 2HDM.  

To provide further intuition into the argument of decoupling and its
close connection to the existence of some non-SSB origin parameter, let us consider 
the following analogy.  It is well known that the top quark in
the SM does not decouple from $h\to\gamma\gamma$. This is because the
top quark receives all its mass from SSB and increasing the its mass
will invariably imply enhancing the Yukawa coupling ($h_t$). Now
suppose that the top quark receives part of its mass ($M$) from some
non-SSB origin, i.e. $m_t=h_tv+M$.  Then the top-loop contribution
will yield a prefactor $h_tv/(h_tv+M)$. In this case, by taking $M \to
\infty$, the top quark contribution can be made to decouple from the
diphoton decay width of the Higgs boson.

\section{Three-Higgs-doublet models} 
$S_3$ or $A_4$ symmetric flavor models are typical examples which
employ three Higgs doublets.  With $\Phi_1$, $\Phi_2$ and $\Phi_3$ as
the three scalar $SU(2)$  doublets, the scalar potential for the $S_3$
symmetric has been already written in \Eqn{potential} (see
e.g. \cite{Barradas-Guevara:2014yoa,Das:2014fea}, and also references
therein for flavor physics discussions both when the $S_3$ symmetry is
exact as well as when it is softly broken). Assuming the lambdas to be real, potential minimization conditions
attribute a relation between two of the three vevs
($v_1=\sqrt{3}v_2$). Using this relation, an alignment limit can be
obtained for this model also \cite{Das:2014fea}.

Now we write the potential satisfying $A_4$ symmetry (see e.g. \cite{Ma:2001dn}),
\begin{eqnarray}
V_{\rm 3HDM}^{A_4} & = &
-\mu^2\left(\Phi_1^\dagger\Phi_1+\Phi_2^\dagger\Phi_2+\Phi_3^\dagger\Phi_3
\right)
+\lambda_1\left(\Phi_1^\dagger\Phi_1+\Phi_2^\dagger\Phi_2+\Phi_3^\dagger\Phi_3
\right)^2 \nonumber
\\ &&+\lambda_2\left(\Phi_1^\dagger\Phi_1\Phi_2^\dagger\Phi_2
+\Phi_2^\dagger\Phi_2\Phi_3^\dagger\Phi_3
+\Phi_3^\dagger\Phi_3\Phi_1^\dagger\Phi_1 \right)
\nonumber\\ 
&& +\lambda_3\left(\Phi_1^\dagger\Phi_2\Phi_2^\dagger\Phi_1
+\Phi_2^\dagger\Phi_3\Phi_3^\dagger\Phi_2
+\Phi_3^\dagger\Phi_1\Phi_1^\dagger\Phi_3 \right) \nonumber
\\ &&+\lambda_4\left[e^{i\epsilon}\left\{\left(\Phi_1^\dagger\Phi_2\right)^2+
\left(\Phi_2^\dagger\Phi_3\right)^2+\left(\Phi_3^\dagger\Phi_1\right)^2
\right\} + {\rm h.c.}\right]
\,.
\label{a4potential}
\end{eqnarray}
In one plausible scenario, the minimization conditions require that all the three
vevs are equal \cite{Toorop:2010ex}. This particular choice
automatically yields a SM-like Higgs as well as two pairs of complex
neutral states with mixed CP properties. Note that for $\epsilon=0$ in
\Eqn{a4potential}, the symmetry of the potential is enhanced to
$S_4$. However, our conclusions do not depend on the value of
$\epsilon$. 

Thus, a 3HDM can provide an SM-like Higgs along with two pairs of
charged scalars, as exemplified with $S_3$ and $A_4$ scenarios.  After
expressing the lambdas in terms of the physical masses, we obtain 
the following expressions for $\kappa_i~(i=1,2)$ in the alignment
limit, which are the same for both $S_3$ and $A_4$:
\begin{eqnarray}
\kappa_i = -\left(1+\frac{m_h^2}{2m_{i+}^2} \right)~~{\rm for}~i=1,2 \,.
\end{eqnarray}
Clearly, the charged scalars do not decouple from the diphoton decay
width, since $\kappa_i$ settles to $-1$ when $m_{i+}$ is very large
compared to $m_h$.  Note, both the charged scalars contribute in the
same direction to reduce $\mu_{\gamma\gamma}$.

\begin{table}[htbp!]
\scriptsize
\begin{center}
\begin{tabular}{|c|c|c|c|c|c|}
\hline
\multicolumn{2}{|c|}{Model} &  Expression for $\kappa_i$ & prediction $\mu_{\gamma\gamma}$ & prediction $\mu_{Z\gamma}$ & Decoupling? \\
\hline\hline
 & Softly broken $Z_2$ & $-\left(1+\frac{m_h^2}{2m_{1+}^2}-\frac{\lambda_5 v^2}{2m_{1+}^2} \right)$ & Depends on $\lambda_5$ & Depends on $\lambda_5$ & Possible \\
\cline{2-6}
2HDM & Exact $Z_2$ & $-\left(1+\frac{m_h^2}{2m_{1+}^2} \right)$ & $\le 0.9$ & $\le 0.96$ &No \\
\cline{2-6}
& Softly broken $U(1)$ & $-\left(1+\frac{m_h^2}{2m_{1+}^2}-\frac{m_A^2}{m_{1+}^2} \right)$ & Depends on $m_A$ & Depends on $m_A$ & Yes \\
\hline
 & Exact $S_3$ & $-\left(1+\frac{m_h^2}{2m_{i+}^2} \right)$ for $i=1,2$ & $\le 0.8$ & $\le 0.93$ & No \\
\cline{2-6}
3HDM & Exact $A_4$ & $-\left(1+\frac{m_h^2}{2m_{i+}^2} \right)$ for $i=1,2$ & $\le 0.8$ & $\le 0.93$ & No \\
\cline{2-6}
& Softly broken $SO(2)$  &
\specialcell{$\kappa_1=-\left(1+\frac{m_h^2}{2m_{1+}^2}-
\frac{m_{h'}^2}{m_{1+}^2} \right)$ \\
$\kappa_2=-\left(1+\frac{m_h^2}{2m_{2+}^2} \right)$}  & Depends on $m_{h'}$ & Depends on $m_{h'}$ & Partial \\
\hline
\end{tabular}
\end{center}
\caption[Decoupling features of multi Higgs-doublet models]{\em Behavior of 2HDM and 3HDM scenarios in the alignment
  limit strictly when all the doublets receive vevs. In the case of
  exact discrete symmetries, every charged scalar pair reduces
  $\mu_{\gamma\gamma}$ approximately by $0.1$. Although explicit
  expression for $\mu_{Z\gamma}$ is not shown in text, its predictions
  in different scenarios are displayed. In the last column where we
  say `Possible', we mean that decoupling can be achieved with some
  tuning, while in the last row `Partial' implies that only the first
  charged scalar decouples.}
\label{BSMtable}
\end{table}

Now we turn our attention to the case of a global continuous symmetry
in 3HDM potential. For illustration, we consider that the symmetry is
$SO(2)$  under which $\Phi_1$ and $\Phi_2$ form a doublet.  The
expression for the scalar potential is similar to \Eqn{potential},
only that now $\lambda_4 =0$ and the potential contains an additional
bilinear term $(-\mu_{12}^2 \Phi_1^\dagger\Phi_2 + {\rm h.c.})$. The
real part of $\mu_{12}^2$ softly breaks the $SO(2)$  symmetry and
prevents the occurrence of any massless scalar in the theory. In any
case, we assume $\mu_{12}^2$ to be real just like any other parameters
in the potential. The relevant minimization conditions are given by
\begin{subequations}
\begin{eqnarray}
v_1\mu_1^2+v_2\mu_{12}^2 &=& v_1(v_1^2+v_2^2)(\lambda_1+\lambda_3)
+\frac{1}{2}v_1v_3^2(\lambda_5+\lambda_6+2\lambda_7)\,, \\
v_2\mu_1^2+v_1\mu_{12}^2 &=& v_2(v_1^2+v_2^2)(\lambda_1+\lambda_3)
+\frac{1}{2}v_2v_3^2(\lambda_5+\lambda_6+2\lambda_7)\,.
\end{eqnarray}
\end{subequations}
Note that nonzero $\mu_{12}^2$ will require $v_1=v_2$. An interchange
symmetry ($1\leftrightarrow 2$) is accidentally preserved even after
spontaneous symmetry breaking.  We will have three CP even scalars
($h',~H,~h$), two pseudoscalars ($A_1,~A_2$) and two pairs of charged
scalars ($H_1^\pm,~H_2^\pm$). Among these, $h'$, $A_1$ and $H_1^\pm$
are odd under the interchange symmetry and the rest are even under
it. Being odd under this interchange symmetry, $h'$ does {\em not}
couple to gauge bosons as $h'VV$ ($V=W,~Z$). Appearance of such an
exotic scalar was noted earlier in the context of an $S_3$ symmetric
3HDM \cite{Bhattacharyya:2010hp,Bhattacharyya:2012ze,Das:2014fea}. The
soft breaking parameter ($\mu^2_{12}$) gets related to the mass of
$h'$ as
\begin{equation}
m_{h'}^2 = 2\mu_{12}^2 \,.
\end{equation}
It is straightforward to express the lambdas in terms of the physical
masses. We then obtain
\begin{subequations}
\begin{eqnarray}
\kappa_1 &=&
-\frac{1}{m_{1+}^2}\left(m_{1+}^2-m_{h'}^2+\frac{m_h^2}{2}\right)
\,, \label{kap1so2} \\ \kappa_2 &=& -\left(1+\frac{m_h^2}{2m_{2+}^2}
\right) \,.
\end{eqnarray}
\label{kapso2}
\end{subequations}
The similarity between \Eqn{kap1so2} and \Eqn{u1kappa} is
striking. Note that $(|m_{1+}^2 - m_{h'}^2|)$ is constrained from
unitarity.  Therefore, when the first charged Higgs mass $m_{1+}$ is
very large, $\kappa_1$ becomes vanishingly small.  However, this
decoupling does not occur in $\kappa_2$ which contains the second
charged Higgs mass $m_{2+}$.  It is not difficult to intuitively argue
that with an extended global symmetry $SO(2)\times U(1)$, together with
an extra soft breaking parameter which is related to $m_{A2}$,
decoupling in $\kappa_2$ can be ensured.  Starting from the softly
broken $SO(2)$  symmetric potential, this additional $U(1)$ extension
($\phi_3 \to e^{i\alpha} \phi_3$) and its soft breaking can be
realized by putting $\lambda_7 = 0$ in Eq.~(\ref{potential}) and
introducing a term that softly breaks this U(1).  A crucial
observation we make in this chapter is that the masses $m_A$ in the 2HDM
context and $m_{h'}$ in the 3HDM context enter into the expressions of
$\kappa_i$ -- e.g. see Eqs.~(\ref{u1kappa}) and (\ref{kap1so2}) --
only when they are related to {\em soft} global symmetry breaking
parameters.

\section{Conclusions and outlook} 
In this chapter we have made an attempt to establish a
connection between decoupling or nondecoupling of charged scalars from
the diphoton decay of the Higgs with the symmetries of the scalar
potential.  We have argued that charged scalars in multi doublet scalar
extensions of the SM do not necessarily decouple from physical processes,
e.g. $\mu_{\gamma\gamma}$ in the context of this chapter, specifically
when the potential has an exact symmetry and all the scalars receive
vevs. In such scenarios, a precisely measured $\mu_{\gamma\gamma}$ can smell
the presence of nonstandard scalars even if
they are super-heavy.  In fact, $\mu_{\gamma\gamma}$ can constrain the
{\em number} of such doublets. Table~\ref{BSMtable} shows that each additional pair
of charged scalars ($H_{i}^\pm$) reduces $\mu_{\gamma\gamma}$
approximately by 0.1 when the potential has an exact discrete
symmetry. Our illustrations are based on two- and three-Higgs-doublet
models which are motivated by flavor symmetries.  We have explicitly
demonstrated how soft breaking of a global $U(1)$ symmetry can ensure
decoupling in 2HDM in the alignment limit. In the case of 3HDM, with a
softly broken global $SO(2)$  symmetry in the potential, decoupling can
be ensured for one pair of charged scalars ($H_{1}^\pm$), while the
second pair ($H_{2}^\pm$) still do not decouple.  Employing the soft
breaking terms of an extended global continuous symmetry, namely,
$SO(2) \times U(1)$, the nondecoupling effects of $H_{2}^\pm$ can be
tamed. If we have more pairs of charged scalars in the theory stemming
from additional scalar doublets, even more enhanced or extended global
continuous symmetries $-$ only softly broken $-$ would be required to
ensure decoupling of all charged scalars from $\mu_{\gamma\gamma}$.
Keeping in mind the expected accuracy in the measurement of the $hhh$
vertex in the high luminosity option of LHC or in the future linear
collider, whose tree level expression in the alignment limit remains
the same as in SM even for multi doublet Higgs structure,
$\mu_{\gamma\gamma}$ may offer a better bet for diagnosing the
underlying layers of the Higgs dynamics.


\begin{savequote}[10cm]  
Look at the end of work, contrast \\
The petty done, the undone vast ...    
 \qauthor{Robert Browning in ``The Last Ride Together''}    
\end{savequote}
%

\chapter{Summary and conclusions} 

\lhead{Chapter 6. \emph{Summary and conclusions}}
\label{Chap6} 

The discovery of a new boson in the July of 2012 at the LHC is undoubtedly the
greatest achievement of this decade in the field of Particle Physics. Precise measurements of the properties
of this new boson will be the major objective of the future high energy collider experiments. Although the
signal strengths of this new boson into various decay channels have been found to be compatible with what is
expected from the SM Higgs scalar, unambiguous determination of the couplings of this new
boson is necessary to make any conclusive remark on its {\em standardness}.

But, besides these fantastic agreements, preliminary data from both CMS and ATLAS showed larger than
$2\sigma$ excess over the SM expectation in the diphoton signal strengths. It is well known that Higgs to diphoton
decay, in the SM, proceeds at the leading order through $W$ and top quark (largest among the fermions) loops.
It is also known that the amplitudes from these two types of loop diagrams interfere destructively with $W$-loop
contribution dominating over the top-loop. Naturally if the sign of the top-Higgs Yukawa coupling is reversed
the interference will be constructive and an enhancement, as was seen in the experiments, can be obtained.
This led many people to speculate that the SM might have been wrong in predicting the sign of the top quark
Yukawa (or, every Yukawa) coupling. But we must remember that in absence of the SM Higgs, the amplitude
of $f\bar{f}\to W_L^+W_L^-$ scattering will grow linearly with the CM energy and the co-efficient of this linear growth is
proportional to the fermion mass, $m_f$ (see Appendix~\ref{AppendixB}). In fact, the requirement of the cancellation of this growth uniquely fixes the $\bar{f}fh$ Yukawa coupling. Thus, tormenting the Yukawa couplings in the SM by hand will lead to an imperfect cancellation of the energy growths spoiling the high energy unitarity of the $f\bar{f}\to W_L^+W_L^-$ scattering and this effect will be most prominent in the case of $t\bar{t}\to W_L^+W_L^-$ scattering because of the large mass of the top 
quark. In Chapter~\ref{Chap2}, it has been showed that any departure from the SM couplings will inevitably introduce a new energy scale at which unitarity will be violated and therefore, some new physics must set in before this energy scale
to restore it. A quantitative analysis of how this energy scale depends on the amount of nonstandard
deviation of the couplings has been demonstrated in this chapter.

However after an updated analysis of the data by ATLAS and CMS, the excess in the diphoton channel
seems to have gone away and all the signal strengths are in agreement with the SM. We have not found any direct evidence for new physics so far. Even if this new boson turns out to
be the SM Higgs particle, there are still some issues, not addressed in the SM, which invite us to take a closer look at the BSM physics.
The major shortcomings of the SM are as follows:
\begin{itemize}
\item Neutrinos are massless in the SM. But oscillation experiments with solar and atmospheric neutrinos have
established that neutrinos have small but finite mass. The neutrino oscillation experiments have raised
new puzzles as the mixing among the neutrinos are very different from that among the quarks.
\item Different cosmological and astrophysical observations have confirmed the existence of {\em dark matter} in
the universe. The fact that we cannot {\em see} it suggests that this must be some new kind of particle
without strong and electromagnetic interactions. On the other hand, the fact that it occurs with a certain
abundance requires that it must have some other kind of interactions which allow this to happen. In the
SM there is no dark matter candidate and one should, therefore, look beyond.
\item All observations in the large as well as the small scale universe have found no evidence of antimatter. There
is only matter in the universe, and the density of matter is only a billionth of the photon density at the
present epoch. There is no way to understand this matter-antimatter asymmetry within the framework
of the SM.
\end{itemize}

Speaking of BSM scenarios, Chapter~\ref{Chap3} is dedicated to the phenomenology of CP conserving 2HDMs -- one of the simplest BSM constructions. The first part of this chapter concentrates on the scalar potential only and the conclusions obtained in this part are independent of the Yukawa structure of the 2HDM. Our motivation here was to explore the scalar potential in view of the observation of a 125 GeV Higgs boson ($h$) at the LHC, using constraints from unitarity
of scalar scattering cross-sections, stability of the potential and electroweak precision tests. These considerations
restrict the spectrum of the non-standard scalars. Since the LHC Higgs data seem to be compatible with the
SM expectations, we restricted ourselves to the alignment limit when the lightest CP even scalar ($h$) resembles
the SM Higgs particle in its gauge and Yukawa couplings. In addition to $h$, a 2HDM contains one heavy CP
even scalar ($H$), one pseudoscalar ($A$) and a pair of charged scalars ($H_1^\pm$). We found that when the potential
has an exact $Z_2$ symmetry, unitarity and stability put severe constraints on $\tan\beta$. This bound, however, is diluted
in the presence of a soft breaking parameter. In the case of a softly broken $U(1)$ symmetry, we find that
unitarity constraints essentially applies to the difference between the nonstandard scalar masses but not on
the individual masses. This means any individual nonstandard mass can be arbitrarily large without violating
unitarity provided the other masses are close by. In particular, it has been shown that unitarity and stability
restricts the quantity $|(m_H^2-m_A^2)(\tan^2\beta+\cot^2\beta)|$. Consequently the difference between $m_H$ and $m_A$ is squeezed with increasing $\tb$ and, for practical purposes,
becomes degenerate for $\tb \gtrsim 5$. On top of this, when we imposed the constraint coming from the $T$-parameter, it
is found that all three nonstandard masses should be degenerate for $\tb \gtrsim 5$.

Later in Chapter~\ref{Chap3} some key features of the BGL models were explored. Note that the BGL models use
a softly broken $U(1)$ symmetry to tame the FCNC couplings by relating them to the off diagonal elements of
the CKM matrix and so all the conclusions of the previous paragraph, in the context of $U(1)$ symmetry, hold for these models too. The crucial
motivation of this chapter was to look for a lighter than conventionally allowed charged scalar which can found
in the next run of the LHC. It should be remembered that the charged scalar mass in the conventional 2HDMs like Type II
2HDM faces stringent constrains mainly from the well measured value of the $b\to s\gamma$ decay width. From this
observable alone the value of $m_{1+}$ in Type II 2HDM is forced to be more than 350~GeV. On the contrary, it has
been found in this chapter that in certain variants of the BGL models a charged scalar in the ballpark of 150-200
GeV can successfully negotiate the major experimental constraints coming from the flavor data. In fact, all the
three nonstandard masses ($m_H$, $m_A$ and $m_{1+}$) can be taken in the same 150-200 GeV range without upsetting
any major theoretical or experimental constrains whatsoever. What’s more interesting is to note that since
the BGL models break the fermionic family universality explicitly, these light nonstandard scalars can have unconventional decay hierarchies which can be observed at LHC and this might be the hallmark signature of
the BGL models.

Multi doublet extensions of the SM scalar sector have been further investigated in Chapter~\ref{Chap4} where
an $S_3$ symmetric 3HDM has been analyzed. Again an alignment limit for this model
has been found when one of the CP even scalars mimics the SM gauge and Yukawa coupling. In addition to
this SM-like scalar ($h$) a 3HDM contains two heavier scalars ($h^0$ and $H$), two pseudoscalars ($A_1$ and $A_2$) and
two pairs of charged scalars ($H_1^\pm$ and $H_2^\pm$). Theoretical constrains coming from the tree-level unitarity of the
scattering amplitudes were derived and the consequences of these constrains on the nonstandard masses have
been explored. It is found that all the nonstandard masses are constrained to lie below 1~TeV. Then the effect
of two pairs of charged scalars on the loop induced Higgs decays has been analyzed. Similar to the case of $Z_2$
symmetric 2HDM, here again we find that none of the charged scalars decouples from diphoton decay and an
enhancement in the diphoton signal strengths is impossible to obtain in the alignment limit. In fact, a sharp
prediction that the diphoton signal strength in this model cannot be more than 80\% of the SM expectation has
been made in this chapter.

In Chapter~\ref{Chap5} it has been shown that decoupling of the charged scalar in the loop induced Higgs decays
like $h\to \gamma\gamma$ is not guaranteed. The role played by symmetries in this context has been emphasized. It is found
that this nondecoupling is a typical characteristic of non-inert multi Higgs-doublet models with an exact discrete symmetry.
It has been suggested that decoupling can be achieved by invoking a soft symmetry breaking term in the scalar potential.

\newpage
\dedicatory{Details are not just details, collectively they make the design.} 

\addtocontents{toc}{\vspace{2em}} 

\appendix 



\chapter{Calculation of the \texorpdfstring{$WW$}{TEXT} scattering amplitude} 

\label{AppendixA} 

\lhead{Appendix A. \emph{Calculation of the WW scattering amplitude}} 
\begin{wrapfigure}{r}{0.45\textwidth}
\centering
 \includegraphics[scale=0.45]{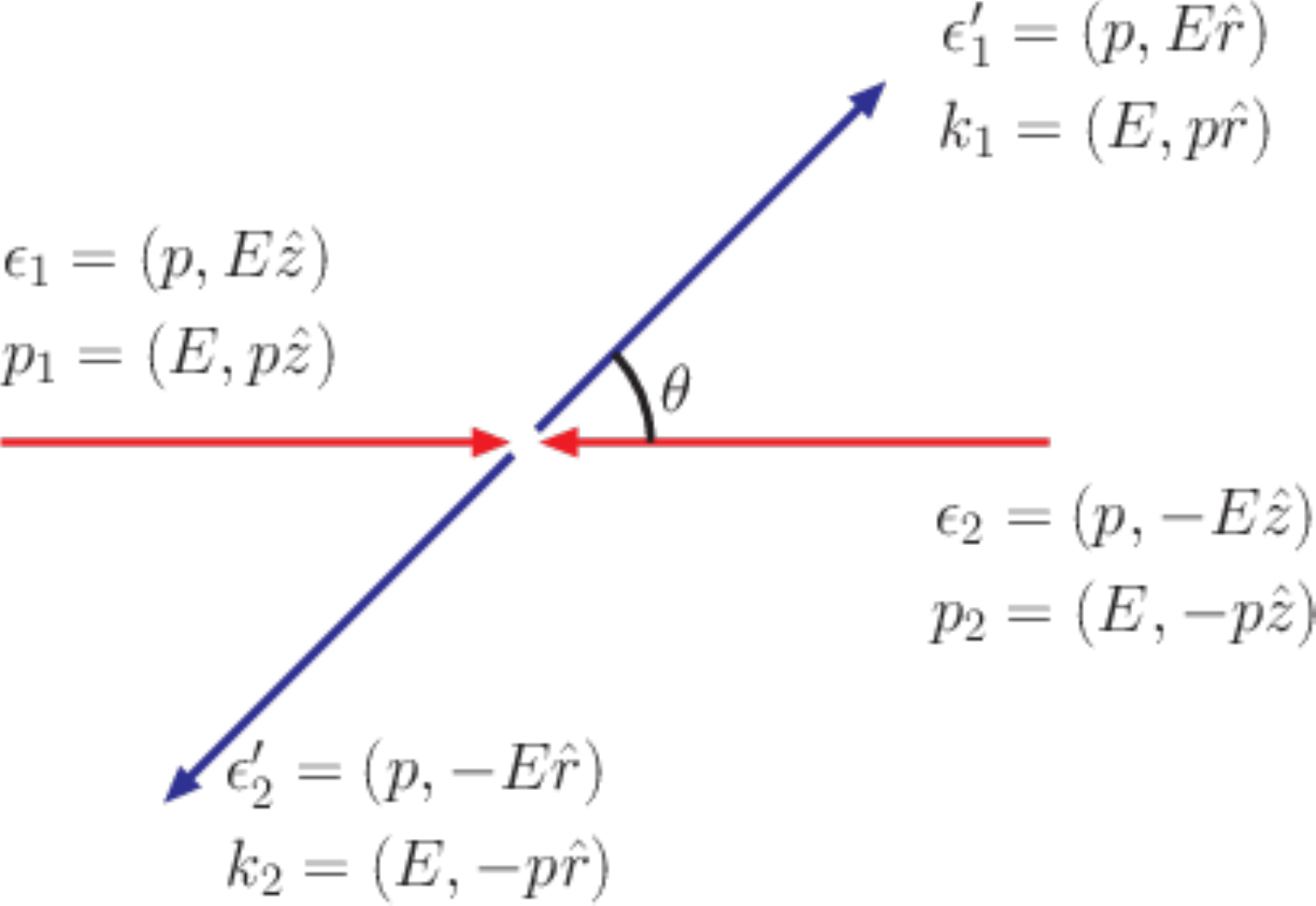}
\end{wrapfigure}%
Consider the following elastic scattering process in the CM frame:
$$W_L^+(p_1)+W_L^-(p_2) \to W_L^+(k_1)+W_L^-(k_2).$$
 We abbreviate $\epsilon_L(p_1)= \epsilon_1/M_W$  etc.
Since $p\cdot\epsilon_L(p)=0$ and $\epsilon_L(p)\cdot\epsilon_L(p)=-1$, we may take, in the CM frame, the momentum and polarization vectors as shown in the adjacent figure.
For our calculation, we shall need the following expressions which follow from the kinematics in the CM frame:
\begingroup
\allowdisplaybreaks
\begin{subequations}
 \begin{eqnarray}
p_1\cdot\epsilon_2 &=& p_2\cdot\epsilon_1=k_1\cdot\epsilon_2^\prime=k_2\cdot\epsilon_1^\prime=2Ep \,,  \\
p_1\cdot\epsilon_1^\prime&=& p_2\cdot\epsilon_2^\prime=k_1\cdot\epsilon_1=k_2\cdot\epsilon_2=Ep(1-\cos\theta) \,, \\
p_1\cdot\epsilon_2^\prime&=&p_2\cdot\epsilon_1^\prime=k_1\cdot\epsilon_2=k_2\cdot\epsilon_1=Ep(1+\cos\theta \,, \\
\epsilon_1\cdot\epsilon_2&=&\epsilon_1^\prime\cdot\epsilon_2^\prime=p^2+E^2 \,, \\
\epsilon_1^\prime\cdot\epsilon_2&=&\epsilon_1\cdot\epsilon_2^\prime=p^2+E^2\cos\theta \,, \\
\epsilon_1\cdot\epsilon_1^\prime&=&\epsilon_2\cdot\epsilon_2^\prime=p^2-E^2\cos\theta \,, \\
p_1\cdot p_2 &=& k_1\cdot k_2=\left(\frac{s}{2}-M_W^2 \right) \,,  \\
t&=&(p_1-k_1)^2=2M_W^2-2p_1\cdot k_1 \nonumber \\
 &=&2M_W^2-2(E^2-p^2\cos\theta) \,, \\
\Rightarrow \frac{t}{2}&=&-p^2(1-\cos\theta) \,, \label{man-t} \\
\Rightarrow p_1\cdot k_1 &=& p_2\cdot k_2=-\left(\frac{t}{2}-M_W^2\right) \,, \\
{\rm Similarly,} ~~~\frac{u}{2}&=&-p^2(1+\cos\theta) \,, \\
\Rightarrow p_1\cdot k_2&=&p_2\cdot k_1=-\left(\frac{u}{2}-M_W^2\right) \,.
 \label{scalarsWW}
 \end{eqnarray}
\end{subequations}
\endgroup
\section{Calculation of the gauge part}
\begin{figure}[h]
\centering
\includegraphics[scale=0.5]{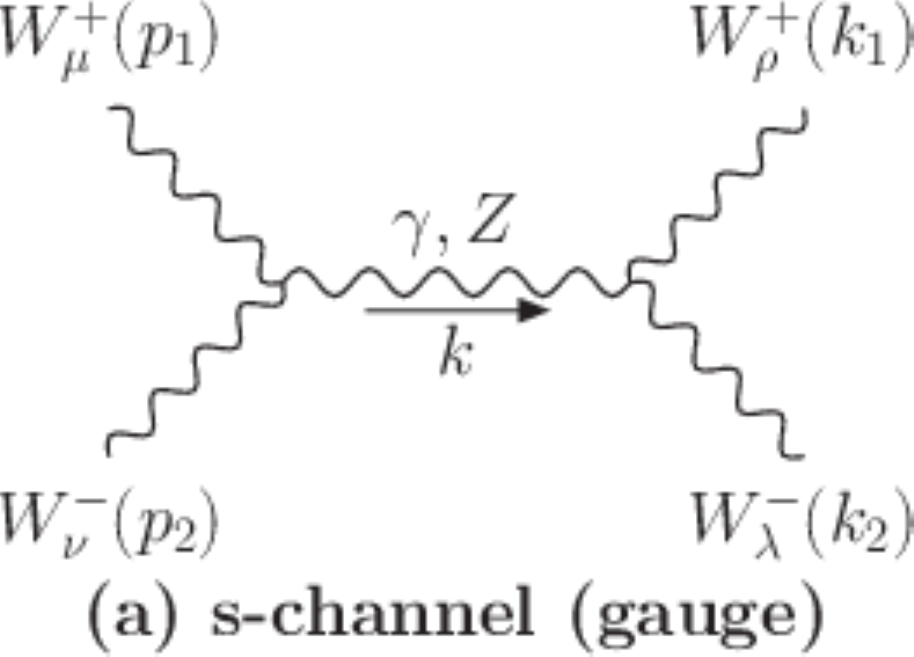}~~~
\includegraphics[scale=0.5]{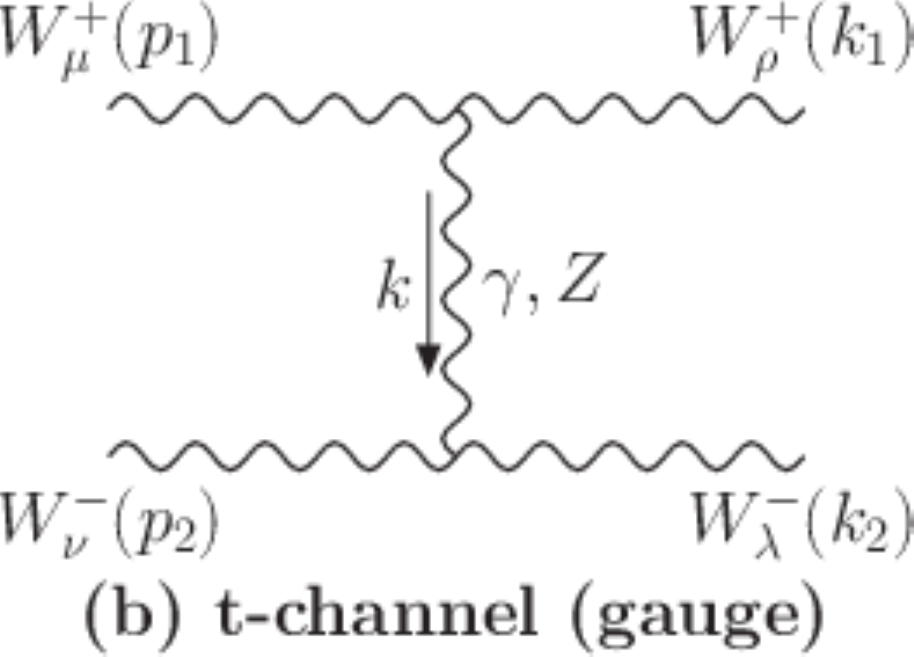}~~~
\includegraphics[scale=0.5]{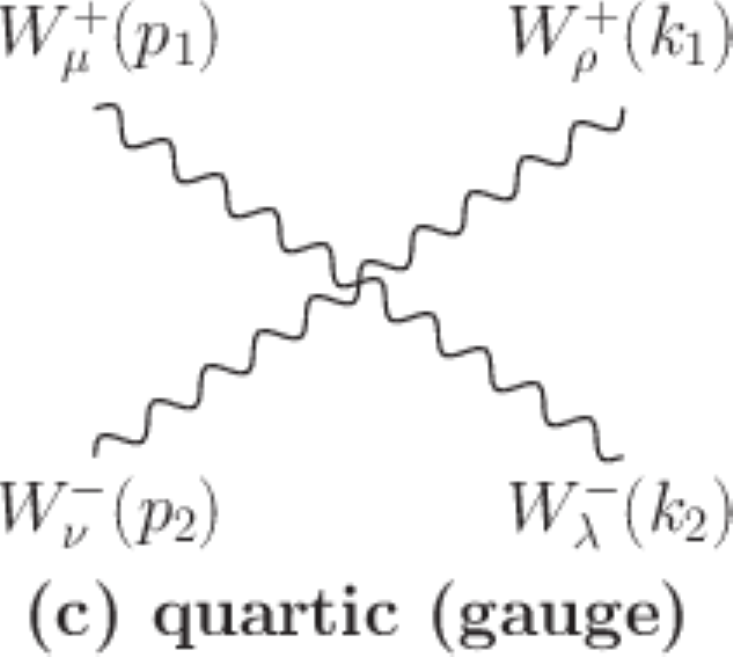}
\caption[Feynman diagrams for WW elastic scattering (gauge part)]{\em Gauge diagrams for $W_L^+(p_1)+W_L^-(p_2) \to W_L^+(k_1)+W_L^-(k_2)$. \nolinebreak}
\label{fWWWW:subfigures}
\end{figure}
\subsection{\texorpdfstring{$s$}{TEXT}-channel photon exchange}
The Feynman amplitude can be written as
\begin{eqnarray}
 i{\cal M}_a^\gamma &=& -g^2\sin^2\theta_w\Big[(p_2-p_1)_\alpha g_{\mu\nu}+(p_1+k)_\nu g_{\mu\alpha}-(k+p_2)_\mu g_{\nu\alpha}\Big]\times\Big(-\frac{ig^{\alpha\beta}}{s}\Big) \nonumber \\
  && \times \Big[(k_2-k_1)_\beta g_{\lambda\rho}+(k_1+k)_\lambda g_{\rho\beta}-(k+k_2)_\rho g_{\lambda\beta}\Big]\times\frac{\epsilon_1^\mu\epsilon_2^\nu\epsilon_1^{\prime\rho}\epsilon_2^{\prime\lambda}}{M_W^4} \,.
  \label{famplitude s photon}
\end{eqnarray}
Rememembering $k=(p_1+p_2)=(k_1+k_2)$ and $p_1\cdot\epsilon_1=p_2\cdot\epsilon_2=k_1\cdot\epsilon_1^\prime=k_2\cdot\epsilon_2^\prime=0$ we may simplify the above equation as
\begin{eqnarray}
{\cal M}_a^\gamma &=& \frac{g^2\sin^2\theta_w}{M_W^4s}\Big[(p_2-p_1)_\alpha(\epsilon_1\cdot\epsilon_2) +2(p_1\cdot\epsilon_2)\epsilon_{1\alpha}-2(p_2\cdot\epsilon_1)\epsilon_{2\alpha}\Big]\times g^{\alpha\beta} \nonumber \\
&& \times \Big[(k_2-k_1)_\beta(\epsilon_2^\prime\cdot\epsilon_1^\prime)-2(k_1\cdot\epsilon_2^\prime) \epsilon_{1\beta}^\prime+2(k_2\cdot\epsilon_1^\prime)\epsilon_{2\beta^\prime}\Big] \,. 
\end{eqnarray}
Using the scalar products of \Eqn{scalarsWW} we may write
\begin{eqnarray}
{\cal M}_a^\gamma&=&\frac{g^2\sin^2\theta_w}{M_W^4s}\Bigg[\Big\{(p_2-p_1)_\alpha(E^2+p^2)+4Ep(\epsilon_1-\epsilon_2)_\alpha\Big\} \nonumber \\
&& \times\Big\{(k_2-k_1)^\alpha(E^2+p^2)+4Ep(\epsilon_1^\prime-\epsilon_2^\prime)^\alpha\Big\}\Bigg] \\
 &=& \frac{g^2\sin^2\theta_w}{M_W^4s} \times T_s ~~~~~~~({\rm say})\,.
 \label{WWsgam}
\end{eqnarray}
\begin{mdframed}
$\blacksquare$ {\bf Terms inside the square bracket:}
\begin{eqnarray}
 T_s &=& \Big[(p_2\cdot k_2-p_2\cdot k_1-p_1\cdot k_2+p_1\cdot k_1)(E^2+p^2)^2+4Ep(E^2+p^2)\times \nonumber \\
  && (p_2\cdot\epsilon_1^\prime-p_2\cdot\epsilon_2^\prime-p_1\cdot\epsilon_1^\prime+p_1\cdot\epsilon_2^\prime
+k_2\cdot\epsilon_1-k_2\cdot\epsilon_2-k_1\cdot\epsilon_1+k_1\cdot\epsilon_2) \nonumber \\
  && +16E^2p^2(\epsilon_1\cdot\epsilon_1^\prime-\epsilon_1\cdot\epsilon_2^\prime-\epsilon_2\cdot \epsilon_1^\prime+\epsilon_2\cdot\epsilon_2^\prime)\Big] \nonumber \\
  &=& \Big[2(-2p^2\cos\theta)(E^2+p^2)^2+32E^2p^2\cos\theta(E^2+p^2)+32E^2p^2(-2E^2\cos\theta)\Big] \nonumber \\
  &=& \Big[-4p^2\cos\theta(E^2+p^2)^2+32E^2p^2\cos\theta(E^2+p^2-2E^2)\Big] \nonumber \\
  &=& -4p^2\cos\theta\Big(\frac{s}{2}-M_W^2\Big)^2-32E^2p^2M_W^2\cos\theta \nonumber \\
  &=& -4E^4s \cos\theta+ \order(M_W^4) \,.
\label{T schannel photon}
\end{eqnarray}
\end{mdframed}

Therefore,
\begin{eqnarray}
 {\cal M}_a^\gamma&=& \frac{g^2\sin^2\theta_w}{M_W^4s}[-4E^4s \cos\theta] + \order(1) \nonumber \\
   &=& -\frac{g^2E^4\sin^2\theta_w}{M_W^4}(4\cos\theta) + \order(1) \,.
\end{eqnarray}
%
\subsection{s-channel \texorpdfstring{$Z$}{TEXT}-boson exchange}
\label{s channel Z}
Firstly, the photon propagator of \Eqn{famplitude s photon} should be replaced by the massive $Z$-boson propagator:
\begin{eqnarray}
 \frac{-g^{\alpha\beta}+k^\alpha k^\beta}{(s-M_Z^2)} \,.
\end{eqnarray}
But it should be noted that the $k^\alpha k^\beta$ term does not contribute to the amplitude because of the following identity:
\begin{eqnarray}
 &&\Big[(p_2-p_1)_\alpha g_{\mu\nu}+(p_1+k)_\nu g_{\mu\alpha}-(k+p_2)_\mu g_{\nu\alpha}\Big]\times k^\alpha k^\beta \nonumber \\ 
 &&\times \Big[(k_2-k_1)_\beta g_{\lambda\rho}+(k_1+k)_\lambda g_{\rho\beta}-(k+k_2)_\rho g_{\lambda\beta}\Big] = 0 \,.
\end{eqnarray}
Another modification is that $WW\gamma$ coupling will be replaced by $WWZ$ coupling, {\it i.e.}, $\sin\theta_w$ should be replaced by $\cos\theta_w$. Keeping this in mind, we can rewrite \Eqn{WWsgam} for s-channel Z-boson mediation as
\begin{eqnarray}
{\cm}_a^{Z} &=& \frac{g^2\cos^2\theta_w}{M_W^4(s-M_Z^2)} \times T_s \simeq\frac{g^2\cos^2\theta_w}{M_W^4(s-M_Z^2)}[-(4E^4s)\cos\theta]  \\
&=& \frac{g^2E^2\cos^2\theta_ws}{M_W^4}\left(1-\frac{M_Z^2}{s}\right)^{-1}\cos\theta \nonumber \\
&=& -\frac{g^2E^4\cos^2\theta_w}{M_W^4}(4\cos\theta)-\frac{g^2E^2}{M_W^2}\cos\theta +\order(1) \,.
\end{eqnarray}
In the last step, we have used $M_W=M_Z\cos\theta_w$. We now get the total amplitude for the s-channel process as
\begin{equation}
{\cm}_a={\cm}_a^\gamma+{\cm}_a^Z=-\frac{g^2E^4}{M_W^4}(4\cos\theta)-\frac{g^2E^2}{M_W^2}(\cos\theta) \,.
\label{1st}
\end{equation}
\subsection{t-channel photon exchange}
The Feynman amplitude for this diagram reads
\begin{eqnarray}
i{\cm}_b^\gamma&=&\left(-g^2\sin^2\theta_w \right)\left[-(p_1+k_1)_\alpha g_{\mu\nu}+(p_1+k)_\nu g_{\mu\alpha}+(-k+k_1)_\mu g_{\nu\alpha}\right]\times\left(-\frac{ig^{\alpha\beta}}{t}\right) \nonumber \\
&& \times\left[(p_2+k_2)_\beta g_{\lambda\rho}-(k_2+k)_\rho g_{\lambda\beta}+(k-p_2)_\lambda g_{\rho\beta}\right] \times
\frac{\epsilon_1^\mu\epsilon_1^{\prime\nu}\epsilon_2^\rho\epsilon_2^{\prime\lambda}}{M_W^4} \,.
\end{eqnarray}
Again remembering $k=p_1-k_1=k_2-p_2$, and using the gauge condition $p.\epsilon(p)=0$ we may simplify the above expression as:
\begin{eqnarray}
{\cm}_b^\gamma&=&\frac{g^2\sin^2\theta_w}{M_W^4t}\left[-(p_1+k_1)_\alpha(\epsilon_1.\epsilon_1^\prime)
+2(p_1.\epsilon_1^\prime)\epsilon_{1\alpha}+2(k_1.\epsilon_1)\epsilon^\prime_{1\alpha}\right] \nonumber \\
&& \times\left[(p_2+k_2)_\beta(\epsilon_2.\epsilon_2^\prime)-2(k_2.\epsilon_2)\epsilon^\prime_{2\beta}
-2(p_2.\epsilon_2^\prime)\epsilon_{2\beta} \right] \times g^{\alpha\beta} \nonumber \\
&=& \frac{g^2\sin^2\theta_w}{M_W^4t} \left[-(p_1+k_1)_\alpha(p^2-E^2\cos\theta)+2Ep(1-\cos\theta)(\epsilon_1+\epsilon_1^\prime)_\alpha \right] \nonumber \\
&& \times\left[(p_2+k_2)^\alpha(p^2-E^2\cos\theta)-2Ep(1-\cos\theta)(\epsilon_2+\epsilon_2^\prime)^\alpha \right]  \\
&=& \frac{g^2\sin^2\theta_w}{M_W^4t} \times T_t ~~~~~~~({\rm say})\,.
\label{WWtgam}
\end{eqnarray}
\begin{mdframed}
$\blacksquare$ {\bf Terms inside the square bracket:}
\begin{eqnarray}
T_t &=& +2Ep(1-\cos\theta)(p^2-E^2\cos\theta)\{(p_2+k_2).(\epsilon_1+\epsilon_1^\prime)+(p_1+k_1).(\epsilon_2+\epsilon_2^\prime)\} \nonumber \\
&&  -(p_1+k_1).(p_2+k_2)(p^2-E^2\cos\theta)^2 \nonumber \\
&& - 4E^2p^2(1-\cos\theta)^2(\epsilon_1+\epsilon_1^\prime).(\epsilon_2+\epsilon_2^\prime) \nonumber \\
&=& - 2(p^2-E^2\cos\theta)^2(2E^2+p^2+p^2\cos\theta) \nonumber \\
&& + 8E^2p^2(1-\cos\theta)(p^2-E^2\cos\theta)(3+\cos\theta) \nonumber \\
&& -8E^2p^2(1-\cos\theta)^2\{2p^2+E^2(1+\cos\theta)\} \nonumber \\
&=& +8E^2p^2(1-\cos\theta)\left\{(p^2-E^2\cos\theta)(3+\cos\theta)-2p^2(1-\cos\theta)-E^2(1-\cos^2\theta)\right\} \nonumber \\
&&  - 2\{p^2(1-\cos\theta)-M_W^2\cos\theta\}^2\left(3E^2+E^2\cos\theta-M_W^2-M_W^2\cos\theta\right) \,.
\label{Tt1}
\end{eqnarray}
After some algebraic manipulation, the terms inside the curly bracket in the first part of \Eqn{Tt1} 
can be evaluated to be equal to $-M_W^2(1+3\cos\theta)$. Using this in conjunction with \Eqn{man-t}, the terms inside the square bracket becomes:
\begin{eqnarray}
T_t &=& -2\left(\frac{t}{2}+M_W^2\cos\theta \right)^2\left\{E^2(3+\cos\theta)-M_W^2(1+\cos\theta)\right\} \nonumber \\ 
&& +4E^2tM_W^2(1+3\cos\theta) \,, \\
&\approx& -\frac{t^2}{2}E^2(3+\cos\theta)+\frac{t^2}{2}M_W^2(1+\cos\theta) \nonumber \\
&& -2M_W^2tE^2\cos\theta(3+\cos\theta) +4E^2tM_W^2(1+3\cos\theta)+ \order(M_W^4) \,, \\
&=& -\frac{t^2}{2}E^2(3+\cos\theta)+M_W^2tE^2(3+6\cos\theta-\cos^2\theta)+\order(M_W^4) \,.
\end{eqnarray}
\end{mdframed}

Plugging this into \Eqn{WWtgam}, we obtain
\begin{eqnarray}
{\cm}_b^\gamma &\approx& \frac{g^2\sin^2\theta_w}{M_W^4t}\left[-\frac{t^2}{2}E^2(3+\cos\theta)+M_W^2tE^2(3+6\cos\theta -\cos^2\theta) \right] \\
&\approx& \frac{g^2E^2p^2\sin^2\theta_w}{M_W^4}(1-\cos\theta)(3+\cos\theta) \nonumber \\ 
&& +\frac{g^2E^2\sin^2\theta_w}{M_W^2} \left(3+6\cos\theta-\cos^2\theta \right) \\
&=& \frac{g^2E^4\sin^2\theta_w}{M_W^4}\left(3-2\cos\theta-\cos^2\theta\right) +\frac{g^2E^2\sin^2\theta_w}{M_W^2}(8\cos\theta)+\order(1) \,.
\end{eqnarray}
\subsection{t-channel \texorpdfstring{$Z$}{TEXT}-boson exchange}
Similar to the s-channel calculation, for t-channel $Z$-boson mediation we may write:
\begin{eqnarray}
{\cm}_b^Z &=& \frac{g^2\cos^2\theta_w}{M_W^4(t-M_Z^2)}\times T_t \nonumber \\
&\approx& \frac{g^2\cos^2\theta_w}{M_W^4t}\left(1+\frac{M_Z^2}{t}\right)\times T_t \nonumber \\
&=& \left(\frac{g^2\cos^2\theta_w}{M_W^4t}+\frac{g^2}{M_W^2t^2}\right) \times T_t \nonumber \\
&\approx& \frac{g^2E^4\cos^2\theta_w}{M_W^4}\left(3-2\cos\theta-\cos^2\theta\right) +\frac{g^2E^2\cos^2\theta_w}{M_W^2}(8\cos\theta) \nonumber \\
&& +\frac{g^2}{M_W^2t^2} \left[-\frac{t^2}{2}E^2(3+\cos\theta)\right] \\
&=& \frac{g^2E^4\cos^2\theta_w}{M_W^4}\left(3-2\cos\theta-\cos^2\theta\right) +\frac{g^2E^2\cos^2\theta_w}{M_W^2}(8\cos\theta) \nonumber \\
&& -\frac{g^2E^2}{M_W^2}\left(\frac{3}{2}+\frac{1}{2}\cos\theta\right) +\order(1) \,.
\end{eqnarray}
Therefore, the total amplitude for the t-channel process becomes:
\begin{eqnarray}
{\cm}_b &=& {\cm}_b^\gamma+{\cm}_b^Z \nonumber \\
&=& \frac{g^2E^4}{M_W^4}\left(3-2\cos\theta-\cos^2\theta\right) +\frac{g^2E^2}{M_W^2}\left(\frac{15}{2}\cos\theta-\frac{3}{2}\right) +\order(1) \,.
\label{2nd}
\end{eqnarray}
\subsection{The quartic gauge vertex}
The Feynman amplitude is given by
\begin{eqnarray}
&& i{\cm}_c=ig^2(2g_{\mu\rho}g_{\nu\lambda}-g_{\mu\nu}g_{\rho\lambda}-g_{\mu\lambda}g_{\nu\rho})
\times \frac{\epsilon_1^\mu\epsilon_2^\nu\epsilon_1^{\prime\lambda}\epsilon_2^{\prime\rho}}{M_W^4} \\
&\Rightarrow& {\cm}_c=\frac{g^2}{M_W^4}
\left[2(\epsilon_1.\epsilon_2^\prime)(\epsilon_2.\epsilon_1^\prime)
-(\epsilon_1.\epsilon_2)(\epsilon_1^\prime.\epsilon_2^\prime) -(\epsilon_1.\epsilon_1^\prime)(\epsilon_2.\epsilon_2^\prime) \right] \\
&\Rightarrow& {\cm}_c =\frac{g^2}{M_W^4}\left[2(p^2+E^2\cos\theta)^2-(p^2+E^2)^2-(p^2-E^2\cos\theta)^2 \right] \\
&\Rightarrow& {\cm}_c=\frac{g^2E^4}{M_W^4}(6\cos\theta-2-\sin^2\theta)+\frac{g^2E^2}{M_W^2}(2-6\cos\theta)  +\order(1) \,.
\label{3rd}
\end{eqnarray}

Adding Eqs.~(\ref{1st}), (\ref{2nd}) and (\ref{3rd}) together one can see that the quartic growth is already canceled. The remaining quadratic growth in the total gauge part of the amplitude is given below:
\begin{eqnarray}
{\cm}^{\gamma+Z} &=& {\cm}_a+{\cm}_b+{\cm}_c \,, \nonumber \\
&=& \frac{g^2E^2}{2M_W^2}(1+\cos\theta)+\order(1) \,, \\
&=& \frac{g^2s}{8M_W^2}(1+\cos\theta)+\order(1)
\equiv \frac{G_Fs}{\sqrt{2}}(1+\cos\theta)+\order(1) \,.
\label{final_gauge}
\end{eqnarray}
\section{Calculation of the Higgs part}
\begin{figure}[h]
\centering
\includegraphics[scale=0.6]{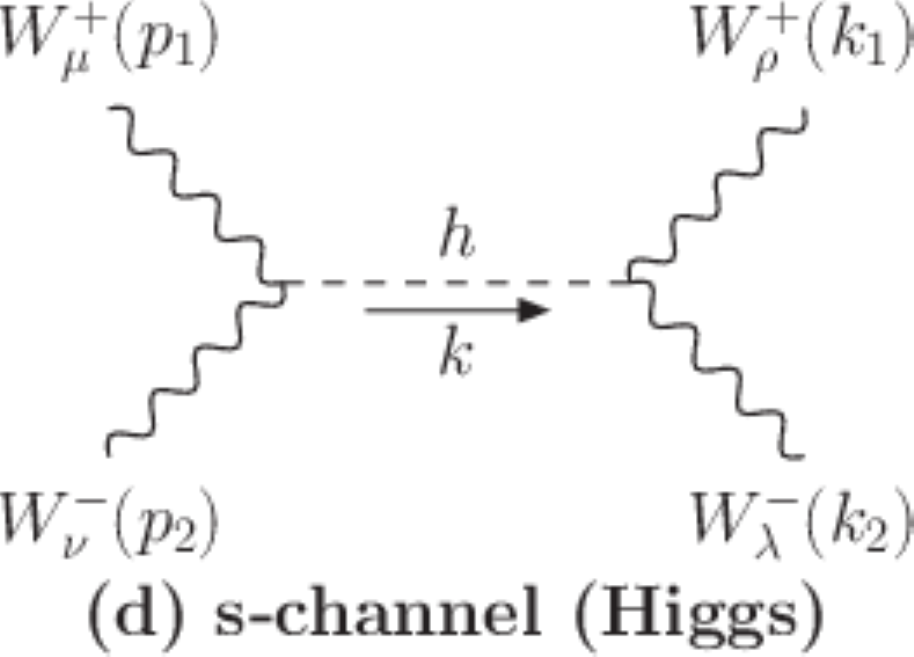}~~~~~
\includegraphics[scale=0.6]{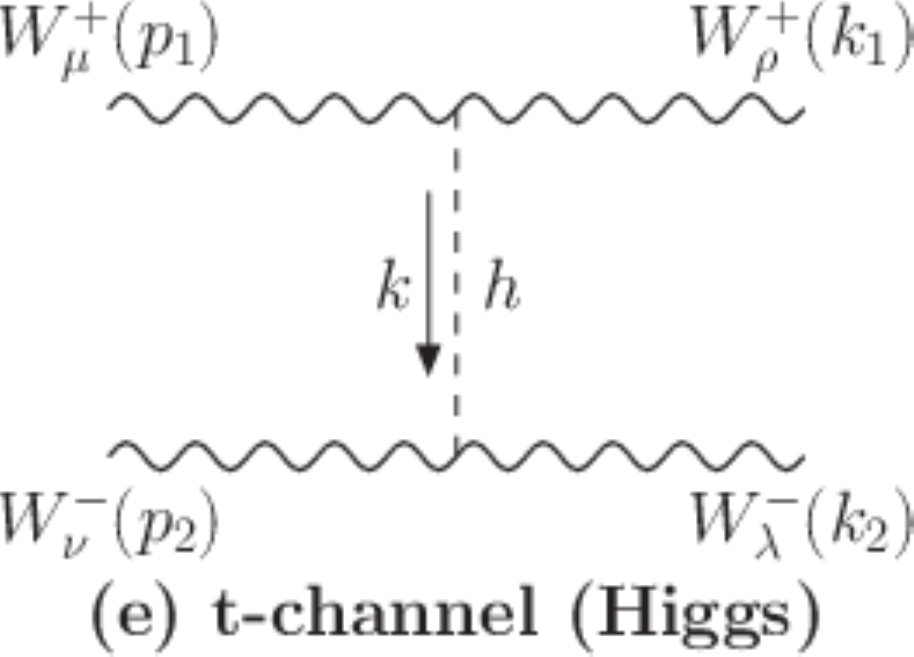}
\caption[Feynman diagrams for WW elastic scattering (Higgs part)]{\em Higgs mediated diagrams for $W_L^+(p_1)+W_L^-(p_2) \to W_L^+(k_1)+W_L^-(k_2)$. \nolinebreak}
\label{fWWWW-higgs}
\end{figure}
\subsection{s-channel Higgs exchange}
The Feynman amplitude for diagram~\ref{fWWWW-higgs}d can be written as follows:
\begin{eqnarray}
&& i{\cm}_d=\left(igM_Wg_{\mu\nu}\right) \frac{i}{\left(s-m_h^2\right)} \left(igM_Wg_{\alpha\beta}\right)
\frac{\epsilon_1^\mu\epsilon_2^\nu\epsilon_1^{\prime\alpha}\epsilon_2^{\prime\beta}}{M_W^4} \nonumber \\
&\Rightarrow& {\cm}_d=-\frac{g^2}{M_W^2(s-m_h^2)}p_1^\mu p_2^\nu k_1^\alpha k_2^\beta g_{\mu\nu}g_{\alpha\beta} \,.
\label{e:WWWW-h-s}
\end{eqnarray}
Note that unlike in the gauge part of the calculation, here we already have $M_W^2$ in the denominator. 
Therefore, any modification of the $\order(M_W^2/E^2)$ in the numerator will lead to a  constant term term independent of the CM energy. For this reason, here we can approximate
\begin{eqnarray}
\epsilon_L^\mu(p)\approx\frac{p^\mu}{M_W} \,.
\end{eqnarray}
Using this, we may rewrite \Eqn{e:WWWW-h-s} as
\begin{eqnarray}
{\cm}_d &=& -\frac{g^2(p_1\cdot p_2)(k_1\cdot k_2)}{M_W^2\left(s-m_h^2\right)} \nonumber \\
 &=& -\frac{g^2s^2}{4M_W^2\left(s-m_h^2\right)}\nonumber \\
&=& -\frac{g^2}{4M_W^2}\left[s+\frac{m_h^2s}{\left(s-m_h^2\right)}\right] \,.
\end{eqnarray}
\subsection{t-channel Higgs exchange}
Similar to the s-channel calculation, for diagram~\ref{fWWWW-higgs}e we can write
\begin{eqnarray}
{\cm}_e=-\frac{g^2}{4M_W^2}\left[t+\frac{m_h^2t}{(t-m_h^2)}\right] \,.
\end{eqnarray}
Thus, the total amplitude for the Higgs part becomes
\begin{eqnarray}
&& {\cm}^h={\cm}_d+{\cm}_e= -\frac{g^2}{4M_W^2}(s+t)-\frac{g^2m_h^2}{4M_W^2}\left( \frac{s}{s-m_h^2}+\frac{t}{t-m_h^2}\right) \, \\
&\Rightarrow& {\cm}^h \approx -\frac{g^2s}{8M_W^2}(1+\cos\theta)-\frac{g^2m_h^2}{4M_W^2}\left( \frac{s}{s-m_h^2}+\frac{t}{t-m_h^2}\right) \,.
\label{final_higgs}
\end{eqnarray}
In the last step we have used the following identity:
\begin{eqnarray}
-(s+t)=u=-2p^2(1+\cos\theta)\approx -2E^2(1+\cos\theta)=-\frac{s}{2}(1+\cos\theta) \,.
\end{eqnarray}
\section{Total amplitude and Lee-Quigg-Thacker limit}
Adding \Eqs{final_gauge}{final_higgs} together, we can easily see that the remnant quadratic energy growth in the gauge part of the amplitude is exactly canceled by the corresponding growth in the Higgs part of the amplitude. After the cancellation of the bad energy growth, the total amplitude reads
\begin{eqnarray}
&& {\cm}(\theta)={\cm}^{\gamma+Z}+{\cm}^h=-\frac{g^2m_h^2}{4M_W^2}\left( \frac{s}{s-m_h^2}+\frac{t}{t-m_h^2}\right) \,, \\
&\Rightarrow& {\cm}(\theta)= -\frac{g^2m_h^2}{4M_W^2} \left[ \frac{s}{s-m_h^2} + \frac{\frac{s}{2}(1-\cos\theta)}{\frac{s}{2}(1-\cos\theta)+m_h^2} \right] \,, \\
&& ~~~ =-\frac{g^2m_h^2}{4M_W^2}\left[{\cm}_1(\theta)+{\cm}_2(\theta)\right]  ~~~{\textrm{(say)}} \,.
\end{eqnarray}
Now, let us define the following:
\begin{eqnarray}
\mathscr{I}_1&=& \int_{-1}^{+1}{\cm}_1(\theta)d(\cos\theta) = \frac{2s}{s-m_h^2} \,, \\
\mathscr{I}_2&=& \int_{-1}^{+1}{\cm}_2(\theta)d(\cos\theta) = \int_{-1}^{+1}\frac{\frac{s}{2}(1-\cos\theta)}{\frac{s}{2}(1-\cos\theta)+m_h^2}d(\cos\theta) \nonumber \\
 &=& \left[2-\frac{2m_h^2}{s}\ln\left(1+\frac{s}{m_h^2}\right) \right] \,.
\end{eqnarray}
One may recall from \Eqn{partial_extraction}
\begin{eqnarray}
a_0 &=& \frac{1}{32\pi}\int_{-1}^{+1} {\cm}(\theta) d(\cos\theta) \nonumber\\
&=& \frac{1}{32\pi}\left( -\frac{g^2m_h^2}{4M_W^2}\right) \left(\mathscr{I}_1+\mathscr{I}_2 \right) \\
&=& -\frac{G_Fm_h^2}{8\pi\sqrt{2}}\left[2+\frac{m_h^2}{s-m_h^2}-\frac{m_h^2}{s}\ln\left(1+\frac{s}{M_H^2}\right)  \right] \,.
\end{eqnarray}
At the limit $s>>m_h^2$, the above equation can be approximated as:
\begin{eqnarray}
a_0\approx -\frac{G_Fm_h^2}{4\pi\sqrt{2}} \,.
\end{eqnarray}
Now, if we apply the unitarity condition of \Eqn{unitarity}, we can get an upper limit on the Higgs mass as:
\begin{eqnarray}
m_h^2\leq \frac{4\pi\sqrt{2}}{G_F} =1.26~ \textrm{TeV} \,, 
\end{eqnarray}
where, in the last step, we have used $G_F=1.12\times 10^{-5}$ GeV$^{-2}$.
\chapter{Calculation of the \texorpdfstring{$e^-e^+ \to WW$}{TEXT} scattering amplitude} 
\label{AppendixB} 
\lhead{Appendix B. \emph{Calculation of the $e^-e^+ \to WW$ scattering amplitude}}
\begin{wrapfigure}{r}{0.45\textwidth}
\centering
 \includegraphics[scale=0.45]{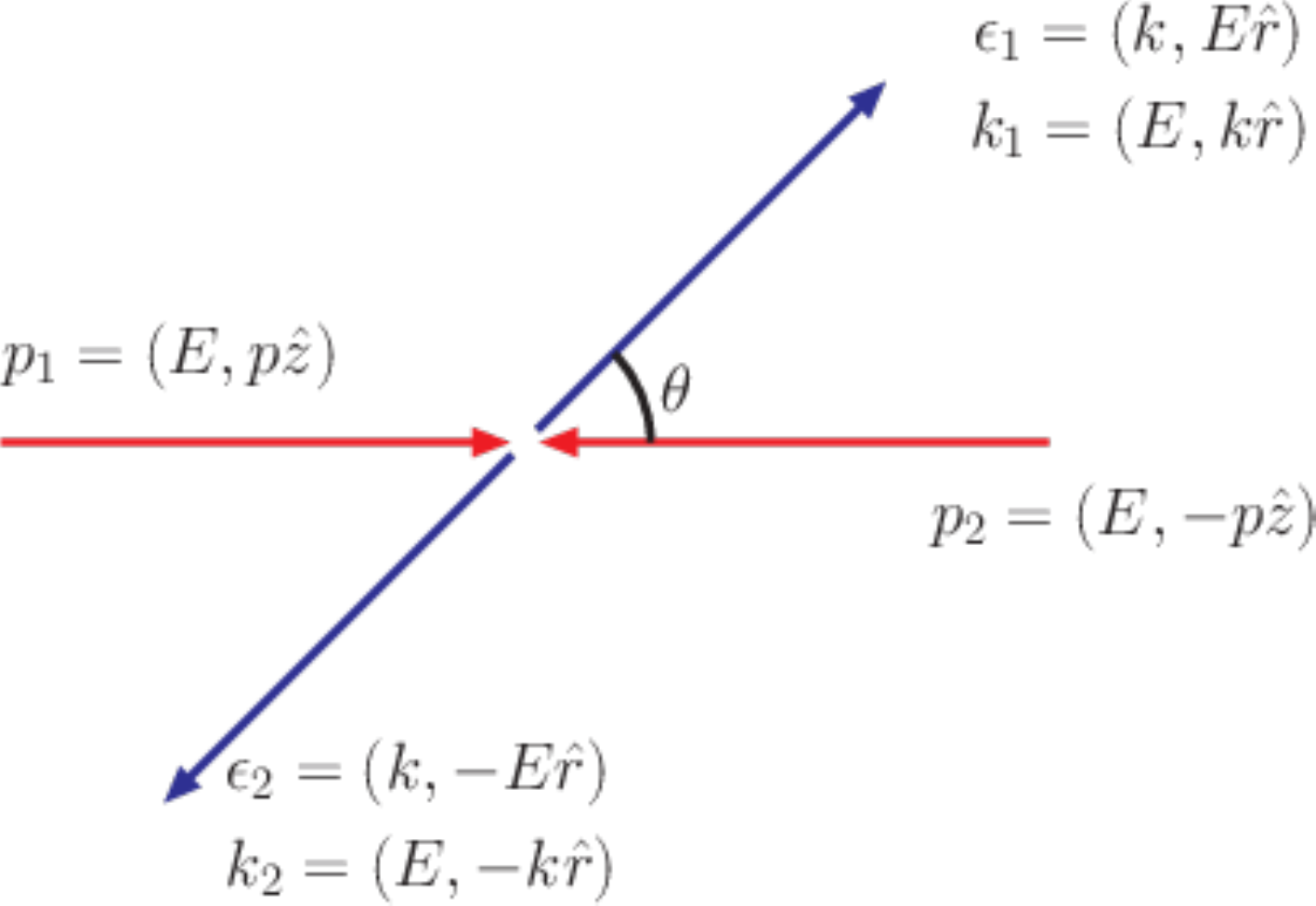}
\end{wrapfigure}%
Consider the following elastic scattering process in the CM frame:
$$e^+(p_1)+e^-(p_2) \to W_L^-(k_1)+W_L^+(k_2).$$
Although we display the calculation for $e^+e^-$ pair, it can be easily generalized for any fermion-antifermion pairs in the SM. As in Appendix~\ref{AppendixA}, here also we abbreviate $\epsilon_L(k_1)= \epsilon_1/M_W$  and $\epsilon_L(k_2)= \epsilon_2/M_W$.
Since $k\cdot\epsilon_L(k)=0$ and $\epsilon_L(k)\cdot\epsilon_L(k)=-1$, we may take, in the CM frame, the momentum and polarization vectors as shown in the adjacent figure.
For our calculation, we shall need the following expressions which follow from the kinematics in the CM frame:
 \begin{subequations}
 \begin{eqnarray}
\epsilon_1^\mu &=& k_1^\mu +\frac{M_W^2}{2E}X_1^\mu ~~~{\rm with} ~~ X_1^\mu =(-1,\hat{r}) \,, \\
\epsilon_2^\mu &=& k_2^\mu +\frac{M_W^2}{2E}X_2^\mu ~~~{\rm with} ~~ X_2^\mu =(-1,-\hat{r}) \,, \\
(\epsilon_1-\epsilon_2)^\mu &=& (k_1-k_2)^\mu +\frac{M_W^2}{E}X^\mu ~~~{\rm with} ~~ X^\mu =(0,\hat{r}) \,, \\
k_1\cdot k_2 &=& \epsilon_1\cdot \epsilon_2 = E^2+k^2 = \frac{s}{2}-M_W^2 \,, \\
k_2\cdot \epsilon_1 &=& k_1\cdot \epsilon_2 = 2Ek = \frac{s}{2}-M_W^2 +\order(\frac{M_W^4}{E^2}) \,.
 \end{eqnarray}
 \end{subequations}
Thus, in the high energy limit when $s >> M_W^2$, we can approximate
\begin{eqnarray}
k_1\cdot k_2 = \epsilon_1\cdot \epsilon_2=k_2\cdot \epsilon_1 = k_1\cdot \epsilon_2 =  \frac{s}{2}-M_W^2 \,.
\end{eqnarray}
The Feynman diagrams for the scattering process in consideration is displayed below.
\begin{figure}[h]
\centering
\includegraphics[scale=0.5]{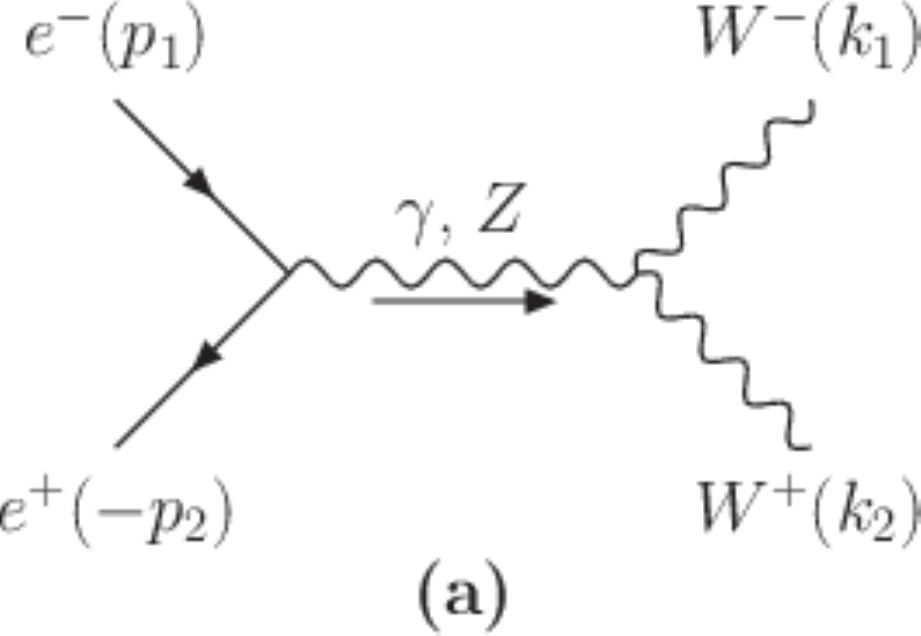}~~~
\includegraphics[scale=0.5]{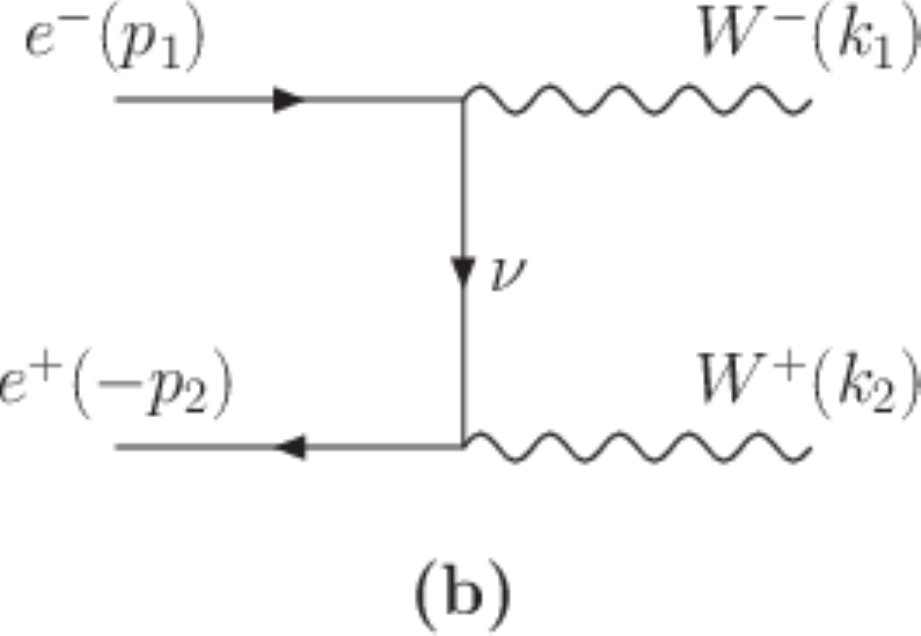}~~~
\includegraphics[scale=0.5]{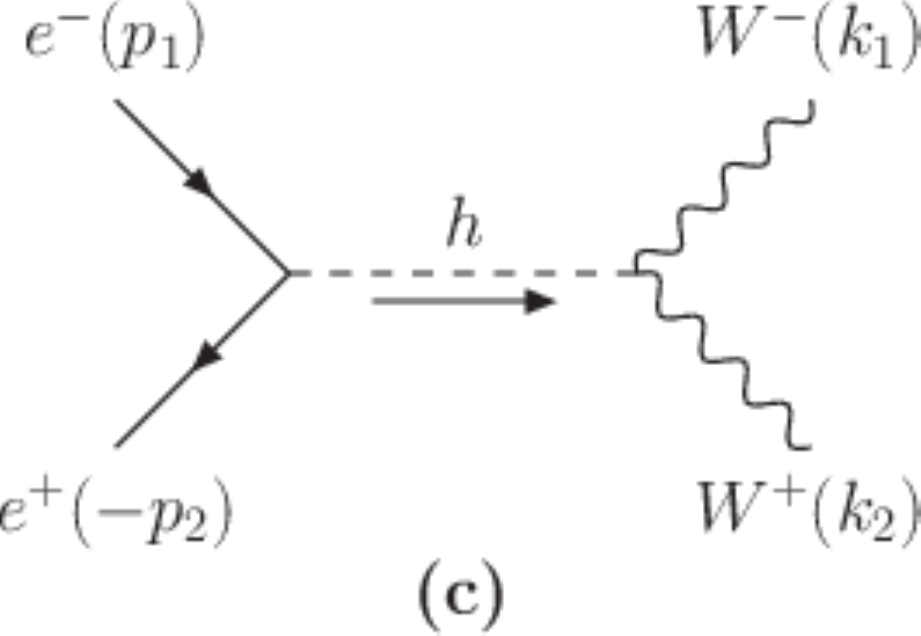}
\caption[Feynman diagrams for $e^-(p_1)+e^+(p_2)\to W_L^-(k_1)+W_L^+(k_2)$]{\em Feynman diagrams for $e^-(p_1)+e^+(p_2) \to W_L^-(k_1)+W_L^+(k_2)$. \nolinebreak}
\label{f:eeWW}
\end{figure}

\section{\texorpdfstring{$s$}{TEXT}-channel photon exchange}
The amplitude for the relevant Feynman diagram can be written as
\begin{eqnarray}
i{\cm}_a^\gamma &=& \bar{v}(p_2)(ie\gamma_\alpha) u(p_1) \times \left(-\frac{ig^{\alpha\beta}}{s} \right) \nonumber \\
&& \times ie\left[(-k_1+k_2)_\beta g_{\mu\nu} +(-k_2-q)_\mu g_{\nu\beta}+(q+k_1)_\nu g_{\beta\mu} \right] \times \frac{\epsilon_1^\mu\epsilon_2^\nu}{M_W^2} \,,
\end{eqnarray}
where, $q=p_1+p_2$ is the four momentum of the intermediate particle. After some simplification we obtain
\begin{subequations}
\begin{eqnarray}
{\cm}_a^\gamma &=& \frac{e^2}{M_W^2s} \bar{v}(p_2) \gamma^\beta u(p_1) \left[(k_2-k_1)_\beta(\epsilon_1\cdot\epsilon_2) -2(k_2\cdot\epsilon_1)\epsilon_{2\beta} +2(k_1\cdot\epsilon_2)\epsilon_{1\beta} \right] \\
&=& \frac{e^2}{M_W^2s} \bar{v}(p_2) \gamma^\beta u(p_1) \left[(k_2-k_1)_\beta-2(\epsilon_2-\epsilon_1)_\beta \right]  \left\{\left(\frac{s}{2}-M_W^2 \right)+\order(\frac{M_W^4}{E^2}) \right\} \\
&=& \frac{e^2}{M_W^2s} \bar{v}(p_2) \gamma^\beta u(p_1) \left[-(k_2-k_1)_\beta+\frac{M_W^2}{E}X_\beta \right] \left\{ \left(\frac{s}{2}-M_W^2 \right)+\order(\frac{M_W^4}{E^2}) \right\} \label{gam-analog} \\
&=& \frac{e^2}{M_W^2s} \bar{v}(p_2) \left[(\slashed{k_1}-\slashed{k_2})+\frac{M_W^2}{E}\slashed{X} \right]u(p_1) \times \left(\frac{s}{2}-M_W^2 \right) \,.
\end{eqnarray}
\end{subequations}
Now, using the conservation of four momentum we get $k_1-k_2 = 2k_1-(p_1+p_2)$ and Dirac equation implies
$\bar{v}(p_2)(\Slash{p_1}+\Slash{p_2})u(p_1)=0$. Hence we may write
\begin{eqnarray}
\bar{v}(p_2)(\Slash{k_1}-\Slash{k_2})u(p_1) = 2~ \bar{v}(p_2)~\Slash{k_1}u(p_1) \,.
\end{eqnarray}
Using this, we get the final expression as
 \begin{subequations}
 \begin{eqnarray}
 {\cm}_a^\gamma &=& \frac{e^2}{M_W^2s} \bar{v}(p_2) \left[2~\slashed{k_1}+\frac{M_W^2}{E}\slashed{X} \right]u(p_1) \times \left(\frac{s}{2}-M_W^2 \right) \\
&=& \underbrace{\frac{e^2}{M_W^2} \bar{v}(p_2)~\Slash{k_1}u(p_1)}_{\order(E^2)} -\underbrace{\frac{2e^2}{s} \bar{v}(p_2)~\Slash{k_1}u(p_1)+\frac{e^2}{2E} \bar{v}(p_2)~\Slash{X}u(p_1) }_{\order(1)} \,.
 \label{gam-final}
 \end{eqnarray}
 \end{subequations}
We have learned that the term involving $X_\beta$ in \Eqn{gam-analog} contributes to $\order(1)$ terms in the amplitude and therefore will not be significant for the high energy behavior. We shall drop such terms hereafter.

\section{\texorpdfstring{$s$}{TEXT}-channel \texorpdfstring{$Z$}{TEXT} exchange}
To begin with, we will assume a general coupling of the following form:
\begin{eqnarray}
\bar{e}\gamma^\mu (g_LP_L+g_RP_R) e~Z_\mu \equiv \frac{1}{2} \bar{e}\gamma^\mu \left\{(g_R+g_L)+(g_R-g_L)\gamma_5 \right\} e~Z_\mu \,.
\end{eqnarray}
Next let us be convinced that the $q^\alpha q^\beta$ term in the $Z$-boson propagator does not contribute to the amplitude. For this, we note the following:
 \begin{subequations}
 \begin{eqnarray}
&& q^\alpha q^\beta \left[(-k_1+k_2)_\beta g_{\mu\nu} +(-k_2-q)_\mu g_{\nu\beta}+(q+k_1)_\nu g_{\beta\mu} \right] \times \epsilon_1^\mu\epsilon_2^\nu \\
&=& q^\alpha q^\beta \left[(k_2-k_1)_\beta (\epsilon_1\cdot \epsilon_2) -2(k_1\cdot \epsilon_2)(\epsilon_2-\epsilon_1)_\beta \right] = 0 \,.
 \end{eqnarray}
 \end{subequations}
The last step follows if we substitute $q=k_1+k_2$ and remember that $k_1^2=k_2^2=M_W^2$ because the external $W^\pm$s are on-shell and $k_1\cdot\epsilon_1 = k_2\cdot\epsilon_2 =0$ due to gauge invariance.

Now denoting the $WWZ$ coupling strength by $g_{WWZ}$, we can write the relevant amplitude by slightly modifying \Eqn{gam-analog} as:
 \begin{subequations}
 \begin{eqnarray}
{\cm}_a^Z &=& \frac{g_{WWZ}}{2M_W^2(s-M_Z^2)} \bar{v}(p_2) \gamma^\beta \left[(g_R+g_L)+(g_R-g_L)\gamma_5 \right] u(p_1) \left[-(k_2-k_1)_\beta \right] \nonumber \\
&& ~~~ \times \left\{ \left(\frac{s}{2}-M_W^2 \right)+\order(\frac{M_W^4}{E^2}) \right\} \\
&=& \frac{g_{WWZ}}{2M_W^2(s-M_Z^2)} \bar{v}(p_2) (\Slash{k_1}-\Slash{k_2}) \left[(g_R+g_L)+(g_R-g_L)\gamma_5 \right] u(p_1) \nonumber \\
&& ~~~ \times \left\{ \left(\frac{s}{2}-M_W^2 \right)+\order(\frac{M_W^4}{E^2}) \right\} \,.
 \label{eeWW1}
 \end{eqnarray}
 \end{subequations}
As has been calculated previously, here we have the following:
 \begin{subequations}
\label{eeWW2}
 \begin{eqnarray}
\bar{v}_2 (~\Slash{k_1}-\Slash{k_2}) u_1 &=& 2~\bar{v}_2~\Slash{k_1} u_1 \,, \\
{\rm and,}~~~ \bar{v}_2 (~\Slash{k_1}-\Slash{k_2})\gamma^5 u_1 &=& \bar{v}_2 \left\{2~\Slash{k_1} - (~\Slash{p_1}+\Slash{p_2}) \right\} \gamma^5 u_1 \nonumber \\
&=& 2\bar{v}_2~\Slash{k_1}\gamma^5u_1 -\bar{v}_2(~ \Slash{p_1}+\Slash{p_2}) \gamma^5u_1 \\
&=& 2\bar{v}_2~\Slash{k_1}\gamma^5u_1 + m_e\bar{v}_2\gamma^5u_1 +\bar{v}_2\gamma^5\Slash{p_1} u_1 \\
&=& 2\bar{v}_2~\Slash{k_1}\gamma^5u_1 +2m_e \bar{v}_2\gamma^5u_1 = 2\bar{v}_2(~\Slash{k_1}+m_e)\gamma^5 u_1 \,,
 \end{eqnarray}
 \end{subequations}
where we have introduced the shorthands $\bar{v}_2$ and $u_1$ for $\bar{v}(p_2)$  and $u(p_1)$ respectively. Now we rewrite \Eqn{eeWW1} as follows:
 \begin{subequations}
 \begin{eqnarray}
{\cm}_a^Z &=& \frac{g_{WWZ}}{2M_W^2s} \times \left(1+\frac{M_Z^2}{s}+ \dots \right) \nonumber \\
&& \times \bar{v}_2 (~\Slash{k_1}-\Slash{k_2}) \left\{(g_R+g_L)+(g_R-g_L)\gamma^5 \right\}u_1 \times \left(\frac{s}{2}-M_W^2 \right) +\order(1)  \\
&\approx& \frac{g_{WWZ}}{2M_W^2s} \times \bar{v}_2 (~\Slash{k_1}-\Slash{k_2}) \left\{(g_R+g_L)+(g_R-g_L)\gamma^5 \right\}u_1 \times \frac{s}{2} +\order(1) \\
&=& \frac{g_{WWZ}}{4M_W^2}\bar{v}_2 (~\Slash{k_1}-\Slash{k_2}) \left\{(g_R+g_L)+(g_R-g_L)\gamma^5 \right\}u_1 + \order(1) \\
&=& \frac{g_{WWZ}}{2M_W^2}\bar{v}_2 \left\{(g_R+g_L)~\Slash{k_1} +(g_R-g_L)(~\Slash{k_1}+m_e) \gamma^5 \right\}u_1 + \order(1) \,.
 \end{eqnarray}
 \end{subequations}
In the last step we have used the results of \Eqn{eeWW2}. The final expression for the $Z$ mediated amplitude is given below:
\begin{eqnarray}
{\cm}_a^Z &=& \underbrace{\frac{g_{WWZ}}{2M_W^2}(g_R+g_L)\bar{v}_2~\Slash{k_1}u_1+ \frac{g_{WWZ}}{2M_W^2}(g_R-g_L)\bar{v}_2~\Slash{k_1}\gamma^5 u_1}_{\order(E^2)} \nonumber \\
&& +\underbrace{\frac{g_{WWZ}m_e}{2M_W^2}(g_R-g_L)\bar{v}_2\gamma^5 u_1}_{\order(E)} +\order(1) \,.
\label{Z-final}
\end{eqnarray}

\section{\texorpdfstring{$t$}{TEXT}-channel \texorpdfstring{$\nu$}{TEXT} exchange}
We assume the $W$-boson coupling with the fermions to be of the following form
$$-\frac{g}{\sqrt{2}}\bar{\nu}\gamma^\mu P_L eW_\mu^+ +~{\rm h.c.} $$
We now write the Feynman amplitude for the diagram \ref{f:eeWW}b as follows:
 \begin{subequations}
 \begin{eqnarray}
&& i{\cm}_b^\nu = \bar{v}_2 \left\{-\frac{ig}{2\sqrt{2}}\gamma_\nu(1-\gamma^5) \right\} \frac{i(\slashed{q}+m_\nu)}{q^2-m_\nu^2} \left\{-\frac{ig}{2\sqrt{2}}\gamma_\mu(1-\gamma^5) \right\}u_1 \times \frac{\epsilon_1^\mu\epsilon_2^\nu}{M_W^2} \\
&\Rightarrow& {\cm}_b^\nu = -\frac{g^2}{8M_W^2(t-m_\nu^2)} \bar{v}_2 \gamma_\nu(1-\gamma^5)(\slashed{q}+m_\nu) \gamma_\mu(1-\gamma^5) u_1 \times \epsilon_1^\mu\epsilon_2^\nu \,,
 \label{eeWW3}
 \end{eqnarray}
where, $q=p_1-k_1=k_2-p_2$ is the four momentum of the intermediate fermion. Note that the term proportional to $m_\nu$ at the numerator of \Eqn{eeWW3} do not contribute because $(1-\gamma^5)\gamma_\mu(1-\gamma^5)$ $=0$. Using $(1-\gamma^5)^2=2(1-\gamma^5)$ we now rewrite \Eqn{eeWW3} as 
\begin{eqnarray}
{\cm}_b^\nu = -\frac{g^2}{4M_W^2(t-m_\nu^2)} \bar{v}_2 \gamma_\nu \slashed{q} \gamma_\mu(1-\gamma^5) u_1 \times \epsilon_1^\mu\epsilon_2^\nu \,.
\label{eeWW4}
\end{eqnarray}
It is worth noting at this point that any modification of $\order(M_W^2/E)$ in $\epsilon_1$ or $\epsilon_2$ will lead to $\order(1)$ contribution to the amplitude. Hence, using $\epsilon_1^\mu\approx k_1^\mu$ and $\epsilon_2^\mu\approx k_2^\mu$ in \Eqn{eeWW4}, we obtain
\begin{eqnarray}
{\cm}_b^\nu = -\frac{g^2}{4M_W^2(t-m_\nu^2)} \left[\bar{v}_2 \slashed{k_2}(\slashed{p_1}-\slashed{k_1}) \slashed{k_1} (1-\gamma^5) u_1 \right] +\order(1) \,.
\label{eeWW5}
\end{eqnarray}
 \end{subequations}

\begin{mdframed}
Now note that
 \begin{subequations}
 \begin{eqnarray}
(\slashed{p_1}-\slashed{k_1})\slashed{k_1} u_1 &=& (\slashed{p_1}\slashed{k_1} - k_1^2)u_1 = (2p_1\cdot k_1 - \slashed{k_1}\slashed{p_1}-k_1^2) u_1 \nonumber \\
&=& (2p_1\cdot k_1-k_1^2)u_1 - m_e\slashed{k_1}u_1 = (-t+m_e^2)u_1-m_e\slashed{k_1}u_1 \,.
 \end{eqnarray}
Similarly, one can easily obtain 
\begin{eqnarray}
(\slashed{p_1}-\slashed{k_1})\slashed{k_1}(1-\gamma^5) u_1=  (-t+m_e^2)(1-\gamma^5) u_1-m_e\slashed{k_1}(1+\gamma^5)u_1 \,.
\end{eqnarray}
Let us now proceed to simplify the expression that appears inside the square bracket in \Eqn{eeWW5}:
\begin{eqnarray}
\bar{v}_2 \slashed{k_2}(\slashed{p_1}-\slashed{k_1}) \slashed{k_1} (1-\gamma^5) u_1 = \underbrace{-(t-m_e^2)\bar{v}_2 \slashed{k_2}(1-\gamma^5) u_1}_{T_1} \underbrace{ -m_e\bar{v}_2 \slashed{k_2} \slashed{k_1}(1+\gamma^5)u_1}_{T_2} \,.
\label{t1t2}
\end{eqnarray}

$\blacksquare$ {\bf Evaluation of $T_1$:}
\begin{eqnarray}
T_1 &=& -(t-m_e^2)\bar{v}_2 \slashed{k_2}(1-\gamma^5) u_1 = -(t-m_e^2)\bar{v}_2 (\slashed{p_1}+\slashed{p_2} -\slashed{k_1})(1-\gamma^5) u_1 \nonumber \\
&=& (t-m_e^2)\bar{v}_2 \slashed{k_1}(1-\gamma^5) u_1 -(t-m_e^2)\bar{v}_2 (\slashed{p_1}+\slashed{p_2})(1-\gamma^5) u_1 \nonumber \\
&=& (t-m_e^2)\bar{v}_2 \slashed{k_1}(1-\gamma^5) u_1 -2m_e(t-m_e^2)\bar{v}_2 \gamma^5 u_1 \,,
\label{t1}
\end{eqnarray}
where, in the last step, we have used the following identity:
\begin{eqnarray}
\bar{v}_2 (\slashed{p_1}+\slashed{p_2})(1-\gamma^5) u_1 &=& \bar{v}_2\slashed{p_2}(1-\gamma^5) u_1 + \bar{v}_2 (1+\gamma^5)\slashed{p_1} u_1 \nonumber \\
&=& -m_e\bar{v}_2 (1-\gamma^5) u_1 +m_e \bar{v}_2(1+\gamma^5) u_1 \nonumber \\
 &=& 2m_e\bar{v}_2\gamma^5 u_1 \,.
\end{eqnarray}

$\blacksquare$ {\bf Evaluation of $T_2$:}
For this, let us first note the following:
\begin{eqnarray}
\slashed{k_2} \slashed{k_1} &=& (\slashed{p_1}+\slashed{p_2} -\slashed{k_1})\slashed{k_1} = \slashed{p_1} \slashed{k_1} + \slashed{p_2} \slashed{k_1} -k_1^2 \nonumber \\
&=& 2(p_1\cdot k_1) -\slashed{k_1} \slashed{p_1} + \slashed{p_2} \slashed{k_1} -k_1^2 =-(t-m_e^2) -\slashed{k_1} \slashed{p_1} + \slashed{p_2} \slashed{k_1} \,.
\end{eqnarray}
Now we can write
\begin{eqnarray}
T_2 &=& -m_e\bar{v}_2 \slashed{k_2} \slashed{k_1}(1+\gamma^5)u_1 \nonumber \\
&=& m_e(t-m_e^2) \bar{v}_2(1+\gamma^5) u_1 +m_e \bar{v}_2 \slashed{k_1} \slashed{p_1}(1+\gamma^5) u_1 -m_e \bar{v}_2\slashed{p_2} \slashed{k_1}(1+\gamma^5) u_1 \nonumber \\
&=& m_e(t-m_e^2) \bar{v}_2(1+\gamma^5) u_1 +m_e^2 \bar{v}_2 \slashed{k_1} (1-\gamma^5) u_1 +m_e^2 \bar{v}_2 \slashed{k_1}(1+\gamma^5) u_1 \,.
\label{t2}
\end{eqnarray}
Therefore, adding \Eqs{t1}{t2} together, we obtain
\begin{eqnarray}
T_1+T_2 &=& t\bar{v}_2 \slashed{k_1}(1-\gamma^5) u_1 +m_e(t-m_e^2)\bar{v}_2 (1-\gamma^5) u_1 +m_e^2\bar{v}_2 \slashed{k_1}(1+\gamma^5) u_1 \nonumber \\
&=& \underbrace{t\bar{v}_2 \slashed{k_1}(1-\gamma^5) u_1}_{{\rm leads~to}~\order(E^2)} +\underbrace{m_et\bar{v}_2 (1-\gamma^5) u_1}_{{\rm leads~to}~\order(E)} \nonumber \\
&& +\underbrace{m_e^2\bar{v}_2 \slashed{k_1}(1+\gamma^5) u_1}_{{\rm leads~to}~\order(1)} -\underbrace{m_e^3 \bar{v}_2 (1-\gamma^5) u_1}_{{\rm leads~to}~\order(1/E)} \,.
\label{t1+t2}
\end{eqnarray}
\end{subequations}
\end{mdframed}

Now we recast \Eqn{eeWW5} as follows:
\begin{eqnarray}
{\cm}_b^\nu &=& -\frac{g^2}{4M_W^2t}\left(1+\frac{m_\nu^2}{t}+\dots \right) \left[T_1+T_2 \right] +\order(1) \nonumber \\
&=&  -\frac{g^2}{4M_W^2} \bar{v}_2 \slashed{k_1}(1-\gamma^5) u_1 -\frac{g^2m_e}{4M_W^2} \bar{v}_2 (1-\gamma^5) u_1 +\order(1) \,,
\label{nu-final}
\end{eqnarray}
where we have used the result of \Eqn{t1+t2}. One interesting thing to note from \Eqn{nu-final} is that the mass of the intermediate fermion ($m_\nu$ in this case) does not appear in the coefficients of the terms that grow with energy.

\section{Amplitude without the Higgs}
We now add Eqs.~(\ref{gam-final}), (\ref{Z-final}) and (\ref{nu-final}) together to obtain
\begin{eqnarray}
{\cm}_{a+b} &=& \left[\frac{e^2}{M_W^2}+\frac{g_{WWZ}}{2M_W^2}(g_R+g_L)-\frac{g^2}{4M_W^2} \right] \underbrace{\bar{v}_2 \slashed{k_1} u_1}_{\order(E^2)} +\left[\frac{g_{WWZ}}{2M_W^2}(g_R-g_L)+\frac{g^2}{4M_W^2} \right] \underbrace{\bar{v}_2 \slashed{k_1}\gamma^5 u_1}_{\order(E^2)} \nonumber \\
&& +m_e\left[\frac{g_{WWZ}}{2M_W^2}(g_R-g_L)+\frac{g^2}{4M_W^2} \right] \underbrace{\bar{v}_2 \gamma^5 u_1}_{\order(E)} -\frac{g^2m_e}{4M_W^2}\underbrace{\bar{v}_2 u_1}_{\order(E)} +\order(1) \,.
\label{wh-final}
\end{eqnarray}
Clearly, the cancellation of $\order(E^2)$ growth requires
 \begin{subequations}
 \label{con}
 \begin{eqnarray}
e^2+\frac{g_{WWZ}}{2}(g_R+g_L)-\frac{g^2}{4} &=& 0 \,, \\
\frac{g_{WWZ}}{2}(g_R-g_L)+\frac{g^2}{4} &=& 0 \,.
 \label{con2}
 \end{eqnarray}
 \end{subequations}
Note that the condition (\ref{con2}) also guarantees the cancellation of $\bar{v}_2 \gamma^5 u_1$ term in \Eqn{wh-final}. Thus no additional new physics contribution is required to cancel the $\order(E)$ growth carried by the $\bar{v}_2 \gamma^5 u_1$ term. Remember that in the case of SM
 \begin{subequations}
 \begin{eqnarray}
&& g_L+g_R = \frac{g}{\cos\theta_w}\left(\frac{1}{2}-2\sin^2\theta_w \right) \,,  ~ g_L-g_R = \frac{g}{2\cos\theta_w} \,, \\
&& e=g\sin\theta_w \,,~~ {\rm and} ~~ g_{WWZ}=g\cos\theta_w \,,
 \end{eqnarray}
 \end{subequations}
so that the conditions (\ref{con}) are trivially satisfied. Thus only the $\order(E)$ growth carried by the term proportional to $\bar{v}_2 u_1$ remains to be canceled in \Eqn{wh-final}. This necessitates the introduction of a new scalar particle with suitable interactions.

\section{\texorpdfstring{$s$}{TEXT}-channel Higgs exchange}
We assume the relevant couplings to be of the form
\begin{eqnarray}
g_{eeh} \bar{e}(\cos\delta +i\sin\delta\gamma^5)eh +g_{WWh}W_\mu^+W^{\mu -}h \,.
\end{eqnarray}
With this, one can write the Feynman amplitude for the diagram~\ref{f:eeWW}c as follows:
 \begin{subequations}
 \begin{eqnarray}
i{\cm}_c^h &=& \bar{v}_2\left\{ig_{eeh}(\cos\delta+i\sin\delta\gamma^5) \right\} u_1 \frac{i}{s-m_h^2}(ig_{WWh}g_{\mu\nu}) \frac{\epsilon_1^\mu\epsilon_2^\nu}{M_W^2} \\
\Rightarrow~~ {\cm}_c^h &=& -\frac{g_{eeh}g_{WWh}}{M_W^2(s-m_h^2)} \bar{v}_2(\cos\delta+i\sin\delta\gamma^5)u_1 (\epsilon_1\cdot \epsilon_2) \\
&=& -\frac{g_{eeh}g_{WWh}}{M_W^2s} \left(1+\frac{m_h^2}{s}+\dots \right) \left(\frac{s}{2}-M_W^2 \right) \times \bar{v}_2(\cos\delta+i\sin\delta\gamma^5)u_1 \\
&\approx& -\frac{g_{eeh}g_{WWh}}{2M_W^2}\cos\delta~ \bar{v}_2u_1 -\frac{ig_{eeh}g_{WWh}}{2M_W^2}\sin\delta ~\bar{v}_2\gamma^5 u_1 +\order(\frac{1}{E}) \,.
 \label{h-final}
 \end{eqnarray}
 \end{subequations}
In the case of SM 
\begin{eqnarray}
g_{eeh} = -\frac{gm_e}{2M_W} \,, ~~ \delta=0 \,, ~~{\rm and} ~~ g_{WWh} =gM_W \,.
\end{eqnarray}
Hence,
\begin{eqnarray}
\left({\cm}_c^h\right)^{\rm SM} = \frac{g^2m_e}{4M_W^2} \bar{v}_2u_1 +\order(\frac{1}{E}) \,,
\end{eqnarray}
which is exactly opposite to the residual $\order(E)$ growth in \Eqn{wh-final}.

\section{Explicit evaluation of the energy growths}
To begin with, we write down the Dirac spinors for free fermions as
\begin{eqnarray}
u^r(\vec{p})= \sqrt{E+m} \begin{pmatrix}
\chi^r \\ \frac{\vec{\sigma}\cdot\vec{p}}{E+m} \chi^r
\end{pmatrix} \,, &&
v^r(\vec{p})= -\sqrt{E+m} \begin{pmatrix}
\frac{\vec{\sigma}\cdot\vec{p}}{E+m} \chi^r \\ \chi^r
\end{pmatrix} \,,
\end{eqnarray}
where the superscript $r$ (= 1, 2) represents the helicity of the fermion. The column matrices, $\chi^r$, are given by
\begin{eqnarray}
\chi^1= \begin{pmatrix} 1 \\ 0 \end{pmatrix} \,, && \chi^2= \begin{pmatrix} 0 \\ 1 \end{pmatrix} \,.
\end{eqnarray}
\begin{mdframed}
The orthonormality conditions satisfied by the spinors $u$ and $v$ are given below:
 \begin{subequations}
 \begin{eqnarray}
&& u^{r\dagger}(\vec{p})u^s(\vec{p}) = 2E\delta^{rs} ~~~ {\rm or,}~~ \bar{u}^r(\vec{p})u^s(\vec{p})= 2m\delta^{rs} \,, \\
&& v^{r\dagger}(\vec{p})v^s(\vec{p}) = 2E\delta^{rs} ~~~ {\rm or,}~~ \bar{v}^r(\vec{p})v^s(\vec{p})= -2m\delta^{rs} \,, \\
&& \bar{u}^r(\vec{p})v^s(\vec{p}) = \bar{v}^r(\vec{p})u^s(\vec{p}) = 0 \,.
 \end{eqnarray}
But note that,
\begin{eqnarray}
&& u^{r\dagger}(\vec{p})v^s(\vec{p}) \ne 0 ~~~ {\rm and} ~~ v^{r\dagger}(\vec{p})u^s(\vec{p}) \ne 0 \,, \\
{\rm however,} && u^{r\dagger}(\vec{p})v^s(-\vec{p}) = 0 ~~~ {\rm and} ~~ v^{r\dagger}(-\vec{p})u^s(\vec{p}) \ne 0 \,.
\end{eqnarray}
 \end{subequations}
\end{mdframed}
Taking the direction of $\vec{p}_1$ to be the positive $z$-axis and assuming that the scattering takes place in the $x$-$z$ plane,
we may take the different momentum four vectors as follows:
 \begin{subequations}
 \begin{eqnarray}
&& p_1^\mu = (E,~0,~0,~p) \,,~~~~ p_2^\mu = (E,~0,~0,~-p) \,, \\
&& k_1^\mu = (E,~k\sin\theta,~0,~k\cos\theta) \,, ~~~~ k_2^\mu = (E,~-k\sin\theta,~0,~-k\cos\theta) \,.
 \end{eqnarray}
 \end{subequations}
In the Dirac-Pauli representation, different gamma matrices take the following form:
\begin{eqnarray}
\gamma^0=\gamma_0 = \begin{pmatrix} 1 & 0 \\ 0 & -1 \end{pmatrix} \,, ~~
\gamma^i=-\gamma_i = \begin{pmatrix} 0 & \sigma_i \\ -\sigma_i & 0 \end{pmatrix} \,,
\gamma^5=\gamma_5 = \begin{pmatrix} 0 & 1 \\ 1 & 0 \end{pmatrix} \,.
\end{eqnarray}
Using these one may easily obtain
\begin{eqnarray}
\slashed{k_1} = k_1^0\gamma_0+k_1^i\gamma_i = \begin{pmatrix} E & -k\vec{\sigma}\cdot\hat{e} \\ k\vec{\sigma}\cdot\hat{e} & -E \end{pmatrix} \,,
\end{eqnarray}
where $\hat{e}=(\sin\theta,~0,~\cos\theta)$ is the unit vector in the direction of $\vec{k}_1$. Similarly, one can also get
\begin{eqnarray}
\slashed{k_1}\gamma^5 = \begin{pmatrix}  -k\vec{\sigma}\cdot\hat{e} & E \\ -E & k\vec{\sigma}\cdot\hat{e}  \end{pmatrix} \,.
\end{eqnarray}
Then clearly,
\begin{eqnarray}
\gamma^0\slashed{k_1} = \begin{pmatrix} E & -k\vec{\sigma}\cdot\hat{e} \\ -k\vec{\sigma}\cdot\hat{e} & E  \end{pmatrix} \,,
~~~~~
\gamma^0\slashed{k_1}\gamma^5 = \begin{pmatrix}  -k\vec{\sigma}\cdot\hat{e} & E \\ E & -k\vec{\sigma}\cdot\hat{e}  \end{pmatrix} \,.
\end{eqnarray}
Again note that,
\begin{eqnarray}
\frac{\vec{\sigma}\cdot\vec{p}_1}{E+m} \approx \sigma_z +\order(\frac{1}{E}) \,, ~~~~~~ \frac{\vec{\sigma}\cdot\vec{p}_2}{E+m} \approx -\sigma_z +\order(\frac{1}{E}) \,.
\end{eqnarray}
Now using $\sigma_z\chi^1=\chi^1$ and $\sigma_z\chi^2=-\chi^2$ one may easily show the following:
 \begin{subequations}
 \begin{eqnarray}
&& u^1(\vec{p}_1) = \sqrt{E} \begin{pmatrix} 1 \\ 1 \end{pmatrix} \chi^1 + \order(\frac{1}{\sqrt{E}}) \,, ~~~~
u^2(\vec{p}_1) = \sqrt{E} \begin{pmatrix} 1 \\ -1 \end{pmatrix} \chi^2 + \order(\frac{1}{\sqrt{E}}) \,,  \\
&& v^1(\vec{p}_2) = \sqrt{E} \begin{pmatrix} 1 \\ -1 \end{pmatrix} \chi^1 + \order(\frac{1}{\sqrt{E}}) \,, ~~
v^2(\vec{p}_1) = -\sqrt{E} \begin{pmatrix} 1 \\ 1 \end{pmatrix} \chi^2 + \order(\frac{1}{\sqrt{E}}) \,, \\
\Rightarrow && \bar{v}^1(\vec{p}_2) =v^{1\dagger}(\vec{p}_2)\gamma^0 = \sqrt{E}\chi^{1\dagger} \begin{pmatrix} 1 & 1 \end{pmatrix}  \,, ~~
\bar{v}^2(\vec{p}_1) = -\sqrt{E}\chi^{2\dagger} \begin{pmatrix} 1 & -1 \end{pmatrix}  \,.
 \end{eqnarray}
 \end{subequations}
 
\subsection{The same helicity case}
$\blacksquare$ {\bf Case-I ($r_1=r_2=1$):}
\begin{subequations}
\begin{eqnarray}
&& \bar{v}^1(\vec{p}_2)u^1(\vec{p}_1) = E \chi^{1\dagger} \begin{pmatrix}1 & 1 \end{pmatrix} \begin{pmatrix}1 \\ 1 \end{pmatrix} \chi^1 = 2E \,. \\
&& \bar{v}^1(\vec{p}_2)\gamma^5 u^1(\vec{p}_1) = E \chi^{1\dagger} \begin{pmatrix}1 & 1 \end{pmatrix} \begin{pmatrix} 0 & 1 \\ 1 & 0 \end{pmatrix} \begin{pmatrix}1 \\ 1 \end{pmatrix} \chi^1 = 2E \,. \\
&& \bar{v}^1(\vec{p}_2) \slashed{k_1} u^1(\vec{p}_1) = E \chi^{1\dagger} \begin{pmatrix}1 & 1 \end{pmatrix} \begin{pmatrix} E & -k\vec{\sigma}\cdot\hat{e} \\ k\vec{\sigma}\cdot\hat{e} & -E \end{pmatrix} \begin{pmatrix}1 \\ 1 \end{pmatrix} \chi^1 = 0 \,. \\
&& \bar{v}^1(\vec{p}_2) \slashed{k_1}\gamma^5 u^1(\vec{p}_1) = E \chi^{1\dagger} \begin{pmatrix}1 & 1 \end{pmatrix} \begin{pmatrix}  -k\vec{\sigma}\cdot\hat{e} & E \\ -E & k\vec{\sigma}\cdot\hat{e}  \end{pmatrix} \begin{pmatrix}1 \\ 1 \end{pmatrix} \chi^1 = 0 \,.
\end{eqnarray}
\end{subequations}

$\blacksquare$ {\bf Case-I ($r_1=r_2=2$):}
\begin{subequations}
\begin{eqnarray}
&& \bar{v}^2(\vec{p}_2)u^2(\vec{p}_1) = E \chi^{2\dagger} \begin{pmatrix}1 & -1 \end{pmatrix} \begin{pmatrix}1 \\ -1 \end{pmatrix} \chi^2 = 2E \,. \\
&& \bar{v}^2(\vec{p}_2)\gamma^5 u^2(\vec{p}_1) = E \chi^{2\dagger} \begin{pmatrix}1 & -1 \end{pmatrix} \begin{pmatrix} 0 & 1 \\ 1 & 0 \end{pmatrix} \begin{pmatrix}1 \\ -1 \end{pmatrix} \chi^2 = -2E \,. \\
&& \bar{v}^2(\vec{p}_2) \slashed{k_1} u^2(\vec{p}_1) = E \chi^{2\dagger} \begin{pmatrix}1 & -1 \end{pmatrix} \begin{pmatrix} E & -k\vec{\sigma}\cdot\hat{e} \\ k\vec{\sigma}\cdot\hat{e} & -E \end{pmatrix} \begin{pmatrix}1 \\ -1 \end{pmatrix} \chi^2 = 0 \,. \\
&& \bar{v}^2(\vec{p}_2) \slashed{k_1}\gamma^5 u^2(\vec{p}_1) = E \chi^{2\dagger} \begin{pmatrix}1 & -1 \end{pmatrix} \begin{pmatrix}  -k\vec{\sigma}\cdot\hat{e} & E \\ -E & k\vec{\sigma}\cdot\hat{e}  \end{pmatrix} \begin{pmatrix}1 \\ -1 \end{pmatrix} \chi^2 = 0 \,.
\end{eqnarray}
\end{subequations}

\subsection{The opposite helicity case}
$\blacksquare$ {\bf Case-I ($r_1=1,~r_2=2$):}
\begin{subequations}
\begin{eqnarray}
&& \bar{v}^2(\vec{p}_2)u^1(\vec{p}_1) = -E \chi^{2\dagger} \begin{pmatrix}1 & -1 \end{pmatrix} \begin{pmatrix}1 \\ 1 \end{pmatrix} \chi^1 = 0 \,. \\
&& \bar{v}^2(\vec{p}_2)\gamma^5 u^1(\vec{p}_1) = -E \chi^{1\dagger} \begin{pmatrix}1 & -1 \end{pmatrix} \begin{pmatrix} 0 & 1 \\ 1 & 0 \end{pmatrix} \begin{pmatrix}1 \\ 1 \end{pmatrix} \chi^1 = 0 \,. \\
&& \bar{v}^2(\vec{p}_2) \slashed{k_1} u^1(\vec{p}_1) = -E \chi^{2\dagger} \begin{pmatrix}1 & -1 \end{pmatrix} \begin{pmatrix} E & -k\vec{\sigma}\cdot\hat{e} \\ k\vec{\sigma}\cdot\hat{e} & -E \end{pmatrix} \begin{pmatrix}1 \\ 1 \end{pmatrix} \chi^1 \nonumber \\
&& ~~~~~~~~~~~~~ = -2E~ \chi^{2\dagger} \left\{E -k\vec{\sigma}\cdot\hat{e} \right\}   \chi^1 = 2Ek\sin\theta \,. \\
&& \bar{v}^2(\vec{p}_2) \slashed{k_1}\gamma^5 u^1(\vec{p}_1) = -E \chi^{2\dagger} \begin{pmatrix}1 & -1 \end{pmatrix} \begin{pmatrix}  -k\vec{\sigma}\cdot\hat{e} & E \\ -E & k\vec{\sigma}\cdot\hat{e}  \end{pmatrix} \begin{pmatrix}1 \\ 1 \end{pmatrix} \chi^1 \nonumber \\
&& ~~~~~~~~~~~~~ = -2E~ \chi^{2\dagger} \left\{E -k\vec{\sigma}\cdot\hat{e} \right\}   \chi^1 = 2Ek\sin\theta \,.
\end{eqnarray}
\end{subequations}
In the last two steps we have used the following:
 \begin{subequations}
 \begin{eqnarray}
&& E -k\vec{\sigma}\cdot\hat{e} = E~\mathbf{I}_2 -k\sigma_x\sin\theta -k\sigma_z\cos\theta \,, \\
&& \mathbf{I}_2\chi^1 = \chi^1 \,, ~~ \sigma_z\chi^1=\chi^1 \,, ~~\sigma_x\chi^1=\chi^2 \,, ~~ \chi^{2\dagger}\chi^1=0 \,.
 \end{eqnarray}
 \end{subequations}
 
\chapter{Flavor observables and 2HDM} 
\label{AppendixC} 
\lhead{Appendix C. \emph{Flavor observables and 2HDM}}

\section{Neutral meson mixing}
The dominant one-loop effective Lagrangian for $\Delta F=2$ is
\begin{eqnarray}
{\ml}_{\rm eff}^{\Delta F=2} &=& \frac{G_F^2 M_W^2}{16\pi^2} 
\sum\limits_{a,b = u,c,t}^{} \lambda_a \lambda_b w_a w_b \left[S(w_a,w_b) + \right. \nonumber \\
&& ~~~~~~~\left. X_aX_b \left\{2I_1(w_a,w_b,w_{1+})+X_aX_bI_2(w_a,w_b,w_{1+})\right\} \right] O_F \,.
\end{eqnarray}
Here, the $S(w_a,w_b)$ part is the SM contribution and the rest is due
to the charged Higgs box diagrams.  For $i$-type BGL model, $X_q =
-\cot\beta$ if $q=i$ and $X_q = \tan\beta$ otherwise. Since we have assumed the external particles 
to have zero (four) momenta, the down- type quark masses have been set to zero. Under this approximation,
the charged Higgs part of the Yukawa interaction for Type I and II models are identical and scaled by a $\cot\beta$
factor for all three generations. This means, for Type I and II models, $X_q = \cot\beta$ for $q=u,c,t$.  The dimension-6
operator for $K^0$--$\bar{K}^0$ mixing is
\begin{equation}
O_F = (\bar{s}\gamma^\mu P_L d)^2 \,.
\end{equation}
Similar expressions can be obtained for $B$ systems. The relevant
parameters and functions are defined as follows:
\begin{eqnarray}
\lambda_a = V^*_{ad}V_{as} ~, ~~~~ w_a &=& \frac{m_a^2}{M_W^2} \,,
~~~f(x) = \frac{(x^2-8x+4)\ln x +3(x-1)}{(x-1)^2} \,,
\nonumber\\ S(w_a,w_b) &=& \frac{f(w_a)-f(w_b)}{w_a-w_b} \,,
~~~ g(x,y,z) = \frac{x(x-4)\ln x}{(x-1)(x-y)(x-z)} \,,
\nonumber\\ I_1(w_a,w_b,w_c) &=& [g(w_a,w_b,w_c) +
  g(w_b,w_c,w_a)+g(w_c,w_a,w_b)] \,,
\nonumber\\ I_2(w_a,w_b,w_c) &=&
\frac{1}{w_a-w_b}\left[\frac{w_a^2\ln w_a}{(w_c-w_a)^2}
  -\frac{w_b^2\ln w_b}{(w_c-w_b)^2} \right] \nonumber \\ && +
\frac{w_c [(w_c-w_a)(w_c-w_b)+\{2w_aw_b-w_c(w_a+w_b)\}\ln
    w_c]}{(w_c-w_a)^2(w_c-w_b)^2}\,.
\end{eqnarray}
Obtaining $M_{12}$ from the effective Lagrangian is
straightforward. As an example, for $K$-meson system (with $B_K$ as
bag parameter),
\begin{eqnarray}
&& M_{12}^K = -\frac{1}{2m_K} \Bra{K^0}{\mathscr L}_{\rm eff}^{\Delta
    F=2} \Ket{\bar{K^0}} \,, \\ && \Bra{K^0}O_F\Ket{\bar{K^0}} =
  \frac{2}{3}f_K^2 m_K^2 B_K \,.
\end{eqnarray}

\section{Expressions for \texorpdfstring{$b\to s\gamma$}{TEXT}}
The effective Lagrangian for $b\to s\gamma$ can be written as
\begin{eqnarray}
{\cal L}_{\rm eff} &=& \sqrt{\frac{G_F^2}{8\pi^3}}V_{ts}^*V_{tb} m_b 
\left[\sqrt{\alpha}\left\{C_{7L} \bar{s}_L \sigma^{\mu\nu}b_R +
C_{7R} \bar{s}_R \sigma^{\mu\nu}b_L\right\}F_{\mu\nu}\right. \nonumber\\
&&~~~~~\left.\sqrt{\alpha_s} \left\{C_{8L} \bar{s}_L T_a \sigma^{\mu\nu}b_R +
C_{8R} \bar{s}_R T_a \sigma^{\mu\nu}b_L\right\}G^a_{\mu\nu}\right] +
\rm{h.c.} \, ,
\label{bsgam-effl}
\end{eqnarray}
where $F_{\mu\nu}$ and $G^a_{\mu\nu}$ are field strength tensors for photon
and gluon, respectively, and $T^a$s are the SU(3) generators. 
The branching ratio ${\rm Br}(b\to s\gamma)$ is given by 
\begin{eqnarray}
\frac{{\rm Br}(b\to s\gamma)}{{\rm Br}(b\to c e \bar{\nu})} = 
\frac{6\alpha}{\pi B}\left\vert 
\frac{V^*_{ts}V_{tb}}{V_{cb}}\right\vert^2 \left[  \left\vert C_{7L}^{\rm eff}\right\vert^2 
+\left\vert C_{7R}^{\rm eff}\right\vert^2 \right] \,,
\label{b2sgam}
\end{eqnarray}
where, we have taken $B=0.546$ \cite{Deschamps:2009rh}. 
The effective Wilson coefficients are
\begin{subequations}
\begin{eqnarray}
C_{7L}^{\rm eff} &=& \eta^{16/23} C_{7L} +
\frac{8}{3}\left(\eta^{14/23} -\eta^{16/23} \right) C_{8L} + {\cal C}
\,, \\ C_{7R}^{\rm eff} &=& \eta^{16/23} C_{7R} +
\frac{8}{3}\left(\eta^{14/23} -\eta^{16/23} \right) C_{8R} \,.
\end{eqnarray}
\label{effective Wilson}
\end{subequations}
In the above equations, $\eta = \alpha_s(M_Z)/\alpha_s(\mu)$, where
$\mu$ is the QCD renormalization scale; ${\cal C}$ corresponds to the
leading log QCD corrections in SM.  In the expression for the
effective Wilson co-efficients (Eq.\ (\ref{effective Wilson})), the
correction term is given by
\begin{eqnarray}
{\cal C} &=& \sum\limits_{i=1}^{8} h_i \eta^{a_i} \,,
\end{eqnarray}
where,
\begin{subequations}
\begin{eqnarray}
a_i &=&
\left(\frac{14}{23},~\frac{16}{23},~\frac{6}{23},~-\frac{12}{23},~0.4086,~-0.4230,~-0.8994,~0.1456\right)\,,
\\ h_i &=&
\Bigg(\frac{626126}{272277},~-\frac{56281}{51730},~-\frac{3}{7},~-\frac{1}{14}, \nonumber \\
&&  ~-0.6494,~-0.0380,~-0.0186,~-0.0057\Bigg)
\,.
\end{eqnarray}
\end{subequations}
These values of $h_i$ and $a_i$ can be found in \cite{Gambino:2001ew}
[see Eq.\ (2.3) and Table 1 of Ref.~\cite{Gambino:2001ew}].

\begin{center}
 \begin{figure}
 \includegraphics[scale=0.45]{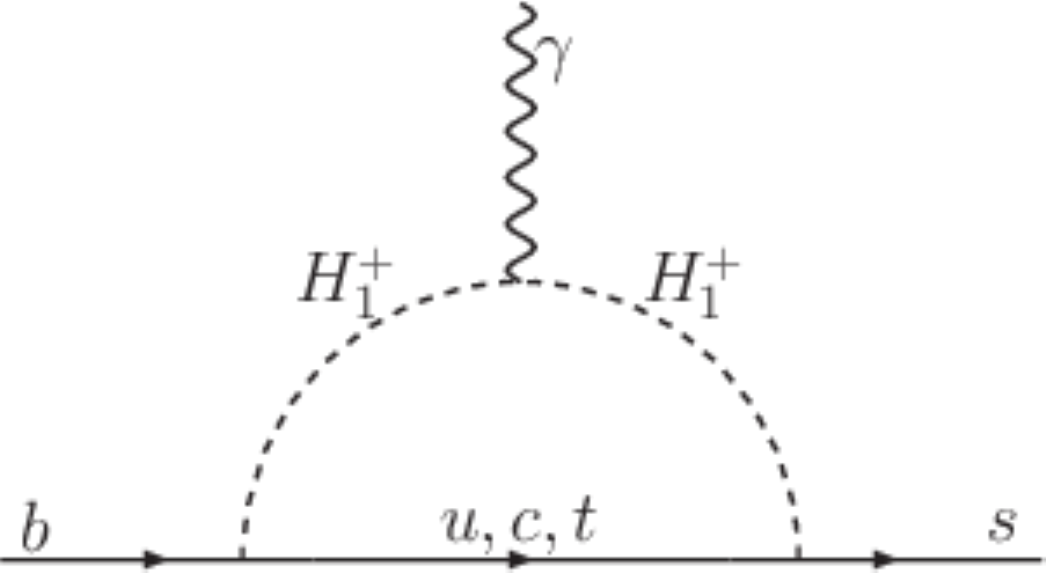}~~~~~
\includegraphics[scale=0.45]{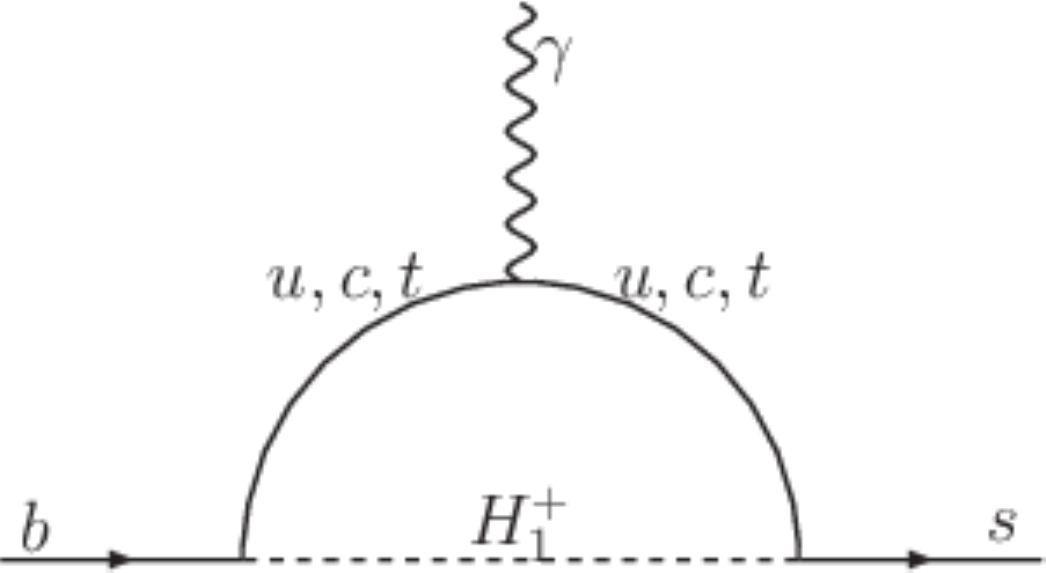}~~~~~
\includegraphics[scale=0.45]{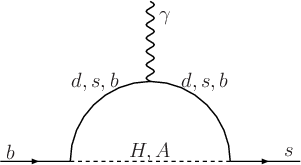}
\caption[NP contributions to $b\to s\gamma$]{\em Feynman diagrams involving nonstandard scalars
  contributing to $b\to s\gamma$ amplitude.}
\label{f:b2sg}
\end{figure}
\end{center}
To understand the above expressions, we first define the following functions:
\begin{subequations}
\begin{eqnarray}
\mathscr{F}_0(t) = \int\limits_{0}^{1} dx \frac{1-x}{x+(1-x)t} &=&
-\frac{1}{1-t}-\frac{\ln t}{(1-t)^2}\,, \\ 
\mathscr{F}_1(t) = \int\limits_{0}^{1} dx \frac{(1-x)^2}{x+(1-x)t} &=&
\frac{-3+4t-t^2}{2(1-t)^3}-\frac{\ln t}{(1-t)^3}\,, \\ 
\mathscr{F}_2(t) = \int\limits_{0}^{1} dx
\frac{(1-x)^3}{x+(1-x)t} &=& \frac{-11+18t-9t^2+2t^3-6\ln
  t}{6(1-t)^4}\,, \\ 
\overline{\mathscr{F}}_0(t) =
\int\limits_{0}^{1} dx \frac{x}{x+(1-x)t} &=& \frac{1-t+t\ln
  t}{(1-t)^2}\,, \\ 
\overline{\mathscr{F}}_1(t) =
\int\limits_{0}^{1} dx \frac{x^2}{x+(1-x)t} &=&
\frac{1-4t+3t^2-2t^2\ln t}{2(1-t)^3}\,, \\
\overline{\mathscr{F}}_2(t) = \int\limits_{0}^{1} dx
\frac{x^3}{x+(1-x)t} &=& \frac{2-9t+18t^2-11 t^3+6t^3\ln
  t}{6(1-t)^4}\,.
\end{eqnarray}
\end{subequations}
Let us further define $x_t={m_t^2}/{m_W^2}$, $y_q={m_q^2}/{m_{1+}^2}$,
$z_q={m_q^2}/{m_H^2}$, $z'_q={m_q^2}/{m_A^2}$. Now the
expressions for $C_{7 L},~C_{7 R},~C_{8 L},~C_{8 R}$ read
\begin{subequations}
\begin{eqnarray}
C_{7L} &=& A_{\gamma}^{\rm SM} +A_{\gamma L}^{+}+
\frac{Q_b}{V_{ts}^*V_{tb}}\sum\limits_{q=b,s}^{}\left[
  A_{L}^H(z_q)+A_{L}^A(z'_q)\right] \,, \\ 
 C_{7R} &=&
\frac{m_s}{m_b} A_{\gamma}^{\rm SM} + A_{\gamma R}^{+}
+\frac{Q_b}{V_{ts}^*V_{tb}}\sum\limits_{q=b,s}^{}\left[A_{R}^H(z_q)+A_{R}^A(z'_q)\right]
\,, \\
C_{8L} &=& A_{g}^{\rm SM} +A_{gL}^{+}+\frac{1}{V_{ts}^*V_{tb}}\sum
\limits_{q=b,s}^{}\left[ A_{L}^H(z_q)+A_{L}^A(z'_q)\right] \,, \\
C_{8R} &=& \frac{m_s}{m_b} A_{g}^{\rm SM} + A_{gR}^{+} +
\frac{1}{V_{ts}^*V_{tb}}\sum\limits_{q=b,s}^{}\left[A_{R}^H(z_q)+A_{R}^A(z'_q)\right] \,.
\end{eqnarray}
\end{subequations}
In the above expressions, $A^+$ and $A^{H,A}$ represent the (nonstandard) charged and neutral scalar contributions respectively (See Fig.~\ref{f:b2sg}). Note that $A^{H,A}=0$ for Type I and II 2HDMs because of the absence of tree-level FCNC in these models. They only become relevant in case of BGL models.

The SM and the new physics contributions are given below:
\paragraph*{$\blacksquare$ SM:}
\begin{subequations}
\begin{eqnarray}
A_{\gamma}^{\rm SM} &=&
\frac{1}{2}\left[\overline{\mathscr{F}}_1(x_t)+\overline{\mathscr{F}}_2(x_t) +
  \frac{1}{2}x_t\overline{\mathscr{F}}_2(x_t)
  -\frac{3}{2}x_t\overline{\mathscr{F}}_1(x_t) +
  x_t\overline{\mathscr{F}}_0(x_t) \right. \nonumber \\ &&~~~
  \left. +\frac{4}{3}\mathscr{F}_0(x_t) -2\mathscr{F}_1(x_t) +
  \frac{2}{3}\mathscr{F}_2(x_t)+\frac{1}{3}x\mathscr{F}_1(x_t)
  +\frac{1}{3}x_t\mathscr{F}_2(x_t) \right] -\frac{23}{36} \,, \\
A_{g}^{\rm SM} &=& \frac{1}{2} \left[2\mathscr{F}_0(x_t) -
3\mathscr{F}_1(x_t) +\mathscr{F}_2(x_t)+\frac{1}{2}x_t\mathscr{F}_1(x_t) +
\frac{1}{2}x_t\mathscr{F}_2(x_t)\right] - \frac{1}{3} \,, 
\end{eqnarray}
\end{subequations}

\paragraph*{$\blacksquare$ Charged Higgs: \\} 
{\bf BGL Models:}
\begin{subequations}
\begin{eqnarray}
A_{gL}^{+} &=&
\frac{1}{4V_{ts}^*V_{tb}}\sum\limits_{q=u,c,t}^{}V_{qs}^*V_{qb}X_q^2
\left[y_q\mathscr{F}_1(y_q) + y_q\mathscr{F}_2(y_q) \right] \,, \\
A_{\gamma L}^{+} &=&
\frac{1}{V_{ts}^*V_{tb}}\sum\limits_{q=u,c,t}^{}V_{qs}^*V_{qb}X_q^2
C(y_q) \,, \\
A_{gR}^{+} &=& \frac{m_s}{m_b} A_{gL}^{+} \,, ~~~~ A_{\gamma R}^{+} = \frac{m_s}{m_b} A_{\gamma L}^{+} \,,
\end{eqnarray}
\end{subequations}
with 
\begin{equation}
 C(y) = \frac{1}{2}\left[\frac{1}{2}y\overline{\mathscr{F}}_2(y) 
 -\frac{3}{2}y\overline{\mathscr{F}}_1(y) +y\overline{\mathscr{F}}_0(y)
 +\frac{1}{3}y\mathscr{F}_1(y) +\frac{1}{3}y\mathscr{F}_2(y)\right] \,,
\end{equation}
and for $i$-type model, $X_q = -\cot\beta$ if $q=i$ and $X_q =
\tan\beta$ otherwise (e.g. for t-type model, $X_u=X_c=\tan\beta$,
$X_t=-\cot\beta$).

{\bf Type I and II Models:}
\begin{subequations}
\begin{eqnarray}
A_{\gamma L}^{+} &=& \frac{1}{V_{ts}^*V_{tb}}\sum\limits_{q=u,c,t}^{} V_{qs}^*V_{qb}\left[C_{1L}(y_q)+\frac{2}{3} C_{2L}(y_q)\right] \,, \\
A_{\gamma R}^{+} &=& \frac{1}{V_{ts}^*V_{tb}}\sum\limits_{q=u,c,t}^{} V_{qs}^*V_{qb}\left[C_{1R}(y_q)+\frac{2}{3} C_{2R}(y_q)\right] \,, \\
A_{g L}^{+} &=& \frac{1}{V_{ts}^*V_{tb}}\sum\limits_{q=u,c,t}^{} V_{qs}^*V_{qb} C_{2L}(y_q) \,, \\
A_{g R}^{+} &=& \frac{1}{V_{ts}^*V_{tb}}\sum\limits_{q=u,c,t}^{} V_{qs}^*V_{qb} C_{2R}(y_q) \,,
\end{eqnarray}
\end{subequations}
with,
\begin{subequations}
\begin{eqnarray}
C_{1L}(y_q)&=& \frac{y_q}{2}\Bigg[\frac{1}{2}\left\{\overline{\mathscr{F}}_2(y_q)-\overline{\mathscr{F}}_1(y_q)\right\} \left(\frac{m_s^2}{m_q^2}Y^2+X^2\right) \nonumber \\
&& ~~~~~~~~~~+XY\left\{\overline{\mathscr{F}}_1(y_q)-\overline{\mathscr{F}}_0(y_q)\right\} \Bigg] \,, \\
C_{1R}(y_q)&=& \frac{y_q}{2}\Bigg[\frac{1}{2}\left\{\overline{\mathscr{F}}_2(y_q)-\overline{\mathscr{F}}_1(y_q)\right\} \left(\frac{m_b^2}{m_q^2}Y^2+X^2 \right) \nonumber \\
&& ~~~~~~~~~~+XY\left\{\overline{\mathscr{F}}_1(y_q)-\overline{\mathscr{F}}_0(y_q)\right\} \Bigg] \,, \\
C_{2L}(y_q)&=& \frac{y_q}{2}\left[\frac{1}{2}\left\{{\mathscr{F}}_2(y_q)-{\mathscr{F}}_1(y_q)\right\} \left(\frac{m_s^2}{m_q^2}Y^2+X^2 \right) - XY{\mathscr{F}}_1(y_q)  \right] \,, \\
C_{2R}(y_q)&=& \frac{y_q}{2}\left[\frac{1}{2}\left\{{\mathscr{F}}_2(y_q)-{\mathscr{F}}_1(y_q)\right\} \left(\frac{m_b^2}{m_q^2}Y^2+X^2 \right) - XY{\mathscr{F}}_1(y_q)  \right] \,.
\end{eqnarray}
\label{type12bsg}
\end{subequations}
In \Eqn{type12bsg}, we have to substitute $X=\cot\beta$, $Y=-\cot\beta$ for Type I model and $X=\cot\beta$, $Y=\tan\beta$ for Type II model.

\paragraph*{$\blacksquare$ CP-even Higgs (BGL Models):} 
%
\begin{subequations}
\begin{eqnarray}
A_{L}^H(z_b) &=& -\frac{1}{8}\left[\left\{z_b\mathscr{F}_1(z_b) -
  z_b\mathscr{F}_2(z_b)\right\} \left(\frac{AD}{m_b^2} +
  \frac{BC}{m_b^2} \frac{m_s}{m_b} \right) +2z_b\mathscr{F}_1(z_b)
  \frac{AC}{m_b^2} \right] \,, \\ 
 A_{R}^H(z_b) &=&
-\frac{1}{8}\left[\left\{z_b\mathscr{F}_1(z_b) -
  z_b\mathscr{F}_2(z_b)\right\} \left(\frac{AD}{m_b^2} \frac{m_s}{m_b}
  + \frac{BC}{m_b^2} \right) +2z_b\mathscr{F}_1(z_b) \frac{BD}{m_b^2}
  \right] \,,
\end{eqnarray}
\end{subequations}
with $A=(N_d)_{sb} \,, B=(N_d)_{bs}^*\,, C=(N_d)_{bb}\,, D=(N_d)_{bb}^*\,.$ 
\begin{subequations}
\begin{eqnarray}
A_{L}^H(z_s) &=& -\frac{1}{8}\left[\left\{z_s\mathscr{F}_1(z_s) -
  z_s\mathscr{F}_2(z_s)\right\} \left(\frac{AD}{m_s^2} +
  \frac{BC}{m_s^2} \frac{m_s}{m_b} \right) +2z_s\mathscr{F}_1(z_s)
  \frac{AC}{m_s^2} \frac{m_s}{m_b} \right] \\
 A_{R}^H(z_s) &=&
-\frac{1}{8}\left[\left\{z_s\mathscr{F}_1(z_s) -
  z_s\mathscr{F}_2(z_s)\right\} \left(\frac{AD}{m_s^2}\frac{m_s}{m_b}
  + \frac{BC}{m_s^2} \right) +2z_s\mathscr{F}_1(z_s) \frac{BD}{m_s^2}
  \frac{m_s}{m_b} \right] 
\end{eqnarray} 
\end{subequations}
with $A=(N_d)_{ss} \,, B=(N_d)_{ss}^*\,, C=(N_d)_{sb}\,, D=(N_d)_{bs}^*\,.$ 

\paragraph*{$\blacksquare$ CP-odd Higgs (BGL Models):}
\begin{subequations}
\begin{eqnarray}
A_{L}^A(z'_b) &=& \frac{1}{8}\left[\left\{z'_b\mathscr{F}_1(z'_b) -
  z'_b\mathscr{F}_2(z'_b)\right\} \left(\frac{AD}{m_b^2} +
  \frac{BC}{m_b^2} \frac{m_s}{m_b} \right) +2z'_b\mathscr{F}_1(z'_b)
  \frac{AC}{m_b^2} \right] \,, \\ 
A_{R}^A(z'_b) &=&
\frac{1}{8}\left[\left\{z'_b\mathscr{F}_1(z'_b) -
  z'_b\mathscr{F}_2(z'_b)\right\} \left(\frac{AD}{m_b^2}
  \frac{m_s}{m_b} + \frac{BC}{m_b^2} \right) +2z'_b\mathscr{F}_1(z'_b)
  \frac{BD}{m_b^2} \right] \,,
\end{eqnarray}
\end{subequations}
with $A=(N_d)_{sb} \,, B=-(N_d)_{bs}^*\,, C=(N_d)_{bb}\,, D=-(N_d)_{bb}^*\,.$
\begin{subequations}
\begin{eqnarray}
A_{L}^A(z'_s) &=& \frac{1}{8}\left[\left\{z'_s\mathscr{F}_1(z'_s) -
  z'_s\mathscr{F}_2(z'_s)\right\} \left(\frac{AD}{m_s^2} +
  \frac{BC}{m_s^2} \frac{m_s}{m_b} \right) +2z'_s\mathscr{F}_1(z'_s)
  \frac{AC}{m_s^2} \frac{m_s}{m_b} \right]  \\ 
A_{R}^A(z'_s) &=&
\frac{1}{8}\left[\left\{z'_s\mathscr{F}_1(z'_s)
  -z'_s\mathscr{F}_2(z'_s)\right\}
  \left(\frac{AD}{m_s^2}\frac{m_s}{m_b} + \frac{BC}{m_s^2} \right) +
  2z'_s\mathscr{F}_1(z'_s) \frac{BD}{m_s^2} \frac{m_s}{m_b} \right]
\end{eqnarray}
\end{subequations}
with $A=(N_d)_{ss} \,, B=-(N_d)_{ss}^*\,, C=(N_d)_{sb}\,, D=-(N_d)_{bs}^*\,.$ 

\section{Leptonic and semileptonic \texorpdfstring{$B$}{TEXT} decays}
We shall quote the formula used in \cite{Crivellin:2012ye}. The effective Hamiltonian is written as
\begin{eqnarray}
{\cal H}_{\rm eff} = C_{\rm SM}^{qb}O_{\rm SM}^{qb} + C_R^{qb}O_R^{qb} + C_L^{qb}O_L^{qb} \,, 
\end{eqnarray}
with
\begin{eqnarray}
O_{\rm SM}^{qb} &=& (\bar{q}\gamma_\mu P_L b)(\bar{\tau}\gamma^\mu P_L\nu_\tau) \,, \\
O_{R}^{qb} &=& (\bar{q} P_R b)(\bar{\tau} P_L\nu_\tau) \,, \\
O_{L}^{qb} &=& (\bar{q} P_L b)(\bar{\tau} P_L\nu_\tau) \,, 
\end{eqnarray}
In the above equations, $q=u$ for $B\to \tau\nu$ and $q=c$ for $B\to D^{(*)}\tau\nu$.
\begin{eqnarray}
\frac{R(D)}{R_{\rm SM}(D)} &=& 1 +1.5~ {\rm Re}\left(\frac{C_R^{cb}+C_L^{cb}}{C_{\rm SM}^{cb}}\right) + 1.0 \left|\frac{C_R^{cb}+C_L^{cb}}{C_{\rm SM}^{cb}}\right|^2 \,, \\
\frac{R(D^*)}{R_{\rm SM}(D^*)} &=& 1 +0.12~ {\rm Re}\left(\frac{C_R^{cb}-C_L^{cb}}{C_{\rm SM}^{cb}}\right) + 0.05 \left|\frac{C_R^{cb}-C_L^{cb}}{C_{\rm SM}^{cb}}\right|^2 \,, \\
\frac{{\rm Br}(B\to\tau\nu)}{{\rm Br}(B\to\tau\nu)_{\rm SM}} &=& \left|1+\frac{m_B^2}{m_b m_\tau}\frac{(C_R^{ub}-C_L^{ub})}{C_{\rm SM}^{ub}}\right|^2 \times \frac{\Gamma^{\rm SM}_B}{\Gamma_B}\,.
\end{eqnarray}
$\Gamma_B$ is the total decay width of the $B$ meson. If the NP only affects the rare decay modes, then we can take $\Gamma_B \approx \Gamma^{\rm SM}_B$. Now, let us proceed to find the Wilson coefficients. For the SM part, the relevant Lagrangian should be 
\begin{eqnarray}
{\mathscr L}_{\rm SM} = \frac{g}{\sqrt{2}} V_{qb} (\bar{q}\gamma^\mu P_L b) W_\mu^+  + \frac{g}{\sqrt{2}}(\bar{\tau}\gamma^\mu P_L \nu_\tau) W_\mu^- \,. 
\end{eqnarray}
Hence at low energy, we may write
\begin{eqnarray}
C_{\rm SM}^{qb} = \frac{g^2}{2M_W^2}V_{qb} =\frac{2}{v^2}V_{qb} = 2\sqrt{2}G_F V_{qb} \,.
\end{eqnarray}
For the NP contributions, $C_L$ and $C_R$, we need to look at
\begin{eqnarray}
{\mathscr L} = \frac{\sqrt{2}}{v}\left[\bar{u}(VN_d)P_R d -\bar{u}(N_u^\dagger V)P_L d\right] H_1^+ +\frac{\sqrt{2}}{v} \bar{e} N_e^\dagger P_L \nu H_1^- \,. 
\end{eqnarray}
According to the definition, we may write
\begin{eqnarray}
-C_R^{qb} &=& \frac{2}{v^2m_\xi^2} (VN_d)_{qb} (N_e^\dagger)_{\tau\tau} \,, \\
-C_L^{qb} &=& -\frac{2}{v^2m_\xi^2} (N_u^\dagger V)_{qb} (N_e^\dagger)_{\tau\tau} \,.
\end{eqnarray}
Note the occurrence of an extra negative sign compared to $C^{qb}_{\rm SM}$. This is because the spin-1 propagator differs from a spin-0 one by a relative negative sign ($-ig^{\mu\nu}/(k^2-M_W^2)$ compared to $i/(k^2-m_{1+}^2)$). Since $N_u$ is diagonal with real entries, we have
\begin{eqnarray}
 (N_u^\dagger V)_{qb} = (N_u)_{qq} V_{qb} \,.
\end{eqnarray}
and,
\begin{eqnarray}
(VN_d)_{qb} &=& \sum_a V_{qa}(N_d)_{ab} \nonumber \\
&=& \tan\beta \sum_a V_{qa}\delta_{ab} m^d_a - m_b (\tan\beta +\cot\beta)\left( \sum_a V_{qa}V_{ia}^* \right) V_{ib}  \\
&=& \tan\beta~ m_b V_{qb} -(\tan\beta +\cot\beta) m_bV_{ib}\delta_{iq} \,.
\end{eqnarray}

Now, the question is, what should we use for the matrix $N_e$ in the leptonic sector? According to our assumption, the leptonic sector will be the exact replica of the quark sector except that neutrinos are massless. This allows us to choose same rotation matrices for $e_L$ and $\nu_L$ which will make the PMNS matrix, $V=I_{3\times 3}$ in the leptonic sector. So the quark sector couplings will apply also to the leptonic sector with $V=I_{3\times 3}$ which means that there will be no flavor violating processes in the leptonic sector. Thus, for the leptonic sector we may write 
\begin{subequations}
\begin{eqnarray}
(N_e)_{ab} &=&  \tan\beta~ m^e_a \delta_{ab} -(\tan\beta+\cot\beta)\delta_{ia} \delta_{ib} m^e_b \,,  \\
 &=& \tan\beta~ {\rm diag}\{0,m_\mu,m_\tau\} -\cot\beta~{\rm diag}\{m_e,0,0\} ~~~~ {\rm for}~~ i=u \,, \\
 &=& \tan\beta~ {\rm diag}\{m_e,0,m_\tau\} -\cot\beta~{\rm diag}\{0,m_\mu,0\}   ~~~~ {\rm for}~~ i=c \,,\\
 &=& \tan\beta~ {\rm diag}\{m_e,m_\mu,0\} -\cot\beta~{\rm diag}\{0,0,m_\tau\}   ~~~~ {\rm for}~~ i=t \,, \\
N_\nu &=& 0 \,, 
\end{eqnarray}
\end{subequations}
where the change of notation is obvious.

Now it is straightforward to evaluate the matrix elements for specific models:
\paragraph*{u-model:}
\begin{subequations}
 \begin{eqnarray}
&& (N_e)_{\tau\tau} = m_\tau \tan\beta \,, \\
&& (N_u^\dagger V)_{ub} = -m_u \cot\beta~ V_{ub} \,; ~~~~(N_u^\dagger V)_{cb} = m_c\tan\beta~ V_{cb} \,, \\
&& (VN_d)_{ub} = -m_b\cot\beta~ V_{ub} \,; ~~~~ (VN_d)_{cb} = m_b \tan\beta~ V_{cb} \,.
\end{eqnarray}
\end{subequations}
\paragraph*{c-model:}
\begin{subequations}
 \begin{eqnarray}
&& (N_e)_{\tau\tau} = m_\tau \tan\beta \,, \\
&& (N_u^\dagger V)_{ub} = m_u \tan\beta~ V_{ub} \,; ~~~~(N_u^\dagger V)_{cb} = -m_c\cot\beta~ V_{cb} \,, \\
&& (VN_d)_{ub} = m_b\tan\beta~ V_{ub} \,; ~~~~ (VN_d)_{cb} = -m_b \cot\beta~ V_{cb} \,.
\end{eqnarray}
\end{subequations}
\paragraph*{t-model:}
\begin{subequations}
 \begin{eqnarray}
&& (N_e)_{\tau\tau} = -m_\tau \cot\beta \,, \\
&& (N_u^\dagger V)_{ub} = m_u \tan\beta~ V_{ub} \,; ~~~~(N_u^\dagger V)_{cb} = m_c\tan\beta~ V_{cb} \,, \\
&& (VN_d)_{ub} = m_b\tan\beta~ V_{ub} \,; ~~~~ (VN_d)_{cb} = m_b \tan\beta~ V_{cb} \,.
\end{eqnarray}
\end{subequations}
Note that, for t-models neither of the above mentioned decay widths will depend on $\tan\beta$.

\section{\texorpdfstring{$B_s\to\mu^+\mu^-$}{TEXT}}
The effective Hamiltonian is
\begin{eqnarray}
{\cal H}_{\rm eff} = C_{A}^{bs}O_{A}^{bs} + C_S^{bs}O_S^{bs} + C_P^{bs}O_P^{bs} \,, 
\end{eqnarray}
with
\begin{eqnarray}
O_{A}^{bs} = (\bar{b}\gamma_\alpha P_L s)(\bar{\mu}\gamma^\alpha
\gamma_5\mu) \,, ~~ O_{S}^{bs} = m_b(\bar{b} P_L
s)(\bar{\mu} \mu) \,, ~~ O_{P}^{bs} = m_b(\bar{b} P_L
s)(\bar{\mu} \gamma_5 \mu) \,.
\end{eqnarray}
Note that in addition to the above operators, there will be operators
of the form $(\bar{b}P_R s)(\bar{\mu}\mu)$ and $(\bar{b} P_R
s)(\bar{\mu} \gamma_5 \mu)$.  But the Wilson coefficients
corresponding to these operators will be proportional to $m_s$
(instead of $m_b$) and their contribution can be neglected ($m_b\gg
m_s$) as argued in \cite{Logan:2000iv}.  With this assumption we can
write
\begin{equation}
\frac{{\rm Br}(B_s\to\mu^+\mu^-)}{{\rm Br}(B_s\to\mu^+\mu^-)_{\rm SM}}
= \left\{\left|1-m^2_{B_s}\frac{C_P^{bs}}{2m_\mu C_A^{bs}}\right|^2
+m_{B_s}^4\left(1-\frac{4m_\mu^2}{m_{B_s}^2}\right) \left|
\frac{C_S^{bs}}{2m_\mu C_A^{bs}}\right|^2 \right\} 
\frac{\Gamma^{\rm SM}_B}{\Gamma_B}\,.
\end{equation}
The relevant part of the Lagrangian (for BGL models) to evaluate $C_S^{bs}$ and $C_P^{bs}$ is
\begin{eqnarray}
{\mathscr L}_{\rm quark} &=& \frac{R}{v} \bar{d}(N_dP_R + N_d^\dagger
P_L)d +i\frac{A}{v}\bar{d}(N_dP_R - N_d^\dagger P_L)d \nonumber \\ &=&
(N_d^\dagger)_{bs}\frac{R}{v} {\bar{b}P_Ls }
-i(N_d^\dagger)_{bs}\frac{A}{v} {\bar{b}P_Ls } \nonumber\\ &=&
(N_d^\dagger)_{bs}\frac{h}{v}\cos(\beta-\alpha) {\bar{b}P_Ls} -
(N_d^\dagger)_{bs}\frac{H}{v} \sin(\beta-\alpha){\bar{b}P_Ls } -
i(N_d^\dagger)_{bs}\frac{A}{v} {\bar{b}P_Ls }\,, \\ {\mathscr L}_{\rm
  lepton}&=& -\frac{H^0}{v}\bar{e}D_e e + \frac{R}{v} \bar{e}(N_eP_R +
N_e^\dagger P_L)e +i\frac{A}{v}\bar{e}(N_eP_R - N_e^\dagger P_L)e
\nonumber\\ &=& -\frac{m_\mu}{v}{ \bar{\mu}\mu H^0}
+\frac{(N_e)_{\mu\mu}}{v} {\bar{\mu}\mu R} + \frac{i(N_e)_{\mu\mu}}{v}
{\bar{\mu}\gamma_5\mu A} \nonumber\\ &=&\Bigg[
  \frac{h}{v}\left\{-\sin(\beta-\alpha)m_\mu+\cos(\beta-\alpha)(N_e)_{\mu\mu}\right\} \nonumber \\
&&~  + \frac{H}{v}\left\{-\cos(\beta-\alpha)m_\mu -
  \sin(\beta-\alpha)(N_e)_{\mu\mu}\right\}\Bigg] \bar{\mu}\mu
 +i \frac{A}{v}(N_e)_{\mu\mu}{\bar{\mu}\gamma_5\mu} \,.
\end{eqnarray}
Note that terms involving $\bar{b}P_Rs$ have not been displayed.
Their coefficients are proportional to $(N_d)_{bs}$, which is
proportional to $m_s$, and are therefore neglected.
\begin{subequations}
\begin{eqnarray}
(N_d)_{bs}&=& -(\tan\beta+\cot\beta)V^*_{ib}V_{is}m_s \,, \\
(N_d^\dagger)_{bs}&=& (N_d)^*_{sb}= -(\tan\beta+\cot\beta)V^*_{ib}V_{is}m_b \,.
\end{eqnarray}
\end{subequations}
The Wilson coefficients are
\begin{eqnarray}
C_S^{bs} &=& (\tan\beta+\cot\beta)\frac{V_{ib}^*V_{is}}{v^2}\left\{
\frac{\cos(\beta-\alpha)}{m_h^2}[-\sin(\beta-\alpha)m_\mu+\cos(\beta-\alpha)(N_e)_{\mu\mu}]
\right. \nonumber \\ && \left. ~~~~~~~~~~~~~~~~~~~
+\frac{\sin(\beta-\alpha)}{m_H^2}[\cos(\beta-\alpha)m_\mu+
  \sin(\beta-\alpha)(N_e)_{\mu\mu}]\right\}\,,
\end{eqnarray}
and
\begin{eqnarray}
C_P^{bs} &=& (\tan\beta+\cot\beta)\frac{V_{ib}^*V_{is}}{v^2}\frac{(N_e)_{\mu\mu}}{m_A^2} \,.
\end{eqnarray}
The SM Wilson coefficient is \cite{Logan:2000iv} 
\begin{equation}
C_A^{bs} =\frac{\alpha G_F}{2\sqrt{2}\pi \sin^2\theta_w} V_{tb}^*V_{ts} 2 Y(x_t)\,,\ \ \ 
Y(x_t)= 0.997 \left[\frac{m_t(m_t)}{166~{\rm GeV}}\right]^{1.55} \approx 1.0 \,. 
\end{equation}

\section{Current experimental numbers}
\begin{enumerate}
\item {\bf $b\to s\gamma$:} SM = $(3.15\pm 0.23)\times 10^{-4}$ \cite{Gambino:2001ew,Misiak:2006zs,Misiak:2006ab}, EXP=$(3.55\pm 0.26)\times 10^{-4}$ \cite{Amhis:2012bh}, ${\rm Br}(b\to c e \nu)_{\rm exp} =(10.51\pm 0.13)\%$ \cite{pdg1}.
Note that we have taken ${\rm Br}(b\to c e \nu) \approx {\rm Br}(b\to c \ell \nu)$.

\item {\bf $R(D)$, $R(D^*)$:} The experimental numbers are \cite{Fajfer:2012jt}:
\begin{eqnarray}
R(D) &\equiv& \frac{{\rm Br}(B\to D\tau\nu)}{{\rm Br}(B\to D\ell\nu)} = 0.440\pm 0.072 \,, \\
R(D^*) &\equiv& \frac{{\rm Br}(B\to D^*\tau\nu)}{{\rm Br}(B\to D^*\ell\nu)} = 0.332\pm 0.030 \,, \\
R &\equiv& \frac{\tau(B^0)}{\tau(B^-)} \frac{{\rm Br}(B^-\to \tau^-\bar{\nu})}{{\rm Br}(\bar{B}^0\to \pi^+\ell^-\bar{\nu})} = 0.73\pm 0.15 \,.
\end{eqnarray}
But, for analysis purpose, the following numbers will be directly usable \cite{Fajfer:2012jt}:
\begin{eqnarray}
\frac{R(D)_{\rm exp}}{R(D)_{\rm SM}} &=& 1.49\pm 0.26 \,, \\
\frac{R(D^*)_{\rm exp}}{R(D^*)_{\rm SM}} &=& 1.32\pm 0.12 \,, \\
\frac{R_{\rm exp}}{R_{\rm SM}} &=& 2.38\pm 0.66 \,.
\end{eqnarray}

\item {\bf $B_s\to \mu^+\mu^-$:} SM = $(3.65\pm 0.30)\times 10^{-9}$ \cite{DeBruyn:2012wj, Fleischer:2012bu}, SM(updated)= $(3.54\pm 0.23)\times 10^{-9}$ \cite{Bobeth:2013uxa}, EXP= $(3.2\pm 1.0)\times 10^{-9}$ \cite{HFAGbs}.

\item For $\Delta M_K$, $\Delta M_d$ and $\Delta M_s$, we have used the central values given in PDG because the errors are very small.
\end{enumerate}

\addtocontents{toc}{\vspace{2em}} 

\backmatter


\lhead{\emph{Bibliography}} 
\bibliographystyle{JHEP}
\bibliography{Bibliography} 
%
%
%
%

\end{document}